\documentclass[a4paper, twoside,12pt]{Latex/Classes/PhDthesisPSnPDF}

\usepackage{lineno}
\usepackage{amsbsy}
\usepackage{xspace}
\usepackage{wtmmPkg}
\usepackage{natbib}
\usepackage{multirow}
\usepackage{paralist}
\usepackage{titlesec}
\usepackage{lscape}
\usepackage{quotchap}
\usepackage{epstopdf}
\usepackage{fancyhdr}

   \usepackage{Latex/StyleFiles/backrefx}
       \renewcommand{\backrefpagesname}{Cited on page~}
       \renewcommand{\backrefpagesnames}{Cited on pages~}

\DeclareRobustCommand{\ion}[2]{%
\relax\ifmmode
\ifx\testbx\f@series
{\mathbf{#1\,\mathsc{#2}}}\else
{\mathrm{#1\,\mathsc{#2}}}\fi
\else\textup{#1\,{\mdseries\textsc{#2}}}%
\fi}

\renewcommand{\vec}[1]{ {\mathbf #1} }

\newcommand{\vol}{ {\mathcal V} }
\newcommand{\bndry}{ {\mathcal S} }

\graphicspath{{images/}}

\usepackage{setspace}
\begin{document}

\renewcommand\baselinestretch{1.2}
\baselineskip=18pt plus1pt


\newcommand{\titlefont}{\bfseries \fontsize{22}{26.42pt}\selectfont}
\newcommand{\largetitlefont}{\bfseries \fontsize{29.88}{35.88pt}\selectfont}
\newcommand{\othertitlefont}{\fontsize{14.4}{17.28}\selectfont}
\newcommand{\authorfont}{\bfseries \fontsize{14.4}{17.28}\selectfont}
\newcommand{\informationfont}{\fontsize{10}{12}\selectfont}
\newcommand{\dedicationfont}{\slshape \fontsize{14.4}{17.28}\selectfont}

\newcommand{\thisyear}{\number\year}
\def\thismonth{\ifcase\month\or January\or February\or March\or
  April\or May\or June\or July\or August\or September\or October\or November\or December\fi}
\newcommand{\todaysdate}{\thismonth\space \thisyear}

\renewcommand{\baselinestretch}{1}
\newpage \thispagestyle{empty}
\vspace*{1.5cm}
\begin{flushright}
\Huge{\textbf{EG Andromedae:\\A Symbiotic System as an Insight\\into Red Giant Chromospheres}}

\end{flushright}

\vspace*{4cm}
\begin{flushright}
A dissertation submitted to the University of Dublin \\
for the degree of Doctor of Philosophy
\end{flushright}

\vspace*{\fill}
\begin{flushright}
{\authorfont {Joseph Roche} \\[1mm]
{\othertitlefont {Trinity College Dublin}, January 2012}\\[.5mm]
\rule{0.9\textwidth}{0.5mm}\\[4mm]

\begin{minipage}[b][15mm][t]{12.5cm}
\raggedleft \sc
School of Physics\\
University of Dublin\\
Trinity College\\
\end{minipage}
\hspace*{1mm}
\begin{minipage}[b][15mm][t]{1.15cm}
\includegraphics[height=16mm]{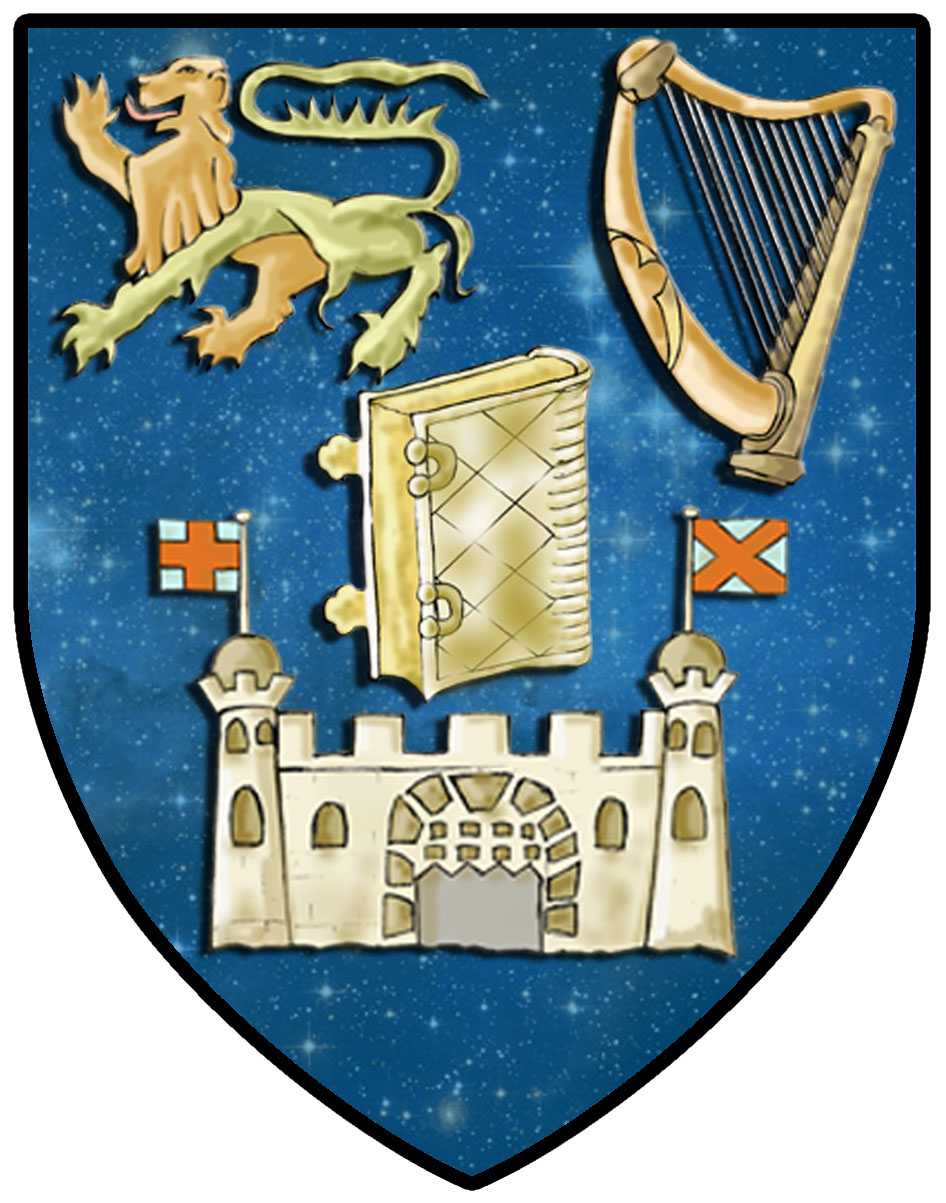}
\end{minipage}}

 \end{flushright}


\newpage
\thispagestyle{empty}
\mbox{}

\frontmatter



\begin{declaration}        

I declare that this thesis has not been submitted as an exercise for a degree at this or
any other university and it is entirely my own work.

\vspace{10mm}

I agree to deposit this thesis in the University's open access institutional repository or
allow the library to do so on my behalf, subject to Irish Copyright Legislation and
Trinity College Library conditions of use and acknowledgement.

\vspace{30mm}

\textbf{Name:} Joseph Roche	

\vspace{15mm}

\textbf{Signature:}  ........................................		\textbf{Date:}  ..........................

\end{declaration}

\newpage
\thispagestyle{empty}
\mbox{}



\begin{abstracts} 

Symbiotic systems are interacting binary stars consisting of both hot and cool components. This results in a complex environment that is ideal for studying the latter stages of stellar evolution along with interactions within binary systems. As a star approaches the end of its life, in particular the red giant phase, it exhausts its supply of core hydrogen and begins burning its way through successively heavier elements. Red giants lose mass in the form of a dense wind that will replenish the interstellar medium with chemical elements that are formed through nuclear processes deep in the stellar interior.  When these elements reach the interstellar medium they play a central role in both stellar and planetary evolution, as well as providing the essential constituents needed for life. The undoubted significance of these cool giants means the study of their atmospheres is necessary to help understand our place in the Universe. 

This thesis presents Hubble Space Telescope ({\sl HST}) observations of the symbiotic system EG Andromedae as an insight into red giant stars. EG And is one of the brightest and closest symbiotic systems and consists of a red giant primary along with a white dwarf. The presence of the white dwarf in the system allows spatially resolved examination of the red giant primary. The benefits of using such a system to better understand the base of red giant chromospheres is shown. Although EG And can help further our comprehension of red giant stars, some of the physical properties governing the system are not known to a high degree of accuracy. New measurements of interstellar extinction (0.05), distance (568 pc),  and wind velocity (70 km s$^{-1}$) are provided. The onset of TiO lines in spectral subclasses is identified. Along with the observations of EG And, new {\sl HST} observations of an isolated red giant spectral standard HD148349 are described.  The similarity between the isolated spectral standard and the red giant primary of EG And is demonstrated, showing that much of the information gleaned from a symbiotic system can be applied to the general red giant population.  Using both ultraviolet and optical spectroscopy, the atmosphere of EG And and HD148349 are investigated and contrasted. EG And is found to have an electron density of $7\times10^{8}$ cm$^{-3}$, while for HD148349 it is found to be $4\times10^{8}$ cm$^{-3}$.  The evidence of spectral line asymmetries in the photosphere arising from stellar granulation is discussed as a possible energy source for a wind-driving mechanism. Finally, a complete energy budget for the atmosphere of EG And is presented which shows the total energy needed to heat the chromosphere and drive the wind is 1.3$\times$10$^{6}$ erg cm$^{-2}$ s$^{-1}$.

\end{abstracts}



\begin{dedication} 

\large{\emph{For Mom and Dad}}

\end{dedication}

\newpage
\thispagestyle{empty}
\mbox{}



\begin{acknowledgements}      

I would like to thank Brian~Espey and Cian~Crowley for getting me started with this research and for all of their help, especially in the initial stages.

I wish to express my gratitude to Peter~Gallagher for his support and encouragement throughout. I am eternally grateful to Graham~Harper for advising me, motivating me and having faith in me.

I am thankful to the academic, administrative and technical staff in both the School~of~Physics and Trinity~Hall as well as the members of the Astrophysics~Research~Group, past and present, for their kindness over the years.

Finally, I would like to thank my friends and family for being there for me always.

\end{acknowledgements}


\newpage
\thispagestyle{empty}
\mbox{}


\chapter{List of Publications}
\label{chapter:publications}

\begin{enumerate}

\item \textbf{Roche, J.}, Espey, B.~R., \& Crowley, C.\\
``Symbiotic Stars and the Origin of Red Giant Winds'', \\
Proceedings of the 16th Cambridge Workshop on Cool Stars, Stellar Systems and the Sun.  
Astronomical Society of the Pacific Conference Series, vol. 448, pp. 713 (2011)
 
\item \textbf{Roche, J.}, Espey, B.~R., \& Crowley, C.\\
``Exploring the Origin of Red Giant Winds'', \\
Proceedings of the 15th Cambridge Workshop on Cool Stars, Stellar Systems and the Sun.  American Institute of Physics
Conference Series, vol. 1094, pp. 888 (2009)

\item Crowley, C., Espey, B.R., Harper, G.M. \& \textbf{Roche, J.}\\
``Winds and Chromospheres of Cool (Super-) Giants'', \\
Proceedings of the 15th Cambridge Workshop on Cool Stars, Stellar Systems and the Sun.  American Institute of Physics
Conference Series, vol. 1094, pp. 267 (2009)

\end{enumerate}

\thispagestyle{empty}


\setcounter{secnumdepth}{3} 
\setcounter{tocdepth}{3}    
\tableofcontents            


\newpage
\thispagestyle{empty}
\mbox{}
\listoffigures	
\newpage
\thispagestyle{empty}
\mbox{}
\listoftables  
\newpage
\thispagestyle{empty}
\mbox{}


\mainmatter

\pagestyle{fancy}

\chapter{Introduction}
\label{chapter:intro}

As sentient beings we have long considered our place in the Universe. Who we are and why we are here are musings that are natural extensions of our capacity for introspection. Science, as a way of thinking more than as a body of knowledge \citep{sagan1997demon}, has helped us unravel some of the mysteries of our existence. It accomplishes this by being, at its core, a self-correcting process. It does not pretend to show the final truth, but rather to test and refine hypotheses until they approach what we think is true of the natural world around us \citep{prothero2007evolution}. Science has led us to understand that everything on the Earth is composed of a handful of essential elements. The Earth is predominantly iron, oxygen, silicon and magnesium. Nitrogen, oxygen and hydrogen largely account for the elements of the atmosphere and the oceans, while the human body is mostly carbon, nitrogen, oxygen and hydrogen. These elements are created deep in the interior of stars and eventually reach the stellar atmosphere where they are dispelled into the interstellar medium to become new stars and planets.  It is during the red giant phase of a star's life that most of this material is lost. Unfortunately, atmospheres of red giants remain poorly understood. Observations of red giant stars are disk-averaged over the stellar surface. Eclipsing symbiotic systems provide a means to probe red giant atmospheres. Symbiotic systems consist of a compact white dwarf orbiting a red giant star. Observing the white dwarf through periodic eclipses allows spatial investigation of the atmosphere and unveils more about its structure and the physical processes taking place within it. The more that is known about red giant atmospheres, the better our comprehension of stellar/planetary evolution and the likely emergence of life.

\section{Stellar Evolution}\label{sec:stellar_evo}

To comprehend stellar evolution it is first instructive to discuss the most well-studied star. The Sun is a G2V main sequence star with a mass of $M_{\odot}= 1.99\times10^{33}$ g. It has a radius of $R_{\odot}=6.96\times10^{10}$ cm and a luminosity of $L_{\odot} = 3.85\times10^{33}$ erg s$^{-1}$ \citep{prialnik2000introduction}. The luminosity is given by:
\begin{equation}
L_{\odot} = 4\pi R_{\odot}^2 \sigma T_{e}^{4}
\end{equation}
where $\sigma$ is the Stefan-Boltzmann constant and  $T_{e}$ is the effective temperature. The Sun is about halfway through its lifecycle and is expected to enter the red giant phase in five billion years \citep{schroder_smith_2008}. 

\citet{phillips1999physics} describes how gravity is the driving force behind stellar evolution. A star could arise from the interstellar gas  between the spiral arms of the galaxy. This gas, predominantly hydrogen, is confined to the plane of the galaxy due to its own pressure. The gas is extremely tenuous and hot due to a balance between heating by X-ray emitting objects in the galaxy, and radiative cooling. If a small pressure increase were to occur, it could result in a sudden increase in cooling due to ionic recombination. At this point, the gas could break up into clouds of relatively dense, cool gas. Increasing pressure will cool the clouds to extremely low temperatures due to infrared radiation emitted by the hydrogen molecules. If the internal pressure of a cloud is not strong enough to resist its own gravitational force, it will collapse into itself, forming a protostar. The criteria for such a cloud to collapse was outlined by \citet{jeans_1902_instability} who showed that for a cloud of given temperature $T$ and radius $r$, the cloud will collapse if its mass exceeds `Jeans Mass':
\begin{equation}
M_J = \frac{3kT}{G\bar{m}}r
\label{eqn:jeans_mass}
\end{equation}
where $G$ is the gravitational constant, $k$ is the Boltzmann constant and $\bar{m}$ is the mean mass of the particles in the cloud. Equivalent limits on radius and density follow from Jeans' criterion. As the protostar contracts it is heated up by the release of gravitational energy. Energy from the centre of the star is transported to the surface via convection. The onset of hydrogen burning is delayed by deuterium burning. While the protostar is still accreting material, deuterium burning keeps the temperature constant until the dominant mode of energy transport changes from convective to radiative. The cessation in convection reduces the supply of deuterium and the subsequent rise in temperature results in hydrogen burning \citep{bally2006birth}. The specific processes that a protostar undergoes are governed by its mass and are described in Section \ref{sec:hr_diagram}. If the protostar's mass is extremely low ($<0.08\ M_\odot$) then it will remain purely convective and thus a sub-stellar object, such as a brown dwarf. The protostar is deemed a star proper when the interior is hot enough for hydrogen fusion to occur and become the driving energy-generating mechanism. 

\section{Stellar Structure}

Stellar evolution allows us to understand the physical and chemical properties of stars and their development with time. Several assumptions are necessary to interpret the standard idea of stellar evolution. These assumptions, as outlined in \citet{salaris_2005_stellar_evo_book}, include  considering stars as spherically symmetric systems made up of matter plus radiation, while ignoring the effects of magnetic fields and rotation.

With these assumptions in place, stars are simply governed by the equations of stellar structure which allow the pressure, temperature, luminosity, radius and chemical element abundance to be described in terms of mass and evolution with time. The force of gravity is balanced by an inwardly increasing pressure gradient. This prevents the star from collapsing in on itself or breaking apart, effectively rendering a state of hydrostatic equilibrium:
\begin{equation}
\frac{dP(r)}{dr}= -  \frac{GM(r)\rho(r)}{r^2}
\label{eqn:hdyr_equilib}
\end{equation}
where $P$ is pressure and $\rho$ is density. 
The density can also be given by the ideal gas law:
\begin{equation}
\rho = \frac{P \mu m_H}{kT}
\label{eqn:ideal_gas1}
\end{equation}
where $\mu$ is the average particle mass, $m_H$ is the mass of hydrogen, $k$ is the Boltzmann constant and $T$ is temperature. Mass conservation provides the equation of mass-continuity:
\begin{equation}
\frac{dM(r)}{dr}= 4\pi r^2 \rho(r)
\label{eqn:mass_contin}
\end{equation}
The temperature, $T(r)$, is governed by energy flows through the star, i.e.\ the luminosity as a function of radius $L(r)$. The path of photons through the star needs to be taken into account by including the opacity $\kappa$, where:
\begin{equation}
\kappa=\frac{1}{\rho l}
\label{eqn:opacity}
\end{equation}
and $l$ is the mean free path of a photon in the star. $T(r)$ is subsequently defined by the equation of radiative energy transport:
\begin{equation}
\frac{dT(r)}{dr}= -  \frac{3L(r)\kappa(r)\rho(r)}{4\pi r^2 a c T^3 (r)}
\label{eqn:energy_trans}
\end{equation}
where $a$ is a constant:
\begin{equation}
a =  \frac{8\pi^5 k^4}{15c^3 h^3} 
\label{eqn:a_const}
\end{equation}
which is $7.6\times10^{-15}$ erg cm\textsuperscript{-3}K \textsuperscript{-4} \citep{maoz_2007_astrophysics}. 
As well as heat flow being transported by electromagnetic radiation, convection is also a transport mechanism for cool stars. This is because the outer envelopes of cool stars have temperatures that are low enough for hydrogen to not be ionized. This increases opacity and makes convection a more efficient energy transport than radiation. The convective envelope obeys the Schwarzschild criterion for convection instability \citep{schrijver2008solar}. The Schwarzschild Criterion determines whether a rising or sinking globule of gas will continue to rise/sink or if it will return to its original depth. The derivation of this criterion follows that described by Schwarzschild in 1906.

If a globule of gas were lifted to a height $\delta r$ by a disturbance, the thermodynamic conditions in the globule would be initially the same as those outside the globule: 
\begin{equation}
\rho_{in}(r) =\rho (r) 
\label{eqn:schwarz_dens}
\end{equation}
However the rise would be sufficiently fast that no heat exchange would occur between the globule and its surroundings. Although being transported adiabatically, the internal pressure, $p_{in}$, would be balanced by the ambient pressure, $p$. These circumstances are feasible in a cool star due to the high opacity and short travel time for sound waves across the globule. If the internal density of the globule were to remain smaller than the external density, it would continue to rise and the condition for instability would be:
\begin{equation}
\rho_{in}(r + \delta r) - \rho(r + \delta r)  = \delta r \left[\left(\frac{d\rho}{dr}\right)_{ad} - \frac{d \rho}{dr}\right] < 0
\label{eqn:schwarz_convec_instab}
\end{equation}
where $(d\rho/dr)_{ad}$ is the density gradient under adiabatic conditions. Using Equation \ref{eqn:ideal_gas1} and Equation \ref{eqn:schwarz_dens}, the instability condition can be rewritten as:
\begin{equation}
\left(\frac{dT}{dr}\right)_{ad} - \frac{T}{\mu}\left(\frac{d\mu}{dr}\right)_{ad} > \frac{dT}{dr} - \frac{T}{\mu}\frac{d\mu}{dr}
\label{eqn:schwarz_convec_unstab_temp}
\end{equation}
The mean molecular weight, $\mu$, is a function of $p$ and $T$, but because of the pressure equilibrium, $(dp/dr)_{ad} = dp/dr$, the gradients $(d\mu/dr)_{ad}$ and $d\mu/dr$ differ only in terms of the gradient $dT/dr$, and:
\begin{equation}
\left|\frac{dT}{dr}\right| > \left| \left( \frac{dT}{dr} \right)_{ad}  \right| 
\label{eqn:convec_schwarz}
\end{equation}
which is the Schwarzschild criterion for convection instability.

The luminosity of the star can be described by the equation of energy conservation:
\begin{equation}
\frac{dL(r)}{dr}=4\pi r^2 \rho(r)\epsilon(r)
\label{eqn:energy_conserv}
\end{equation}
where $\epsilon$ is the power produced per unit mass of stellar material.

\section{Hertzsprung-Russell Diagram}\label{sec:hr_diagram}
 
Having defined the equations that govern stellar structure, the number of stars of a particular mass can be described by the initial mass function \citep{salpeter_1955_inital_mass_function}. This is a power law that  predicts the number of stars of mass $M$ per unit mass interval, and is be given by:
\begin{equation}
\frac{dN}{dM}\propto M^{-\alpha}
\label{eqn:initial_mass}
\end{equation}
where $\alpha = 2.35$ over most of the mass range \citep{maoz_2007_astrophysics}.  

The equations of stellar structure depend on the star remaining in equilibrium. Given the star's mass and composition it is possible to determine its structure. In reality, the assumption of equilibrium cannot hold indefinitely. Nuclear reactions in the stellar core synthesize hydrogen into helium which will alter the elemental composition over time. Mixing due to convection will also change the composition. Eventually the star will exhaust its fuel supply and lose the energy supply necessary to resist gravitational collapse. Consequently, it is inevitable that stars will evolve with time. To help understand how mass and stellar evolution are intertwined, the use of a Hertzsprung-Russell diagram (hereafter HR diagram) is practical. Between 1911 and 1913, Ejnar Hertzsprung and Henry Norris Russell independently devised a plot to show how the absolute magnitude of stellar populations changed with spectral type \citep{zeilik_1992_introductory}. 

\begin{figure}[ht!]
\includegraphics[width=\textwidth]{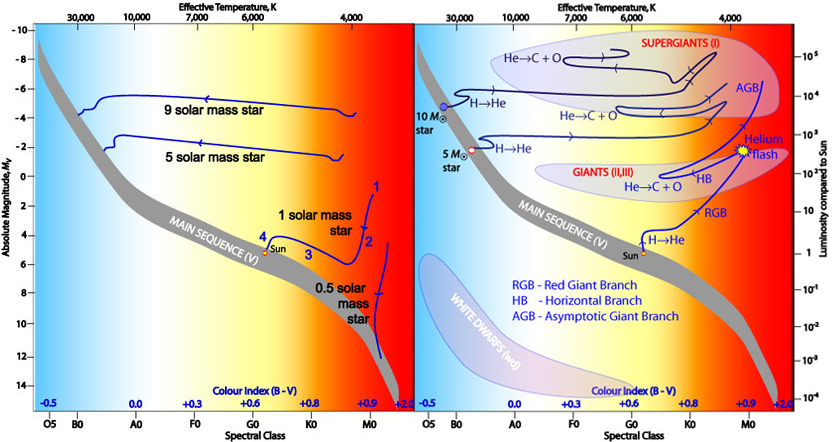}
\caption[Pre- and Post-Main-Sequence Stellar Evolution]{
Pre- and Post-Main-Sequence Stellar Evolution. The left panel shows a HR diagram with Henyey and Hayashi Tracks for protostars of different masses highlighted. Four pre-main-sequence stages are labelled. (1) The initial cloud collapse to form a protostar. (2) The temperature increases due to gravity but the decrease in surface area causes a reduction in luminosity and a vertical drop on the HR Diagram. (3) The onset of  nuclear fusion causes an increase in surface temperature. (4) The outward gas and radiation pressures match the inward gravitational force and the star attains hydrostatic equilibrium and settles onto the main sequence. The right panel is a HR diagram showing post-main-sequence evolutionary paths. \emph{Image Credit: Australian Telescope National Facility.}
\label{fig:combined_tracks}}
\end{figure}

Figure \ref{fig:combined_tracks} shows how the initial stages of stellar birth (outlined above) result in a star debuting on the main sequence of a HR diagram. The HR diagram itself is a plot of logarithmic luminosity (relative to solar luminosity; $3.8\times10^{33}$ erg s$^{-1}$) against  surface temperature (and/or colour, spectral type). Sometimes stellar magnitudes are plotted instead of luminosity. Traditionally, the temperature increases from right to left along the plot. On a HR diagram, certain regions are more densely populated than others. Hydrogen burning stars should give rise to the most densely populated region of the plot as stars spend most of their lives in this phase. This is known as the main sequence, and as pointed out above, is the region on the plot that a star makes its debut. It encompasses a diagonal swathe across the plot from blue supergiants to red dwarves. It is estimated that 80 - 90\% of observed stars are main sequence stars \citep{phillips1999physics}. 

The point that the star debuts on the main sequence is given by its initial mass. The left panel of Figure \ref{fig:combined_tracks} shows the pre-main sequence tracks taken by protostars of different masses. For protostars with masses $< 0.5\ M_{\odot}$, \citet{hayashi_1961} showed that when hydrostatic equilibrium is reached the protostar is fully convective. The star will continue to contract but its temperature will remain constant as it moves vertically along a track which became known as the Hayashi track. \citet{henyey_1955} showed that protostars with masses $> 0.5\ M_{\odot}$ will reach the end of the Hayashi track when radiative energy transport becomes more efficient than convective transport. The protostar will then take a near horizontal path across the HR diagram maintaining radiative equilibrium and almost constant luminosity. This path became known as the Henyey track. 

The length of time a star spends on the main sequence will be determined by its mass. The star's luminosity  at this stage is proportional to its mass, while the total energy radiated away is also mass-dependent. This leads to the time spent on the main sequence being given by:
\begin{equation}
t_{\mbox{ms}} \sim M^{1 - \alpha}
\label{eqn:main_sequence}
\end{equation}
For intermediate mass stars  $\alpha \sim 3$ \citep{maoz_2007_astrophysics} and:
\begin{equation}
t_{\mbox{ms}} \sim \frac{1}{M^2}
\label{eqn:main_sequence_ims}
\end{equation}
While higher mass stars will exhaust their hydrogen more quickly, solar mass stars could expect to spend around $\sim 10^{10}$ years on the main sequence. Once the star exhausts its supply of hydrogen, having spent most of its life on the main sequence, it then takes one of several paths off of the main sequence. These post-main-sequence tracks are again governed by the mass of the star. Several examples are shown in Figure \ref{fig:combined_tracks}.

\section{Evolution to the Red Giant Stage}\label{sec:rg_evo}

\begin{figure}[ht!]
\includegraphics[width=\textwidth]{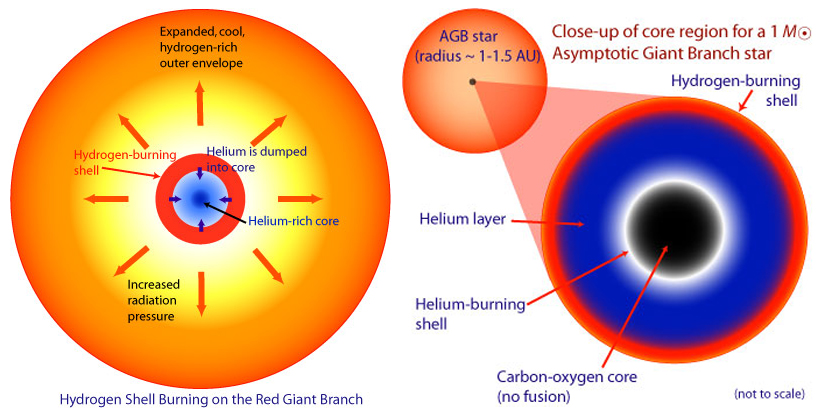}
\caption[Red Giant and Asymptotic Giant Branch Stars]{
Red Giant and Asymptotic Giant Branch Stars. The left side of the figure shows the basic structure of a star on the giant branch of the HR diagram, while the right side shows a similar star after it has evolved to ascend the asymptotic giant branch. \emph{Image Credit: Australian Telescope National Facility.}
\label{fig:rg_shells.jpg}}
\end{figure}

Less massive stars (i.e.\ $< 8\ M_{\odot}$)  can spend billions of years on the main sequence. Eventually, the supply of hydrogen in the core will be spent and it will contract, causing the temperature of the stellar interior to rise. This results in shells of burning hydrogen in the less-processed regions around the core. At this point, the outer layers of the star undergo huge expansion leading to increased luminosity, but diminished effective surface temperature. The star is now giant-size ($\sim$ 10 - 100 R$_{\odot}$) and cooler, changes that move the star upwards and to the right on the HR diagram onto the `red giant branch' (See the right panel of Figure \ref{fig:combined_tracks}). The basic structure of a red giant is shown in the left panel of Figure \ref{fig:rg_shells.jpg}. 
A dredge-up will occur, which causes the products of fusion to be mixed into the stellar atmosphere. The next stage in stellar evolution occurs when the helium in the core has been exhausted and contractions cause helium burning in a shell beneath the hydrogen burning shell. This double-burning shell around a core of inert carbon/oxygen (right-hand side of Figure \ref{fig:rg_shells.jpg}) results in the star ascending the asymptotic giant branch. A second dredge-up causes some of the elements newly synthesised in the core to be dispersed into the outer layers of the star. These stages of stellar evolution are pivotal as evolved stars undergo large mass loss, especially on the giant and asymptotic giant branches\footnote{Very little mass is lost ascending the red giant branch, but $\sim$ 0.1 - 0.2 M$_{\odot}$ can be shed at the top of the branch.}.  Red giants make up as much as 10\% of the stars in The Bright Star Catalogue \citep{percy_harrett_2004}.

The dense matter in the core will remain as a degenerate electron gas supported against further contractions. The outer layers will continue to expand and be blown off, creating firstly a preplanetary nebula and then a planetary nebula. The exposed core is now considered a white dwarf. More massive stars ($> 8\ M_{\odot}$) deplete their supply of hydrogen in a few million years as they try to resist gravitational contraction. The mass of the core might be such that gravity causes further contractions, leading to the formation of a neutron star or black hole \citep{salaris_2005_stellar_evo_book}. This study is most interested in the red giant phase of stellar evolution for stars with lower initial masses, although the presence of a white dwarf in the target system is also important.

\section{Symbiotic Stars}\label{sec:symb_stars}

\begin{figure}[ht!]
\centering 
\includegraphics[width=0.6\textwidth]{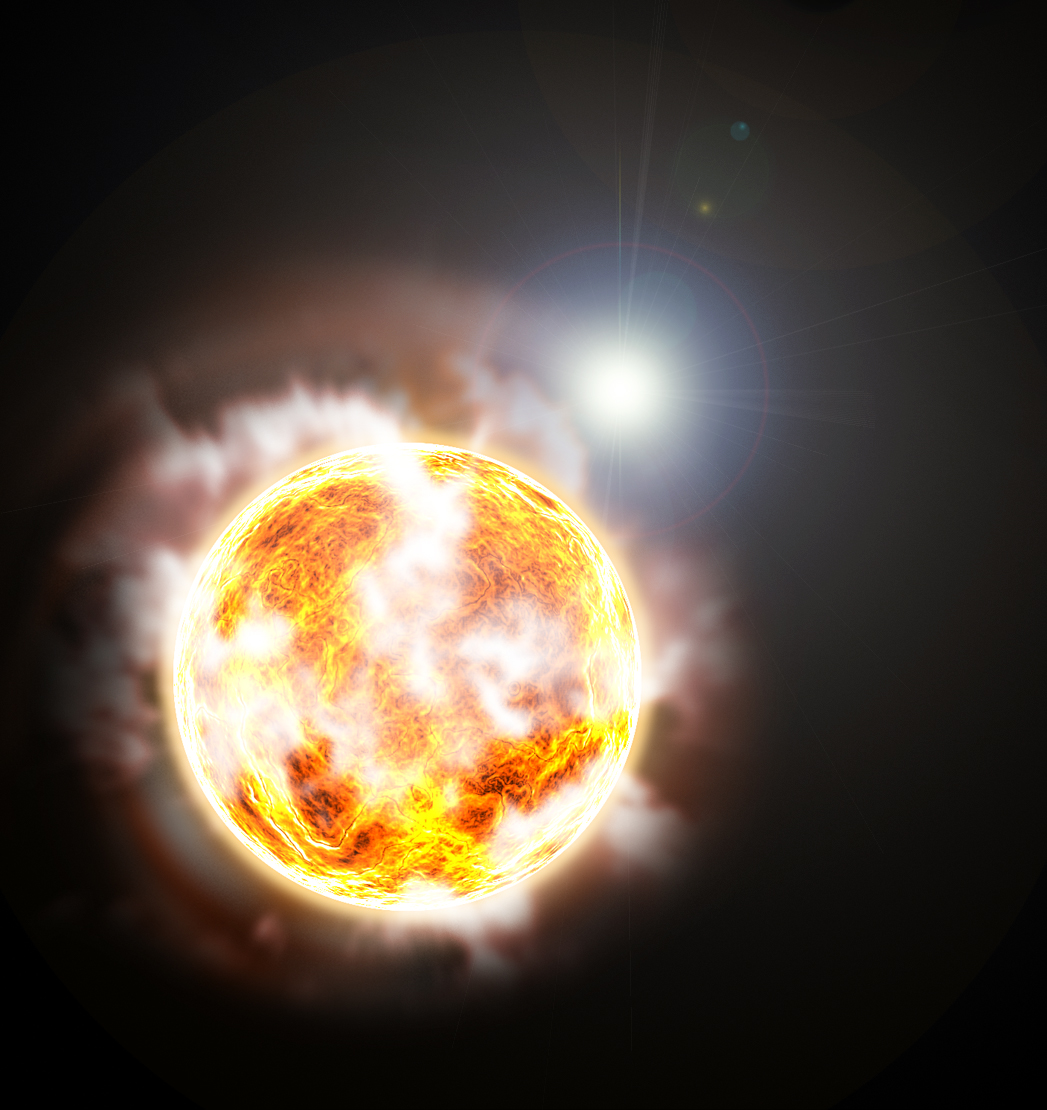} 
\caption[Author's Impression of a Symbiotic Star]{
Author's impression of a symbiotic star, depicting the compact white dwarf orbiting behind the giant and  illuminating its atmosphere.
\label{fig:symb}}
\end{figure}

Symbiotic binaries consist of a cool giant star experiencing large scale mass-loss and a hot, compact  component, usually a white dwarf (Figure \ref{fig:symb}). As well as the two stars there is also an extended nebula that owes its presence to the red giant wind \citep{leed_2006}.  
Stellar symbiosis in stars was first tackled early in the twentieth century. The first examples were noted by their combination spectra and are thought to have been CI Cyg, RW Hya and Ax Per \citep{percy2007understanding}. \citet{berman_1932} and \citet{hogg_1936} first suggested the binarity of the stars, while \citet{merrill_1944_symb} attempted  to group the stars into a different class based on the visibility of emission lines. Merrill's class was made up of 12 stars showing combination spectra and  would become the first group of symbiotic stars. This group originally consisted of Z And, R Aqr, UV Aur, T CrB, BF Cyg, CI Cyg, AG Dra, RW Hya, SY Mus, AG Peg, AX Per and RX Pup. EG And was not suggested as a possible symbiotic star until the 1950s, and was later confirmed as one in the 1980s (this is discussed further in Chapter 2). While initially it was difficult to know if these stars were long period variables or binary stars, Merrill was convinced of their ``symbiotic'' nature \citep{merrill_1950_symb,merrill_1958}. It was not until the launch of The International Ultraviolet Explorer ({\sl IUE}) in the 1970s that the presence of the ultraviolet component (the white dwarf) was truly confirmed.  Both the Space Telescope Imaging Spectrograph (STIS) aboard the Hubble Space Telescope ({\sl HST}), and the Far Ultraviolet Spectroscopic Explorer ({\sl FUSE}) bolstered UV and FUV observations of symbiotic stars. There are currently more than 180 observed symbiotic stars and a further 30 that are suspected symbiotics \citep{belczy_2000}. 

\begin{figure}[ht!]
\centering
\includegraphics[width=\textwidth]{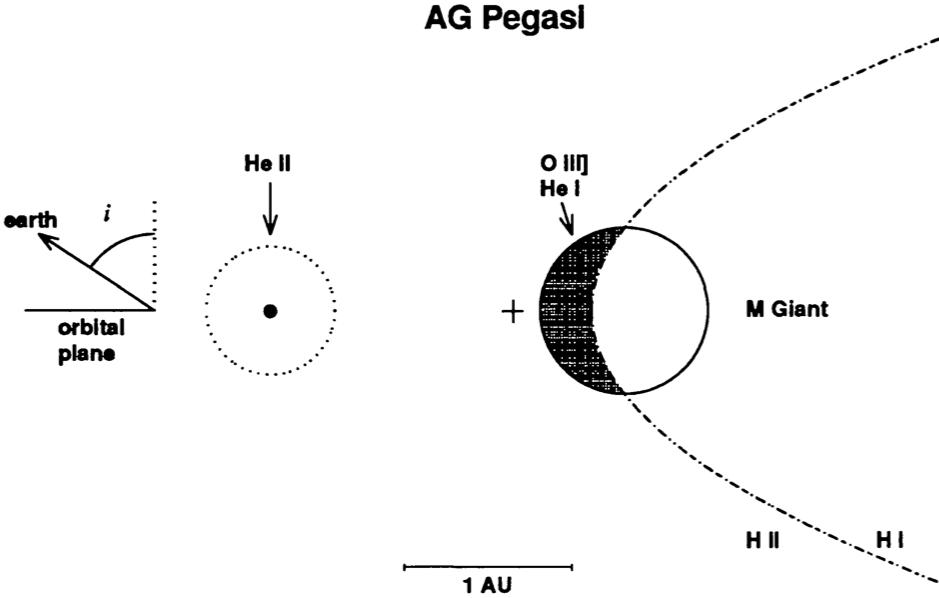} 
\caption[AG Peg Symbiotic Star]{
A schematic illustration of a symbiotic star, in this case AG Peg  \citep{kenyon_1993_ag_peg} viewed in the orbital plane of the binary. The cross marks the centre of mass and the line of sight to the Earth is inclined at an angle of $50^{\circ}$. The dotted line surrounding the hot component shows the \ion{He}{II} recombination zone. This heats the hemisphere of the red giant facing the hot component and produces \ion{He}{I} and \ion{O}{III]}. The dashed-dot line around the giant shows the \ion{H}{II} region.
\label{fig:ag_peg}}
\end{figure}

While the three components required for symbiosis (the cool star, hot star and ionised nebula) appear straightforward at first, there are several different combinations of hot and cool stars that can produce the required symbiotic spectrum. It is estimated that up to 80\% of all symbiotic stars are made up of a red giant star under-filling its Roche lobe\footnote{Orbiting material within the Roche lobe of a star in a binary system is gravitationally bound to that star \citep{eggleton_1983_rochelobe}.} by a factor of two or three, a hot compact star similar to the central star of a planetary nebula and a large ionised nebula that is responsible for many of the emission lines in the spectrum \citep{1984_kenyon_webbink}. A schematic of this type of system is shown in Figure \ref{fig:ag_peg}. A second group of symbiotics is characterised by wider binaries with separations of up to 100AU and a Mira variable instead of a red giant as the primary star \citep{whitelock_1987_symb_mira}. IR observations allow these two groups to be distinguished and help define them as either short orbital period {\it S-type} (stellar) or longer orbital period {\it D-type} (dusty) systems. The {\it S-type} symbiotics have IR colour temperatures of $\sim 2500 - 3500$ K while the {\it D-type} symbiotics have temperatures of $\sim 1000$ K. The {\it S-type} systems also have higher density nebulae ($n_e \sim 10^{10}$ cm$^{-3}$) compared to the {\it D-type} systems ($n_e \sim 10^6$ cm$^{-3}$).

The following description of symbiotic formation largely follows the discussion in \citet{kenyon_1994_symb_evo}. The red giant primary constrains the possible evolutionary paths of a symbiotic system. It sets the size of the system. It must be large enough to allow a main sequence star to evolve to radii greater than 50 R$_{\odot}$, but small enough to allow the red giant to transfer sufficient mass to its companion. This results in symbiotic stars being interacting binaries with the longest orbital period and the largest component separation \citep{mikolajewska_2007}. For {\it S-type} symbiotics their orbital period is limited to between 1 and 10 years, while for the longer period {\it D-type} systems it can be up to 100 years.  In the case of {\it S-type} symbiotics, the two stars start out as main sequence stars in a binary system. One of the stars will follow an evolutionary path similar to a single star, evolving to a red giant phase, ejecting a planetary nebula and becoming a white dwarf. The system then approaches the symbiotic phase as the original secondary star evolves to ascend the red giant or asymptotic giant branch to become the red giant primary of the system. In the process, it will lose mass and rejuvenate its companion star. The lifetime of the system is also limited by the red giant. Symbiotic red giants have masses of 1 - 3 M$_{\odot}$ and as the two stars evolve at different rates the system can only remain in the symbiotic stage for approximately a few million years. Most {\it S-type} systems have circular orbits and seem to interact by wind-driven mass-loss (although some of them may show Roche lobe overflow). The primary is usually an M2 - 6 giant and are often quite similar to single M giants. The main difference is that symbiotic red giants have higher mass-loss rates than single giants which may be a trigger for symbiotic activity \citep{mikolajewska_2002}. 

This thesis is chiefly concerned with a specific type of {\it S-type} symbiotic known as an eclipsing symbiotic. While isolated giants provide only disk-averaged information on the structure of their atmospheres, eclipsing symbiotics allow ultraviolet observations of the compact white dwarf through periodic eclipsing, defined in terms of our sightline. By comparing the UV spectrum of the white dwarf when it is outside of eclipse (and therefore less affected by the extended atmosphere or the wind) to observations as it passes behind the giant's atmosphere, we can obtain spatially resolved information. The vast differences in spectral energy distribution means the giant and dwarf star can be readily deblended from one another. The white dwarf contributes mostly to the UV continuum, while the optical region is dominated by the giant. As the white dwarf goes into eclipse, its UV continuum will gradually have more absorption features imposed upon it by the red giant atmosphere, giving detailed spatial information about the atmosphere close to the giant. Observing at different stages of eclipse is discussed further in Chapter 2.

\section{Stellar Atmospheres}

Mass-loss during the red giant phase is a crucial event in the star's life. Post-main-sequence stars are responsible for almost all of the elements in the Universe that are heavier than helium. Low gravity giants  are specially important for the CNO abundances in the solar neighbourhood \citep{schroder_2001_gal}. Understanding red giants  in terms of their atmospheres and mass-loss is essential to interpreting both stellar and galactic chemical evolution. 

\begin{figure}[ht!]
\includegraphics[width=\textwidth]{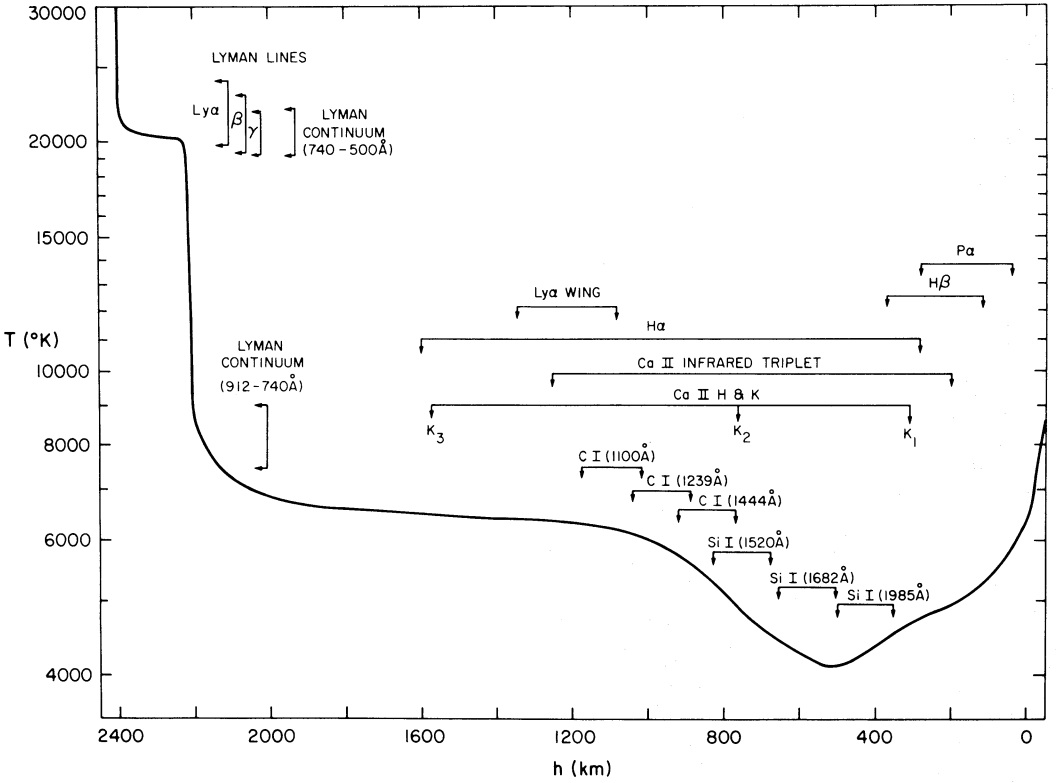}
\caption[Temperature Distribution of the Solar Atmosphere]{
A simplified representation of the temperature distribution of the solar atmosphere from \citet{vernazza_1973_val}. The figure shows where different ultraviolet spectral features are formed in the solar atmosphere. 
\label{fig:val_plot}}
\end{figure}

\citet{gray_photosphere_book} defines a stellar atmosphere as `\emph{a transition region from the stellar interior to the interstellar medium}'. Figure \ref{fig:val_plot} shows a simplified one-dimensional representation of the solar atmosphere from \citet{vernazza_1973_val} which helps to identify the different regions of the atmosphere. The sub-photosphere is below the solar surface (less than 0km on the plot) and marks the end of the convective envelope. The region from 0 to $\sim 500$km is the photosphere. This is the visible surface of a star; the part of the star at which the plasma of the star becomes transparent to photons of visible light (i.e.\  the optical depth reaches unity). At this point, any energy generated in the stellar interior becomes free to propagate out into space. Typically, the temperature drops by more than a factor of 2 from the top to the bottom of the photosphere.  Solar-like stars have reasonably well-defined photospheric layers. For cool giants,  their lower surface-gravities and higher mass-loss rates make defining the photosphere much more difficult. As their photospheres are cooler they are heavily influenced by absorption bands of molecular species. Figures showing optical spectra of cool giants being dominated by TiO bands are shown in Chapter 4, while the photosphere of cool giants is discussed further in Chapter 6.

\begin{figure}[ht!]
\includegraphics[width=\textwidth]{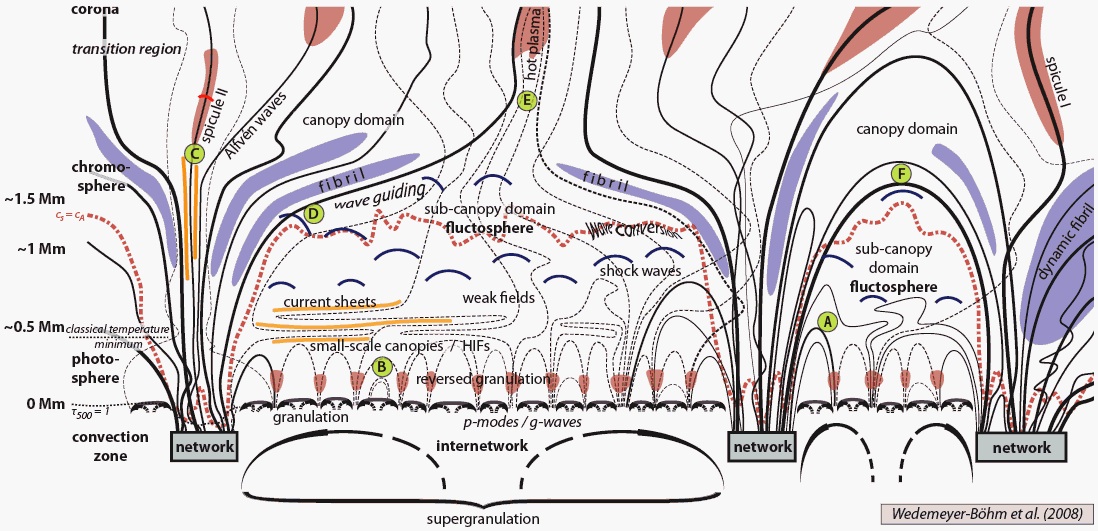}
\caption[Complexity of the Solar Atmosphere]{
A more realistic illustration of the complexity of the solar atmosphere \citep{wedemeyer_2009}.
\label{fig:wede_bohm}}
\end{figure}

The simplified temperature distribution in Figure \ref{fig:val_plot} also shows a temperature rise around $500-2000$\ km, where $T_e \sim 4000-20,000$\ K. This is the solar chromosphere. It demonstrates a rise in temperature despite decreasing density. A further sharp rise shows the transition region and finally the corona which is an extremely hot, tenuous and extended layer. In reality, the atmosphere is far more complex than the simplified picture shown in Figure \ref{fig:val_plot}. \citet{wedemeyer_2009} show that the solar atmosphere is characterized by a complex interplay of competing physical processes which include convection, radiation, conduction, and magnetic fields (Figure \ref{fig:wede_bohm}). Cool giants could be expected to have similar intricacy in their atmospheric structure as well as having an inhomogeneous extended atmosphere that becomes a strong outflow of gas and sometimes dust. 

Cool stars can be grouped into two broad categories separated by those that have corona and those that have not. \citet{carpenter_1998_coolstars} describes how coronal stars include cool dwarfs, giants earlier than K2, and supergiants earlier than G8.  These stars are thought to have atmospheres similar to the Sun in that above the photosphere they display a chromosphere of 5000 - 10,000\ K, a hotter transition region at around $10^5$\ K and an even hotter corona at approximately $10^6$\ K. These stars have fast winds upwards of $50 - 100$\ km\ s$^{-1}$ and relatively low mass-loss rates of around $10^{-13} - 10^{-14}$\ M$_{\odot}$\ yr$^{-1}$ \citep{dupree_reimers_1987}. The non-coronal stars encompass giants later than K2 and supergiants later than G8. They show a similar temperature distribution in the atmosphere out to the chromosphere but do not exhibit transition regions or coronal material. They show slower winds of about $10 - 50$\ km s$^{-1}$ but higher mass-loss rates, $10^{-10} - 10^{-8}$ for giants, $10^{-7} - 10^{-5}$\ M$_{\odot}$ yr$^{-1}$ for supergiants\footnote{This is a broad categorisation, as ``hybrid stars'' exist near the Linsky-Haisch dividing-line \citep{linsky_haisch_1979} that can display properties from both groups i.e.\ evidence for coronae along with high mass-loss rates \citep{hartmann_1980}.}.

\section{Stellar Winds}

Mass-loss, in the form of stellar wind, is different for both the coronal and non-coronal stars. Historically, observations of massive hot stars and novae proved that mass outflows existed as far back as the 19\textsuperscript{th} Century. \citet{campbell_1892} found that the width of spectroscopic lines, $\Delta \lambda$, was proportional to its velocity, confirming that the Doppler effect was responsible for broadening: \begin{equation}\Delta \lambda = \frac{\lambda v}{c}\end{equation} Although broadening could arise from expansion or some other form of turbulent motion, photographs of novae showed shells around outbursts. The shell diameters increased with time, proving that the they were expanding away from the central star. \citet{deutsch_1956}, using observations of the $\alpha$ Her system (M5Iab giant primary), was the first to show that material could escape the gravitational pull of a red giant surface, indicating the presence of a wind. Deutsch noted how the earlier-type secondary star\footnote{The secondary star is now thought to be a multiple star system.} could be used to probe the outer regions of the giant's wind. 

Stellar winds are now described as outflows of material (of both radiation and particles) from almost all stars. The mass-loss rate, $\dot{M}$, describes the amount of material shed by a star per unit time, and is usually given in terms of solar masses per year (M$_{\odot}$ yr$^{-1}$). The other crucial parameter when characterizing stellar winds is the terminal velocity of the wind, $v_{\infty}$. The mass-loss rate will affect the evolutionary path of the star. By multiplying both sides of  the equation of mass-continuity (Equation \ref{eqn:mass_contin}) by velocity, a mass-loss rate for a stationary spherically symmetric wind can be obtained:
\begin{equation}
\dot{M}= 4\pi r^2 \rho(r)v(r)
\label{eqn:mass_loss}
\end{equation} 
This simplified equation merely states that no material is created or lost in the wind and that the same amount of gas flows per second through a sphere at any radial distance from the centre of the star. 

\citet{lamers_book} also show how gas that escapes from the photosphere will asymptotically approach the terminal wind velocity at large distances from the star:
\begin{equation}
v(r)\simeq v_0 + (v_\infty - v_0)\left(1 - \frac{R_\star}{r}\right)^\beta
\label{eqn:beta_law}
\end{equation} 
This is a $\beta${\it -law}, as the parameter $\beta$ describes the steepness of the run of velocity. At the photosphere ($r=R_\star$) the velocity $v_0$ is very small, while at large distances ($r\rightarrow \infty$) the velocity approaches $v_\infty$.
Mass-loss rates were derived for red giants and supergiants by \citet{reimers_1975}. In particular, the mass-loss rate for a subset of red supergiants in binary systems was measured by observing the cool stars wind absorption lines against its hot companion. This led to ``Reimers Relation'':
\begin{equation}
\dot{M} = 4 \times 10^{-13}\eta_{R}\frac{(L_{*}/L_{\odot})(R_{*}/R_{\odot})}{M_{*}/M_{\odot}}
\end{equation}
This equation gives the mass-loss rate in $M_{\odot}$\ yr$^{-1}$, where $\frac{1}{3}<\eta_{R}<3$ is a correction factor for different types of star. Using this relation, it is possible to estimate the mass-loss rate of a star from measurements of its luminosity, radius and mass ($L_{*}, R_{*}, M_{*}$) scaled to those parameters for the solar case ($L_{\odot}, R_{\odot}, M_{\odot}$). 
 
\citet{parker_1958}  assumed a spherically symmetric, isothermal outflow from the Sun. He deduced that  the gas pressure at the surface of the Sun must be much greater than that of the interstellar medium, $P_{\odot}>>P_{\mbox{ISM}}$. If the system were in hydrostatic equilibrium, then the pressure gradient would be balanced by the attraction due to gravity and there would be no wind:
\begin{equation}
\nabla{P} + \rho g = \frac{dP}{dr} + \frac{GM\rho}{r^{2}}= 0
\label{eqn:solar_wind}
\end{equation}
Or:
\begin{equation}
\nabla{P} = - \rho g 
\label{eqn:solar_wind2}
\end{equation}
Assuming an isothermal atmosphere/wind, substituting the ideal gas law (Equation \ref{eqn:igl}) into Equation \ref{eqn:solar_wind}, and integrating from the surface of the star outwards gives:
\begin{equation}
\int_{\rho_{\star}}^{\rho} \frac{d\rho}{\rho} =- \frac{GM\mu m_H}{kT} \int_{r_{\star}}^{r} \frac{1}{r^{2}}dr
\label{eqn:solar_wind_int}
\end{equation}
which becomes:
\begin{equation}
\rho = \rho_{\star}\ exp\left[{-\frac{GM\mu m_H}{kT}}\left(\frac{1}{R_{\star}} - \frac{1}{r} \right)\right]
\label{eqn:solar_expon}
\end{equation}
It can be seen that at the stellar surface, where $r=r_{\star}$, the exponent in Equation \ref{eqn:solar_expon} goes to zero and $\rho = \rho_{\star}$. At large radii, $\rho(\infty) \rightarrow \mbox{Constant}$. Substituting the pressure scale height:
\begin{equation}
H_P = \frac{k T}{\mu m_H g}
\label{eqn:pressure_scale}
\end{equation}
which is the radial distance over which the pressure drops by a factor of $e$ \citep{schrijver2008solar}, and rearranging using the isothermal scale height relationship $H_{\rho} = H_P$, gives:
\begin{equation}
\rho = \rho_{\star}\ exp\left[{-\frac{R_{\star}^2}{H_{\rho}}}\left(\frac{1}{R_{\star}} - \frac{1}{r} \right)\right]
\label{eqn:sol_expl_hp}
\end{equation}
Expressing this in terms of pressure, and for the solar case, gives:
\begin{equation}
P = P_{cor} exp\left[- \frac{R_{\odot}}{H_P}\right]
\label{eqn:pressure_exp}
\end{equation}
Taking the ISM pressure as $P_{ISM}  = 3\times10^{-12}$\ dyn\ cm$^{-2}$ and adding in solar coronal values of $H_P = 7\times10^{9}$\ cm and $P_{cor} = 0.04$\ dyn\ cm$^{-2}$ \citep{mihalis_book}, yields $P_{cor}( \infty) = 1.9\times 10^{-6}$\ dyn\ cm$^{-2}$. This shows that at large distances, the coronal pressure is greater than the ISM pressure, $P_{cor}( \infty) >> P_{ISM}$, and a wind must be present. For giant stars this argument cannot be used to prove the presence of a wind and empirical evidence must be found, such as the $\alpha$ Her method discussed above.

\begin{figure}[ht!]
\centering
\includegraphics[width=\textwidth]{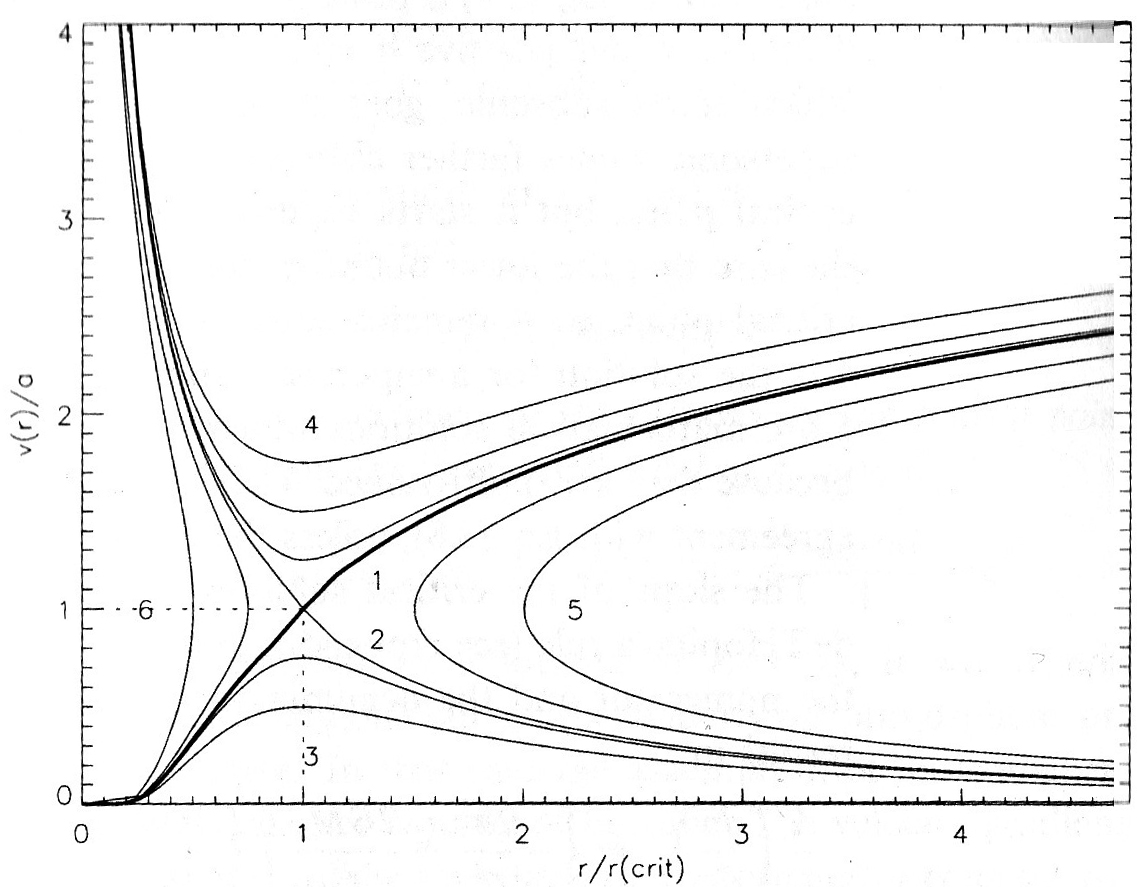}
\caption[Solutions of Momentum Equation for an Isothermal Wind]{
Solutions of the momentum equation (Equation \ref{eqn:parker_wind_solution}) for an isothermal wind from \citet{lamers_book}. The thick curve (labelled Curve 1) shows the critical, transonic,  solution with increasing velocity through the critical point $r_c$, where $v = a$.
\label{fig:parker_final_solution}}
\end{figure}

If the only forces acting on a wind are the gas pressure and gravity, then:
\begin{equation}
v\frac{dv}{dr} + \frac{1}{\rho}\frac{dp}{dr} + \frac{GM_{*}}{r^2}=0
\label{eqn:eqn_motion_1}
\end{equation}
which is the equation of motion for an isothermal wind. The ideal gas law (Equation \ref{eqn:ideal_gas1}) can be re-written as:
\begin{equation}
P = \frac{\rho k T}{\mu m_H}
\label{eqn:igl}
\end{equation}
The pressure gradient can be given by:
\begin{equation}
\frac{1}{\rho}\frac{dp}{dr} = \frac{k\mu}{m_{H}}\frac{dT}{dr} + \frac{kT}{\mu m_{H}\rho}\frac{d\rho}{dr} = \left(\frac{kT}{\mu m_{H}} \right)\frac{1}{\rho}\frac{d\rho}{dr}
\label{eqn:parker_pressure_gradient}
\end{equation}
for an isothermal wind, while the density gradient can be given by:
\begin{equation}
\frac{1}{\rho}\frac{d\rho}{dr} = -\frac{1}{v}\frac{dv}{dr} - \frac{2}{r}
\label{eqn:parker_density_gradient}
\end{equation}
Substituting Equations \ref{eqn:parker_pressure_gradient} and \ref{eqn:parker_density_gradient} into Equation \ref{eqn:eqn_motion_1}:
\begin{equation}
\frac{1}{v}\frac{dv}{dr} = \left\{\frac{2a^2}{r} - \frac{GM_{*}}{r^2} \right\} / \{v^2 - v_c ^2 \}
\label{eqn:parker_wind_solution}
\end{equation}
where:
\begin{equation}
v_c = \left(\frac{kT}{\mu m_{H}}\right)^{\frac{1}{2}}
\label{eqn:parker_wind_speed}
\end{equation}
is the constant isothermal speed of sound. Integrating gives:
\begin{equation}
\left(\frac{v}{v_c}\right)^2 - \log\left(\frac{v}{v_c}\right)^2 = 4\log\left(\frac{r}{r_c}\right) + 4\left(\frac{r}{r_c}\right) + C
\label{eqn:parker_wind_solution_integ}
\end{equation}
Equation \ref{eqn:parker_wind_solution} has several implications for the structure of the wind. In particular, the singularity that occurs where $v(r) = v_c$ implies that the mass loss rate is fixed. This occurs at a critical distance, $r_c$. In order for there to be a positive velocity gradient at all distances, there can be only one solution to the equation that goes through the critical point, the critical solution. Figure \ref{fig:parker_final_solution} displays the various solutions to Equation \ref{eqn:parker_wind_solution}. Curve 1 is the critical solution. It is transonic as it is subsonic initially, passes through the critical point, and is supersonic at larger distances. Curve 2 passes through the critical point but is supersonic initially, while Curve 3 remains subsonic and never reaches the critical point (the ``Solar Breeze Solution''). While Curves 5 and 6 exist mathematically, their multi-valued nature renders them physically meaningless. 

In Parker's steadily expanding wind theory, the forces of Equation \ref{eqn:solar_wind} are not balanced and an outflow is the outcome. From Newton's second law, the resulting force is:
\begin{equation}
\rho \frac{dv}{dt} = -\frac{dP}{dr} -\frac{GM\rho}{r^{2}}
\end{equation}
To expand this simplified theory of mass-loss, for example to add in a magnetic pressure term, $P_B$, it can be added to the sum of the forces on the right-hand side of the equation.
\begin{equation}\rho \frac{dv}{dt} = \sum (F )\end{equation}
\begin{equation} \rho \frac{dv}{dt} =  -\frac{dP}{dr} -\frac{GM\rho}{r^{2}} + \frac{dP_{B}}{dr}\end{equation}

One of the outstanding questions of stellar astrophysics remains the mechanism by which cool giants lose mass \citep{harper_1996_massloss}. While \citet{lamers_book} outline the following main wind-driving mechanisms, none can convincingly match the results shown in cool giant observations. Hot, luminous stars such as OB stars can have line driven winds \citep{kudritzki_2002_linedrivenwinds}. These stars emit radiation in the ultraviolet, where strong resonance transitions of several elements are located. The opacity in these absorption lines is much stronger than in the continuum and the radiation force on the ions effectively drives a stellar wind. This mechanism would not work for cool giants as they do not have strong continuum in the ultraviolet or enough free electrons due to their cooler temperature. 

Cooler stars that do have coronae, have lower mass-loss rates and faster wind speeds. This wind is probably governed by the same wind mechanisms as the Sun. Dissipation of mechanical energy and/or the reconnection of magnetic fields from the convective zone (below the photosphere) result in a temperature rise above the photosphere. This hotter material can produce a gas pressure gradient that can accelerate the wind. This mechanism does not apply to cool giants as it is reliant on the presence of a corona. 

In the absence of a magnetic field, a star with a spherically symmetric and hot corona produces a steady, structureless wind. Additional wind structure is added as a  result of open magnetic field structures. Oscillations induced in the field at the base of the wind by convective motions can generate Alfv\'en waves \citep{alfven_1983, cranmer_2005_alfenwaves}. These are non-compressive, transverse waves consist of travelling oscillations of ions and magnetic field. Alfv\'en wave driven winds have very high terminal velocities as wave damping occurs at large distances from the photosphere.

Finally, cool stars can generate acoustic waves in their photosphere which can propagate outwards carrying wave energy. The resulting wave pressure in the stellar atmosphere can drive a wind from the star. Sound waves \citep{koninx_1992_soundwinds} can only affect the mass-loss rates of low gravity stars and this process is not fully understood.

Dust driven winds \citep{sedlmayr_1995_dustdriven} can occur when temperatures drop low enough for dust to form. The dust can absorb photospheric radiation which heats the dust and results in the energy being radiated isotropically in the infrared. The photons carry momentum which is transferred to the dust particle when the radiation from the star is absorbed by the dust. The radiation field is directional and so an accelerating flow of material moves outwards from the star at low speeds \citep{mihalis_book}:
\begin{equation}
\dot{M} = \frac{L \tau_{\infty}}{v_{\infty}c}
\label{eqn:mihalas_dust}
\end{equation}
where $\tau_\infty$ is the flux weighted mean dust optical depth at the wind acceleration zone. This mechanism only affects stars that are cool enough to allow dust to form close to the photosphere, such as heavily evolved RGB/AGB stars. Pulsating Mira stars can possibly levitate material out to dust forming distances for this mechanism to work. In the atmospheric regions of interest in non-mira cool giants there is no dust present, e.g.\ EG And and $\gamma$ Cru.

It is clear that the structure of cool giant atmospheres and the mechanism of their mass-loss is yet to be resolved. Table \ref{tab:heat_mech} summarises the possible non-radiative heating and wind-driving mechanisms for cool stars.

\begin{table}
\begin{center}
\begin{tabular}{ll}
\hline
\textbf{Source}  & \textbf{Dissipation}   \\
\hline
\hline 
Alfv\'{e}n Waves &  Resonance/Viscous, Landau Damping  \\   
Fast-mode MHD Waves &  Landau Damping  \\ 
Slow-mode MHD Waves & Shock Dissipation  \\  
Magnetoacoustic Surface Waves & Mode coupling, Resonant Absorption \\  
Acoustic  Waves & Shock Dissipation \\  
Pulsation  Waves & Shock Dissipation \\  
Electric Sheets & Reconnection \\  
\hline
\hline
\end{tabular}
\caption[Atmospheric Heating and Wind-Driving Mechanisms]{
Atmospheric heating and wind-driving mechanisms from \citet{schrijver2008solar}, originally from \citet{ulms_1996}.
\label{tab:heat_mech}}
\end{center}
\end{table}

\section{UV Spectroscopy}

Most of what we know about stars comes from electromagnetic radiation \citet{percy2007understanding}. Much of the analysis in this thesis examines the amount of energy as a function of wavelength, where wavelength is given in terms of Angstroms\footnote{1\AA\ is equal to 0.1nm.}. These spectra are observed in the optical and ultraviolet regimes. In terms of the main target star, EG And,  the ultraviolet region corresponds to the radiation emitted by the white dwarf while the optical is dominated by the red giant. Observations made at different stages of eclipse allows probing of the atmospheres close to the star and gives temperature, density and velocity measurements . Fitting spectral features with computationally constructed models often allows us to infer that certain atomic species are present in the wind. The combination of both Doppler and collisional processes leads to a complex change in the equivalent width of a spectral absorption line as the optical depth increases. Optical depth is a measure of transparency, and is characterised by the mean amount of scattering or absorption between a point and an observer. If $I_{0}$ is the intensity of the radiation at the source and $I$ is the observed intensity after it has travelled a given path, then the optical depth, $\tau$ , is defined by the equation: 
\begin{equation}
I/I_{0}=e^{-\tau}\;or\;\tau=-ln(I/I_{0})
\end{equation} 

\citet{aller_atoms_stars_nebulae} describes how when light from a star passes through a stellar wind, the resulting spectrum will have absorption lines which serve to provide us with information about the wind. Each different type of atom can absorb light radiation of distinct wavelengths. Every light quantum will be re-radiated by the atoms but, crucially, the re-radiated light will be emitted in all directions. The beam of energy from the star will be in one direction only and so after passing through the wind, its continuum will be depleted at wavelengths corresponding to the atoms it has encountered. The absorption lines appear as dark lines on the continuous stellar spectrum. To gain information about the wind, both the shapes and the intensities of these dark lines are examined. The intensity is generally given by equivalent width in units of Angstroms. The equivalent width of an absorption feature is given by: 
\begin{equation}
\mbox{EW} = \int_{Line} I_{\nu}(0)-I_{o}(z)d\nu
\end{equation}
where $I_{\nu}(0)$ is the intensity on entering the cloud, $I_{o}(z)d\nu$ is the intensity on exiting, and $z$ is the distance from the observer, $o$, to the source. The area as defined above is equal to the rate at which energy is absorbed from the incident beam. The fractional energy lost is the equivalent width of the line:
\begin{equation}
\mbox{EW} = \int_{line}\frac{I_{\nu}(0)-I_{o}(z)}{I_{\nu}(0)}d\nu = \int_{line}(1-e^{-\tau_{\nu}z})d\nu
\end{equation}

\begin{figure}[ht!]
\centering
\includegraphics[width=\textwidth]{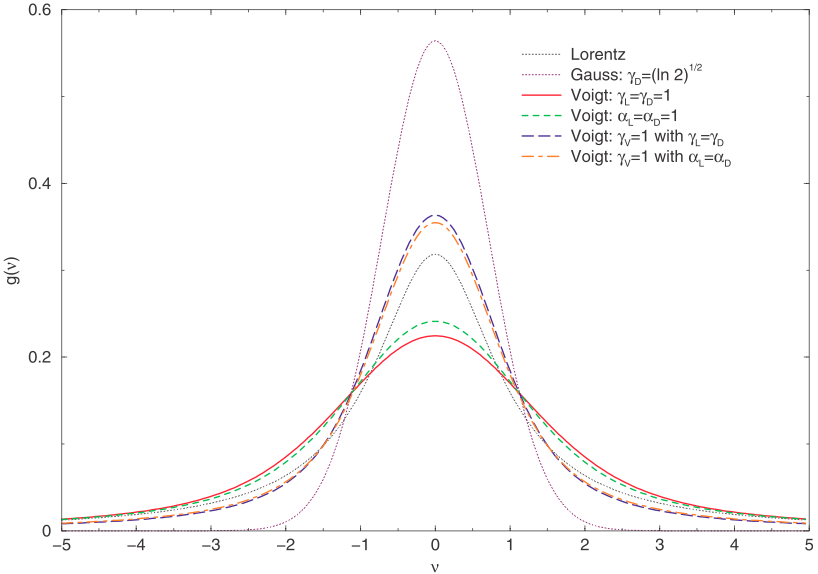}
\caption[Voigt Profile]{
A spectral line will resemble a Voigt profile. The line core will be dominated by the Doppler  part of the profile, but the Lorentzian will govern the damped wings, from \citet{schreier_2009}.
\label{fig:voigt_profile}}
\end{figure}

The line intensities are affected most by three bulk properties of absorbing gases: the chemical composition; the temperature; and the density.  However, the two most important processes that determine the width and shape of astrophysical lines are the natural broadening and the Doppler broadening. Natural line broadening is a property of the energy levels forming that line and, as such, are intrinsic to every spectral line. This results in a Voigt line profile that is a convolution of Doppler and Lorentzian profiles. The Voigt profile is given by:
\begin{equation}
V(\nu, \sigma, \gamma) - \int_{-\infty}^{\infty}G(\nu',\sigma)L(\nu - \nu', \gamma)d\nu'
\label{eqn:voigt}
\end{equation} 
where $\nu$ is frequency from line center, $\gamma$ is the damping constant, $G(\nu,\sigma)$ is the centered Gaussian profile:
\begin{equation}
G(\nu, \sigma) = \frac{1}{\sigma \sqrt{2 \pi}} \, e^{\frac{-\nu^2} {2\sigma^2}}
\label{eqn:gauss_profile}
\end{equation} 
and $L(\nu, \gamma)$ is the centered Lorentzian profile:
\begin{equation}
L(\nu, \gamma) = \frac{\gamma}{\pi(\nu^2 + \gamma^2)}
\label{eqn:lorentz_profile}
\end{equation} 
The Voigt profile is shown in Figure \ref{fig:voigt_profile}.

\section{Atomic Physics}\label{sec:atomic_physics}

The diagnostics used in this research require an understanding of UV emission-line formation. A photon is produced by the following process:
\begin{equation}
X^{m}_j \rightarrow X^{m}_i + h\nu
\label{eqn:photon_production}
\end{equation}
where $X$ is the atomic species, $m$ is the ionisation state and $i$ and $j$ are the lower and upper energy levels respectively. The volume emissivity (in erg cm$^{-3}$ s$^{-1}$) of a plasma is  given by:
\begin{equation}
\epsilon_{ji} = h\nu_{ji}A_{ji}n_{j}
\label{eqn:vol_emiss}
\end{equation}
where $n_i$ is the lower level population density, $n_j$ is the upper level population density and $A_{ji}$ is the Einstein coefficient for spontaneous radiative emission giving the
probability per unit time that the electron in the excited state will spontaneously
decay to the lower state. The flux (in erg cm$^{-2}$ s$^{-1}$) at Earth, observed at a distance $R$ from the star is:
\begin{equation}
F_{ji} = \frac{1}{4 \pi R^2}\int_{\nabla V}h\nu_{ji}A_{ji}n_{j}dV
\label{eqn:flux_basic_earth}
\end{equation}
The number density of ions in the excited level, $j$, can be expressed as:
\begin{equation}
n_j = \frac{n_j}{n_{ion}}\frac{n_{ion}}{n_{el}}\frac{n_{el}}{n_{H}}\frac{n_H}{n_e}n_e
\label{eqn:}
\end{equation}
where $n_{j}/n_{ion}$ is the relative population of the excited level, $n_{ion}/n_{el}$ is the relative abundance of the ionic species, $n_{el}/n_{H}$ is the abundance of the element relative to hydrogen, and $n_{H}/n_e$ is the number density of hydrogen relative to the number density of electrons. Using this expression with Equation \ref{eqn:flux_basic_earth} yields:
\begin{equation}
F_{ji} = \frac{h\nu_{ji}A_{ji}}{4 \pi R^2}\int_{\nabla V}\frac{n_j}{n_{ion}}\frac{n_{ion}}{n_{el}}\frac{n_{el}}{n_{H}}\frac{n_H}{n_e}n_edV
\label{eqn:flux_basic_earth_new}
\end{equation}
The main atomic processes for UV emission line spectroscopic diagnostics are summarized in Table \ref{tab:marisk_atom_procs}. Collisional transitions from level $i$ to level $j$ are governed by the collisional rate coefficient $C_{ij}$. 
\begin{table}
\begin{center}
\begin{tabular}{ll}
\hline
\textbf{Process}  & \textbf{Rate}   \\
& (cm$^{-3}$ s$^{-1}$) \\
\hline
\hline 
Collisional Excitation &  $n_in_eC_{ij}$  \\   
Collisional De-excitation &  $n_jn_eC_{ji}$  \\   
Spontaneous Radiative Decay &  $n_jA_{ji}$  \\   
\hline
\hline
\end{tabular}
\caption[Important Atomic Processes]{
Important atomic processes for UV emission line diagnostics \citep{mariska1992solar}.
\label{tab:marisk_atom_procs}}
\end{center}
\end{table}
Collisional de-excitation is less important as its characteristic time is much longer than that of spontaneous radiative decay. For each level in the ion, the rate equation is described by:
\begin{equation}
\frac{dn_i}{dt} = \sum_{j\ne i}n_jn_eC_{ji} - n_i\sum_{j\ne i}n_e C_{ij} + \sum_{j > i}n_j A_{ji} - n_i \sum_{j<i}A_{ij}
\label{eqn:atomic_procs_rate}
\end{equation}
The first two terms on the right-hand side of Equation \ref{eqn:atomic_procs_rate} show how level $i$ is populated and depopulated by collisions while the second two terms show how level $i$ is populated and depopulated by radiative decays. The processes involved have short timescales meaning that the left-hand side of Equation \ref{eqn:atomic_procs_rate} is zero. The electron collision rate coefficient $C_{ij}$ can be calculated by integrating the cross-section for excitation by collisions with electrons of velocity $v$ over the electron velocity distribution $f(v)$. The collision rate between the lower and  upper levels is then: 
\begin{equation}
n_e n_i C_{ij} = n_e n_i \int_{v_{0}}^{\infty}\sigma_{ij}(v)f(v)\, v\, dv
\label{eqn:atomic_procs_collis_rate}
\end{equation}
where $\sigma_{ij}$ is the electron excitation cross-section and $v_{0}$ is the threshold energy  velocity for the transition. The Maxwellian distribution is:
\begin{equation}
f(v) = 4\pi \left(\frac{m}{2\pi kT}\right)^{3/2} v^2 \exp\left( \frac{-mv^2}{2kT}\right)
\label{eqn:max_dist}
\end{equation}
where $m$ is the mass of the electron and $k$ is once again Boltzmann's constant. The collision cross-section is often expressed in terms of the collision strength $\Omega_{ij}(E)$:
\begin{equation}
\sigma_{ij} = \frac{\pi a_{0}^2 \Omega_{ij}(E)}{\omega_{i} E}
\label{eqn:collis_strength}
\end{equation}
where $a_{0}$ is the Bohr radius and $\omega_i$ is the statistical weight of level $i$. If the collision strength is independent of the incident energy then:
\begin{equation}
C_{ij} = \frac{8.63\times10^{-6}\Omega_{ij}}{\omega_i T^{1/2}}\exp\left(\frac{- \Delta E_{ij}}{kT} \right)
\label{eqn:collis_coeff}
\end{equation}
where $\Delta E_{ij}$ is the threshold energy of the transition. 

The two-level atom approximation uses only the ground level and excited level responsible for the line to calculate the line flux. In this case the collisional excitations from the ground level balance spontaneous radiative decay from the excited level:
\begin{equation}
n_e n_1 C_{12} = n_2 A_{21}
\label{eqn:coronal_approx}
\end{equation}
This is known as the \emph{coronal approximation}. The populations of other excited levels are small enough to assume all ions are in the ground level. The expression for the line flux becomes:
\begin{equation}
F_{21} = \frac{h\nu_{21}}{4\pi R^2}\int_{\Delta V} n_e n_{ion} C_{12} dV
\label{eqn:line_flux_a}
\end{equation}
Substituting Equation \ref{eqn:collis_coeff} gives:
\begin{equation}
F_{21} = \frac{h\nu_{21}}{4\pi R^2} \frac{8.63\times10^{-6}\Omega_{12}0.8A_{el}}{\omega_1}\times \int_{\Delta V} n_{e}^2 \frac{n_{ion}}{n_{el}}T^{-1/2}\exp\left(\frac{-h\nu}{kT} \right) dV
\label{eqn:line_flux_b}
\end{equation}
where the hydrogen-to-electron number density ratio ($n_H/N_e$) is taken to be 0.8. The temperature-dependent terms are often grouped together into the contribution function:
\begin{equation}
G(T) = \frac{n_{ion}}{n_{el}}T^{-1/2}\exp\left(\frac{-h\nu}{kT}\right)
\label{eqn:contribution_fucntion}
\end{equation}
Incorporating this gives:
\begin{equation}
F_{21} = \frac{2.2\times10^{-15}}{4\pi R^2}f A_{el} \int_{\Delta V} gG(T) n_{e}^2 dV
\label{eqn:line_flux_c}
\end{equation}

The temperature of the emitting plasma enters the emission line flux equation due to the temperature dependence of the relative ion abundance, $n_{ion}/n_{el}$, along with the temperature dependence of the collisional excitation rates. Determining the temperature of the plasma from the emission line fluxes observed is possible due to this latter dependence. An ion with emission lines originating from energy levels 3 and 2 that are exited from and decay to the lowest energy level 1, will have a ratio of fluxes (in an isothermal plasma):
\begin{equation}
\frac{F_{31}}{F_{21}}= \frac{\Delta E_{13}}{\Delta E_{12}}\frac{C_{13}}{C_{12}}
\label{eqn:temp_diag_1}
\end{equation}
Using Equation \ref{eqn:collis_coeff} for the collision rates:
\begin{equation}
\frac{F_{31}}{F_{21}}= \frac{\Delta E_{13}}{\Delta E_{12}}\frac{\Omega_{13}}{\Omega_{12}}\exp\left(\frac{\Delta E_{12} - \Delta E_{13}}{kT} \right)
\label{eqn:temp_diag_2}
\end{equation}
If $(\Delta E_{13} - \Delta E_{12})/kT \ge 1$, the flux ratio is a temperature sensitive diagnostic. 

The electron density affects the emission-line flux equation through the number density of ions in the excited level as well as the density dependence of the collision rates. In order to understand density diagnostics it is necessary to look beyond the two-level atom approximation.  The two-level atom is usually restricted by the ions having lines produced by electric dipole transitions that does not warrant a change in spin. This leads to these \emph{allowed} transitions having large transition probabilities and any collisional excitation is instantly followed by a spontaneous radiative decay. There are some occasions when the spontaneous transition probabilities are much smaller. This is the case for transitions from metastable levels that require a spin change, known as \emph{intercombination} transitions, or magnetic dipole transitions called \emph{forbidden} transitions. In order for an electron sensitive diagnostic to exist an ion needs at least three levels. One of the excited levels is always populated by collisions and depopulated by radiative decays. The second excited level is populated by collisions but at high enough densities because of a small $A$-value is depopulated by both collisions and and spontaneous radiative decays. The ratio of the two emission lines produced by spontaneous decays will therefore be sensitive to density \citep{mariska1992solar}.

\begin{figure}[ht!]
\centering
\includegraphics[width=\textwidth]{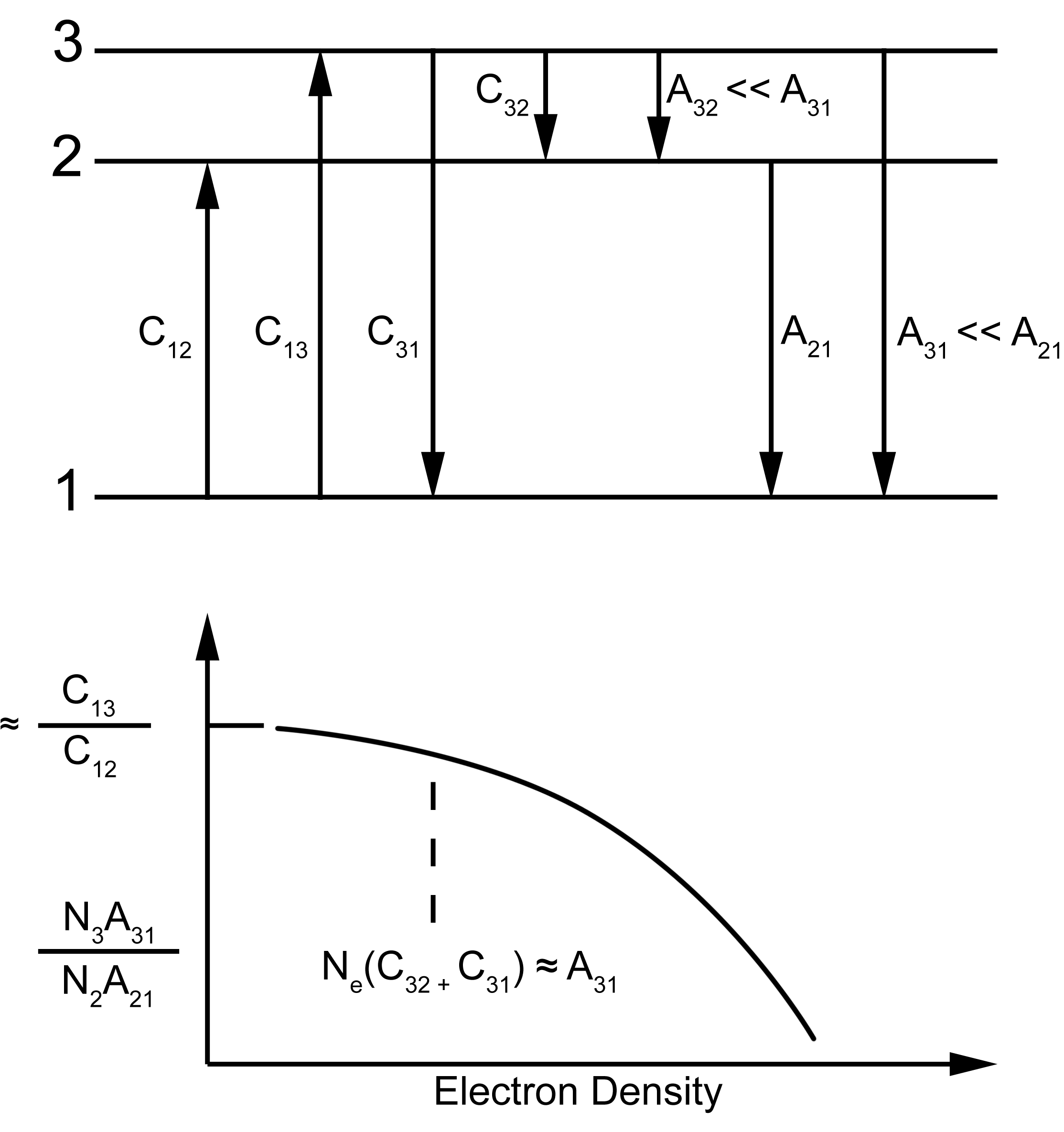}
\caption[Model Three-Level Atom]{
A model three-level atom with a metastable state leading to density sensitivity \citep{mariska1992solar}.
\label{fig:mariska_electron_dens_diag_3level}}
\end{figure}

The statistical equilibrium equations for levels 2 and 3 in a three-level atom case (shown in Figure \ref{fig:mariska_electron_dens_diag_3level}) can be simplified to:
\begin{equation}
n_3(A_{32} + A_{31} + n_e C_{32} + n_e C_{31}) = n_1 n_e C_{13}
\label{eqn:three_level_atom_stat_eq1}
\end{equation}
and:
\begin{equation}
n_2 A_{21} =  n_1 n_e C_{12} +  n_3(A_{32} + n_e C_{32})
\label{eqn:three_level_atom_stat_eq2}
\end{equation}

\begin{figure}[ht!]
\centering
\includegraphics[width=\textwidth]{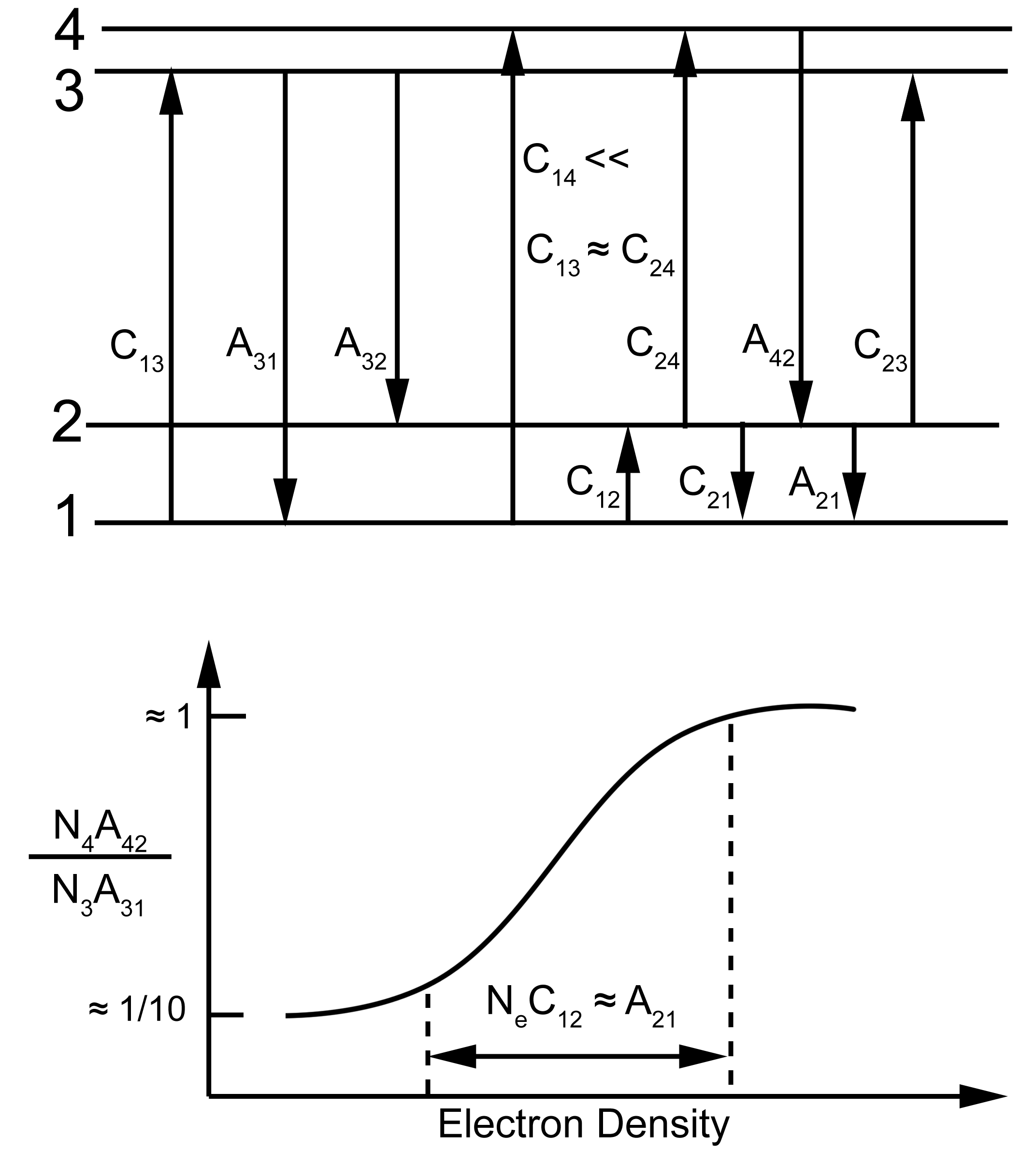}
\caption[Model Four-Level Atom]{
A model four-level atom with a metastable state leading to density sensitivity \citep{mariska1992solar}. It can be seen that between the very low and very large density extremes, the  ratio is sensitive to changing electron density.
\label{fig:mariska_electron_dens_diag}}
\end{figure}

Combining these equations gives:
\begin{equation}
\frac{n_3}{n_2} = \frac{C_{13}A_{21}}{C_{12}(A_{31} + n_e C_{31})}
\label{eqn:three_level_atom_1}
\end{equation}
and the line ratio is:
\begin{equation}
R = \frac{n_3 A_{31}}{n_2 A_{21}}
\label{eqn:three_level_atom_ratio1}
\end{equation}
which implies:
\begin{equation}
R = \frac{C_{13} A_{31}}{C_{12}(A_{31}+n_e C_{31})}
\label{eqn:three_level_atom_ratio2}
\end{equation}
Very low values of electron density will result in collisional de-excitation from level 3 to level 1 becoming small relative to spontaneous radiative transitions:
\begin{equation}
R = \frac{C_{13}}{C_{12}}
\label{eqn:three_level_atom_ratio3}
\end{equation}
but when the electron density is large enough, collisional de-excitation from level 3 to 1 becomes more dominant than spontaneous radiative transitions and:
\begin{equation}
R = \frac{A_{31}C_{13}}{C_{12}n_e C_{31}}
\label{eqn:three_level_atom_ratio4}
\end{equation}
This leads to inverse proportionality to the electron density with only a very weak temperature dependence. Figure \ref{fig:mariska_electron_dens_diag} shows a four-level atom containing a metastable state which leads to density sensitivity. 

This can be shown by the statistical equilibrium equations for levels 4, 3 and 1:
\begin{equation}
n_4(A_{42} + A_{41}) = n_1 n_e C_{14} + n_2 n_e C_{24}
\label{eqn:four_level_atom_1}
\end{equation}
\begin{equation}
n_3(A_{32} + A_{31}) = n_1 n_e C_{13} + n_2 n_e C_{23}
\label{eqn:four_level_atom_2}
\end{equation}
\begin{equation}
n_1 n_e(C_{12} + C_{13} + C_{14}) = n_4 A_{41} + n_3 A_{31} + n_2 (n_4 A_{21} +  n_e C_{21})
\label{eqn:four_level_atom_3}
\end{equation}
The density sensitive ratio is defined as:
\begin{equation}
R = \frac{n_4 A_{42}}{n_3 A_{31}}
\label{eqn:four_level_atom_ratio_1}
\end{equation}
Using the statistical equilibrium equations this becomes:
\begin{equation}
R = \frac{\alpha_{42}[C_{14} + (n_2 / n_1 ) C_{24}]}{\alpha_{31} [C_{13} + (n_2 / n_1 ) C_{23}]}
\label{eqn:four_level_ratio_2}
\end{equation}
where:
\begin{equation}
\alpha = \frac{A_{ji}}{A_{j1} + A_{j2}}
\label{eqn:four_level_ratio_3}
\end{equation}
The only density dependence in Equation \ref{eqn:four_level_ratio_2} is the ratio $n_2 / n_1$, which is:
\begin{equation}
\frac{n_2}{n_1} = \frac{C_{12} + C_{13}\alpha_{32} + C_{14}\alpha_{42}}{A_{21}/n_e + C_{21} + C_{23}\alpha_{31} + C_{24}\alpha_{41}}
\label{eqn:four_level_ratio_4}
\end{equation}
This ratio is small for low densities and leads to:
\begin{equation}
R = \frac{\alpha_{42} C_{14}}{\alpha_{31} C_{13}}
\label{eqn:four_level_ratio_5}
\end{equation}
At large densities the density-dependent term in $n_2 / n_1$ becomes small and the ratio is once again constant. Between the very low and very large density extremes the ratio is density sensitive as show in Figure \ref{fig:mariska_electron_dens_diag}.

\section{Motivation for Research}

The complexity of cool evolved stellar atmospheres, coupled with the lack of a convincing wind-driving mechanism, has meant that these stars are not well understood. The expanse of the chromosphere, and the physical conditions within it, are poorly established. It is the aim of this work to address this uncertainty, by looking at the conditions at the base of the outflow from two targets, and examining their implications for energy transport. Isolated giants provide only disk-averaged information, whereas symbiotic stars allow spatially resolved probing of the regions of interest - particularly the base of the chromosphere. Observations of EG And, an eclipsing symbiotic binary, provide a means to advance this field of red giant study. The white dwarf in the system, coupled with a knowledge of its orbital parameters, can be utilised as an orbiting ultraviolet backlight \citep{cian_thesis}. Ultraviolet observations at different stages of eclipse will give an insight into how the atmosphere of the giant evolves both spatially and temporally.

\begin{figure}[ht!]
\centering
\includegraphics[width=\textwidth]{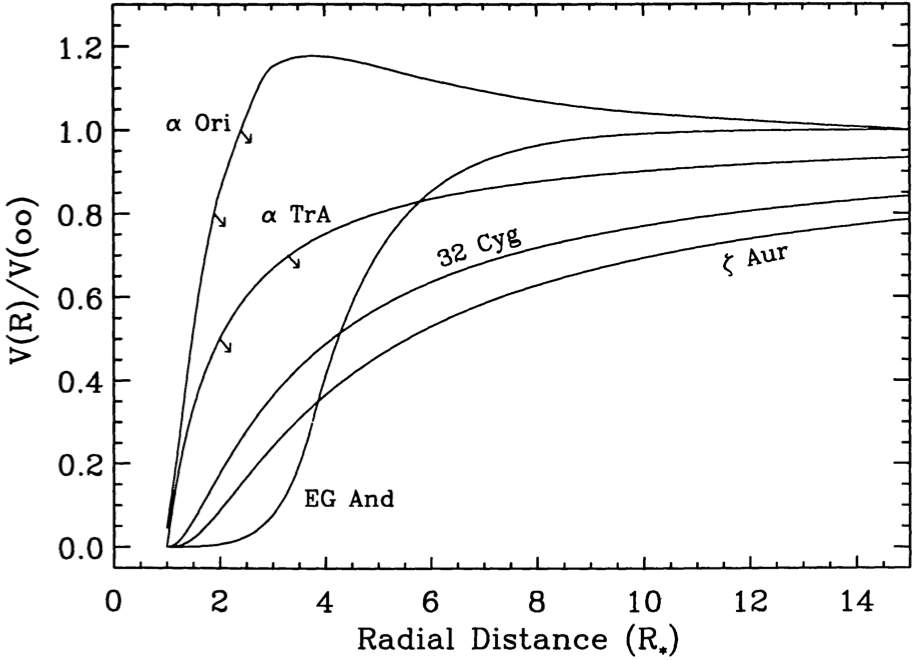}
\caption[Wind Velocity Laws]{
Normalised wind velocity laws for single stars and eclipsing systems EG And and $\zeta$ Aur  \citep{harper_1996_massloss}.
\label{fig:zeta_aur_windobs}}
\end{figure}

Previous analysis of other binary systems, $\zeta$ Aur (consisting of a K4Iab primary and a B5V secondary) and VV Cep (consisting of a M2Iab primary and a B-type main sequence secondary),  allowed information to be obtained on temperature structure, radiative processes and the physical distribution of circumstellar material  \citep{reimers_1987_binary, baade_1990_binary}. Unfortunately, the spectra of these binary stars were often difficult to disentangle and the presence of the secondary seemed to alter the wind acceleration profile (Figure \ref{fig:zeta_aur_windobs}). Symbiotic binary systems have the advantage that the spectra of the two components display vastly differing spectral characteristics and are quite easy to disentangle. The giant, dominant in the optical, offers low-velocity and low-ionisation features while the the dwarf contributes high-velocity and high-ionisation features. Comparing the bright UV continuum from the dwarf  when it  is in front of (and unattenuated by) the red giant, to observations as it passes behind the wind, will highlight any absorption features superimposed on it due to the red giant atmosphere.

To make these UV observations, it is necessary to use a space-based observatory, as ground-based observations would be hindered by the Earth's atmosphere. The Space Telescope Imaging Spectrograph (STIS) aboard the Hubble Space Telescope ({\sl HST}) is the only spaced-based instrument that currently covers the UV with the resolution and sensitivity needed for this work \citep{ihb, hst_primer}.

Symbiotic stars themselves warrant study in their own right as complex systems with companion stars at different stages of evolution. EG And is a well-studied object but uncertainty still remains. This study aims to improve some of the parameters of the system. The reasons for choosing EG And ahead of other symbiotic systems is dealt with more in Chapter 2.

A further goal of this work is to improve understanding of the similarities and differences between the red giant of EG And and isolated giants. This is addressed by using observations of an M3 giant spectral standard, HD148349. As well as looking at the base of the chromosphere of both stars, synthetic photometry and extinction analysis on their optical spectra will allow their variability to be examined. By comparing HD148349 to EG And in this manner, it is hoped that the effect of the white dwarf on EG And can be understood and the results can be extended to isolated giants. 

Lastly, efforts will be made to interpret the photospheric motions of giants, specifically line asymmetries in the optical regime. By measuring the changing asymmetry across the photosphere, a greater insight into the possible role of stellar granulation in mass-loss can be obtained.

 \section{Outline of Thesis}\label{sec:outline_thesis}

Chapter 2 summarises the main target of this study, the symbiotic eclipsing binary EG Andromedae. The earliest observations are recounted, along with all of the significant new results and parameters that have been discovered. Its importance as a tool for studying base outflows of red giant atmospheres is outlined, along with the reasons why it is utilised ahead of other symbiotic systems. An observational history and a list of known parameters is also included for the spectral standard, HD148349. Instrumental details for the STIS and {\sl HST} are recorded, with particular attention paid to the echelle gratings that were used for the observations in this thesis. The observations themselves are then listed, along with all of the observational information.

Chapter 3 illustrates the route taken by the data after observation, through the processing pipeline and into the archive. Manual calibration and data reduction steps that were taken outside of the pipeline are explored along with some of the main pitfalls of STIS echelle data reduction.

Chapter 4 compares EG And to the spectral standard HD148349.  A new method of obtaining low-resolution optical spectra from perviously unusable acquisition images is discussed at length. A method of preforming synthetic photometry on those optical spectra is also shown. An extensive study in interstellar extinction is coupled with the photometry results and an analysis of \ion{Mg}{ii} lines in both targets to provide new astrometry values, variabilities and insight into the similarities and differences of the two targets.

Chapter 5 examines the important chromosphere diagnostic multiplet, \ion{C}{ii}] 2325\AA. A technique for simultaneously fitting all of the emission lines in this multiplet is shown and the resultant integrated fluxes are used to draw conclusions on the electron density of both EG And and the spectral standard. Radial velocities, line asymmetries and \ion{Al}{ii}] features are also measured. The influence of the white dwarf on EG And is analysed and some modelling is preformed to interpret the structure of EG And's atmosphere.

Chapter 6 investigates photospheric motions and line asymmetries with a view to better understand their possible role in cool giant mass-loss. A process for obtaining photospheric composite line profiles and subsequent asymmetries is presented.

Chapter 7 summarises the conclusions from the previous chapters as well as discussing the accomplishments of the work. Future directions for the study are suggested in light of these new developments.


\chapter{Targets, Instrumentation and Observations} 
\label{chapter:instrumentation_and_observations}

In this chapter, the target stars, instruments and the observations central to this research are described. In order to compare a symbiotic system to an isolated spectral standard we chose to observe two targets - EG Andromedae (a symbiotic binary consisting of an M2.4 giant and a white dwarf) and HD148349 (an isolated M3III spectral standard). As described in Chapter 1, symbiotic stars can be used to uncover spatially resolved information from the red giant's atmosphere through ultraviolet observations of its orbiting dwarf companion. By observing EG And at several orbital phases and comparing it to observations of HD148349, we can catalogue the similarities and differences caused by the presence of the white dwarf in the symbiotic system. The Earth's atmosphere blocks most UV radiation, resulting in the need to use space-based observatories  to probe the key atomic transitions that lie in the UV. The International Ultraviolet Explorer ({\sl IUE}), launched in 1978 \citep{boggess_1978_iue}, was a precursor for this type of study. Its lifespan was originally supposed to be three years, but it survived until 1996. Its sensitivity in the UV region of the spectrum was invaluable in determining the nature of the hot component in symbiotic systems. {\sl IUE} observed in the spectral range 1150 - 3350\AA\ and following its demise, the UV spectroscopy mantle passed (albeit in the FUV) to the Far Ultraviolet Spectroscopic Explorer ({\sl FUSE}). {\sl FUSE} was launched in 1999 \citep{moos_2000_fuse}.  It had a resolution of 24,000 - 30,000 and a spectral range of 905 - 1195\AA. The satellite was decommissioned on October 18\textsuperscript{th} 2007. In order to obtain new high-resolution ultraviolet observations of the target stars, it was necessary to utilise {\sl HST}, the only space-based observatory in operation with the required UV capabilities \citep{kimble_1998_stis}. Some ground-based observations were also utilized in order to provide photometric observations of the two target stars. Archival data were used  to explore stellar photospheric motions, with a view to understanding granulation and the physics involved in cool giant mass-loss.

\section{EG Andromedae}\label{sec:eg_and}

\begin{figure}[ht!]
\centering
\includegraphics[width=\textwidth]{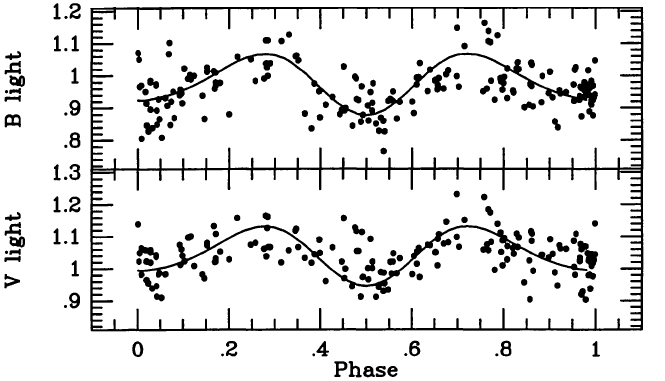}
\caption[EG And Ellipsoidal Light Curve]{
B and V magnitude light curves for EG And from \citet{wilson_vaccaro_1997_egand_photom}. A sinusoidal model fits the light curves when a circular orbit and tidal distortion are taken into account. The fit works less well in the U band (This is discussed further in Chapter 4).\label{fig:wilson_vaccaro_curve}}
\end{figure}

\begin{figure}[ht!]
\centering
\includegraphics[width=\textwidth]{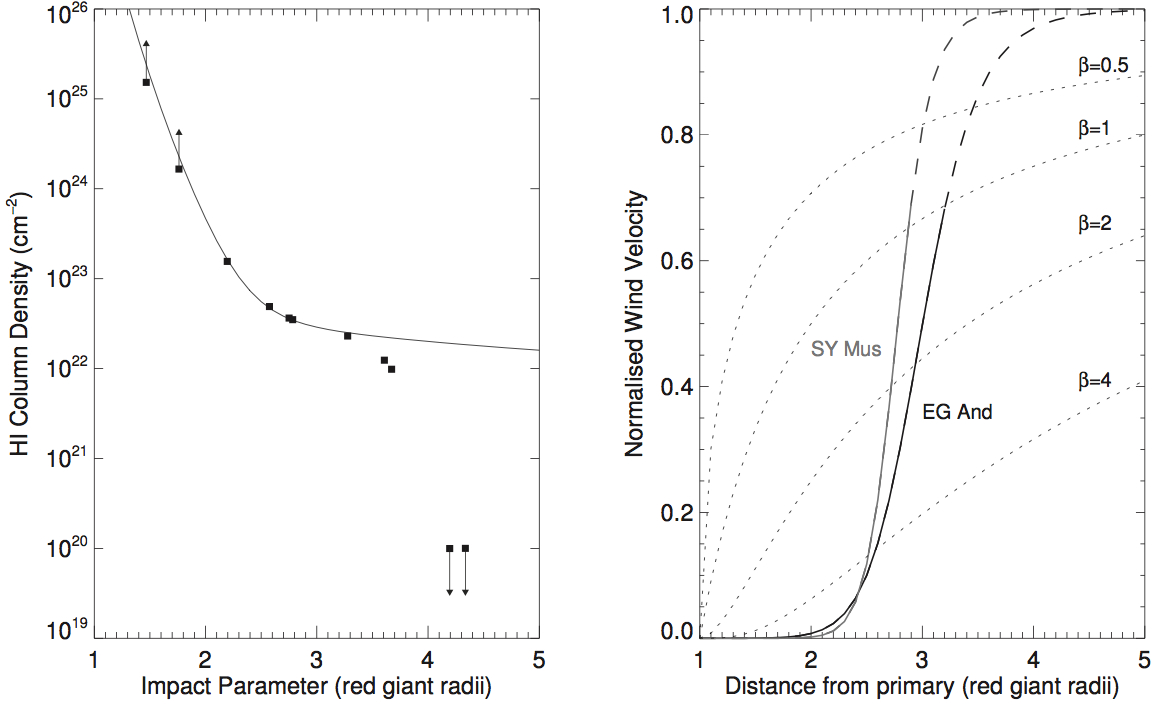}
\caption[EG And Wind Structure]{
Recent work on EG And's wind structure from \citet{cian_2010_passadena}. The left panel shows column density structure of the red giant wind in EG And derived from STIS and FUSE data. Overlayed is a model of the \ion{H}{I} column-density.  The right panels shows the run of the velocity profile of the giant derived from the \ion{H}{I} model on the left using an inversion technique described by \citet{knill_1993}. The right panel shows the wind profile derived for another eclipsing symbiotic, SY Mus, using {\sl IUE} data. Profile sections that are dashed are extrapolated since the wind is ionized at these distances.
\label{fig:cian_egand_plots}}
\end{figure}

EG Andromdae (also known as HD4174 and SAO36618) was chosen as the focus of this work as it is one of the brightest and closest symbiotic systems. Attention was drawn to the star as far back as 1950 when \citet{babcock_1950} and \citet{wilson_1950} noted how interesting the spectrum appeared and suggested that it was a combination (or composite) spectrum. The optical spectrum was indicative of an early M giant but the presence of strong emission lines was not easily explained. \cite{stencel_sahade_1980} suggested that the high-temperature emission lines such as \ion{He}{ii} and \ion{C}{iv} were caused by coronal heating due to the presence of a strong magnetic field of the order of a kilogauss.  Although mentioning that the emission lines could also be due to the presence of a hot companion star, they deemed this was unlikely as there was not enough UV continuum present in the spectrum (this was likely due to the dwarf being eclipsed by the giant during their bluest observations). \cite{smith_1980} compared the two alternate theories of explaining the peculiarities of the EG And UV spectrum - the large magnetic field, and binarity. The expected emission line variations of equivalent width and radial velocity over the time scale of a rising magnetic flux tube were not observed and failed to support the large magnetic field theory. Conversely, the presence of the hot companion was supported by periodic photometric variation and yielded a period of $\sim$470 days. \citet{stencel_1984}, after evidence of a UV continuum was found, supported the binarity model  and EG And was widely accepted as an eclipsing symbiotic binary consisting of a red giant and a hot companion. \citet{wallerstein_1981_olddisk} classified EG And as being an old disk population object. \citet{oliverson_1985} estimated a wind velocity of $75$\ km s$^{-1}$. \citet{kenyon_fernandez_castro_1987} derived a spectral type for the EG And primary star of an M2.4 giant. Using ground-based photometric observations of EG And, \citet{skopal_1988} and \citet{munari_1988} measured an orbital period of 481 days. {\sl IUE} observations of EG And by \citet{munari_1989} showed the effect of the hot component on the atmosphere of the giant primary, causing UV line and continuum emission. Using Equation \ref{eqn:mass_loss} and Equation \ref{eqn:beta_law}, \citet{vogel_1991} derived a way to describe EG And's wind from {\sl IUE} observations. While the $\beta${\it -law} in Equation \ref{eqn:beta_law} shows an early acceleration, Vogel's law showed the velocity remained low farther than two red giant radii, before rising sharply. \citet{muerset_1991}  and \citet{vogel_nussbaumer_1992} employed line diagnostics and eclipse analysis of the {\sl IUE} data to determine values of temperature, luminosity, giant radius and separation of the system. These parameters are listed in Table \ref{tab:target_params}. \citet{van_buren_1994} showed how EG And could not possess a dust-driven wind by calculating the optical depth of the required dust to drive a wind using Equation \ref{eqn:mihalas_dust}. \citet{wilson_vaccaro_1997_egand_photom} modelled the ellipsoidal photometric variation of EG And's light curves to present a case for a circular orbit with tidal distortion (Figure \ref{fig:wilson_vaccaro_curve}). With the arrival of {\sl FUSE} and {\sl HST}, higher resolution observations by \citet{cian_2008} resulted in a clearer understanding of which atomic species are present in the red giant wind and provided clues as to its structure. In particular, \citet{cian_2010_passadena} confirmed that EG And's wind is better described using a {\sl Vogel}-Law than a $\beta$-Law (Figure \ref{fig:cian_egand_plots}).

EG And was chosen as the target of this study as it has a number of advantages over other binary and symbiotic systems. It is an {\it S-type} symbiotic so it has not lost enough mass to form a dusty circumstellar shell. This ensures that its UV photons are not obscured by dust. {\it D-type} systems contain a mira variable with different atmospheric structure to standard giants, so an {\it S-type} system is more useful to this study. EG And is also out of the plane of the galaxy which reduces the amount of interstellar dust along the line of sight. It has an orbital period that is makes it suitable to study (BX Mon, for example, has a period of  $\sim$1401 Days while for  {\it D-type} stars it can be decades). A low ratio of hot to cool star UV luminosity is also desirable so that the hot component's effect on the giant atmosphere is minimised. Existing archival {\sl IUE} observations of EG And are insufficient due to low resolution and low S/N. While \citet{carpenter_1991_prop} observed EG And with the Goddard High Resolution Spectrograph, this permitted only small wavelength regions to be observed (three separate 40\AA\ regions between 1380 - 1670\AA).

\begin{table}
\begin{center}
\begin{tabular}{lcl}
\hline
\textbf{Parameter}  & \textbf{Value} & \textbf{Reference}   \\
\hline
\hline
Orbital Period &  482.6 Days & \citet{fekel_part1}  \\   
Separation & $\sim$ 4.2 R$_{RG}$ & \citet{vogel_nussbaumer_1992}  \\   
RG Radius &  $\sim$ 75 R$_{\odot}$& \citet{vogel_nussbaumer_1992}  \\  
WD Radius &  $\sim$ 0.018 R$_{\odot}$& \citet{vogel_1991}  \\   
Orbit &  Circular & \citet{wilson_vaccaro_1997_egand_photom}  \\   
RG Temperature &  $\sim$ 3,700 K & \citet{kreyes_preblich_2004}  \\  
WD Temperature &  $\sim$ 75,000 K & \citet{muerset_1991}  \\  
RG Mass & $1.5\pm0.6$ M$_{\odot}$ & \citet{mikolajewska_2002}  \\  
WD Mass & $ 0.4\pm0.1$ M$_{\odot}$ & \citet{mikolajewska_2002}  \\  
RG Luminosity &  $\sim$ 950 L$_\odot$ & \citet{vogel_nussbaumer_1992}  \\  
WD Luminosity &  $ \sim$ 16 L$_\odot$ & \citet{vogel_nussbaumer_1992}  \\ 
Spectral Type &  M2.4III & \citet{kenyon_fernandez_castro_1987}  \\   
Distance & $513\pm169$ pc & \citet{vanleeuwen_2007} \\
Right Ascension & $00\ 44\ 37.19$ (h:m:s) & \citet{vanleeuwen_2007} \\
Declination &  $+ 40\ 40\ 45.70$ (d:m:s) & \citet{vanleeuwen_2007} \\
Radial Velocity &$-101\pm3$ km s$^{-1}$ & \citet{wilson_1953_egandradvel}\\
Log(g) &$\sim$ 1.0 & \citet{cian_thesis}\\
Metallicity &$\sim$ -0.4 & \citet{cian_thesis}\\
E(B-V) &  0.05 & \citet{muerset_1991}  \\  
Angular Diameter &$\sim$ 1.27 mas& \citet{angdiam_2001}\\
\hline
\hline
\end{tabular}
\caption[EG And Observational Properties]{
EG And Observational Properties. The Right Ascension is given in units of time; hours, degrees and minutes. The Declination is given in terms of angle; degrees, arcminutes and arcseconds. The parameters have been compiled from several sources and may not be internally consistent.
\label{tab:target_params}}
\end{center}
\end{table}

\section{HD148349}\label{sec:hd48349}

HD148349 (V2105 Oph, SAO141186) was observed as early as 1918 by \citet{lunt_1918} who measured its radial velocity using the 36-inch Lick Telescope at the Mount Wilson Observatory. It is listed as an M3III spectral standard by \citet{keenan_1989} and \citet{1989_hdstand} in their catalogues, based on the MK classification system.  It also features in the Wilson-Bappu sample \citep{wilson_bappu_1957} which is discussed in Chapter 4. \citet{dumm_schild_1998} used {\sl Hipparcos}  data \citep{hipparcos_catalogue_1997} to determine a relationship between the visual surface brightness and the Cousins (V- I) colour index for stars of known angular diameter. They use this relationship to derive stellar parameters for a number of M giants, including HD148349. Stellar masses were then determined by using evolutionary tracks. They list HD148349 as having a temperature of 3720 K, a radius of 83 R$_\odot$ and a mass of 2 M$_\odot$. A list of parameters is provided in Table \ref{tab:target_params_hd}. The similarity of the giant primary of EG And to HD148349 was noted by \citet{cian_thesis} after comparing optical spectra of the two targets. This makes HD148349 an ideal candidate when comparing EG And to isolated giants. Despite (or perhaps because of) the fact that HD148349 is a long-accepted spectral standard, there have been very few observations to determine just how reliable it is as a standard star. This issue will be revisited in Chapter 4. 

\begin{table}
\begin{center}
\begin{tabular}{lcl}
\hline
\textbf{Parameter}  & \textbf{Value} & \textbf{Reference}   \\
\hline
\hline 
Radius &  $\sim$ 83 R$_{\odot}$& \citet{dumm_schild_1998}  \\   
Temperature &  $\sim$ 3720 K & \citet{dumm_schild_1998}  \\  
Mass & $\sim$ 2 M$_{\odot}$ & \citet{dumm_schild_1998}  \\  
Luminosity & $\sim$ 1000 L$_{\odot}$ &    \\  
Spectral Type &  M3III & \citet{keenan_1989} \\  
Distance & $177\pm12$ pc & \citet{vanleeuwen_2007} \\
Right Ascension & $16\ 27\ 43.46$ (h:m:s) & \citet{vanleeuwen_2007} \\
Declination &  $-07\ 35\ 52.56$ (d:m:s) & \citet{vanleeuwen_2007} \\  
Radial Velocity &$99.3\pm0.9$ km s$^{-1}$& \citet{evans_1967_hdradvel}\\
Log(g) &$\sim$ 1.0 & \citet{cian_thesis}\\
Metallicity &$\sim$ -0.4 & \citet{cian_thesis}\\
E(B-V) &  0.33 &   \\  
Angular Diameter &$\sim$ 4.85 mas& \citet{angdiam_2001}\\
\hline
\hline
\end{tabular}
\caption[HD148349 Paramaters]{
HD148349 Parameters. As in Table \ref{tab:target_params}, the Right Ascension is given in units of hours, degrees and minutes, while the Declination is given in degrees, arcminutes and arcseconds. Angular diameter is also given in arcseconds.
\label{tab:target_params_hd}}
\end{center}
\end{table}

\section{Hubble Space Telescope}\label{sec:hst}

The Hubble Space Telescope ({\sl HST}) is a concerted program of the National Aeronautics and Space Administration (NASA) and the European Space Agency (ESA) to operate a long-term, space-based observatory. It was first envisioned in the 1940s when Lyman Spitzer consulted for the RAND Corporation on the scientific merits  of having an Earth-circling satellite \citep{spitzer1997dreams}. Interestingly, the original report in 1946, later reprinted in \citet{spitzer_1990_hst_idea}, highlighted \emph{``Structure of Stellar Atmospheres''}  through spectroscopic UV analysis of resonance lines as an important area of research for the hypothetical telescope. It also mentions \emph{``Analysis of Eclipsing Binaries''} as another possible area of interest. A history of the early development of {\sl HST} (then referred to as the ``Large Space Telescope'') is detailed in \citet{lst_history_1974}. It was designed and built in the 1970s and 1980s as a 2.4-meter reflecting telescope\footnote{The secondary mirror is 30cm in diameter.}. It was deployed in low-Earth orbit (600 kilometers) by the crew of the space shuttle Discovery (STS-31) on 25 April 1990.  {\sl HST} science operations are conducted by the Space Telescope Science Institute (STScI) on the Johns Hopkins University Homewood Campus in Baltimore, Maryland. STScI itself is operated for NASA by the Association of Universities for Research in Astronomy, Inc. (AURA). The STScI maintain a website at \url{http://www.stsci.edu/hst} that provides the most complete and up-to-date information on {\sl HST} and its instruments. 

\begin{figure}[ht!]
\centering
\includegraphics[width=\textwidth]{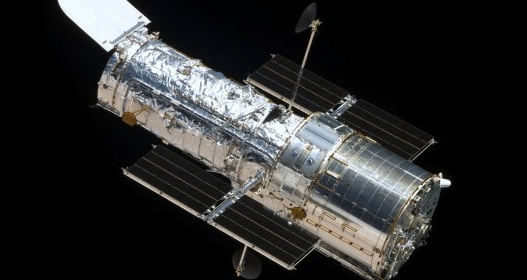}
\caption[{\sl HST} Instrument Design]{
{\sl HST} as seen from the departing Space Shuttle Atlantis, flying Servicing Mission 4 (STS-125), the fifth and final human spaceflight to visit the observatory in May 2009. \emph{Image Credits: NASA}.
\label{fig:hst_design}}
\end{figure}

\begin{figure}[ht!]
\centering
\includegraphics[width=\textwidth]{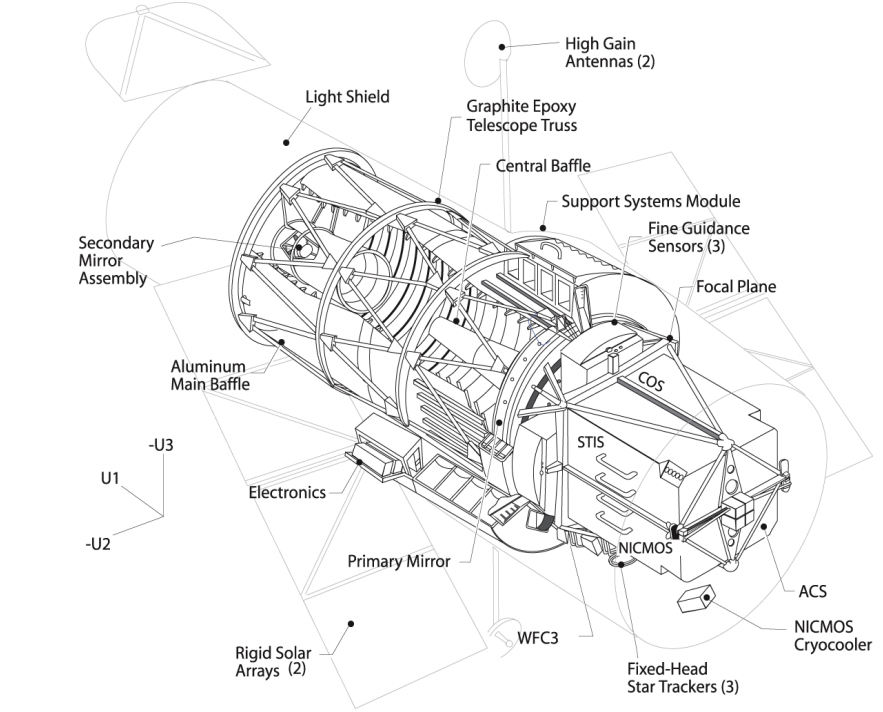}
\caption[{\sl HST} Instrument Design Close-up]{
{\sl HST} Instrument Design Close-up. STIS is located in the axial bays behind the mirror at the end of the telescope.
\emph{Image Credit: NASA}.
\label{fig:hst_design_close}}
\end{figure}

The current suite of {\sl HST} science instruments includes three cameras, two spectrographs, and fine guidance sensors (primarily used for accurate pointing, but also for astrometric observations). The cameras are the Advanced Camera for Surveys (ACS), the Near Infrared Camera and Multi-Object Spectrometer (NICMOS), and Wide Field Camera 3 (WGC3). The spectrographs are the Space Telescope Imaging Spectrograph (STIS) and the Cosmic Origins Spectrograph (COS). Former instruments include the first two Wide Field Planetary cameras (WF/PC-1 and WFPC2), the Faint Object Camera and Spectrograph (FOC and FOS) and the Goddard High Resolution Spectrograph (GHRS). Orbiting above the Earth's atmosphere allows the science instruments to produce high-resolution images of astronomical objects. As discussed in the introduction to this chapter, {\sl HST} can observe ultraviolet radiation, which is blocked by the atmosphere and therefore unavailable to ground-based telescopes. Figure \ref{fig:hst_design} shows {\sl HST} during Servicing Mission 4. A close-up of the telescope design is show in Figure \ref{fig:hst_design_close}.

{\sl HST} is powered primarily by two solar arrays which are turned, and the spacecraft rolled, about its optical access to ensue the panels face the incident sunlight. Its secondary power source consists of nickel-hydrogen batteries which operate during orbital night. A tracking and data relay system is operated with the ground via two high-gain antennae. Although the low-Earth orbit means {\sl HST} is a serviceable spacecraft, there are a number of constraints imposed on its observations as a result. Targets can be occulted by the Earth during the 96-minute orbit. The time available for observing is further reduced by guide star acquisitions and instrument overheads.

\section{Space Telescope Imaging Spectrograph}\label{sec:stis}

\begin{figure}[ht!]
\centering
\includegraphics[width=\textwidth]{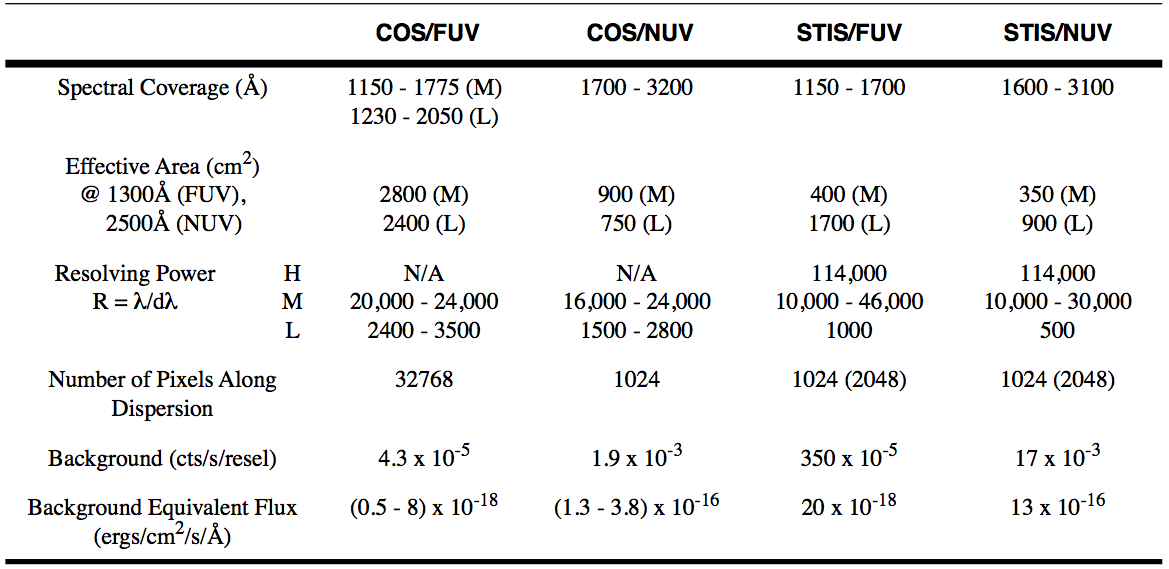}
\caption[Spectroscopy at Ultraviolet Wavelengths]{
Spectroscopy at Ultraviolet Wavelengths using STIS and COS. Table from \citet{hst_primer}.
\label{fig:stis_vs_cos}}
\end{figure}

The Space Telescope Imaging Spectrograph is one of two spectrographs aboard {\sl HST}.  It was installed during servicing mission SM2 in February 1997 and was the natural replacement for the GHRS. In 2004 a failure in the power supply electronics rendered the instrument inoperable. After nearly 5 years of dormancy STIS was repaired during SM4 in May 2009 and has resumed science operations with all channels. For a full description of the instrument see \citet{kimble_1998_stis} and \citet{ihb}. While the other spectrograph (COS) is more sensitive than STIS (by a factor of 10 to 30 in the far-ultraviolet and by a factor of 2 to 3 in the near-ultraviolet) the targets of this study were too bright in the UV to utilise COS. The higher spectral resolution offered by STIS was considered more important to the study of spectral features. A further consideration was that although COS has the capability to observe wavelengths between 900 and 1150\AA, STIS offers additional optical coverage which was deemed more useful. Figure \ref{fig:stis_vs_cos} shows the spectroscopic abilities of STIS compared to those of COS. 

\begin{figure}[ht!]
\centering
\includegraphics[width=\textwidth]{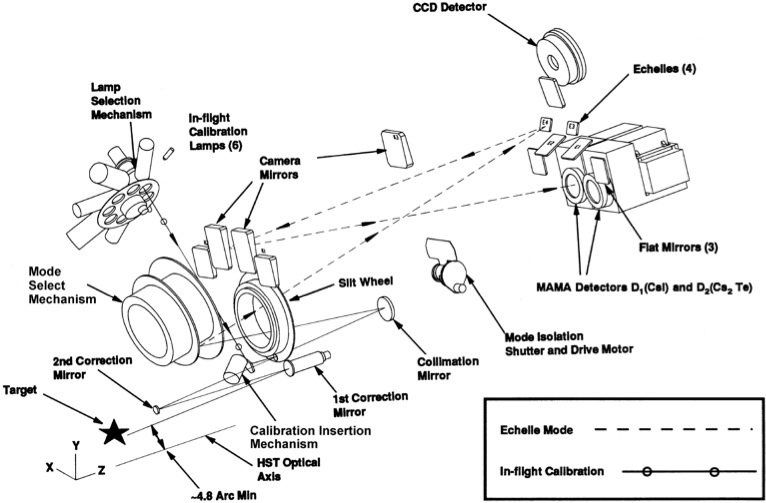}
\caption[STIS Instrument Design]{
The exploded view of STIS shows the optical design of the instrument. It utilises a CCD and two MAMA detectors in conjunction with a prism and first order and echelle gratings to cover the ultraviolet to infrared spectral range. The filter wheel near the lamps was not included in the final design. Figure from \citet{woogdate_stis_1998}.
\label{fig:stis_design}}
\end{figure}

STIS can perform spatially resolved long-slit spectroscopy from the UV to the NIR (1150 - 10,300\AA) at low to medium spectral resolution (R $\sim$ 500 - 17,000) in the first order as well as echelle spectroscopy at medium to high spectral resolution (R $\sim$ 30,000 - 114,000) in the UV (1150 - 3100\AA). It utilizes a CCD detector for visible observations and two Multi-Anode Microchannel Array (MAMA) detectors for the UV. Figure \ref{fig:stis_design} shows the instrument design. Like all current {\sl HST} instruments, STIS' optical design includes corrective optics to compensate for {\sl HST}Õs spherical aberration. It also can be seen in the Figure that a telescope focal plane slit-wheel assembly, collimating optics, a grating selection mechanism, fixed optics, and camera focal plane detectors are all part of STIS. A calibration lamp assembly can illuminate the focal plane with a range of continuum and emission line lamps. The so-called Hole in the Mirror (HITM) system is used to obtain wavelength comparison exposures and to illuminate the slit during target acquisitions. These target acquisitions will be revisited in Chapter 4. The slit wheel allows positioning of the spectroscopic slits, while the grating wheel can be used to position  the first-order gratings, the cross-disperser gratings used with the echelles, the prism, and the mirrors used for imaging. 

\section{Echelle Gratings}

\begin{figure}[ht!]
\centering
\includegraphics[width=\textwidth]{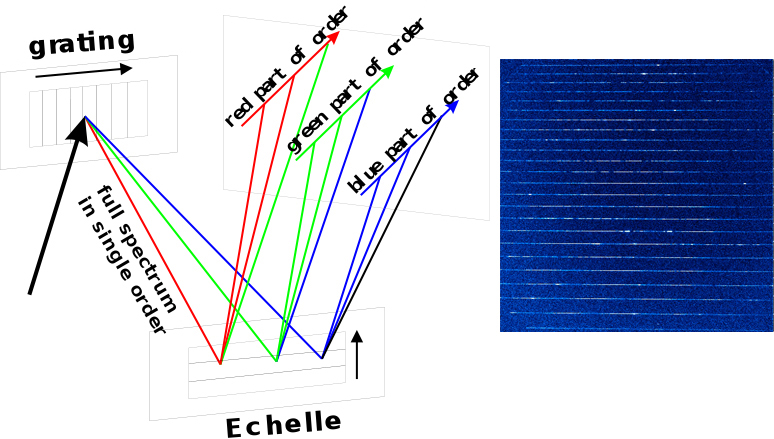}  
\caption[Echelle Grating]{
The echelle grating design (left) is optimized for multiple overlapping higher orders. The first grating is designed for a single lower spectral order while the echelle, mounted orthogonal to the first grating, transversally separates the higher orders. This results in stripes of different but slightly overlapping wavelength ranges. This effect can be seen in the flat-fielded image of EG And (right) which utilised the E140M echelle grating. A full image can be seen in Figure \ref{fig:echelle_raw}. \emph{Echelle Image Credit: Boris Pova{\~z}ay (Cardiff University)}. 
\label{fig:echelle_grating}}
\end{figure}

\begin{figure}[ht!]
\centering
\includegraphics[width=\textwidth]{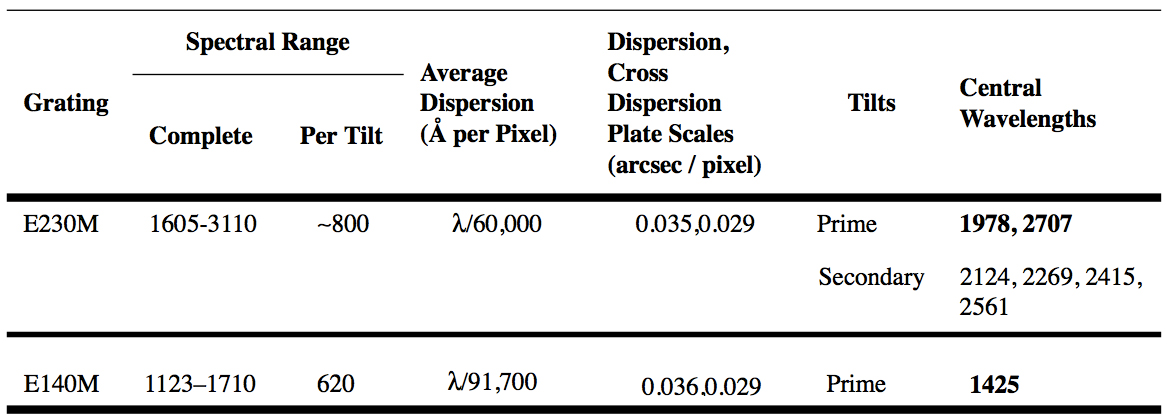}  
\caption[STIS E140M and E230M Echelle Gratings]{
STIS E140M and E230M Echelle Gratings.  Table from \citet{ihb}. 
\label{fig:echelle_grating_details}}
\end{figure}

The observations in this study are mostly obtained using the STIS echelle gratings. Echelle gratings were first suggested by \citet{harrison_1949_echelle} who envisaged that \emph{``Echelles should lend themselves to production of very powerful compact spectrographs giving single-exposure coverage of broad spectral ranges''}. They were designed for spectroscopic devices to give higher resolution and dispersion than ordinary gratings, but with a greater free spectral range than previous types of gratings. Echelles are a special class of gratings, with high groove spacings (i.e.\ a coarse groove pattern), used in high angles in high diffraction orders \citep{palmer2000diffraction}. The echelle gratings allow simultaneous observations of several wavelength orders. This maximises the spectral coverage achieved in a single exposure. An important limitation of echelles is that the orders overlap unless separated optically, for instance by a cross-dispersing element. A prism  is often used for this purpose and the combination leads to an output format well matched to CCD arrays, allowing a large quantity of spectroscopic data to be recorded simultaneously. Figure \ref{fig:echelle_grating} shows an example of an echelle grating and how it produces an image. Both the E140M and E230M gratings were used for the observations during {\sl HST} Cycle 17 (See Table \ref{tab:egand_obs}). The E230M grating is used with the NUV-MAMA and provides echelle spectra at a resolving power of 30,000 from 1570 to 3100\AA. The E140M grating is used with the FUV-MAMA and provides echelle spectra at a resolving power of 45,800 from 1144 to 1710\AA. Spectroscopic details for the two echelle gratings used in this study are shown in Figure \ref{fig:echelle_grating_details}.

\section{Observations of Target Stars}\label{sec:cycle17}

\begin{figure}[ht!]
\includegraphics[width=\textwidth]{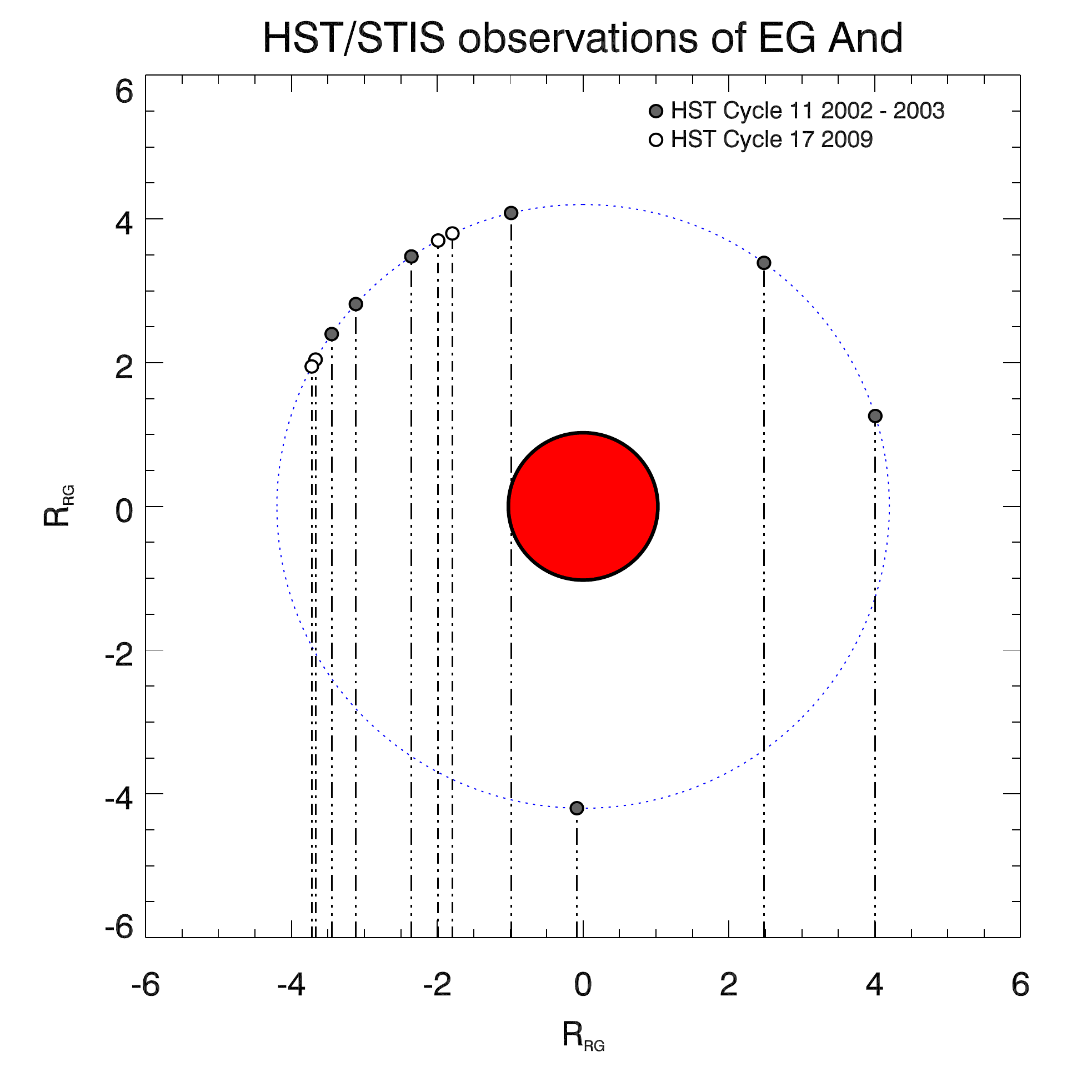}  
\caption[EG And Observations]{
STIS observations of EG And from {\sl HST} Cycle 11 and 17. The view is perpendicular to the orbital plane. The white dwarf size has been exaggerated for clarity, but all other sizes and positions are scaled in terms of red giant radii.
\label{fig:egand_view_orbit}}
\end{figure}

Observations of EG And were made during {\sl HST} Observing Cycle 17. In all, 4 separate UV observations were made of EG And (See Table \ref{tab:egand_obs} and Figure \ref{fig:egand_view_orbit}). An observation of HD148349 was also made on $2^{nd}$ September 2009. For both target stars an acquisition image was obtained using the F25ND3 mode with the CCD as detector and the mirror as the spectral element. The  F25ND3 mode provides imaging with a $25\times25$ arcsecond field of view. These acquisitions took $\sim0.1$s and were followed by a peak-up exposure (taking $\sim4$s) using the $0.2\times0.06$ arcsecond aperture slit. Acquisition exposures are discussed further in Chapter 4. The echelle exposures followed and for each observation used the prime tilt settings. Exposure times ranged from 1200 - 3200s, and used central wavelengths of 1425\AA\ and 2707\AA. These observations compliment previous observations of EG And during Observing  Cycle 11 in 2002 and 2003 (See Table \ref{tab:egand_obs} and Figure \ref{fig:egand_view_orbit}). The original proposal for both observing cycles are \citet{espey_2002_prop} and  \citet{hst_prop}.

\begin{table}[ht!]
\begin{center}
\begin{tabular}{cccc}
\hline
\textbf{Target}  & \textbf{Observation Date} & \textbf{STIS Gratings}  & \textbf{$\phi$ UV} \\
\hline
\hline
EG And &  2002 August 28 & E140M, E230M  & 0.798  \\
EG And &  2002 October 16 & E140M, E230M  & 0.899  \\
EG And &  2002 December 22 & E140M, E230M  & 0.038  \\
EG And &  2003 January 18 & E140M, E230M  & 0.095  \\
EG And &  2003 February 6 & E140M, E230M  & 0.133  \\
EG And &  2003 February 16 & E140M, E230M  & 0.153  \\
EG And &  2003 July 31 & E140M, E230M  & 0.497  \\
\hline
EG And &  2009 August 15 & E140M, E230M  & 0.070  \\ 
EG And & 2009 August 19 & E140M, E230M & 0.078  \\
EG And &  2009 October 2 & E140M & 0.169  \\
EG And & 2009 October 4& E140M, E230M & 0.173  \\ 
\hline
HD148349 & 2009 September 2& E230M &-  \\ 
\hline
\hline
\end{tabular}
\caption[HST Cycle 11 Observations]{
UV phases of observation correspond to the eclipse epoch of the system. $\phi = 0$ corresponds to UV total eclipse when the white dwarf is behind the giant, while $\phi = 1$ is UV maximum. The ephemeris used to calculate the phases was taken from \citet{fekel_part1}.
\label{tab:egand_obs}}
\end{center}
\end{table}

\section{Additional Instrumentation}\label{sec:other_instruments}

Other instruments used in this project included the High Precision Parallax Collecting Satellite, {\sl Hipparcos} \citep{turon_hipparcos_2008}. This was an ESA mission operated between 1989 and 1993. It was dedicated to high-precision astrometry but as a by-product of this principle objective it also recorded photometric data in the visible light passband. 

{\sl ELODIE} - a computer-operated fibre-fed echelle spectrograph permanently situated in a temperature-controlled environment in L'Observatoire de Haute Provence was also utilised. It covered a spectral range 3906 - 6811\AA\ with a spectral resolution of 42,000 using a 1.93 m telescope and 1024$\times$1024 CCD. The full instrument design can be seen in \citet{elodie_baranne}. The purpose of {\sl ELODIE} was to provide a fixed optical system, designed to obtain accurate radial velocity measurements.  It operated from 1993 to 2006 and resulted in a spectral library of 709 stars covering the space of atmospheric parameters: T$_{eff}$ from 370K to 13600K and log(g) from 0.03 to 5.8. The {\sl ELODIE} data are used in Chapter 6.

Photometric data were also gathered from {\sl KAIT}, the Katzman Automatic Imaging Telescope \citep{treffers_1995_bait}. These data were obtained with the assistance of Weidong Li (private communication). The {\sl KAIT} photometric data are used in Chapter 4.


\chapter{Data Processing}
\label{chapter:data_reduction}

In this chapter, the data reduction techniques for {\sl HST} Cycle 11 and 17 observations of EG Andromedae and the spectral standard HD148349 are presented. In the interest of completeness, all of the processes that are carried out on the data from the initial observation on {\sl HST}, to the  pipeline calibration, and any further data reduction techniques are described. In particular, the processes involved with splicing overlapping spectral orders together are described and a statistical analysis of their effect on the data is included. 

\section{Data Transmission}

As mentioned in Chapter 2, {\sl HST} is in contact with the ground through two high-gain
antennae using the Tracking and Data Relay Satellite System (TDRSS). This system consists of a set of satellites in geosynchronous orbit which supports many other spacecraft in addition to {\sl HST}. While {\sl HST} has the resources to allow real-time control of the spacecraft, most observations are scheduled and performed automatically. Likewise, the use of the TDRSS to send instrument commands or to retrieve data must also be scheduled. The command sequences for {\sl HST} are uplinked approximately every 8 hours while data are downloaded between ten to twenty times each day depending on the observing schedule. The TDRSS ground station at White Sands, New Mexico receives the data from the satellites before sending it to the Sensor Data Processing Facility at Goddard Space Flight Center in Greenbelt, Maryland, and then finally to STScI.

\section{Production Pipeline}\label{sec:stsci_pipe}

Once the data arrives at STScI, the production pipeline provides several functions as standard for calibration and product generation. The calibrations include flat fields, wavelength calibrations and background subtraction using the most up-to-date calibration files. The product generation involves converting the data from spacecraft packet format to spectra and images in the form of Flexible Image Transport System (FITS) files \citep{wells_1981_fits}.  FITS files are the resulting data product for almost all STIS reductions. These files generally have two-dimensional image arrays made up of science, error and data quality arrays. STScI maintains a set of tools and support software using the Image Reduction and Analysis Facility (IRAF)\footnote{IRAF is freely available to download at the IRAF homepage: \url{http://iraf.noao.edu/}}. This facility is developed by the National Optical Astronomy Observatories (NOAO) and  contains many applications, called tasks, that are used to calibrate and analyze {\sl HST} data. Complimentary tasks are contained in packages.

One such package called the Space Telescope Science Data Analysis System (STSDAS) contains the software needed to calibrate data from STIS. TABLES is a companion package that provides the tools needed for analysing tabular data. Both packages are layered on IRAF and are accessible through the user interface command language. STSDAS contains the same pipeline tools used by the Space Telescope Science Institute (STScI) to calibrate the data. It is possible to recalibrate raw data by re-running the pipeline and using different reference files and/or calibration switch settings. This approach had to be adopted with a non-standard reduction technique applied to observations using the STIS G430L grating. This process is fully discussed in Chapter 4.

\section{Pipeline Calibration Steps}

\begin{figure}[ht!]
\includegraphics[width=\textwidth]{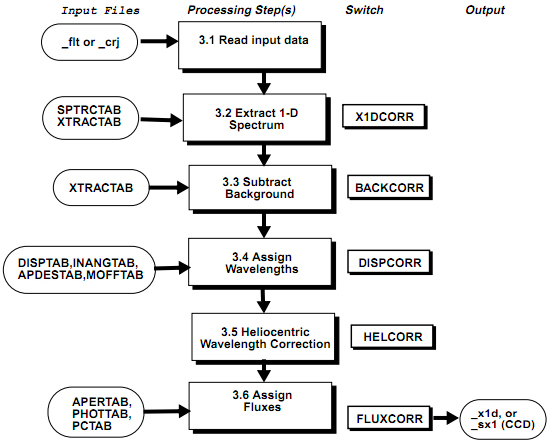}  
\caption[CALSTIS Flow Chart]{
CALSTIS flow chart showing how the extraction of a one-dimensional spectra (\_x1d.fits) from an input flat-fielded (\_flt.fits) or cosmic-ray-rejected (\_crj.fits) takes place through the application of processing steps determined by calibration switches in the file headers. \emph{Image Credit: \citet{stis_isr_calstis_1999}.}
\label{fig:calstis_flow}}
\end{figure}

The pipeline calibration steps are carried out by reading all the input calibration parameters from calibration switches and from reference files whose values and names are indicated in the primary header of the input files. The combination of these steps for STIS is referred to as CALSTIS. All the possible steps that can be carried out are covered in \citet{hodge_1998_calstis0}, \citet{hodge_stis_1998}, \citet{stis_isr_calstis_1999}, and \citet{dhb}. The data processing steps relevant to this work are discussed here and are shown in flow chart form in Figure \ref{fig:calstis_flow}. 

CALSTIS first performs a basic, two-dimensional image reduction to produce a flat-fielded image. The location of the spectrum is found using X1DCORR. This step involves locating the spectrum by performing a cross-correlation between a spectral trace and the science image. The relevant trace is read in from a reference table (SPTRCTAB). The size of the pixel boxes needed to extract the spectrum is also read from a reference table (XTRACTAB). A background subtraction is applied by setting extraction box sizes and offset distances above and below the spectrum on the detector using columns specified in XTRACTAB. A bias subtraction is performed by using a bias reference image. A mean dark current is calculated and subtracted from the image. A dispersion solution (DISPCORR) is applied to the data so that wavelengths are assigned to pixels using dispersion coefficients from the reference table DISPTAB. STIS carries a set of wavelength calibration lamps in order to calibrate the echelle wavelengths. This provides an accuracy of roughly 0.2 pixels for the echelle modes \citep{wavecal_1998_isr}. 
However by taking a distortion map of the calibration lamp into consideration and carrying out a bootstrapping technique it may be possible to improve the calibration even further \citep{ayres_2010_stis_wrinkles}. A further step is applied to transfer the wavelengths to a heliocentric reference frame (HELCORR). Finally, the counts are converted to fluxes (in erg cm$^{-2}$ s$^{-1}$\AA$^{-1}$) using reference files APERTAB, PHOTTAB and PCTAB. This conversion is carried out using a sensitivity scale determined  from observations of spectrophotometric standard stars. For the echelle gratings used in this study the standard stars were G191B2B and BD+28 4211 as discussed in \citet{bohlin_1996_stis_calib}. This absolute flux calibration propagates uncertainty from both the standard star and STIS photometric stability, and leads to an estimated echelle flux uncertainty of $\sim4\%$ \citep{bohin_1998_isr}.

\section{Splicing and Weighting Spectral Orders}\label{sec:splicing_spec_orders}

\begin{figure}[ht!]
\includegraphics[width=\textwidth]{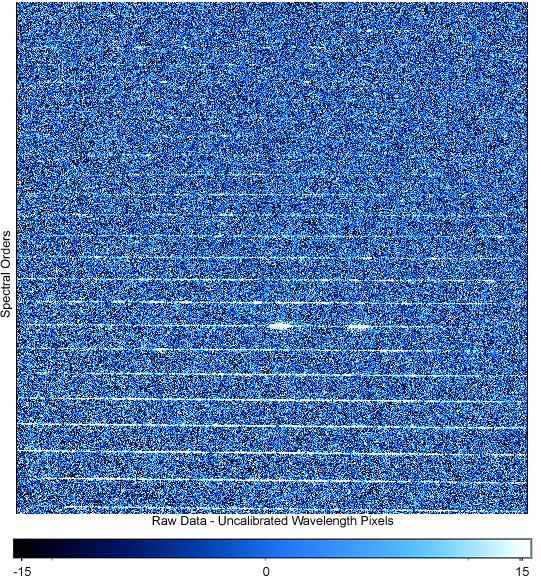}  
\caption[Echelle Raw Data File]{
Raw data file for an E230M observation of HD148349 on 2\textsuperscript{nd} September 2009. The file is displayed using SAOImage DS9, an astronomical data visualization application \citep{joye_2003_ds9}. The number of counts increases from dark to white and has been exaggerated for clarity. The horizontal axis corresponds to the wavelength direction but in this image the raw data is in units of uncalibrated pixels. The spectral orders are seen to be dispersed in the vertical direction and need to be spliced together to form a full spectrum. It can be seen that there are fewer counts visible at lower spectral orders (orders towards the top of the plot) compared to the higher spectral orders nearer the bottom of the plot. This corresponds to the lack of UV flux in HD148349 towards the blue end of the spectrum.
\label{fig:echelle_raw}}
\end{figure}

In this research it is necessary to splice the echelle orders together so that full spectra can be analysed, rather than considering solely the individual orders. Weighting these spliced orders will improve the S/N of the data. The steps outlined in the previous section are automatically applied to the {\sl HST} observations in the pipeline. However the raw, uncalibrated data files are available to download from the archive along with all of the necessary reference files and the individual calibration steps can be applied to the data to try and improve the standard data reduction. STSDAS contains all of the tasks needed to apply the processing steps to the raw data. In order to fully analyse the echelle data, the individual spectral orders (shown in Figure \ref{fig:echelle_raw}) must first be spliced together. 

\begin{figure}[ht!]
\includegraphics[width=\textwidth]{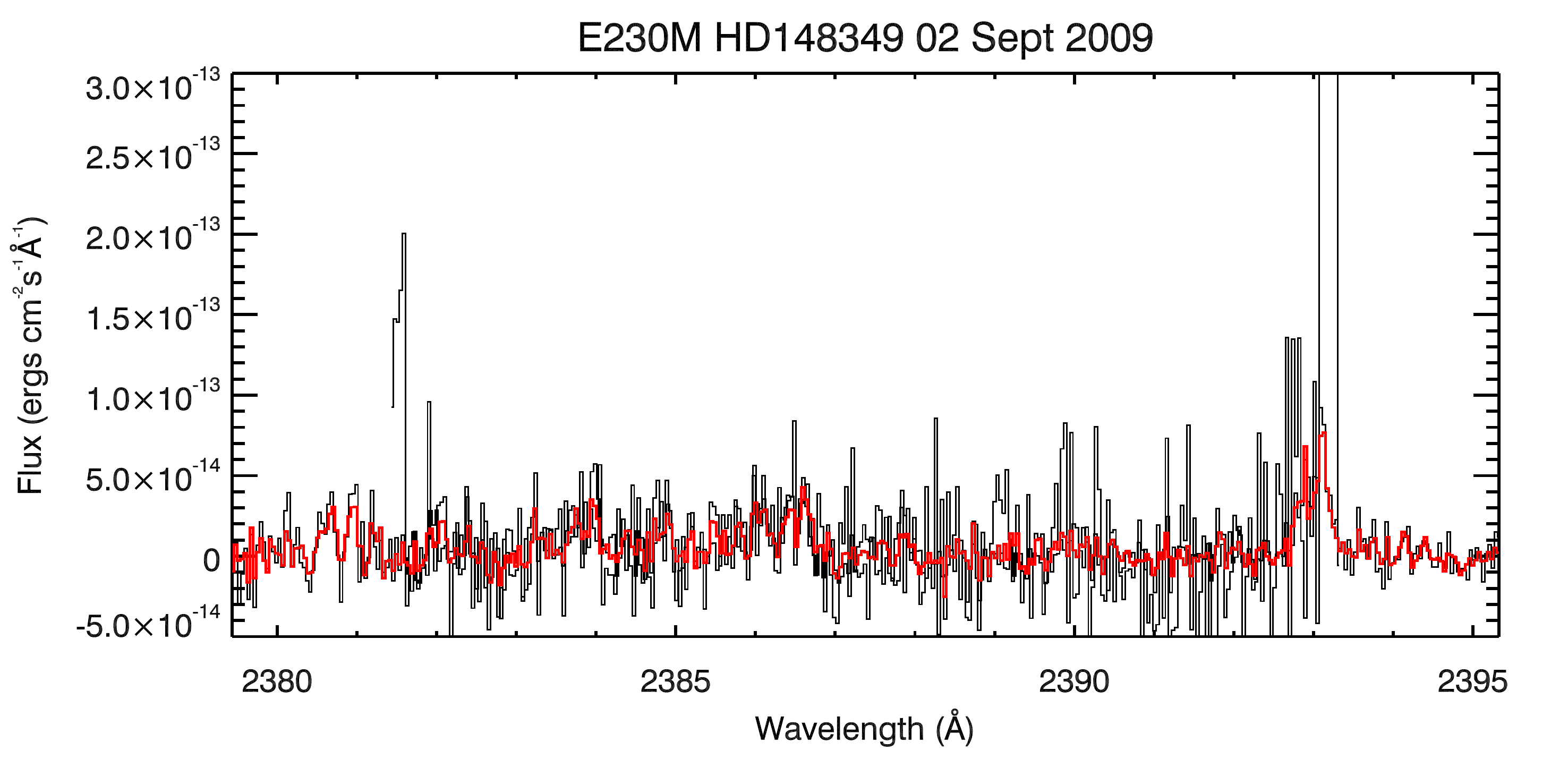}  
\caption[Echelle Data Overlap]{
A region of overlap between spectral orders for an E230M observation of HD148349. The unspliced data shown in black has regions of overlap and noisy data at the edge of each spectral order. The spliced data in red is much cleaner.  
\label{fig:overplot1}}
\end{figure}

The spectral orders can be spliced together using the \textbf{splice} task in STSDAS. Both the error and data quality arrays are taken into account when performing the splicing. This is important as it means that the noisy data at the edge of the detector can be prevented from propagating into the 1-D spectrum when the spectral orders are spliced together. This noise is most evident in the regions of spectral order overlap, as shown in Figure \ref{fig:overplot1}, which shows the difference between the unspliced and spliced data at a region of overlapping spectral orders.

\clearpage

\begin{figure}[ht!]
\includegraphics[width=\textwidth]{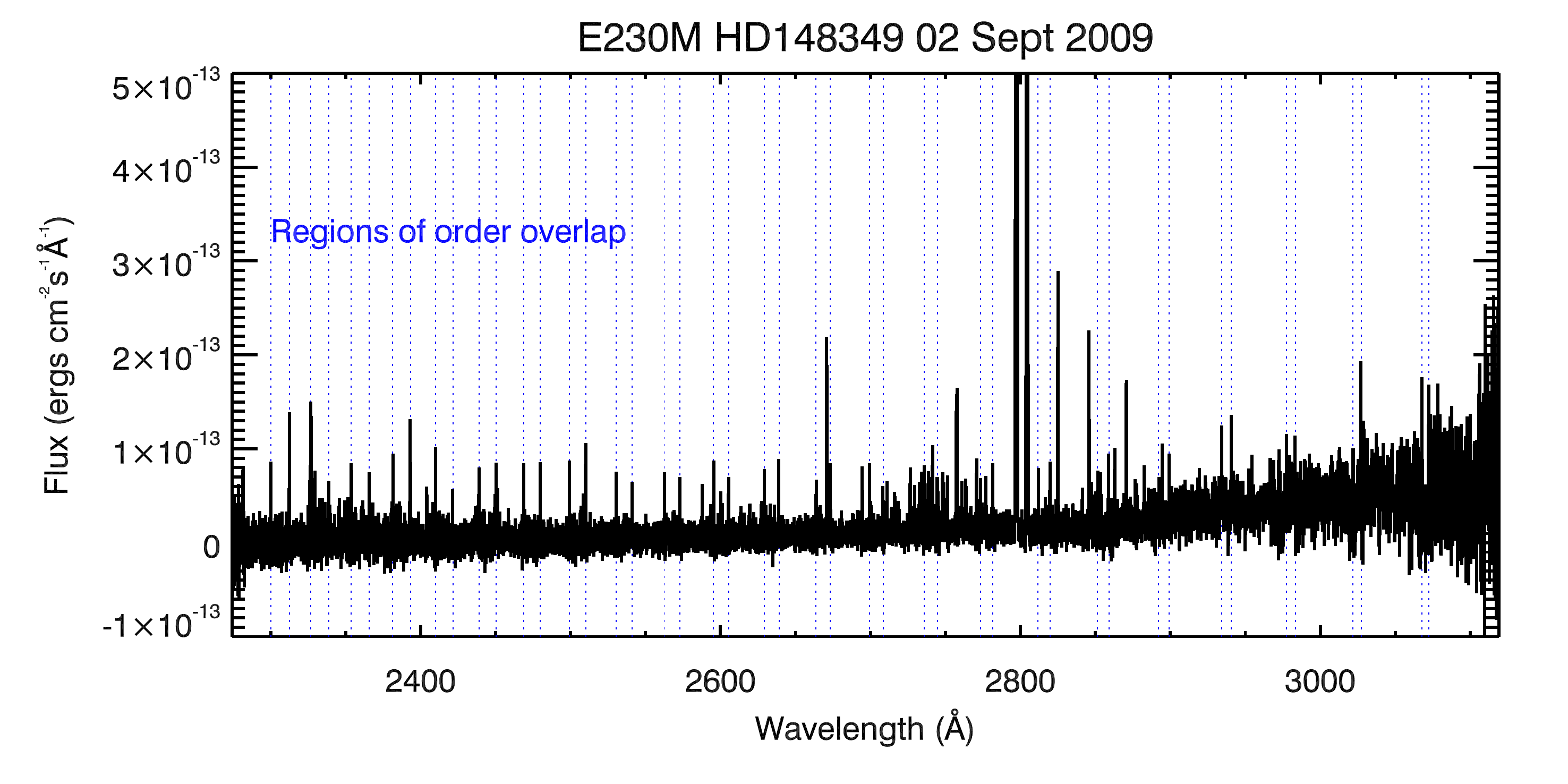}
\includegraphics[width=\textwidth]{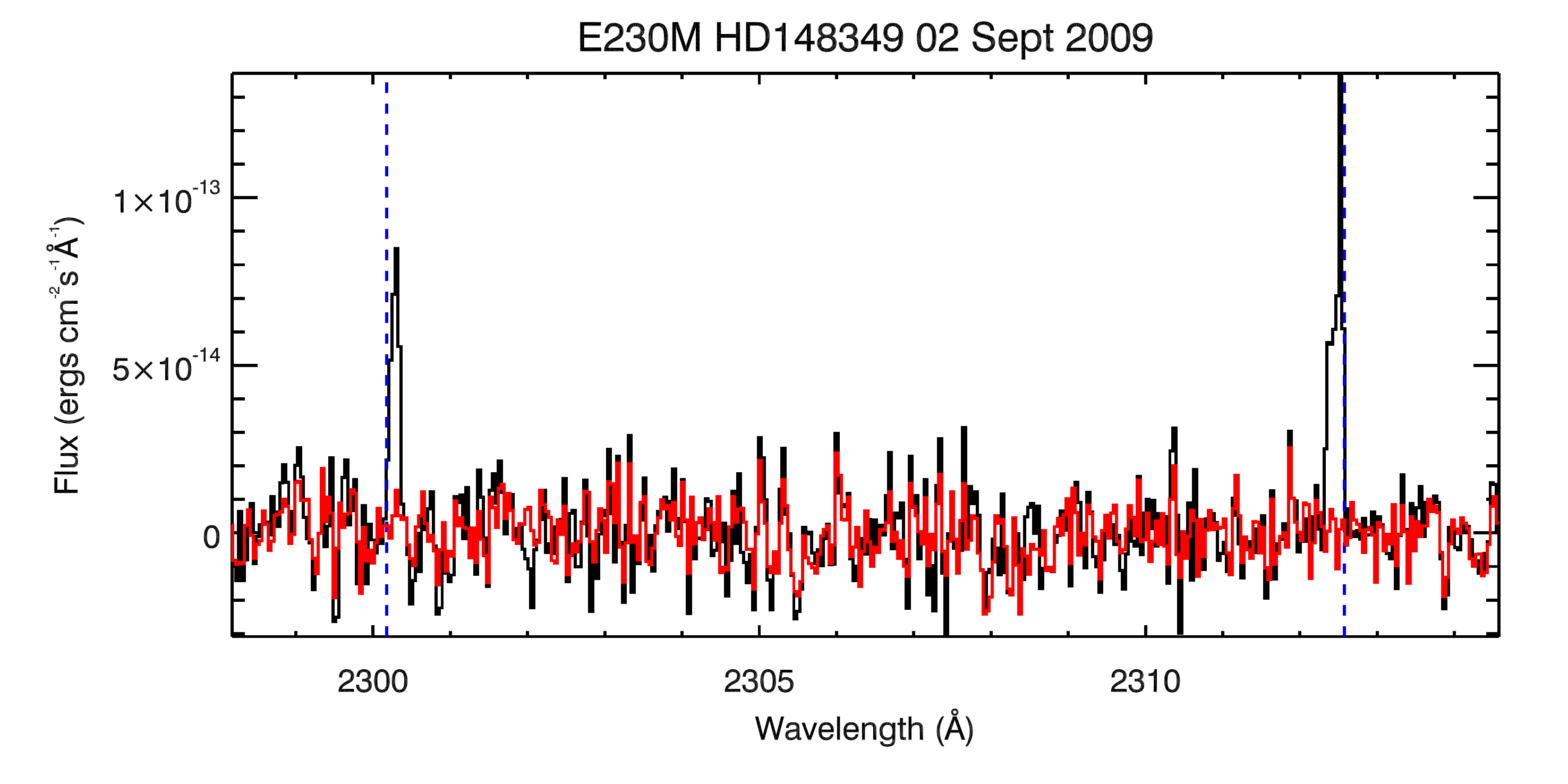}
\caption[STIS Echelle Border Pixels]{
The top plot shows a spliced spectrum of HD148349 that has not had its border pixels properly flagged. The ``emission features'' that increase in number  towards the blue-end of the spectrum line-up with the regions of overlap and are in fact noisy pixels included in the spectrum. This is made obvious by marking the end of each spectral order with a blue dotted line. The bottom plot shows a close-up of one of these overlap regions. The data with no data quality border is shown in black and clearly displays that the noisy pixels at the edge of the detector have been propagated into the spectrum. Inclusion of the data quality border results in a superior clean spectrum, shown in red. 
\label{fig:border}}
\end{figure}

\begin{figure}[ht!]
\includegraphics[width=\textwidth]{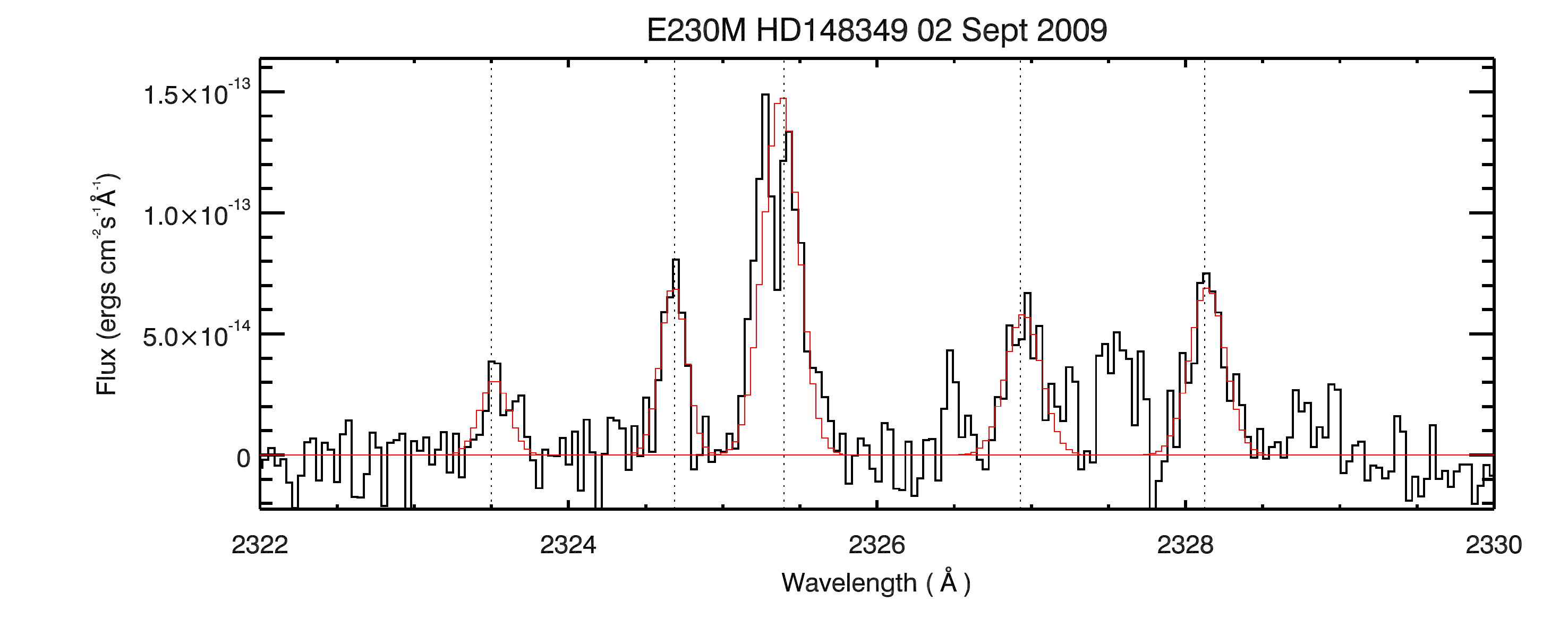}  
\caption[Echelle Splice Issue]{
The dangers of using the default \textbf{splice} settings when manually reducing data are shown in a  \ion{C}{ii}] 2325\AA\ multiplet for HD148349 during {\sl HST} Cycle 17. The central (strongest) emission line is afflicted by noisy pixels from an echelle order edge being propagated into the spliced spectrum. In this case it appears as an erroneous absorption feature that hampers the line fits shown in red.
\label{fig:c_ii_lines_mpcurvefit_nocaptions_1010}}
\end{figure}

The edges need to be masked by including a border in the data quality array of the observation. This is done by setting a border around the flat-fielded image (a $1024\times1024$ image outputted by the pipeline as an intermediary between the raw data and the extracted spectrum). By using the \emph{sdqflags} parameter, a ``bad detector pixels'' value can be assigned to  a border region around the edge of the image where large changes in throughput result in noise. After including this border it is possible to re-extract the spectrum using the \textbf{x1d} task. This extraction will ignore the pixels flagged in the data quality array. When the \textbf{splice} task is utilised manually (outside the pipeline calibration) its default setting is to ignore this flagging step. The reason for this is that the bad pixels change over time for the MAMA detectors. This default setting is one of the main reasons behind data reduction problems with STIS echelle observations. If the data are reduced manually, the \textbf{splice} task may appear to have handled the ends of the echelle orders correctly, but it may have passed some bad data points into the spliced spectrum. Figure \ref{fig:border} shows the effect of the data quality border and the problems that occur with incorrect border flagging over the whole spectrum.  Figure \ref{fig:c_ii_lines_mpcurvefit_nocaptions_1010} shows an even more subtle danger of incorrectly flagging noisy border pixels. The figure shows a manually spliced STIS E230M observation of the \ion{C}{ii}] 2325\AA\ multiplet for HD148349 during {\sl HST} Cycle 17. This feature is important for studying base chromosphere conditions (see Chapter 5). While in most parts of the spectrum the \textbf{splice} task appears to have handled the echelle orders correctly, there is a feature in the central (strongest) emission line that appears to be absorption superimposed on top of the emission line. This `feature' is a remnant of using the default \textbf{splice} settings that fail to take into account enough pixels when applying a border mask.

\clearpage

Along with manually flagging the borders before splicing the spectral orders, another advantage of  reducing the data outside of the pipeline is the option to weight the data. This step will give slightly more weight in the overlapping areas to data points that are close to continuum values. This will protect against spurious emission or absorption lines. The \textbf{fweight} task can be used to compute weights and add them as an additional column to the spectral data.  These calculated weights take signal-to-noise and dispersion into account and will improve the splicing process. To use the \textbf{fweight} task, an input table of wavelengths and smoothed fluxes should be obtained. This table can be generated by using the \textbf{continuum} task. This task fits a one dimensional function to the continuum to produce a continuum normalised spectrum. Using the smoothed flux values reduces the possibility of the weight calculations being affected by emission or absorption lines. The fitted function used for the echelle observations is a $5^{th}$-order legendre polynomial. 

\begin{figure}[ht!]
\centering
\includegraphics[width= \textwidth]{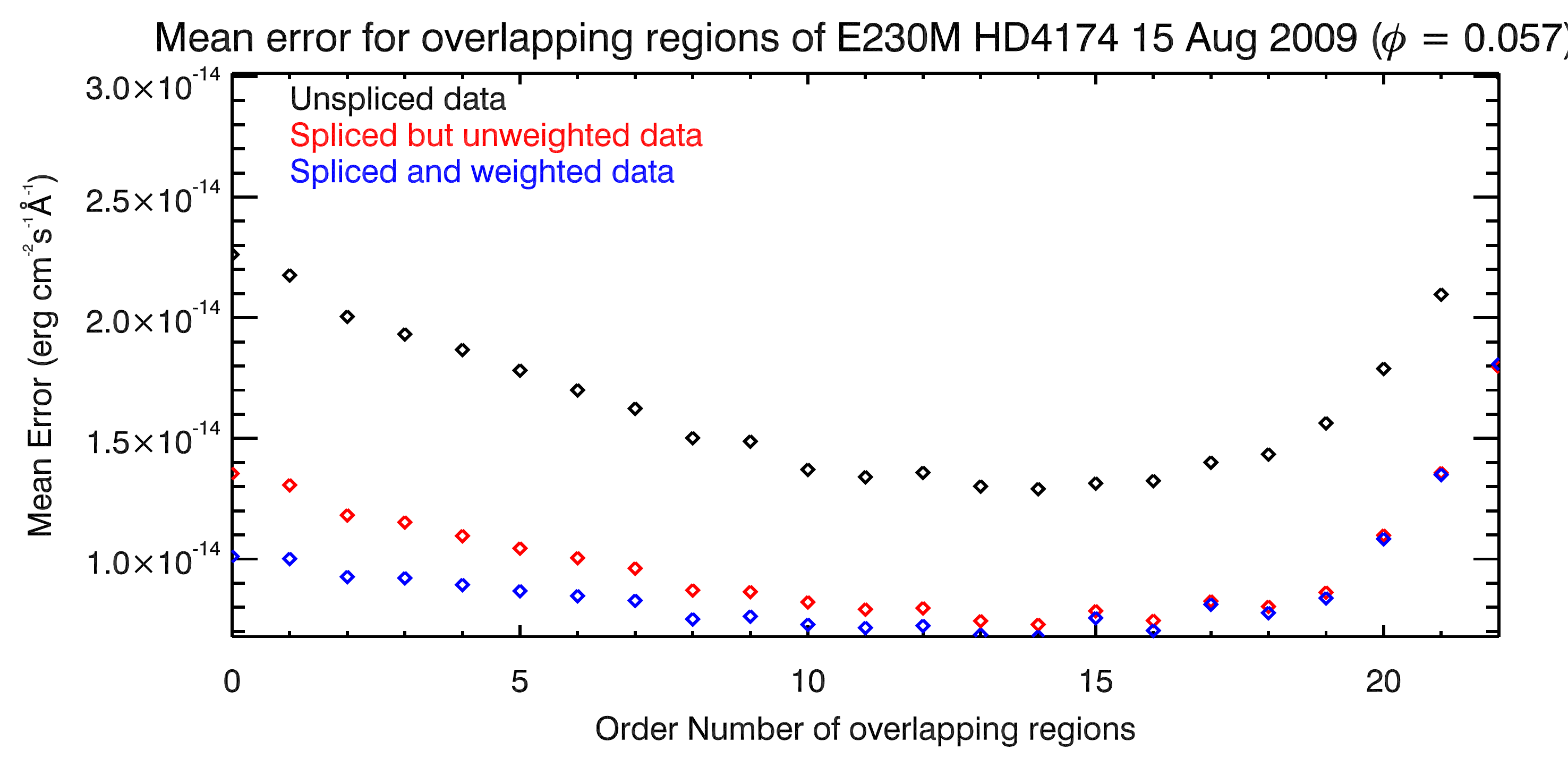}
\includegraphics[width= \textwidth]{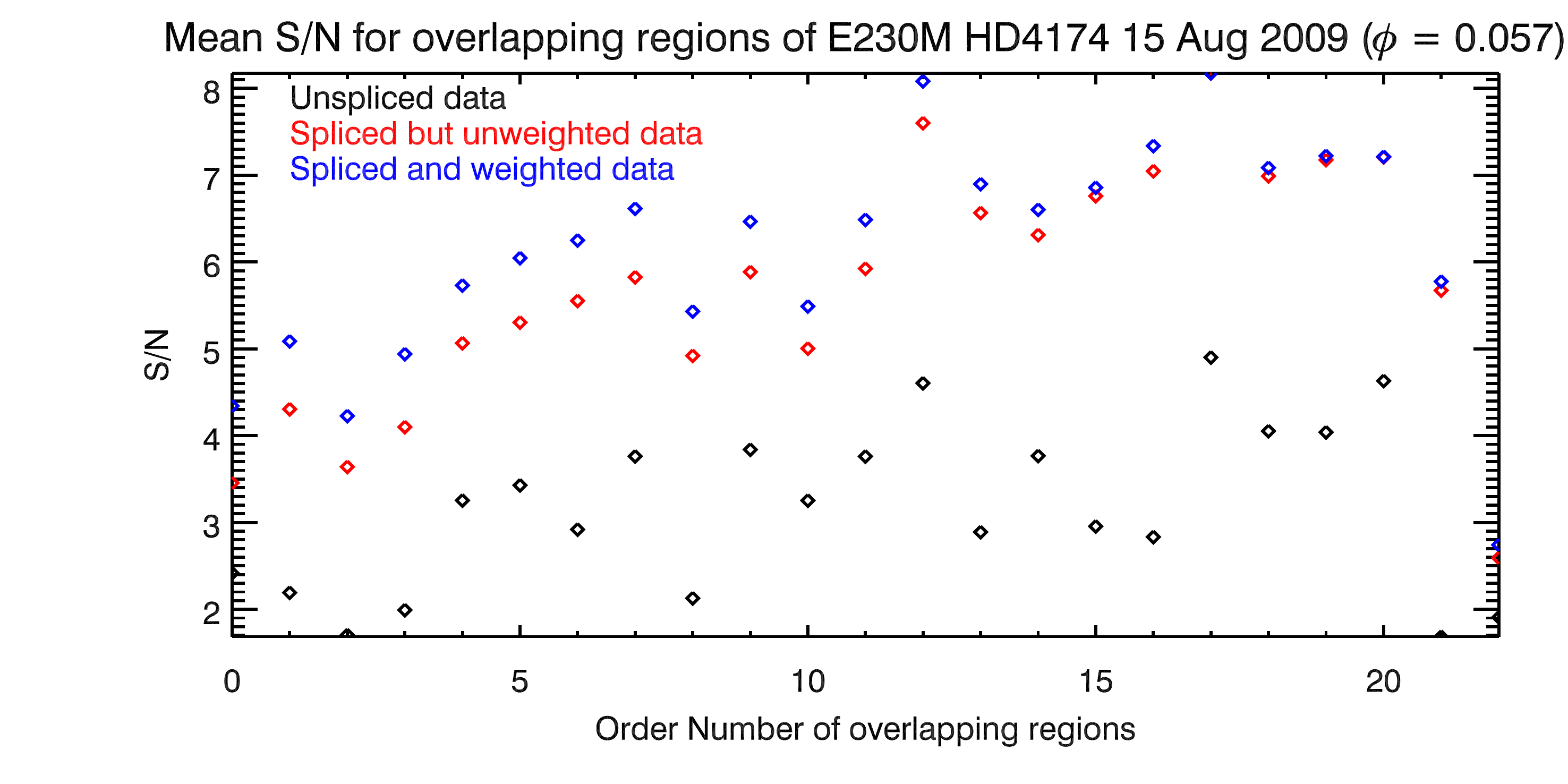}
\caption[Echelle Overlap Statistics: Error and Mean S/N]{
The top plot shows how the mean error changes due to splicing and weighting for the 23 regions of overlap between the spectral orders. The spectrum used is an observation of EG And from the 15\textsuperscript{th} August 2009. The black line corresponds to unspliced data, the red corresponds to spliced but unweighted data, while the blue is both spliced and weighted. The benefits of splicing and weighting the data can be seen in that it reduces the mean error and increases the S/N.
\label{fig:echelle_stats_2}}
\end{figure}

The effect of the data reduction can be appreciated most in the regions of data overlap between the spectral orders. For the E230M observations there are 23 regions of overlap. To demonstrate the effects of splicing and weighting the data, each overlap region of the observations was analysed using three datasets: (1) Unspliced, (2) Spliced but unweighted, (3) Spliced and weighted. At each region of overlap the mean error and the mean signal-to-noise were calculated. A representative sample can be seen in Figure \ref{fig:echelle_stats_2}. It can be seen that the mean error drops drastically due to splicing as the noisy border pixels are ignored. The ratio of signal-to-noise is substantially improved by splicing and boosted further by weighting.

\clearpage

\section{{\sl HST} Data Archive}

The science and calibration data from {\sl HST} eventually comes to be stored in the {\sl HST} data archive. The path of data from {\sl HST} to eventual storage and retrieval from the archive is fully described in \citet{swade_2001_hst_archive}. The archive is available to the public and  can be accessed through the  Multimission Archive at STScI (MAST)\footnote{\url{http://archive.stsci.edu/}}. The data used in this project were from {\sl HST}'s Observing Cycle 17 and Cycle 11. The observations from Cycle 17 can be accessed by entering the proposal ID: 11690 on the MAST web page while the proposal ID: 9487 gives access to the Cycle 11 observations. {\sl HST} archival data were used to supplement this work in Chapter 4.
\newpage
\thispagestyle{plain}
\mbox{}
\chapter{Comparing EG Andromedae and HD148349}
\label{chapter:results_rg_winds}

In this chapter, {\sl HST}/STIS observations of the symbiotic binary system EG Andromedae and an isolated spectral standard HD148349 are presented to compare and contrast a symbiotic system with an isolated red giant. A technique for obtaining contemporaneous low-resolution optical spectra from STIS acquisition images is demonstrated. These optical spectra become fundamental to the comparison of isolated and symbiotic systems. In particular, both targets are investigated from a photometric point of view. Photometric data from several sources are compiled to show the variability of EG And across several passbands, while synthetic photometry is performed on the optical spectra extracted from the STIS acquisition images. The giant's spectrum completely dominates the optical wavelength region and the analysis will establish if the giant behaves like a single star or if the white dwarf component has a significant effect on the system. The periodic variability is discussed and compared to that of the variability seen in the isolated spectral standard. An intensive examination of the effects of interstellar extinction results in reliable extinction values for both targets, along with a confirmation of their spectral type. These extinction values, coupled with an estimate of the brightness ratio between the two target stars, are utilised to produce new astrometry values including a revised estimate of the EG And distance. A thorough analysis of the \ion{Mg}{ii} \emph{h} and \emph{k} lines yields information on the wind from both stars including their terminal wind velocities, while the associated Wilson-Bappu effect provides a different estimate of the distances to the stars. Combining the main results from this chapter allows both of the target stars to be better understood in terms of where they fit into the the larger giant population and shows how the analysis of symbiotic stars can be integral to the understanding of isolated red giant stars. Some of the results in this chapter are presented in \citet{roche_2011_cs16poster}.

\section{G430L 1D Spectral Extraction}\label{sec:g430l_spec_extract}

\begin{figure}[ht!]
\centering
\includegraphics[width=\textwidth]{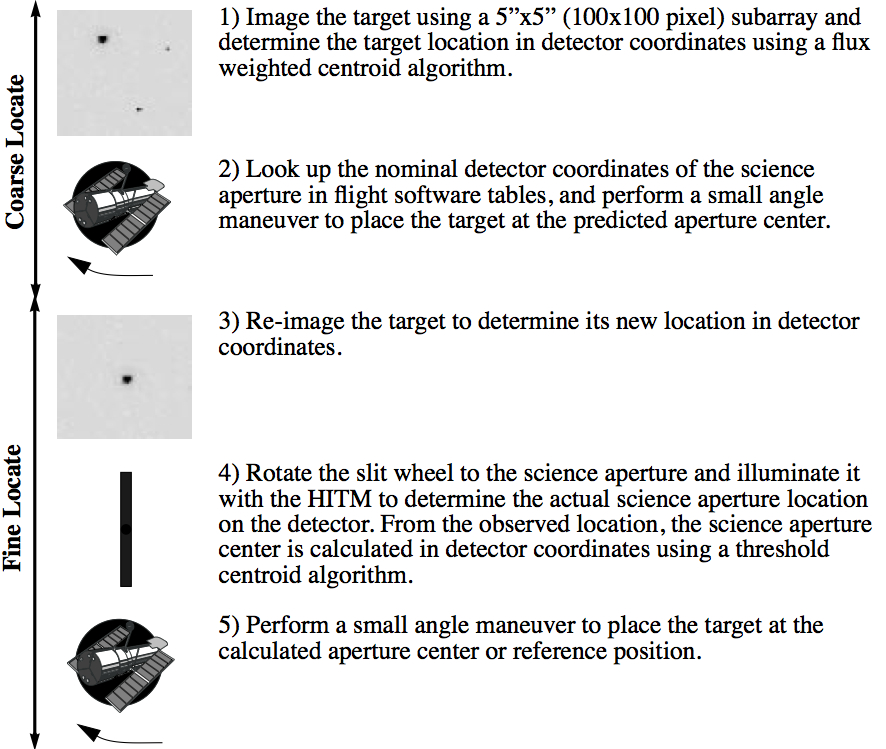}
\caption[STIS Target Acquisition]{
Performing acquisition exposures using STIS as outlined in \citet{stis_acq_isr_1996}. In this figure, HITM refers to the ``Hole-in-the-Mirror'' lamp - one of the onboard calibration lamps on STIS. The threshold centroid algorithm is a flux-weighted algorithm designed to eliminate errors due to non-uniform illumination \citep{kraemer_1997_stic_acq}.
\label{fig:acq_peakup}}
\end{figure}

Small aperture spectroscopy with STIS requires an onboard target acquisition exposure. These exposures are controlled by the {\sl HST} flight software and are necessary to centre the target in the scientific aperture \citep{ihb}. This results in each of the echelle observations having a corresponding and contemporaneous acquisition image. Acquisition exposures traditionally locate a target in the instrument field of view for subsequent science exposures. If a slit of  less than or equal to $0.1^{\prime\prime}$ wide is required for an observation then a second `peakup' acquisition is performed. These exposures are necessary for STIS to centre the target in the spectroscopic apertures more accurately. Following the guide star procurement, STIS  images the target in the acquisition aperture. It then locates the target in the field of view. Next, it determines the spacecraft move needed to correctly position the target in the science aperture. Finally, HST is moved to position the target at the calculated aperture centre. The steps of this sequence are outlined in Figure \ref{fig:acq_peakup} and are explained in detail in \citet{stis_acq_isr_1996}.

\begin{figure}[ht!]
\centering
\includegraphics[width=\textwidth]{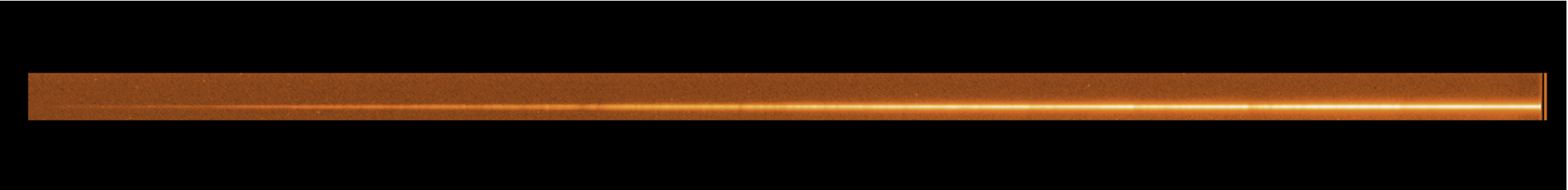}
\caption[G430L Acquisition Strip]{G430L $0.2^{\prime\prime}\times0.06^{\prime\prime}$ acquisition strip on a blank G430L wrapper image. The $1022\times32$ acquisition strip is placed in the center of the blank $1060\times128$ G430L image so that the IRAF tools can extract the 1D spectrum. In this figure the acquisition strip is for EG And taken on the $15^{th}$ August 2009.\label{fig:g430l_acq}}
\end{figure}

The acquisition exposures in this study were carried out using the G430L grating, the $0.2^{\prime\prime}\times0.06^{\prime\prime}$ aperture, with the CCD as the detector. The G430L grating has a range of $2900 - 5700$\AA, with a resolving power of $530 - 1040$. In most cases these exposures are used only for acquisition purposes and are in the form of a $1022\times32$ pixel strip. Consequently, there is no standard reduction procedure in place to extract 1D spectra from these acquisition images. The extraction technique comprises overlaying the $1022\times32$ pixel acquisition strip in the middle of a blank  wrapper file. For this study the wrapper file consists of a blank $1060\times128$ G430L image from an observation using the $52^{\prime\prime}\times0.2^{\prime\prime}$ aperture.  In Figure \ref{fig:g430l_acq} it can be seen that the dimensions of the smaller aperture result in an acquisition strip that is too compact to be processed by the standard pipeline tools. The positioning of the strip on the blank wrapper file is an important manual step, as the extraction requires the location of the likely data pixels to be input by hand. 

\begin{figure}[ht!]
\centering
\includegraphics[width=\textwidth]{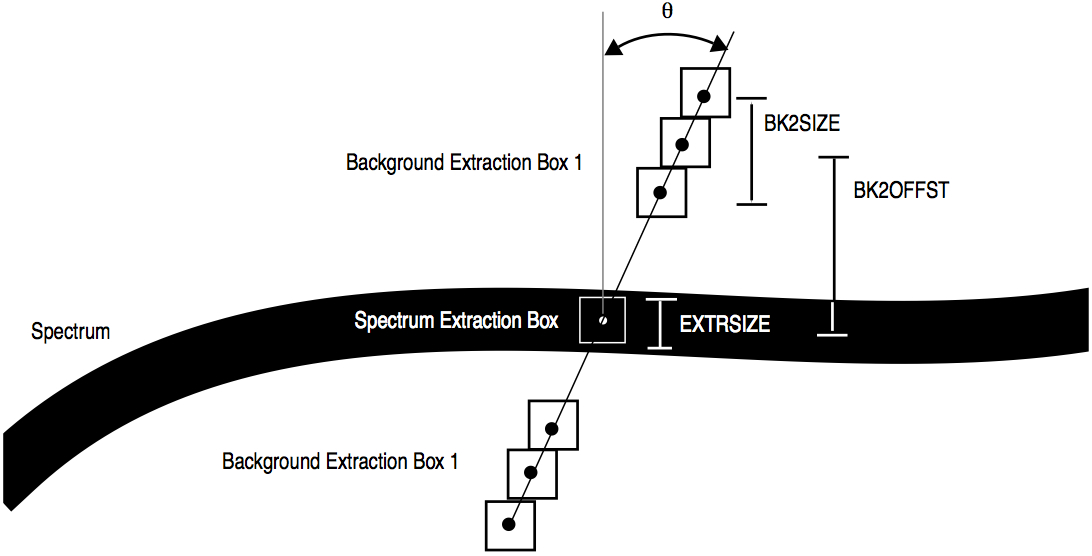}
\caption[1D Spectral Extraction]{
The parameters of the extraction tools need to be set manually before applying them to a spectrum such as that shown in the acquisition strip in Figure \ref{fig:g430l_acq}. The size of the spectrum extraction box, the size of the background extraction box  and the offsets between them (shown in this Figure from \citet{stis_isr_calstis_1999}) have to be optimised to get the best available spectrum within the confines of the narrow acquisition strip. The horizontal direction corresponds to the uncalibrated wavelength direction.
\label{fig:stis_1d_trace_ext}}
\end{figure}

To extract a spectrum, the file header keywords are first edited so that the exposure parameters matches those of the acquisition strip but the image format parameters  remain those of a standard G430L $52^{\prime\prime}\times0.2^{\prime\prime}$ image. Next, the parameter inputs for the CalSTIS \citep{stis_isr_calstis_1999} extraction tools need to be chosen. These parameters are shown in Figure \ref{fig:stis_1d_trace_ext}. The narrowness of the acquisition strip means there is a compromise between the number of pixels used to extract the spectrum and the number used to define the background. The manually determined extraction parameters for the small aperture spectra (coupled with the use of throughput tables for small apertures that are perhaps not maintained as well as tables for the more important apertures) results in the need for a single value multiplicative correction factor being applied to the extracted spectrum. There is a small region of spectral overlap ($2900 - 3110$\AA) between the G430L extracted spectra and the E230M echelle data. This overlap is used to help refine the value of the correction factor and is discussed further in Section \ref{sec:g430l_calib}.

\begin{figure}[ht!]
\centering
\includegraphics[width=\textwidth]{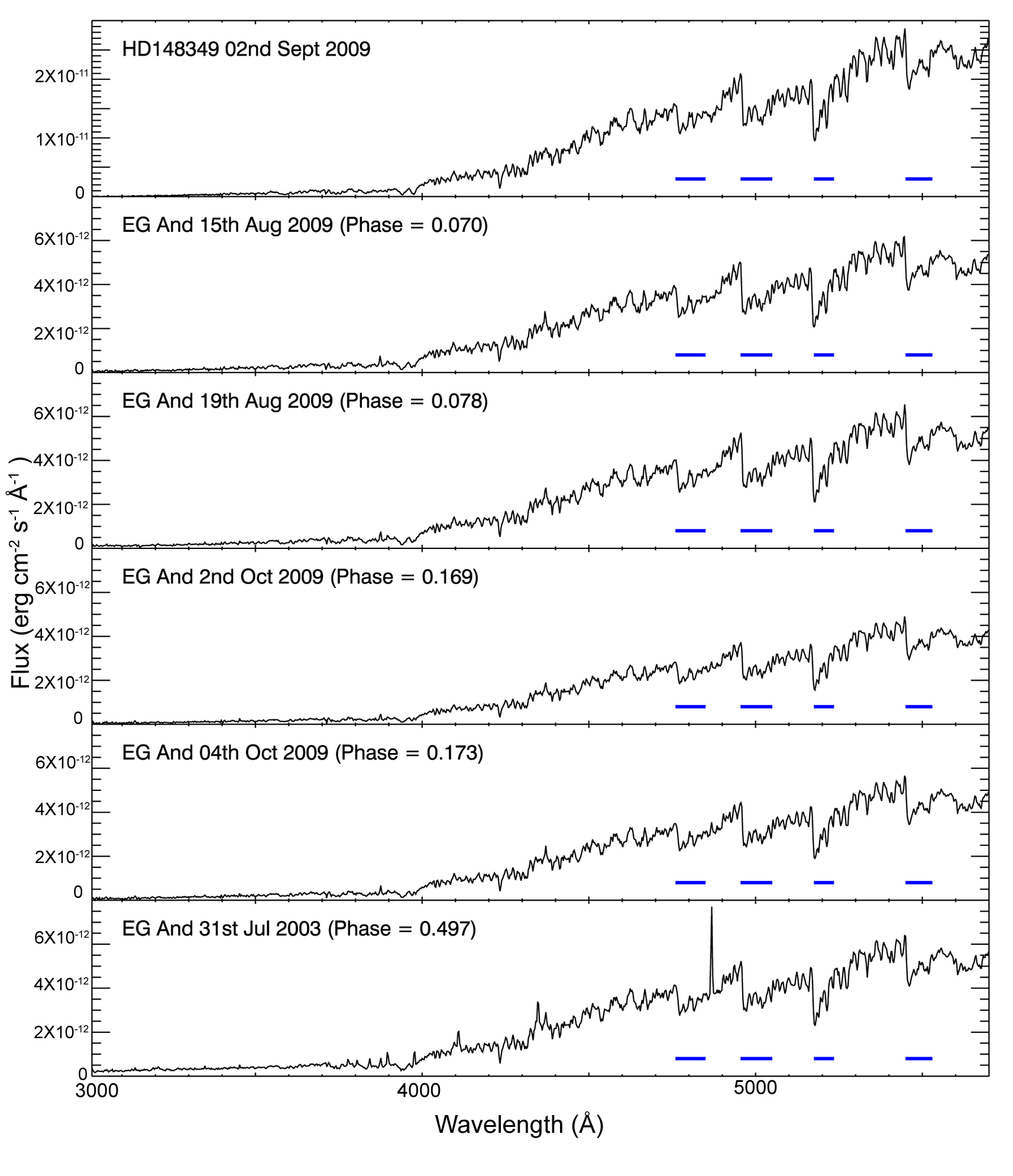}
\caption[G430L 1D Spectra Part 1]{
The G430L 1D extracted spectra for HD148349 is shown in the top panel. The flux range on the y-axis is 0 - $3\times10^{-11}$\ erg\ cm$^{-2}$\ s$^{-1}$\ \AA$^{-1}$. The next 4 panels show the extracted spectra of EG And from Observing Cycle 17. The bottom plot  shows an observation from Observing Cycle 11. The y-axis is 0 - $8\times10^{-12}$\ erg cm$^{-2}$\ s$^{-1}$\ \AA$^{-1}$ for all of the EG And panels, while the x-axis is $3000 - 5700$\AA\ in all panels. The location of the TiO bands are shown as blue dashes at the bottom of the plots \citep{plez_1998}.
\label{fig:g430l_first}}
\end{figure}

\begin{figure}[ht!]
\centering
\includegraphics[width=\textwidth]{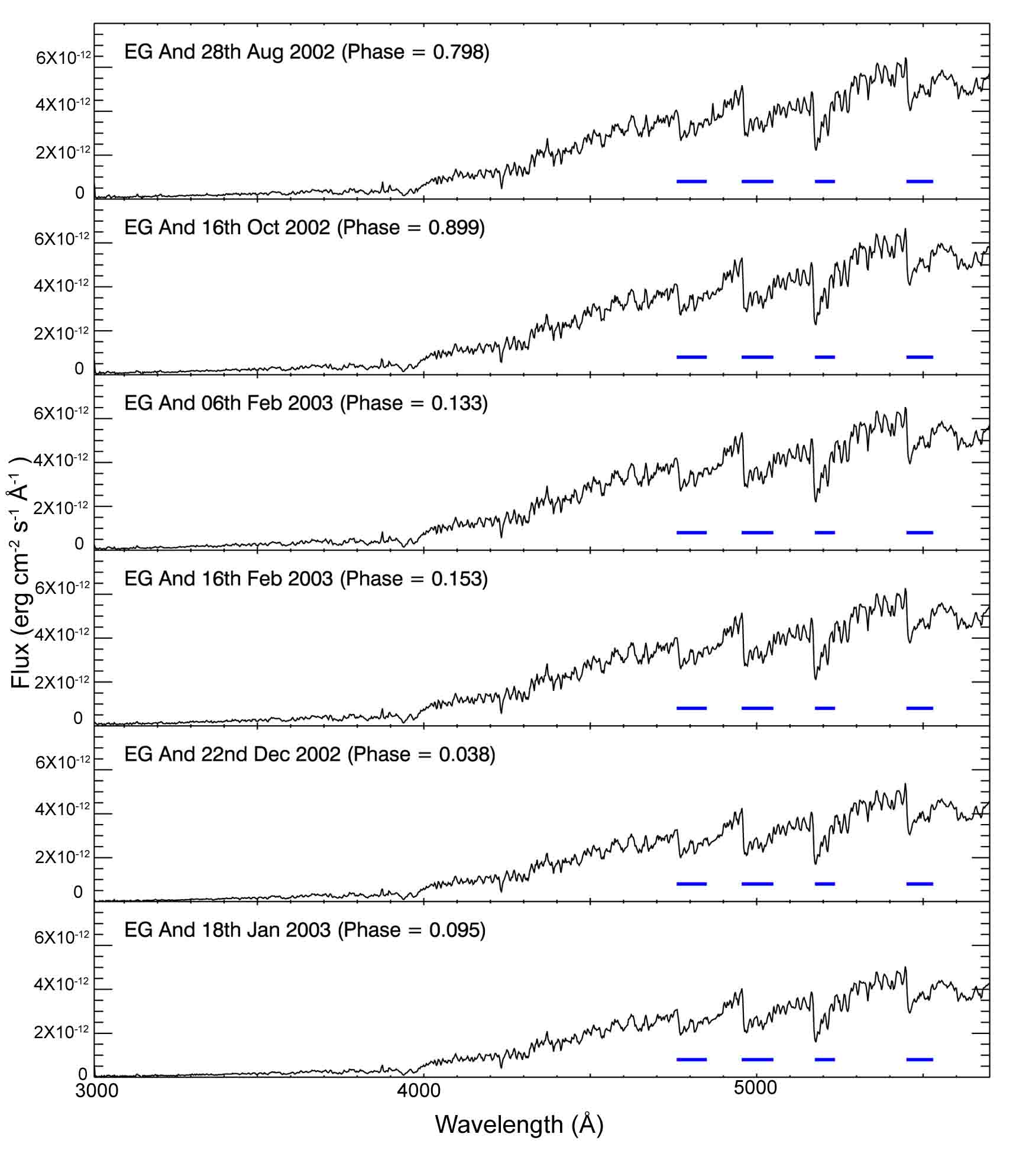}
\caption[G430L 1D Spectra Part 2]{
G430L 1D extracted spectra for observations of EG And during Observing Cycle 11. As in Figure \ref{fig:g430l_first}, the y-axis is 0 - $8\times10^{-12}$ erg cm$^{-2}$ s$^{-1}$ \AA$^{-1}$, while the x-axis is $3000 - 5700$\AA\ in all panels. The location of the TiO bands are shown as blue dashes at the bottom of the plots \citep{plez_1998}.
\label{fig:g430l_second}}
\end{figure}

By carrying out these steps it becomes possible to use the pipeline software developed for a standard $52^{\prime\prime}\times0.2^{\prime\prime}$ exposure and extract a 1D spectrum from the $0.2^{\prime\prime}\times0.06^{\prime\prime}$ acquisition strip using the reduction steps outlined in Chapter 3. This extraction technique means low-resolution optical spectra can be obtained at the same epoch as corresponding UV observations. These spectra can be used to monitor the activity of the EG And giant primary in the optical and compare it to the isolated red giant, HD148349. The extracted spectra are shown in  Figure \ref{fig:g430l_first} and Figure \ref{fig:g430l_second}. 

Figure \ref{fig:g430l_first} shows the extracted spectra of HD148349 in the top panel. The next four panels show the extracted spectra of EG And from Observing Cycle 17. The bottom plot  shows an observation of EG And from Observing Cycle 11 when the dwarf was out of eclipse. The increased intensity of nebular emission lines in this observation was possibly caused by an increase in mass accreted onto the dwarf \citep{cian_thesis}. As discussed in Chapter 1, the optical spectra of cool giants are heavily influenced by the absorption bands of molecular species. These figures are dominated by TiO and VO, the location of which are marked on the plots.

\clearpage

\section{Synthetic Photometry}\label{sec:synth_photom}

Photometry represents a method of measuring the flux of a star over large wavelength bands (passbands). The most common system is the UBV photometric system known as the Johnson-Cousins system. It is a wide band photometric system for classifying stars according to their colors. It was the first and perhaps remains the best known standardized photoelectric photometric system. In this system, UBV stands for ultraviolet, blue and visual respectively. For details of the Johnson-Cousins system see \citet{johnson_morgan_1953_photom}, \citet{johnson_harris_1954_photom} and \citet{cousins_1971_photom}.

\begin{figure}[ht!]
\centering
\includegraphics[width=\textwidth]{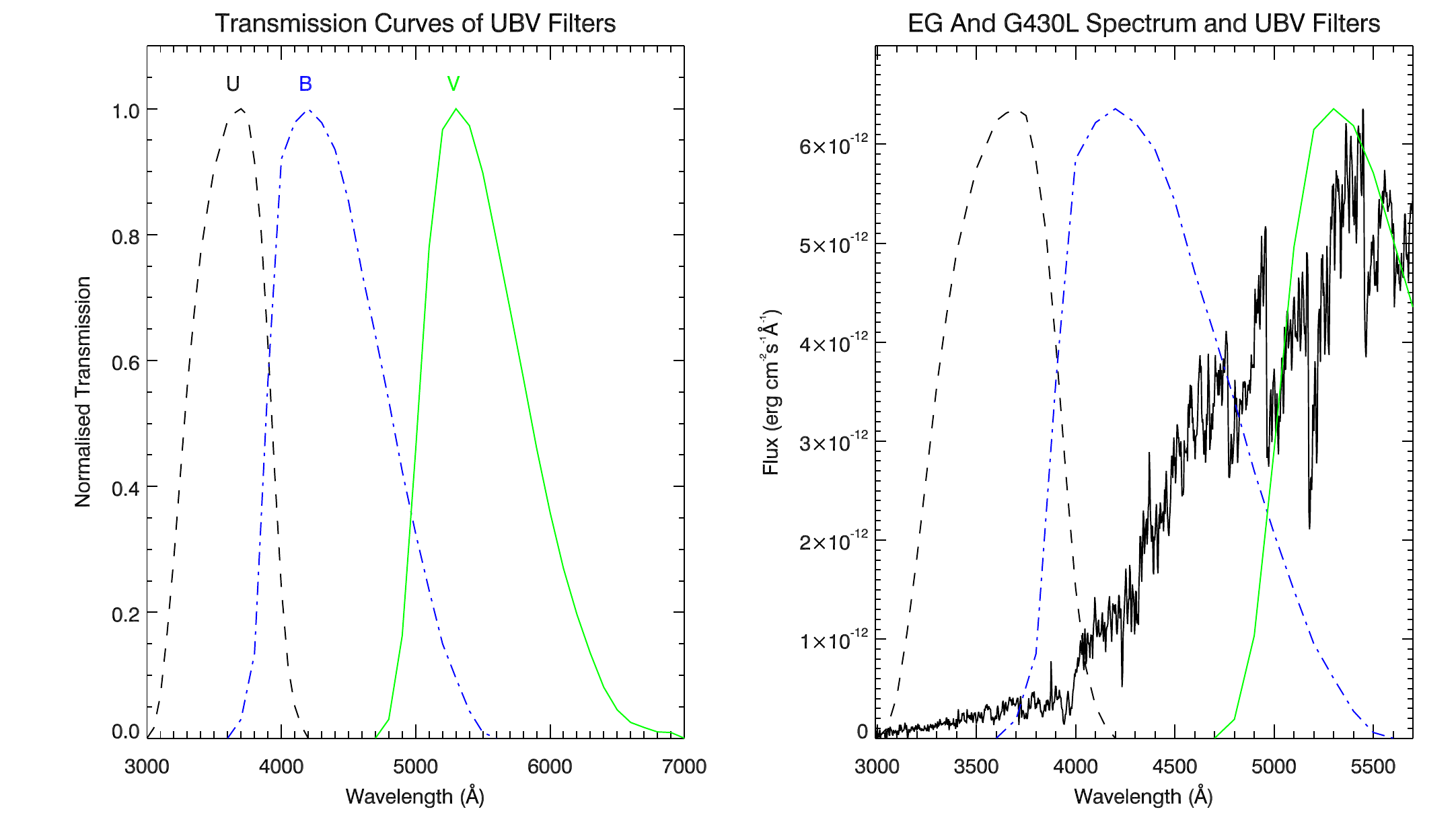}
\caption[Normalised Transmission Curves]
{\textbf{Left:} Normalised transmission curves of Johnson-Cousins UBV Passbands. The passbands span the wavelength region from 3000\AA\ to 7000\AA. This range encompasses most of the G430L wavelength range ($2900 - 5700$\AA) as well as the end of the E230M range ($1570 - 3110$\AA). The U passband, centered around 3700\AA, is shown by the dashed black curve. The B passband, centered around 4200\AA, is shown by the dashed-dot blue curve. The V passband, centered around 5300\AA, is shown by the solid green line. \textbf{Right:} An EG And G430L spectrum with the location of the UBV passbands overplotted. The EG And spectrum corresponds to a Cycle 17 Observation from 15th August 2009 ($\phi=0.070$). The maximum values of UBV transmission curves have been set to match the peak value of the G430L spectrum for clarity.
\label{fig:ubv_transmission_curves}}
\end{figure}

\begin{figure}[ht!]
\centering
\includegraphics[width=\textwidth]{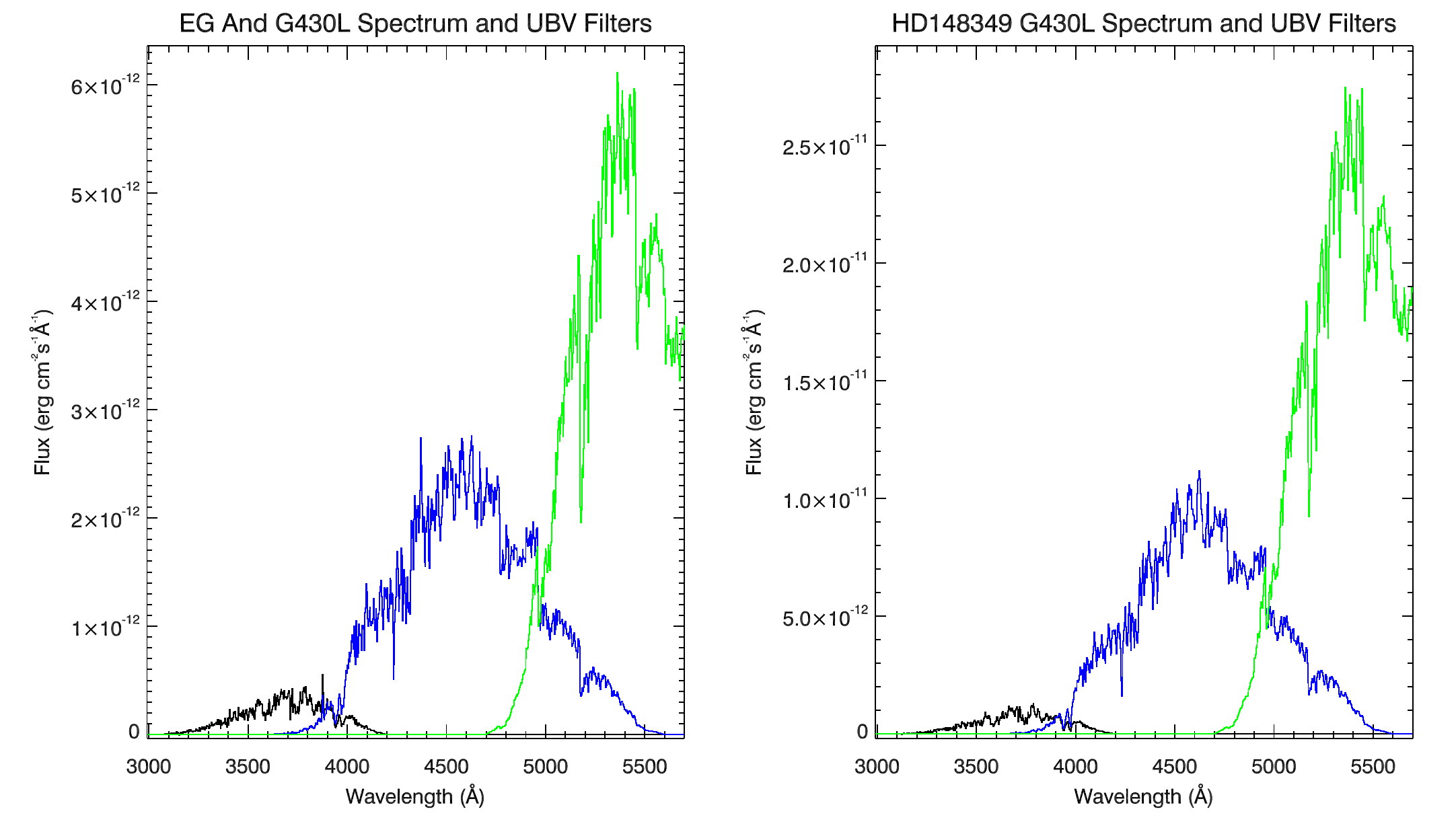}
\caption[Synthetic Photometry for EG And and HD148349]
{Synthetic Photometry for EG And and HD148349. \textbf{Left:} The UBV transmission curves from Figure \ref{fig:ubv_transmission_curves} have been mulitplied by an EG And G430L spectrum corresponding to a Cycle 17 Observation from 15th August 2009 ($\phi=0.070$). The resulting fluxes are shown in black, blue and green, corresponding to the U, B and V passbands respectively. \textbf{Right:} A G430L spectrum of HD148349 from 2nd September 2009 after it has undergone the same process.
\label{fig:eg_hd_transmission_curves}}
\end{figure}

Using the 1D spectral data extracted from the STIS G430L observations (See Section \ref{sec:g430l_spec_extract}) it is possible to perform synthetic photometry and obtain magnitudes in the Johnson-Cousins UBV passbands.  To carry out the synthetic photometry it was necessary to follow several steps. The Johnson-Cousins UBV passbands, as described in \citet{bessell_1990_ubvri_passbands}, were interpolated onto the G430L wavelength region. The normalised transmission curves are shown in Figure \ref{fig:ubv_transmission_curves}.
Each G430L spectrum was multiplied by the individual passbands, $R_{\lambda}$, to calculate the flux, $f_{x}$, that would be detected by a photometric observation with a filter for that passband:
\begin{equation}
f_x= \int{f_{\lambda}R_{\lambda}d\lambda} 
\label{eqn:photom_flux}
\end{equation}
A sample of the fluxes obtained for G430L spectra are shown in Figure \ref{fig:eg_hd_transmission_curves}.
Integrated fluxes can be converted to magnitudes using the following equation:
\begin{equation}
M_{x} = -2.5\log_{10}({f_x}) +  M_{x\mbox{\_offset}}  
\label{eqn:photom_calc}
\end{equation}
where $M_x$ is the magnitude of the passband in question, $f_x$ is the integrated flux from the observed spectrum and $M_{x\mbox{\_offset}}$ is the calibration correction for the system being used.

\begin{figure}[ht!]
\centering
\includegraphics[width=\textwidth]{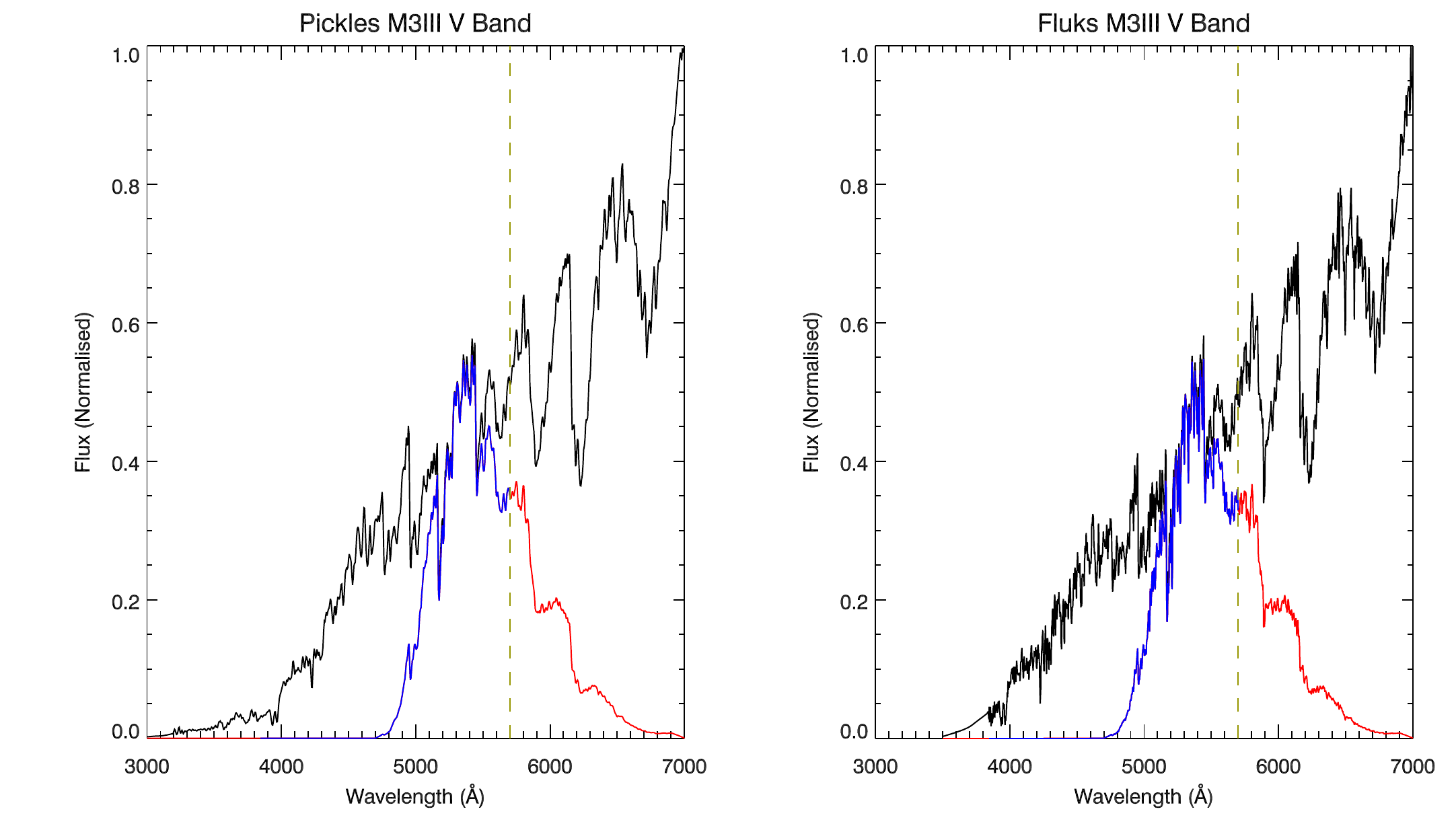}
\caption[Pickles and Fluks V Band Comparison]
{M3III spectra averaged from spectral standards from \citet{pickles_1998_stellar_library} (left) and  \citet{fluks_1994} (right). In both cases the amount of V band flux that is seen in the G430L region is 65\% of the total V band flux. In both plots the G430L V band flux is shown in blue while the red flux is the 35\% that is disregarded. The dashed line shows the G430L cut-off point. 
\label{fig:fluks_pickles_vband}}
\end{figure}

It can be seen in Figure \ref{fig:eg_hd_transmission_curves} that the redward side of the V passband exceeds the end of the G430L spectrum. The G430L data ends at 5700\AA, while the V band extends to 7000\AA. To estimate the flux that is being disregarded by the V band exceeding the G430L wavelength range a number of flux-calibrated averaged spectral standards were analysed. All M giant sub-spectral types were investigated using \citet{pickles_1998_stellar_library} and \citet{fluks_1994} spectra (Section \ref{sec:extinction} discusses these spectral libraries). Across the M0 to M6 sub-spectral types it was found that the G430L region of the V band encompassed between 64\% and 68\% of the total V band flux. Both targets stars are considered M3 giants. On this basis each of the G430L V band fluxes were taken to be 65\% of the total V band flux and increased to the full expected amount accordingly to account for the V band flux falling between 5700\AA\ and 7000\AA. Figure \ref{fig:fluks_pickles_vband} shows both Pickles and Fluks spectra and the V band flux with the G430L limit marked.

\begin{figure}[ht!]
\centering
\includegraphics[width=\textwidth]{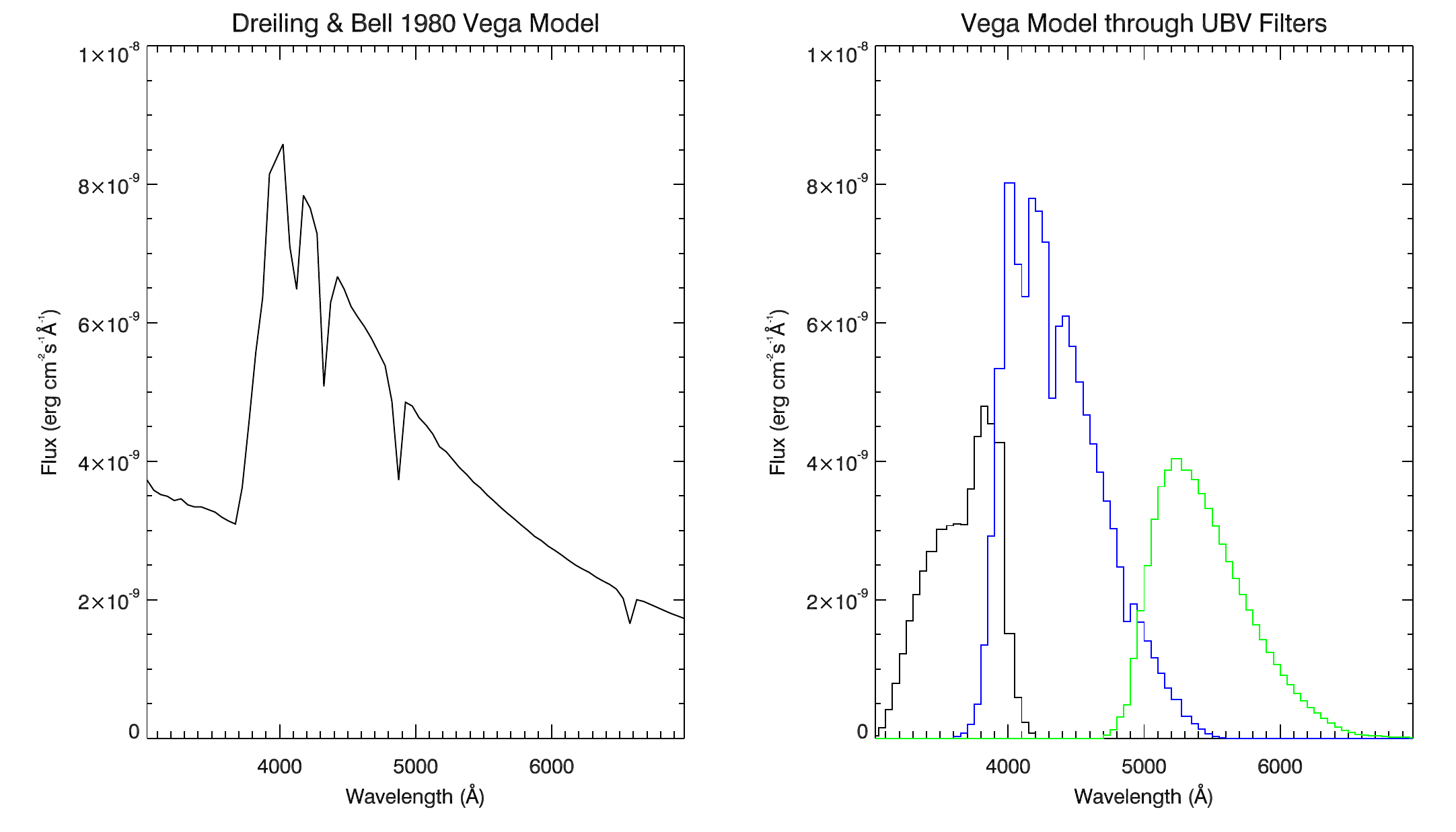}
\caption[Dreiling \& Bell Vega Model and UBV Filters]
{\textbf{Left:} Vega model from \citet{dreiling_bell_1980_vega} shown over the $3000 - 7000$ \AA\ wavelength region. The flux values have been converted to erg s$^{-1}$ cm$^{-2}$ \AA$^{-1}$ and from emergent fluxes to absolute fluxes by dividing by the geometric dilution factor, $(d/R)^2 = 1.62\times10^{16}$. \textbf{Right:} The resulting fluxes after the Vega model has been multiplied by the U, B and V transmission curves from Figure \ref{fig:ubv_transmission_curves}. 
\label{fig:dreiling_bell_vega_model}}
\end{figure}

\begin{figure}[ht!]
\centering
\includegraphics[width=\textwidth]{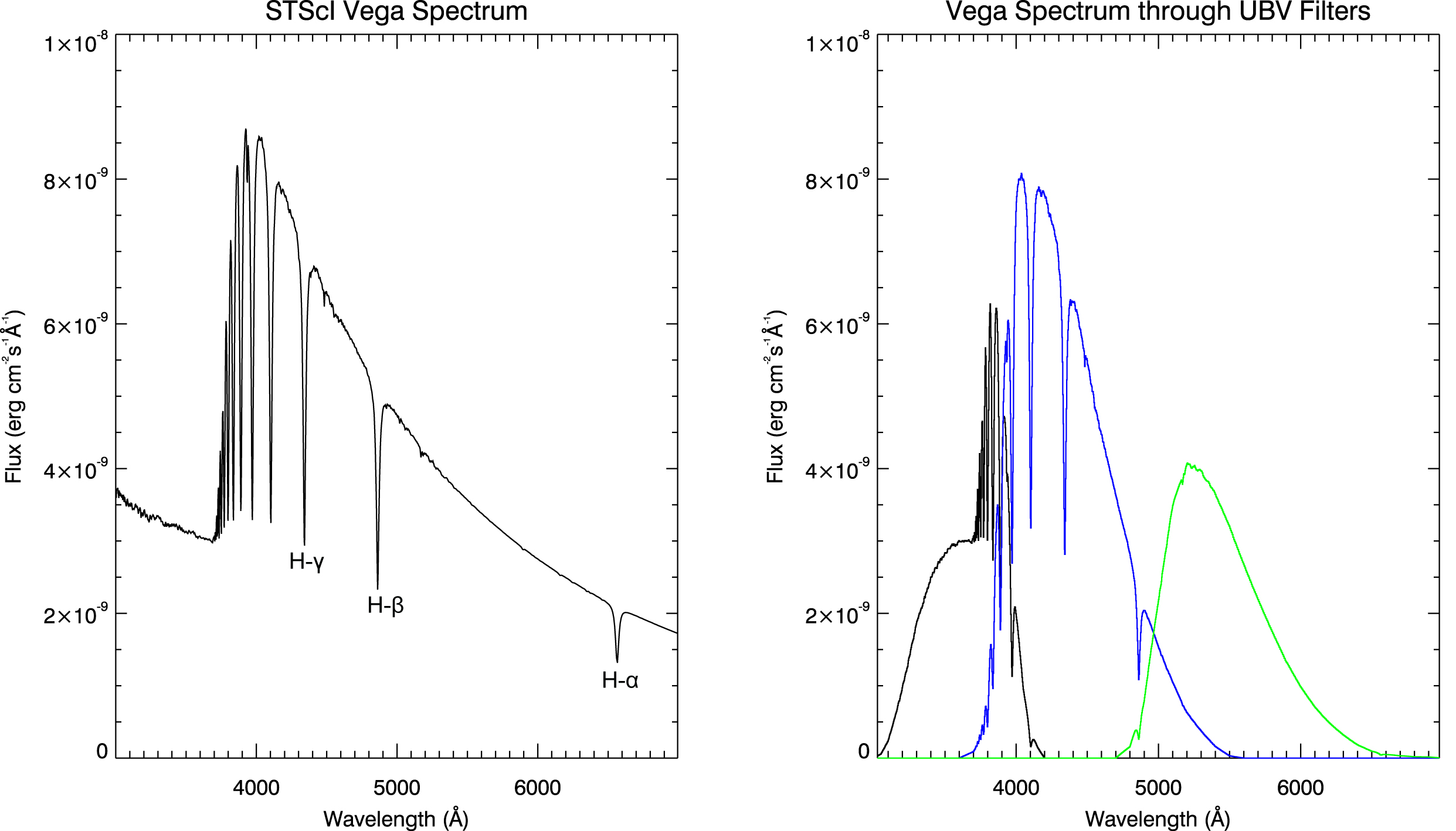}
\caption[STScI Vega Spectrum and UBV Filters]
{\textbf{Left:} STScI calibration spectrum of Vega. The spectrum is a composite of {\sl IUE} data in the FUV, STIS data in the UV to optical with Kurucz models making up the fluxes longward of 5337\AA. \textbf{Right:} The resulting fluxes after the STScI spectrum has been multiplied by the U, B and V transmission curves from Figure \ref{fig:ubv_transmission_curves}. 
\label{fig:stsci_vega_model}}
\end{figure}

A reference spectrum was needed to calculate the U, B, and V magnitudes of both EG And and HD148349. The reference star was chosen as Vega (HD172167, $\alpha$ Lyrae) as it is the primary standard for photometry \citep{gillett_1971_vega}.  \citet{bessell_1990_ubvri_passbands} advises using the model spectrum from \citet{dreiling_bell_1980_vega}. The Dreiling \& Bell spectrum is shown converted to absolute fluxes in Figure \ref{fig:dreiling_bell_vega_model}. Also shown in the figure are the fluxes for each of the U, B and V filters.
As well as the Dreiling \& Bell reference spectrum, a composite spectrum from the STScI calibration database system was also utilised to make sure the technique was tested against two reference spectra. The STScI Vega spectrum combines {\sl IUE} and STIS data along with Kurucz models and is one of the accepted fundamental flux standards for {\sl HST} calibrations. Figure \ref{fig:stsci_vega_model} shows the spectrum and the fluxes after passing through the U, B and V transmission curves. Comparing the STScI spectrum to the Dreiling \& Bell spectra it was found that the fluxes in the B and V bands both differed by less than 1\%. The U band was found to differ by less than 3\%. 

The zero-point magnitudes for the U, B and V filters were taken to be 0.790, -0.104 and 0.008 respectively as these were used in \citet{bessell_1990_ubvri_passbands}. To calibrate the synthetic photometry with the accepted Vega model, the offsets needed for each passband were calculated using:
\begin{equation}
M_{x\mbox{\_offset}} = M_{x\mbox{\_vega}} + 2.5\log_{10}({f_0})
\label{eqn:photom_offset}
\end{equation}
where $x$ is the passband, $M_{x\mbox{\_vega}}$ is the accepted magnitude of Vega in that band and $f_0$ is the integrated flux from the Vega Model spectrum in the $x$ passband. The accepted magnitudes for Vega are 0.02, 0.03 and 0.03 for U, B and V respectively \citep{monet_2003_vega_mags}. The calibration parameters used can be found in Table \ref{tab:ubv_calib}.

\begin{table}[ht!]
\caption[UBV Calibration Parameters]
{To calculate U, B and V magnitudes for STIS G430L observations, the following calibration parameters were used. The offset was calculated for the model from \citet{dreiling_bell_1980_vega} (DB) and also for the STScI Calibration (SC).}
\label{tab:ubv_calib}
\centering
\begin{tabular}{cccc}
\hline\hline
Passband & $M_{\mbox{vega}}$ & $M_{\mbox{offset (DB)}}$ & $M_{\mbox{offset (SC)}}$\\
\hline
     U & 0.02 &  -13.91 &  -13.96 \\
     B & 0.03 &  -13.01 &  -13.02 \\
     V & 0.03 &  -13.70  &  -13.69  \\
\hline
\end{tabular}
\end{table}

\section{Extinction}\label{sec:extinction}

Interstellar extinction is the absorption and scattering of light by interstellar dust grains \citep{fitzpatrick_1999_extinct}. The relationship between interstellar extinction and wavelength is best described by \citet{cardelli_1989}. To continue the comparison of EG And to isolated giants, it was necessary to consider the interstellar reddening suffered by the two target stars of this study. To obtain values of E(B-V), STIS G430L spectra were compared to a range of intrinsic M giant spectra from \citet{fluks_1994} and \citet{pickles_1998_stellar_library}. Two-parameter grid-searches were carried out  to determine which spectral type best matched EG And and HD148349 and the  corresponding E(B-V) value for both stars.

\citet{fluks_1994} presents intrinsic\footnote{The \citet{fluks_1994} Library distinguishes between the spectra averaged from real data as ``intrinsic'', while the modelled spectra are ``synthetic''.} spectra obtained from a study of 97 bright M-giants in the Solar neighbourhood in the spectral range $3800 - 9000$\AA.  The intrinsic spectra span all of the M-spectral subclasses of the MK classification systems and were obtained by averaging the extinction-corrected spectra for each subclass. \citet{pickles_1998_stellar_library} presents a stellar spectral flux library consisting of 131 flux-calibrated spectra, encompassing spectral coverage from $1150 - 10620$\AA. Each spectrum was produced by combining data from several sources overlapping in wavelength coverage. To compare the both the Fluks and Pickles spectra to the G430L observations the data had to be smoothed and rebinned\footnote{The Fluks and Pickles spectral libraries are both available electronically from the Strasbourg Astronomical Data Centre (CDS).}. The data was smoothed using the IDL \textbf{gaussfold}\footnote{From J\"{o}rn Wilms' (Institut f\"{u}r Astronomie und Astrophysik, T\"{u}bingen) AITLIB library at \url{http://astro.uni-tuebingen.de/software/idl/aitlib}.} procedure to smooth the data by convolving with a Gaussian profile. \textbf{rebin\_spectrum}\footnote{From David Schlegel's (Princeton) IDLUTILS library at: \url{http://spectro.princeton.edu/idlutils_doc.html}.} was then used to resample the smoothed data onto the same wavelength grid as the G430L observations.

\begin{figure}[ht!]
\centering
\includegraphics[width=\textwidth]{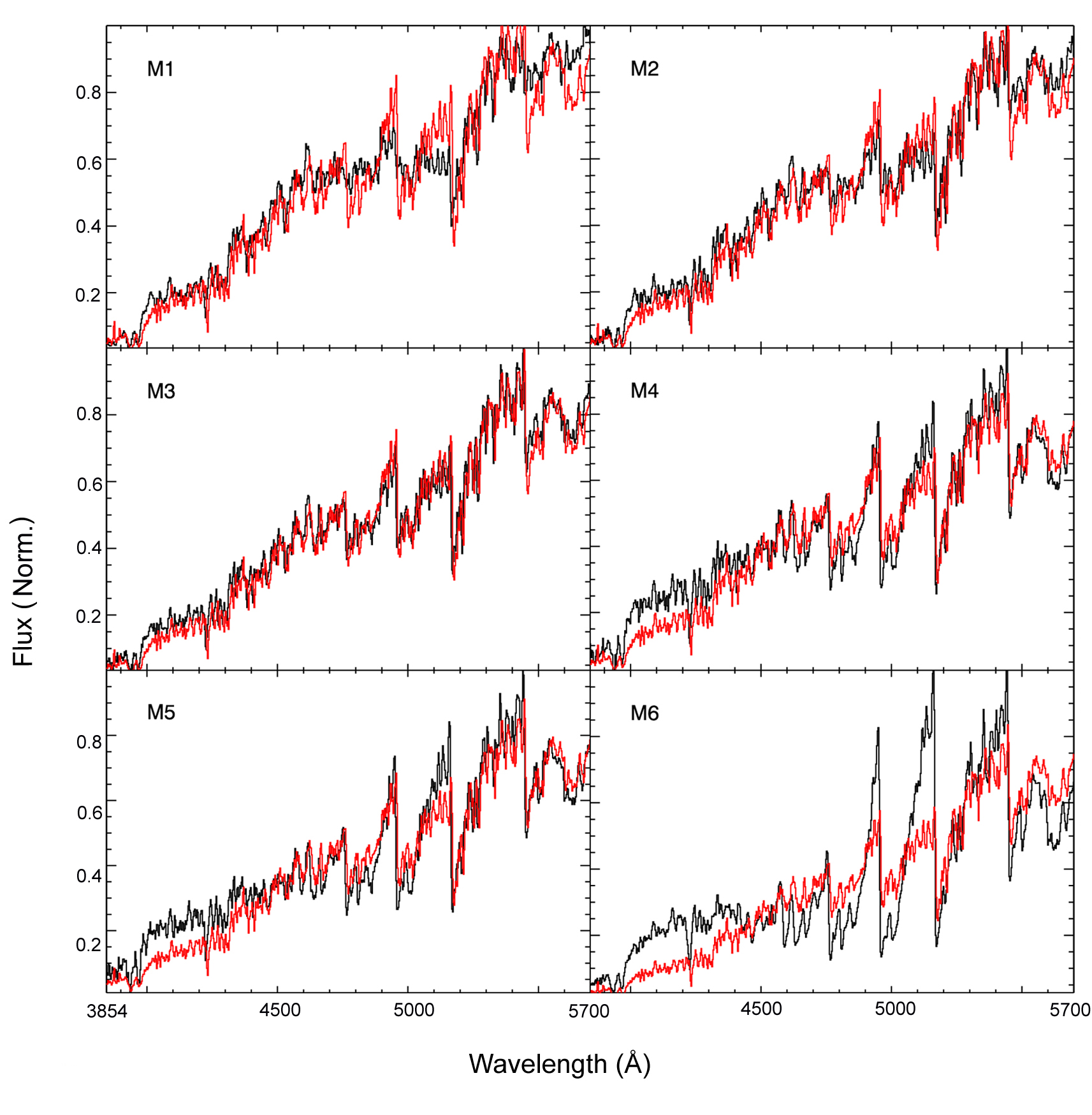}
\caption[Fluks M Giant Comparison to EG And]{
A comparison between EG And and intrinsic M giant spectral subclasses in the \citet{fluks_1994} Library. Fluks M giant spectra are plotted in black while EG And is overplotted in red. In each case the x-axis is the wavelength range $3854 - 5700$\AA\ and the y-axis is normalised flux. The EG And spectra have been scaled and extinction-corrected to produce the best-fits for each comparison. The overall best-fitting spectral-type was the M3 spectrum.
\label{fig:eg_series_extinct_fluks}}
\end{figure}

\begin{figure}[ht!]
\centering
\includegraphics[width=\textwidth]{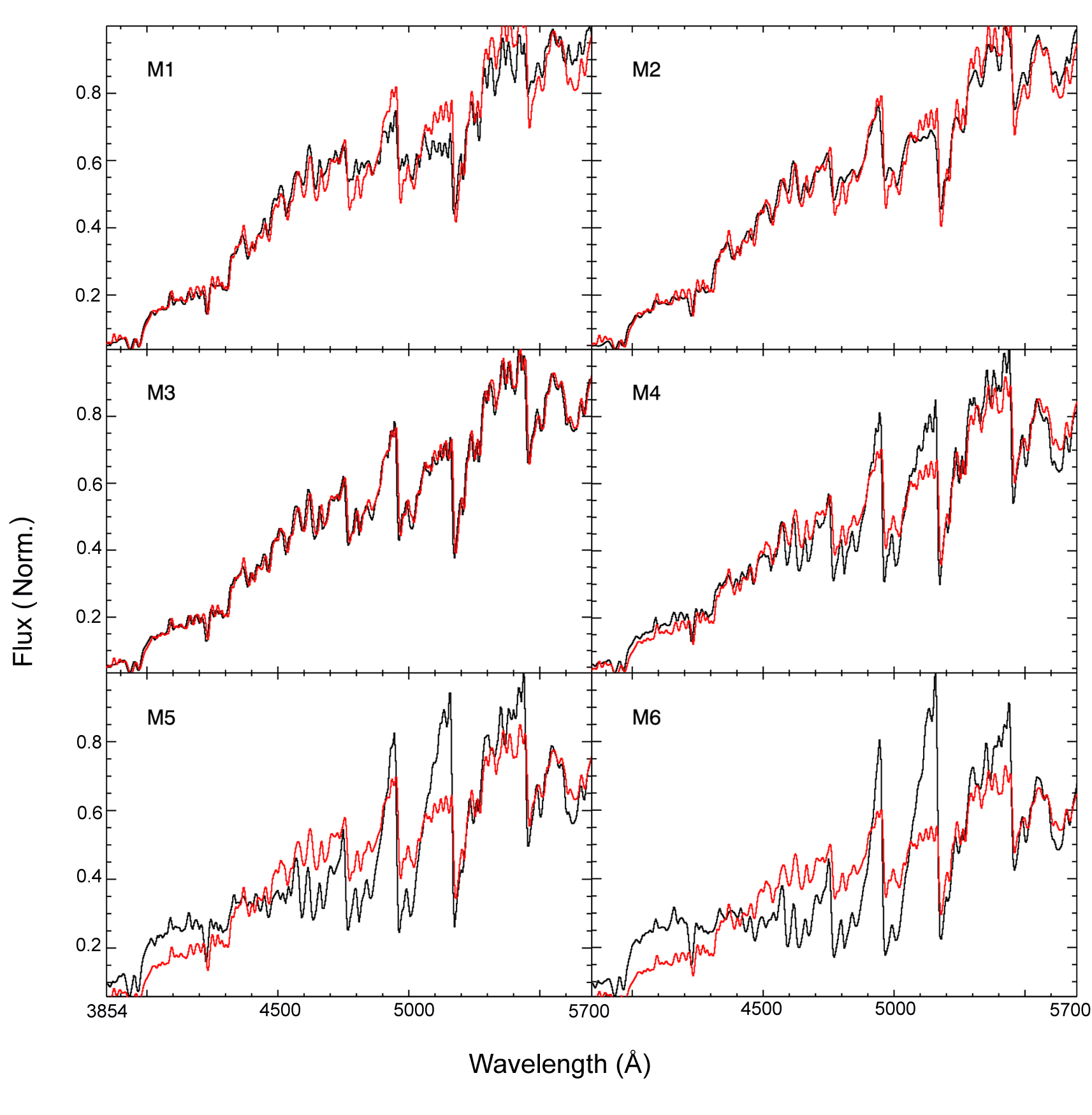}
\caption[Pickles M Giant Comparison to EG And]{
A comparison between EG And and intrinsic M giant spectral subclasses in the \citet{pickles_1998_stellar_library} Library. Pickles M giant spectra are plotted in black while EG And is overplotted in red. In each case the x-axis is the wavelength range $3854 - 5700$\AA\ and the y-axis is normalised flux. The EG And spectra have been scaled and extinction-corrected to produce the best-fits for each comparison. The overall best-fitting spectral-type was the M3 spectrum.
\label{fig:eg_series_extinct_pick}}
\end{figure}

\begin{figure}[ht!]
\centering
\includegraphics[width=\textwidth]{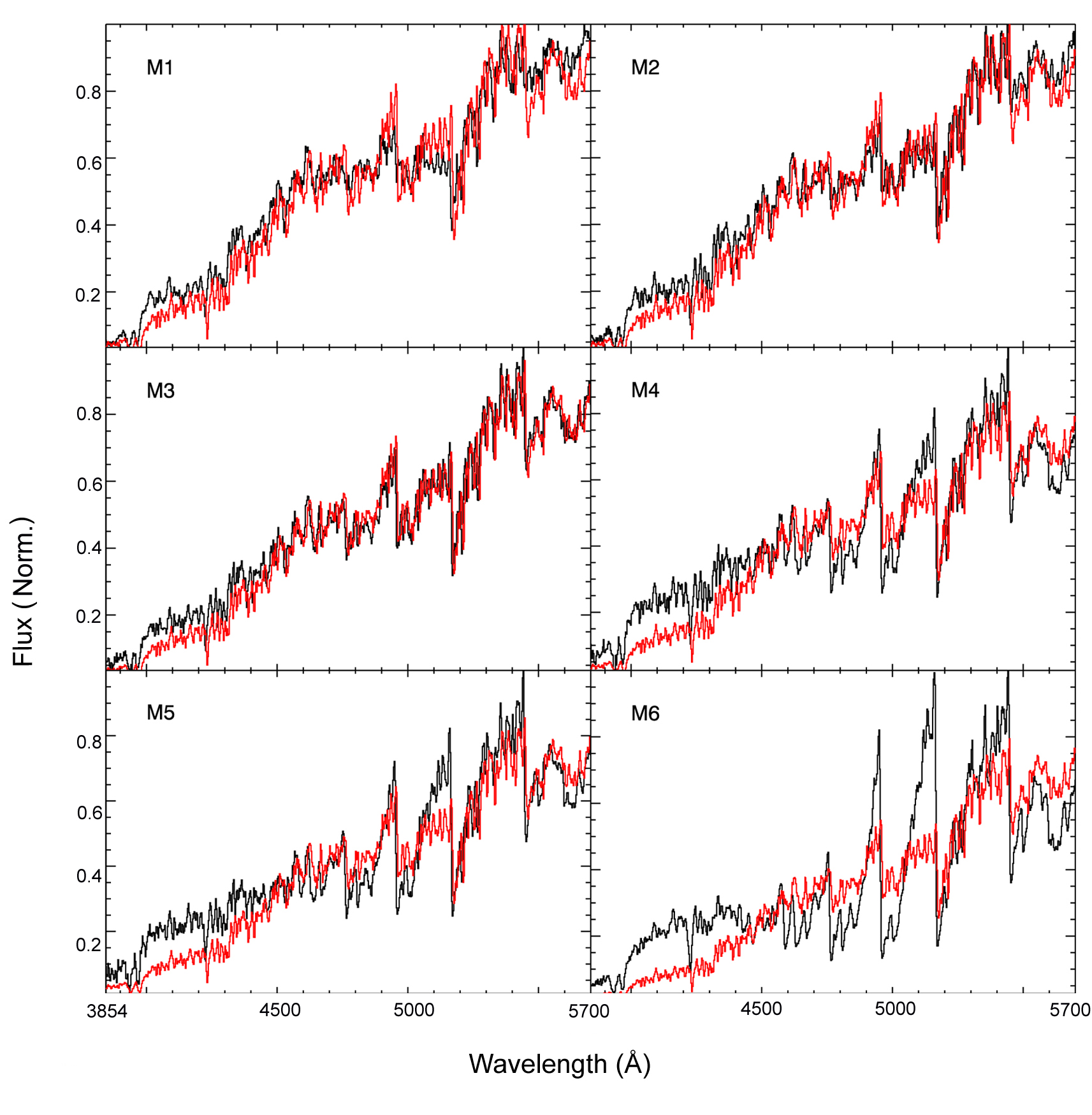}
\caption[Fluks M Giant Comparison to HD148349]{
A comparison between HD148349 and intrinsic M giant spectral subclasses in the \citet{fluks_1994} Library. Fluks M giant spectra are plotted in black while HD148349 is overplotted in red. In each case the x-axis is the wavelength range $3854 - 5700$\AA\ and the y-axis is normalised flux. The HD148349 spectra have been scaled and extinction-corrected to produce the best-fits for each comparison. The overall best-fitting spectral-type was the M3 spectrum.
\label{fig:hd_series_extinct_fluks}}
\end{figure}

\begin{figure}[ht!]
\centering
\includegraphics[width=\textwidth]{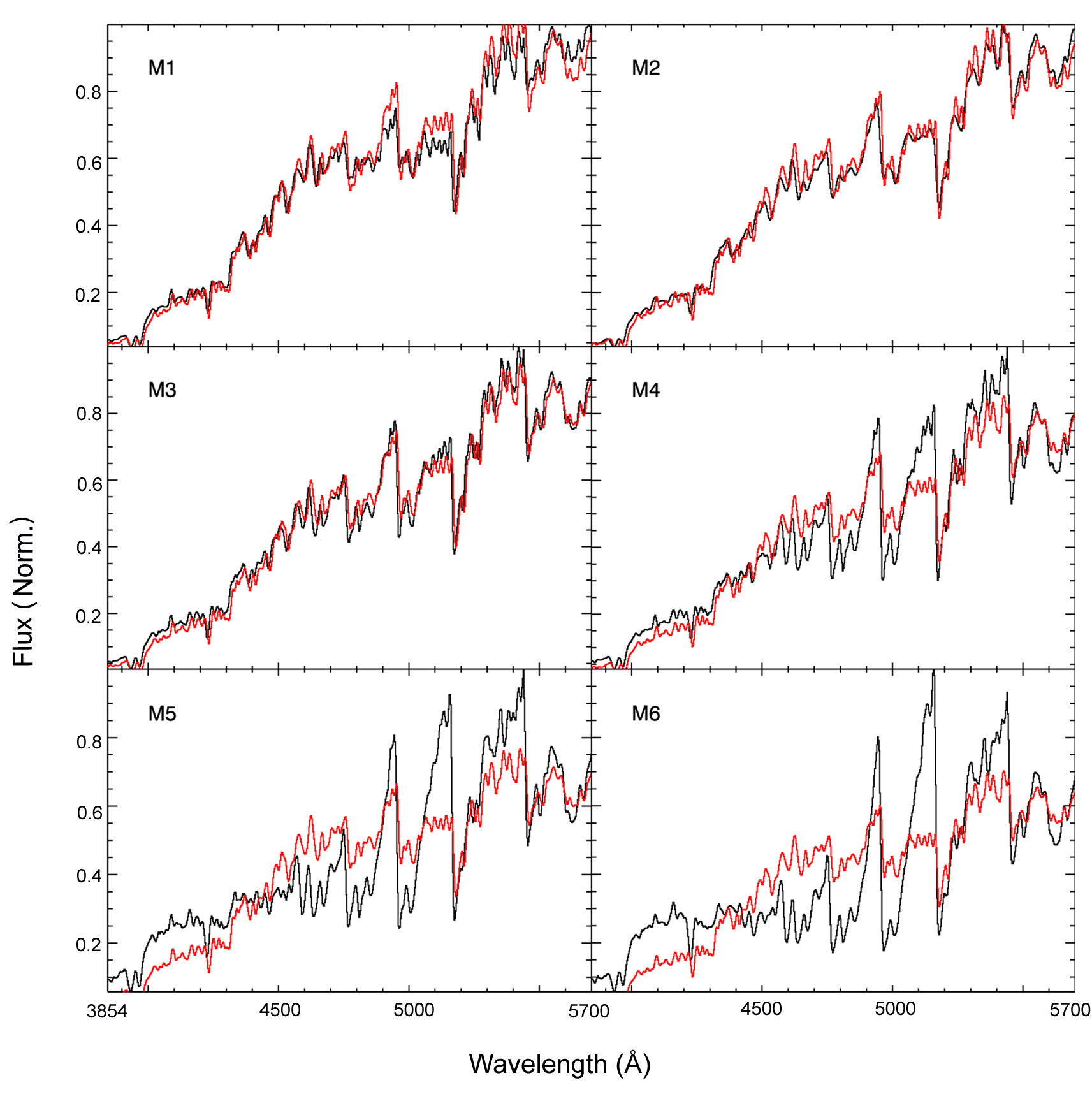}
\caption[Pickles M Giant Comparison to HD148349]{
A comparison between HD148349 and intrinsic M giant spectral subclasses in the \citet{pickles_1998_stellar_library} Library. Pickles M giant spectra are plotted in black while HD148349 is overplotted in red. In each case the x-axis is the wavelength range $3854 - 5700$\AA\ and the y-axis is normalised flux. The HD148349 spectra have been scaled and extinction-corrected to produce the best-fits for each comparison. The overall best-fitting spectral-type was the M3 spectrum.
\label{fig:hd_series_extinct_pick}}
\end{figure}

To find the optimum matching spectral type and extinction value, a 2-parameter grid-search method based on the technique outlined in \citet{bevington_data_reduction}\footnote{The method from \citet{bevington_data_reduction} is originally from \citet{numerical_recipes}.} was employed. For each sub-spectral type in the Fluks and Pickles series, the EG And (or HD148349) spectrum was scaled and extinction-corrected by a range of brightness and E(B-V) values and for each combination of these two parameters the reduced $\chi^2$ statistic was calculated. The procedure calculates the reduced $\chi^2$ values using the \textbf{chisq} IDL routine and:
\begin{equation}
\chi^{2}_{red} = \frac{1}{N_{free}}\sum \frac{(y - y_{lib})^2}{\sigma^2}
\end{equation}
where $y$ is either the EG And spectrum or the HD148349 spectrum, $y_{lib}$ is the library spectrum after being scaled and unreddened, $N_{free}$ is the number of degrees of freedom given by [(no.\ of wavelengths)-2] and $\sigma$ is the flux error array. A grid of all the $\chi^2$ values was constructed and the minimum value and the best-fit was located in parameter space. This grid-search method was tested by extincting a spectrum by a known value and then challenging the code to return the same value. It was found that it returns values less than 1\% away from the expected value. Figure \ref{fig:eg_series_extinct_fluks} shows the best-fits for each spectral sub-type obtained by scaling and extinction-correcting EG And to match the Fluks M giant spectra. Figure \ref{fig:eg_series_extinct_pick} compares EG And to the Pickles M giant spectra, while Figure \ref{fig:hd_series_extinct_fluks} and Figure \ref{fig:hd_series_extinct_pick} compare HD148349 to the Fluks and Pickles spectra respectively. 


\begin{figure}[ht!]
\centering
\includegraphics[width=\textwidth]{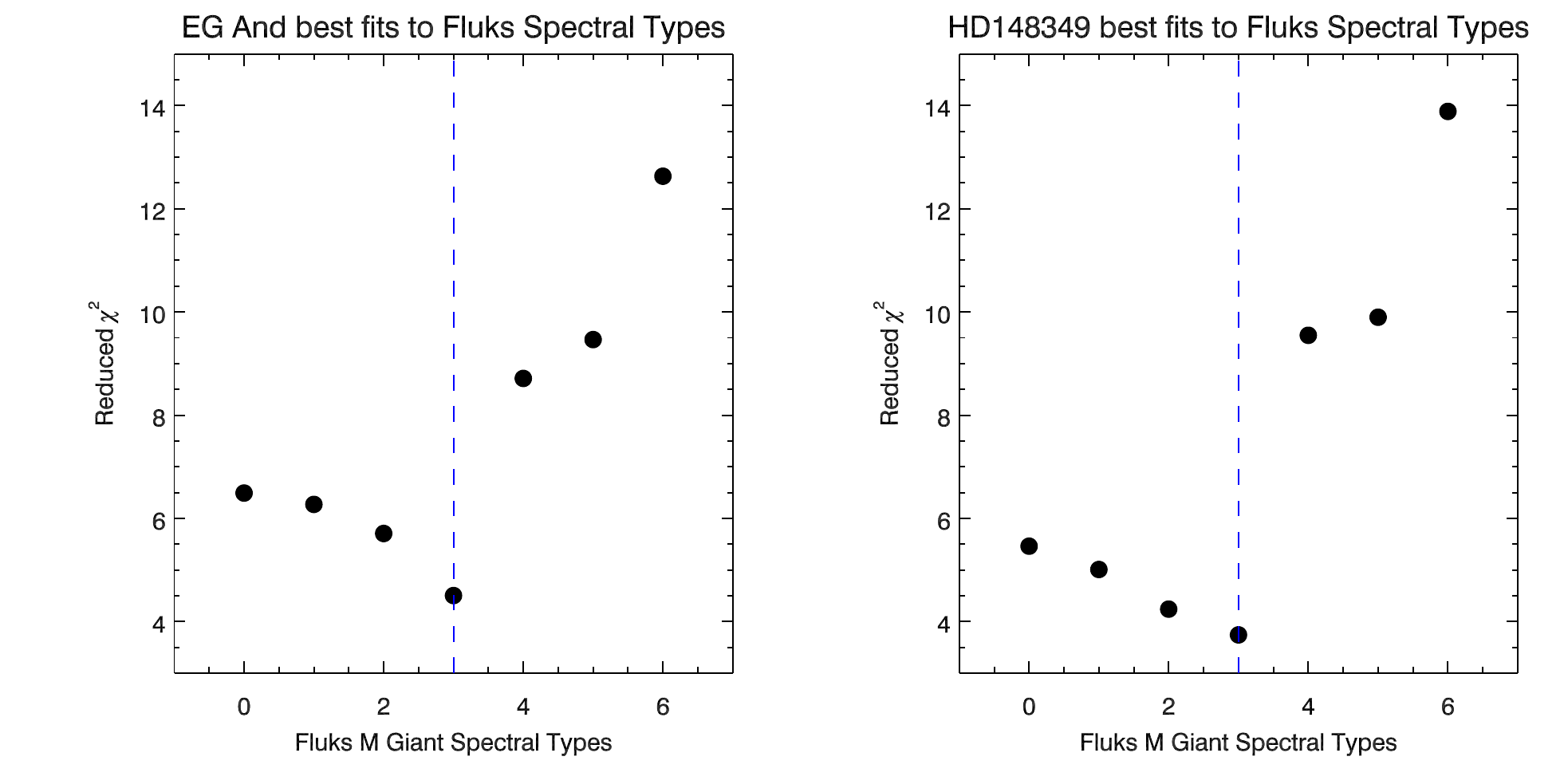}
\caption[Trends across Fluks Spectral Type]{Best-fit trends across Fluks spectral type for EG And (left) and HD148349 (right). The reduced $\chi^2$ was calculated for the best fit to each of the Fluks intrinsic M giant spectral types from M0 up to M6. For both target stars the trend in reduced $\chi^2$ values confirms that they are M3 giants.
\label{fig:extinct_bestfit_fluks}}
\end{figure}

The best-fitting comparisons yielded E(B-V) values of $0.05\pm0.02$  for EG And and $0.32\pm0.02$ for HD148349. Both the Fluks and Pickles comparisons gave best fits for M3-type spectral types for HD148349 and all observations of EG And. The reduced $\chi^2$ value for the best-fitting spectrum was calculated for each target compared to the library spectral types from M0 up to M6. A sample trend, representative of all the comparisons, is shown in  Figure \ref{fig:extinct_bestfit_fluks}. It can be seen that while the spectra produce similar goodness-of-fit values for hotter spectral subclasses, in the cooler subclasses the disparity between the spectra becomes more noticeable even after scaling and extinction-correcting. From M3 to M4 the change in $T_{eff}$ seems to triggers the onset of stronger TiO absorption bands.

\begin{figure}[ht!]
\centering
\includegraphics[width=\textwidth]{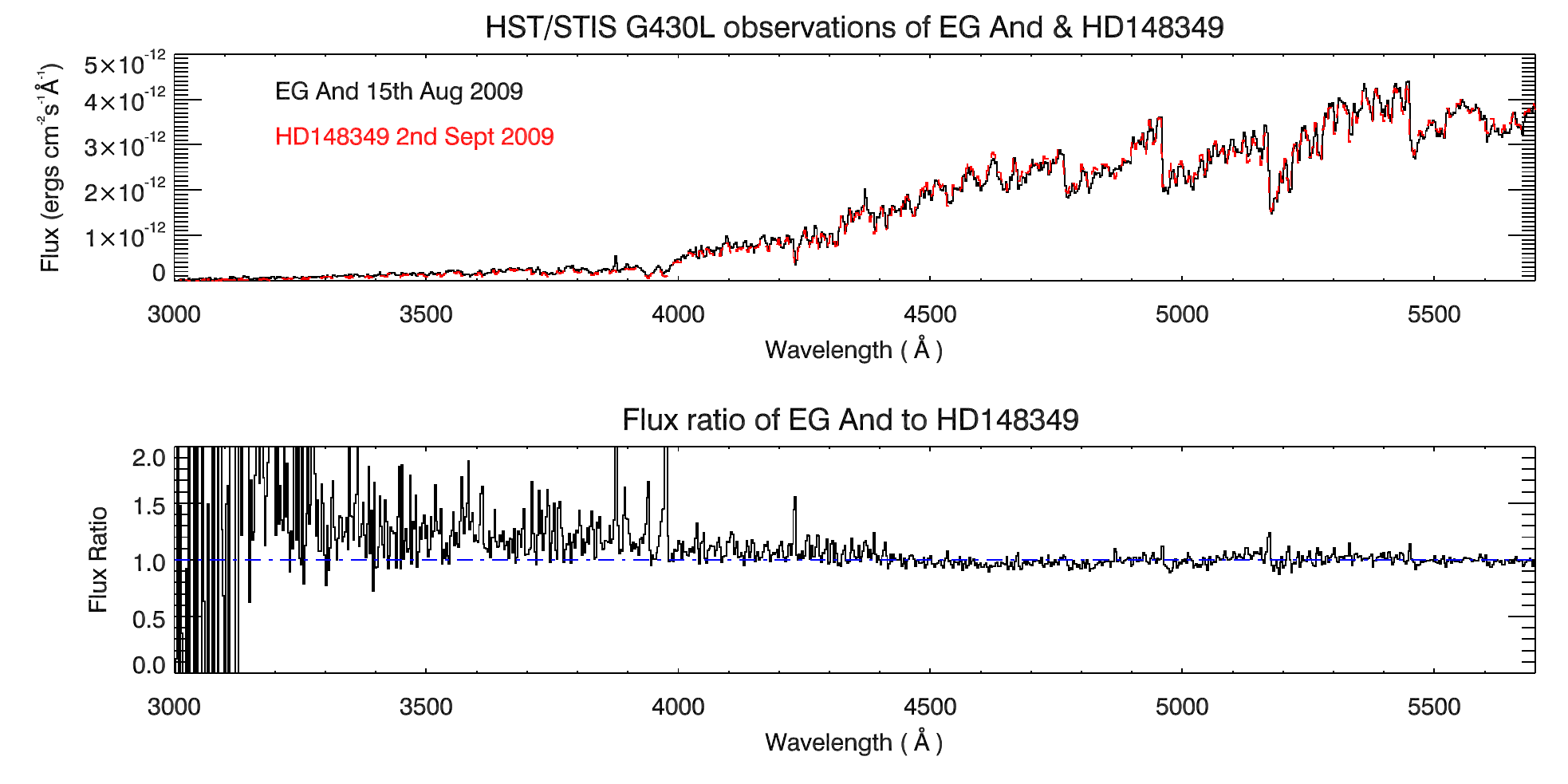}
\caption[{\sl HST}/STIS G430L observations of EG And and HD148349]{
{\sl HST}/STIS G430L observations of EG And and HD148349. Using the 1D spectral extraction technique outlined in Section \ref{sec:g430l_spec_extract}, observations in the optical can be obtained to show the similarity between the two giants. HD148349 (dashed red line) has been scaled by a brightness factor of 0.1 and dereddened using the extinction curve from \citet{cardelli_1989} and an E(B-V) value of 0.27. The bottom panel shows the flux ratio. The EG And emission lines account for a very small portion of the overall flux (around 1\%). 
\label{fig:basic_comp}}
\end{figure}

\begin{figure}[ht!]
\centering
\includegraphics[width=\textwidth]{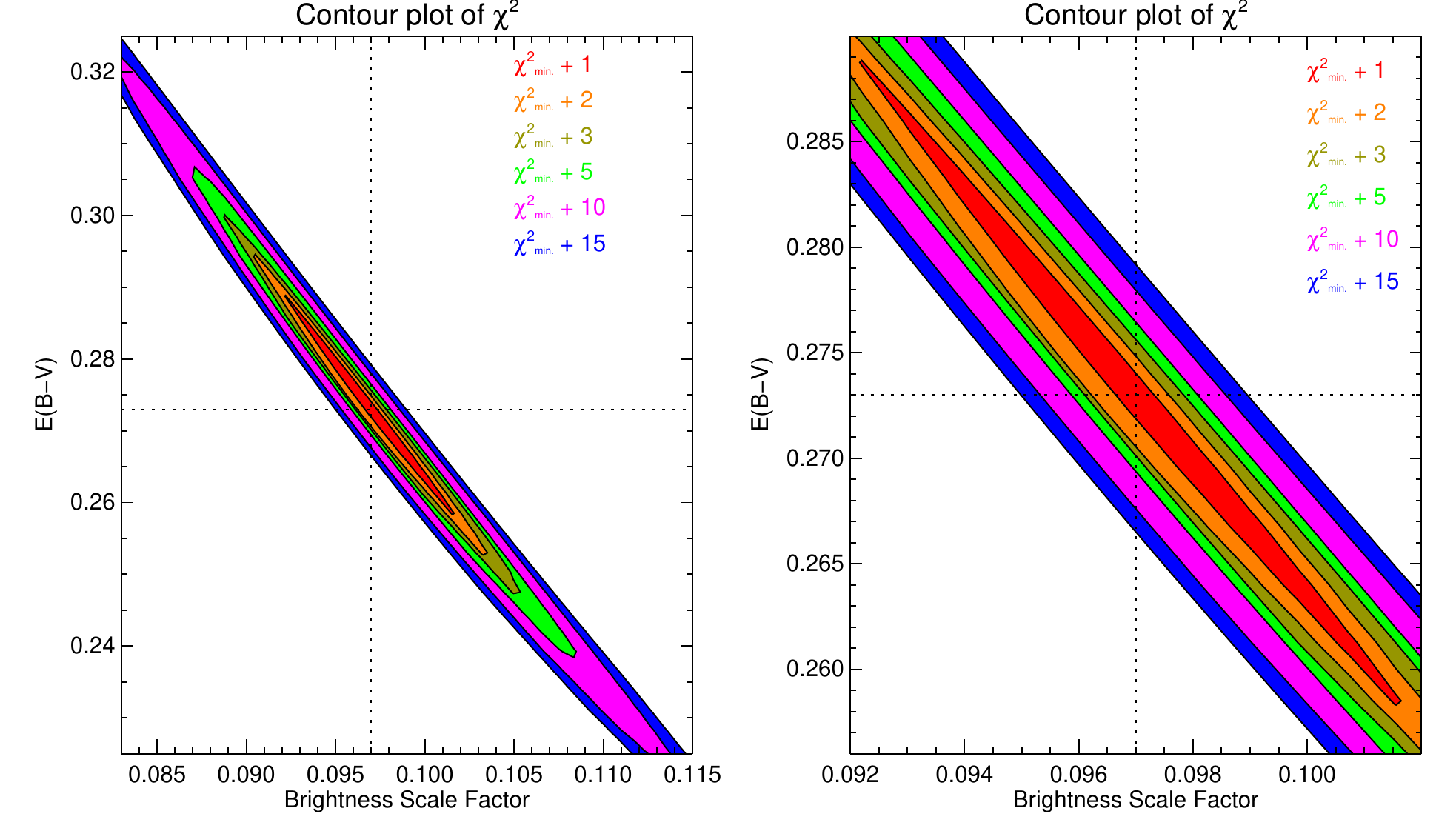}
\caption[Contour Plot of $\chi^2$ in Parameter Space]{
Contour plot of $\chi^2$ in parameter space. In this case the lowest $\chi^2$ value reached is for an E(B-V) value of 0.27 and a brightness scale factor of 0.1. The plot on the right is a close-up of the minimum.
\label{fig:extinct_scale_grid_contour}}
\end{figure}

Comparing the G430L optical spectra of EG And and HD148349 directly against each other shows how similar the EG And giant primary is to an isolated giant (Figure \ref{fig:basic_comp}). The similarity in $T_{eff}$, composition and size means that the only significant differences between the spectra result from differing levels of extinction and flux. The Hipparcos values for the distance to EG And and HD148349 are $513\pm169$ and  $177\pm12$ pc respectively, based on \citet{vanleeuwen_2007}. HD148349 has the higher flux levels but also suffers more light extinction than EG And due to their positions relative to the galactic plane. By using a grid-search method to find the $\chi^2$ minimum in parameter space, the best fitting values of differential extinction and the flux ratio between the two stars are obtained (See Figure \ref{fig:extinct_scale_grid_contour}). The flux ratio can be used to calculate a more precise distance to EG And using  the Hipparcos value for the distance to HD148349 and the inverse-square law of distance and apparent brightness:

\begin{equation}\label{eq:dist_egand_hipp}
D_{\mbox{\scriptsize EG}}  =   D_{\mbox{\scriptsize HD}} \times  \sqrt{B_{\mbox{\scriptsize HD/EG}}}
\end{equation}
where $D_{\mbox{\scriptsize EG}}$ is the distance to EG And, $D_{\mbox{\scriptsize HD}}$ is the Hipparcos distance to HD148349 and $B_{\mbox{\scriptsize HD/EG}}$ is the brightness ratio of HD148349 to EG And. The best fit values yielded E(B-V) of $0.05\pm0.02$ for EG And and $0.32\pm0.02$ for HD148349 and a new distance to EG And of $568\pm41$pc. EG And was found to be about $\sim$10\% as bright as HD148349 in the optical.

\section{G430L 1D Extracted Spectrum Calibration}\label{sec:g430l_calib}

The method of extracting optical spectra from previously unused acquisition images has proven to be integral to the comparison of EG And to HD148349. Due to the non-standard method of extracting a spectrum from the G430L images, it was necessary to apply a calibration scale factor to the extracted spectrum. The value of this scale factor was determined  by using a number of methods. Firstly, the corresponding E230M spectrum was compared  to the G430L extracted spectrum at the region of overlap ($2900 - 3100$\AA). From this it was determined that a correction scale factor of approximately 1.35 was needed. This was a crude method as the region of overlap between E230M and G430L are at the ends of their wavelength coverage regions, where the uncertainty is highest. 

Next, all previous G430L small aperture observations (from all observing cycles and of all targets) were investigated in the MAST archive. It was hoped that by applying the same technique described in Section \ref{sec:g430l_spec_extract} to stars with corresponding E230M observations and with measured photometry values it would be possible to refine the calibration scale factor. Targets that had no corresponding E230M observations or photometry measurements or were considered too variable, for example Mira variables, were discounted. This left 5 stars (HD132475, LTT 7244, LTT 3774, NLTT 26576, and  HD31293). When the G430L acquisition images for these objects were utilised to extract a 1D spectrum in the manner described in Section \ref{sec:g430l_spec_extract}, it was again found that a calibration correction factor was required. By comparing the extracted spectra to the overlap region in the corresponding E230M observations and by performing synthetic photometry on the extracted spectra, UBV magnitudes were found to compare to the accepted values for those targets. While these methods were not precise it resulted in a range of calibration correction factors of 1.4, 1.2, 1.2, 1.3 and 1.45 for the stars in the order listed above. This supported the adopted value of 1.35. 

\begin{figure}[ht!]
\centering
\includegraphics[width=\textwidth]{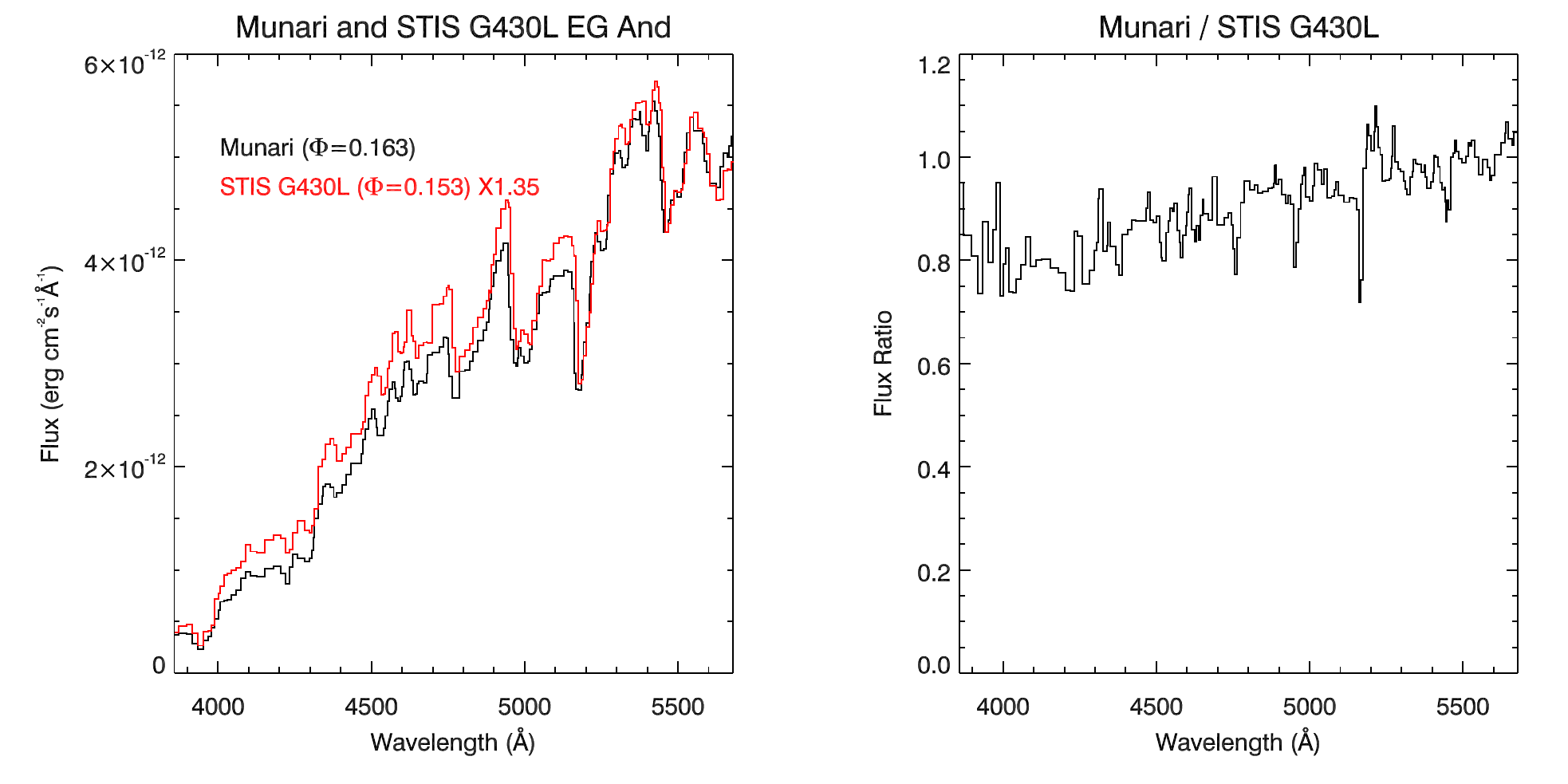}
\includegraphics[width=\textwidth]{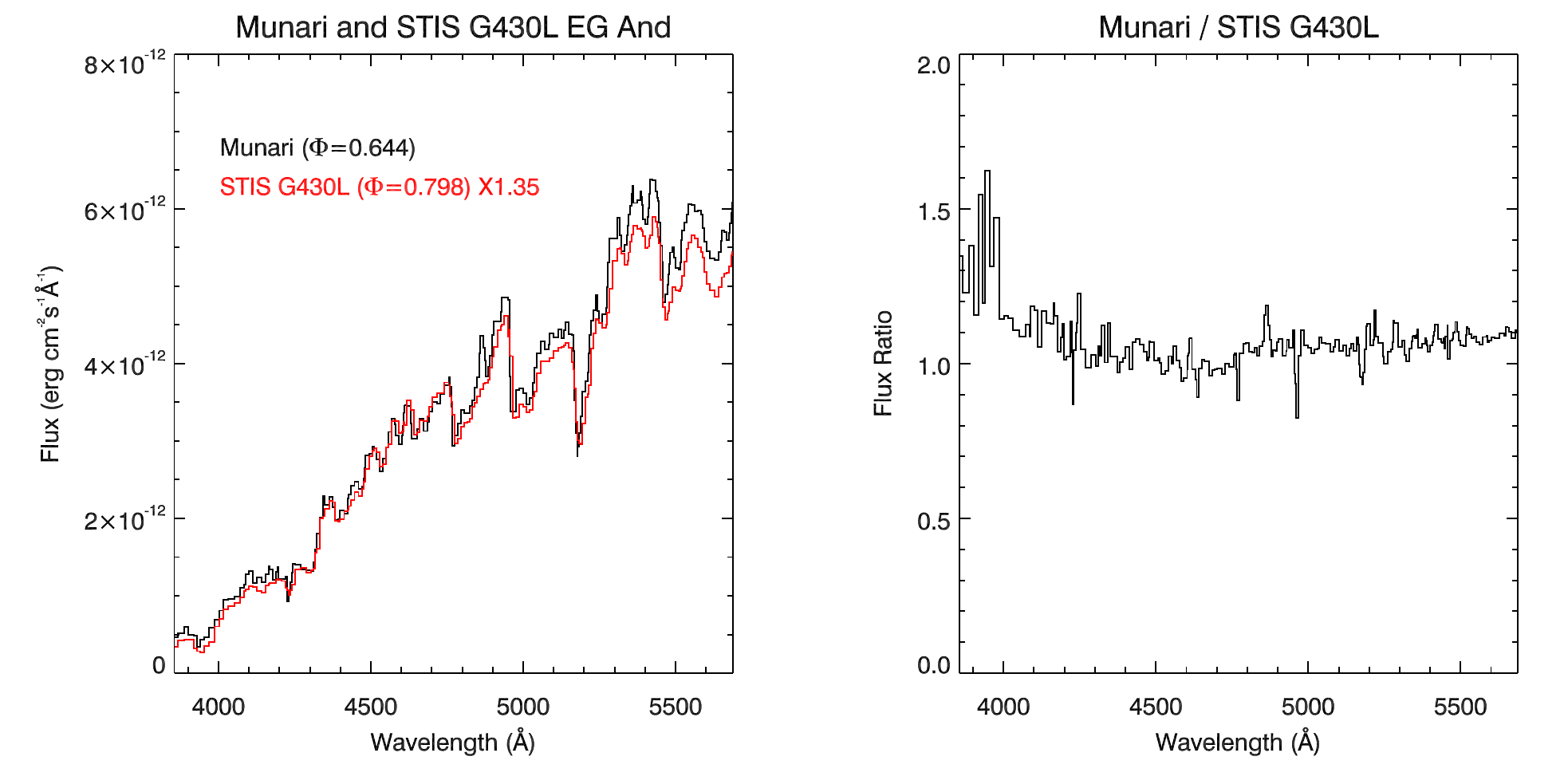}
\caption[G430L Extracted Spectra Compared to Munari Observations]{
G430L 1D extracted spectra (calibration corrected using a value of 1.35) are compared to observations from \citet{munari_zwitter_atlas_2002} at similar orbital phases. The flux ratios are shown on the right to ensure that the extraction is being performed accurately across the entire wavelength region. The agreement between the extracted spectra and the Munari observations is quite good. Any discrepancies can be attributed to the variations discussed in Section \ref{sec:vary}. 
\label{fig:munari_eg_g430l_1}}
\end{figure}

Finally, the Hipparcos transmission curve was applied to the extracted spectra to calculate synthetic H$_p$ magnitudes for the G430L observations and compare them to the V magnitudes.  The curves were taken from \citet{bessell_2000_hipparcos_passbands}. Although the H$_p$ band covers $3400 - 8800$\AA\ the amount of G430L flux was estimated using the same method described in Section \ref{sec:synth_photom} to estimate the amount of total flux that would fall in the H$_p$ band. The range of resulting V-H$_p$ values for both targets stars matched those suggested by \citet{bessell_2000_hipparcos_passbands} for a red giant star.  When the data is extracted and calibrated by 1.35 it can be seen in Figure \ref{fig:munari_eg_g430l_1} that the spectra show agreement with those of similar orbital phases from \citet{munari_zwitter_atlas_2002}, which were based on observations collected with the telescopes of the European Southern Observatory (ESO, Chile) and of the Padova \& Asiago Astronomical Observatories (Italy).

\clearpage

\section{UBV Synthetic Photometry}

\begin{figure}[ht!]
\centering
\includegraphics[width=\textwidth]{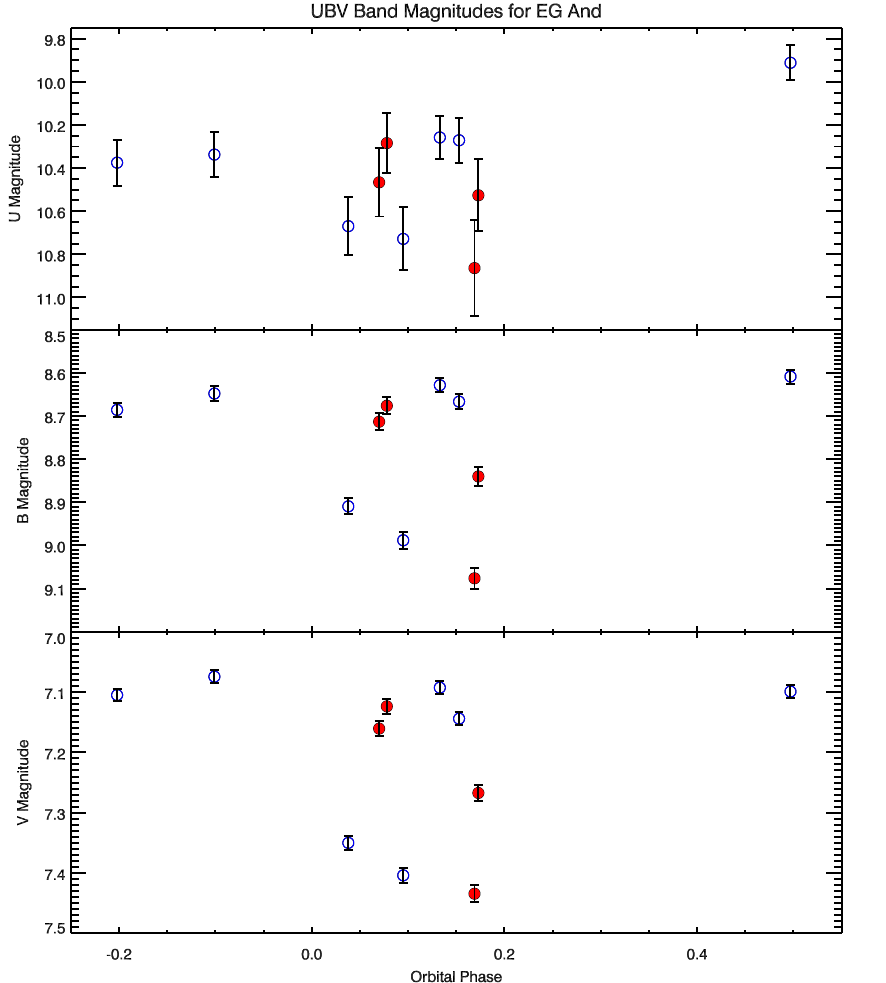}
\caption[UBV Photometry for EG And]
{U, B, and V magnitudes of EG And against orbital phase. The empty blue circles correspond to Cycle 11 observations while the filled red circles correspond to Cycle 17 observations. The uncertainty in the U magnitudes is considerably higher as the amount of flux in this passband region is lower and the error increases towards the blue end of the G430L region. The significant orbit-to-orbit variability can seen in the phase region $\phi = 0.03$ to $\phi = 0.2$.
\label{fig:ubv_egand_photom}}
\end{figure}

For each of the {\sl HST} Cycle 11 and 17 observations there were two G430L spectra extracted (acquisition and peakup as discussed in Section \ref{sec:g430l_spec_extract}). The two spectra were combined for each observation to improve the signal-to-noise ratio.  Synthetic photometry was performed on the combined spectra and the resulting magnitudes were then plotted against orbital phase as shown in Figure \ref{fig:ubv_egand_photom}. These values are also shown in Table \ref{tab:ubv_photom}.

\begin{table}[ht!]
\caption[UBV Magnitudes for G430L EG And and HD148349]{
UBV Magnitudes for G430L EG And and HD148349.
\label{tab:ubv_photom}}
\centering
\begin{tabular}{ccccccccc}
\hline\hline
HJD & Target & Cycle & Orbital Phase & U & B & V & B-V & U-B\\
\hline
     2452515 & EG And &       11 & 0.798 & 10.11 & 8.59 & 7.03 & 1.56 & 1.52 \\
     2452564 & EG And &       11 & 0.899 & 10.07 & 8.55 & 7.00 & 1.55 & 1.52 \\
     2452631 & EG And &       11 & 0.038 & 10.41 & 8.82 & 7.28 & 1.54 & 1.59 \\
     2452658 & EG And &       11 & 0.095 & 10.46 & 8.89 & 7.33 & 1.56 & 1.57 \\
     2452677 & EG And &       11 & 0.133 &  9.99 & 8.53 & 7.02 & 1.52 & 1.46 \\
     2452687 & EG And &       11 & 0.153 & 10.01 & 8.57 & 7.07 & 1.50 & 1.44 \\
     2452852 & EG And &       11 & 0.497 &  9.64 & 8.51 & 7.03 & 1.49 & 1.13 \\
     2455059 & EG And &       17 & 0.070 & 10.21 & 8.62 & 7.09 & 1.53 & 1.59 \\
     2455063 & EG And &       17 & 0.078 & 10.03 & 8.58 & 7.05 & 1.53 & 1.44 \\
     2455107 & EG And &       17 & 0.169 & 10.61 & 8.98 & 7.36 & 1.62 & 1.63 \\
     2455109 & EG And &       17 & 0.173 & 10.27 & 8.75 & 7.19 & 1.56 & 1.53 \\
     \hline
     2455077 & HD148349 &       17 & - & 9.20 & 7.16 & 5.44 & 1.72 & 2.04 \\
\hline
\end{tabular}
\end{table}

\begin{figure}[ht!]
\centering
\includegraphics[width=\textwidth]{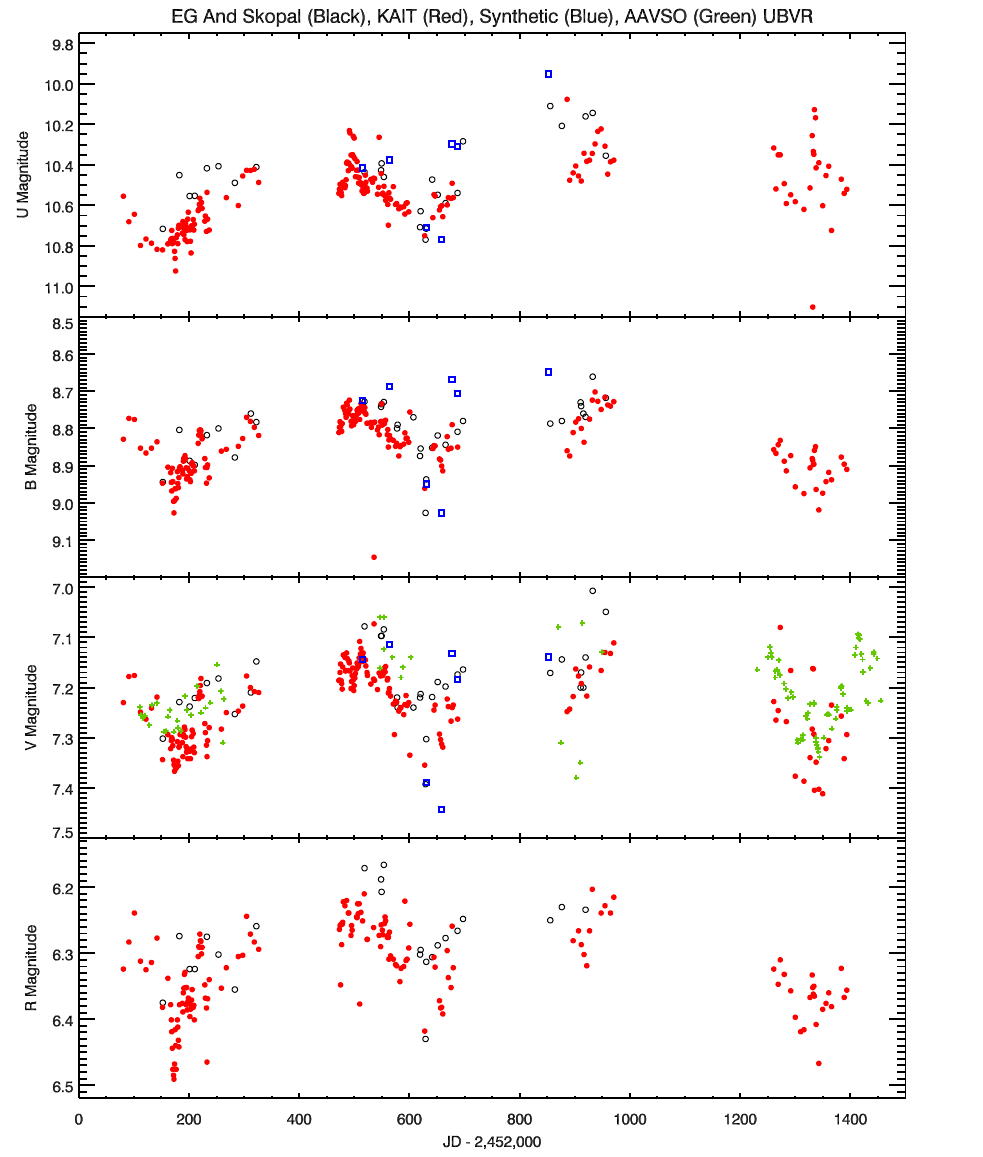}
\caption[EG And Photometry Compilation]{
EG And Photometry from \citet{skopal_2002_egand_photom} covering an observing period from July 1998 to December 2003 (shown as black circles). {\sl KAIT} observations are shown as red filled-circles,  the derived synthetic values from the G430L spectra are shown as blue squares and the AAVSO\footnote{American Association of Variable Star Observers} observations are shown as green triangles.
\label{fig:skopal_kait_photom1}}
\end{figure}

Once again it can be seen that the calibration scale factor is justified when compared to the synthetic magnitudes with those of other observations. Figure \ref{fig:skopal_kait_photom1} shows that the magnitudes compare well with those of \citet{skopal_2002_egand_photom}. Also shown are observations from KAIT\footnote{Discussed in Chapter 2.}. The B-V and U-B colours for HD143849 match the expected values. Previous photometric observations of HD148349 by \citet{smak_1964_hd_photom}, \citet{cousins_1964_hd_photom}, \citet{przybylski_1965_hd_photom} and \citet{mermilliod_1986_hd_photom}, yielded U-B values of 2.13, 2.05, 2.00 and 2.05 respectively. Their B-V values were 1.74, 1.72, 1.73 and 1.75. These compare well with the U-B value of 2.04 and the B-V value of 1.72. The similarity in the values support this method of G430L synthetic photometry for accurately measuring magnitudes contemporaneous to UV observations. 

\clearpage

\clearpage

\section{Comparison of Variabilities}\label{sec:vary}

The main orbital photometric variation is the ellipsoidal distortion of the giant as found by \citet{wilson_vaccaro_1997_egand_photom} and subsequently by \citet{cian_thesis} with a period of half the orbital period due to the circular orbit, as noted in Chapter 2. The effect is difficult to see over the Cycle 11 and Cycle 17 observations but it becomes much clearer when viewed over several cycles (see Figure \ref{fig:skopal_kait_photom1}). The EG And colours show that for B-V, the values expected for for an M3III star match quite well with the values predicted by \citet{bessell_1990_ubvri_passbands}. The U-B colours are lower but this can be attributed to the higher flux in the U band for EG And due to the presence of the white dwarf in comparison with isolated giants. Comparing the variability of the B and V band photometry of EG And to that of an isolated spectral standard, the variability is similar if the ellipsoidal distortion is taken out. \citet{cian_thesis} suggests a variability of 0.1 magnitudes in B and V bands after the effects of ellipsoidal distortion have been removed. 

\begin{figure}[ht!]
\centering
\includegraphics[width=\textwidth]{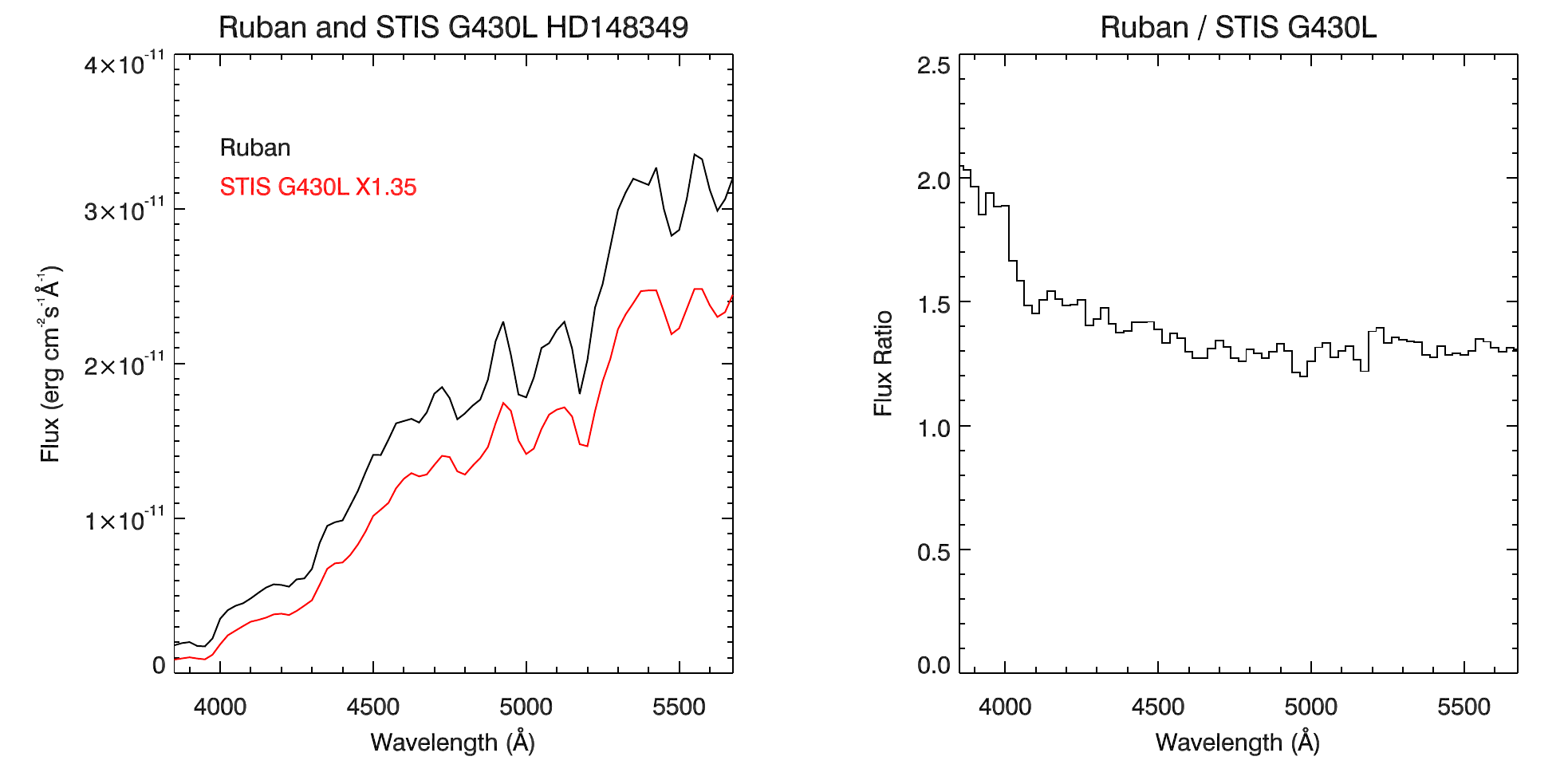}
\caption[G430L HD148349 Spectrum and Ruban Observation]{
The left panel shows the spectrum from \citet{ruban_2006} in black, while the G430L extracted HD148349 spectrum multiplied by 1.35 is shown in red. The right panel shows the ratio of the two spectra.
\label{fig:ruban_hd_g430l}}
\end{figure}

HD148349 is more variable than previously thought. The synthetic photometry magnitudes (see Table \ref{tab:ubv_photom}) indicate the star is less bright than preceding observations. These previous observations gave U, B and V magnitudes of 9.03, 6.99 and 5.27 show that HD148349 was brigher in the past.  When compared to the extracted G430L spectrum of HD148349 (combining both the ACQ and ACQ Peakup to improve the signal-to-noise, and applying the calibration scale factor of 1.35) to a \citet{ruban_2006} spectrum, it can be seen how the flux of the recent observation of HD148349 has fallen significantly since the Ruban observation (See Figure \ref{fig:ruban_hd_g430l}). In particular there is a considerable excess of flux ($\geq 1\times10^{-12}$erg s$^{-1}$ cm$^{-2}$\AA$^{-1}$).

In the recent observation of HD148349 very little flux is seen at wavelengths short of 4000\AA. This is supported in the STIS E230M spectrum where little or no flux is seen at 3000\AA.
Finally, several AAVSO observations of HD148349 show V magnitudes of 5.5 and brighter. This reinforces the point that HD148349 undergoes larger variations in magnitude than previously thought and that the recent STIS observations were taken during an epoch when the star was relatively dimmer than previous observation used to determine its magnitudes. 

\section{\ion{Mg}{ii} \emph{h} and \emph{k} Lines}

\begin{figure}[ht!]
\centering
\includegraphics[width=\textwidth]{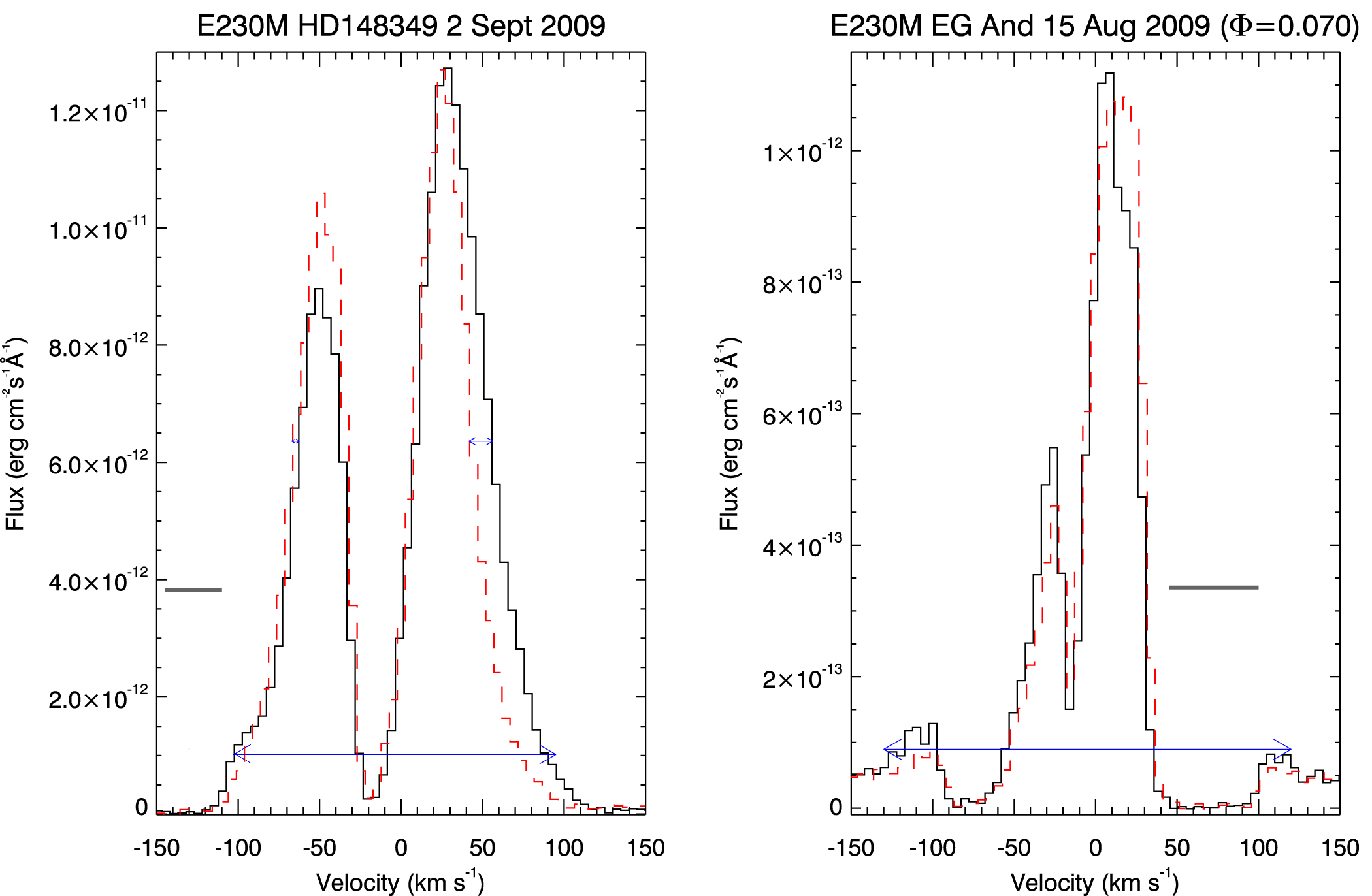}
\caption[\ion{Mg}{ii} \emph{h} and \emph{k} Lines]{
Comparison of the \ion{Mg}{ii} resonance lines for the E230M observations of HD148349 (left)  and EG And (right). The two lines have been overlaid so that their central absoption features conincide, and the weaker \emph{h} line was scaled up to to match the \emph{k} line. The 2803\AA\ \emph{h} line is shown in red, while the 2796\AA\ \emph{k} line is shown in black. The effects of ISM absorption are indicated by thick grey lines and the dotted lines show the wind terminal velocities at $\sim$70km s$^{-1}$ and $\sim$15km s$^{-1}$ for EG And and HD148349 respectively. 
\label{fig:mgii_asymm}}
\end{figure}

The strength of the \ion{Mg}{ii} resonance lines at 2796.352\AA\ \emph{k} line and 2803.531\AA\ \emph{h} line \citep{kaufman1991wavelengths} can be used as chromospheric activity diagnostics, while their asymmetries and blue-shifted absorption features indicate the presence of a wind \citep{stencel_mullan_1980apj}. Figure \ref{fig:mgii_asymm} shows the \ion{Mg}{ii} \emph{h} and \emph{k} lines for both HD148349 and EG And. Measurements of width, asymmetry and offset can quantify the behavior of the line profiles. Using the parameters set out in \citet{robinson_carpenter_1995}, the widths are taken as the width of the \emph{h} and \emph{k} lines at 10\% of the maximum intensity. The asymmetry is taken as the ratio of the maximum fluxes of the blue to red emission peaks. The offset is the difference in velocity space between the wings of the \emph{h} and \emph{k} lines when they the lines are overlaid so that there central absorption features coincide and normalized to the peak intensity (shown as the small blue arrows at 50\% of the peak intensity in the left panel of Figure \ref{fig:mgii_asymm} but not visible in EG And due to absorption).

The width at 10\% of the maximum flux is difficult to measure due to the profiles being mutilated by the ISM at the regions marked by thick grey lines. By comparing the profiles about their centre points and scaling the mutilated side to estimate the flux lost to the ISM (or in the case of EG And, extrapolating from the mutilated wings) an estimate of the line width can be made. The measured line profile parameters are listed in Table \ref{tab:mgii_asymm}. The asymmetries match those expected of giants later than K2. $\gamma$ Cru (M3III) and $\rho$ Per (M4II) have $B/R_h$ values of 0.55 and 0.50 with $B/R_k$ values of 0.62 and 0.90, while EG And and HD148349 were measured to have $B/R_h$ values of 0.4 and 0.85 with $B/R_k$ values of 0.5 and 0.7, respectively.
There are also a couple of discrepancies. The blue wing of the HD148349 lines show very little offset between the \emph{h} and \emph{k} lines where usually there is a shift between the lines of around 2 - 5km s$^{-1}$. In the EG And profile a shift on the red wing would be expected of  10 - 20km s$^{-1}$, but this cannot be seen. For both of these cases absorption due to the ISM can be attributed to causing this discrepancy. A red asymmetry ($B/R < 1$), which both of the targets display, indicates the presence of a massive, slow-moving wind. From the profiles these winds can be estimated to be $\sim$70km s$^{-1}$ and $\sim$15km s$^{-1}$ for EG And and HD148349, respectively. 

\begin{table}
\caption[\ion{Mg}{ii} line profile parameters]{
\ion{Mg}{ii} line profile parameters.  $(B/R)_h$ is the ratio of the maximum intensity of the blue emission peak to the red emission peak for the \emph{h} line.  $(B/R)_k$ is the same ratio but for the \emph{k} line. $\Delta V_r$ is the offset between the two lines on the red wing, while $\Delta V_b$ is the offset on the blue wing. $V_\infty$ is the terminal wind velocity given by the P Cygni profile.  
\label{tab:mgii_asymm}}
\smallskip
\begin{center}
{\small
\begin{tabular}{ccccccc}
\hline
\noalign{\smallskip}
Targets & Line Width & $(B/R)_h$ & $(B/R)_k$ & $\Delta V_r$ & $\Delta V_b$ & $V_\infty$ \\
& (km s$^{-1}$) & & & (km s$^{-1}$) & (km s$^{-1}$) & (km s$^{-1}$)  \\
\noalign{\smallskip}
\hline
\noalign{\smallskip}
EG And & 250 & 0.4 & 0.5 & 0 & 0 & $\sim70$\\
HD148349 & 200 & 0.85 & 0.7 & 5 & 15 & $\sim15$ \\
\noalign{\smallskip}
\hline
\end{tabular}
}
\end{center}
\end{table}

The \ion{Mg}{ii} lines can also be used to estimate absolute magnitude and hence the distance to a star through an empirical relationship between luminosity and the width of the line. A similar relationship between luminosity and the width of \ion{Ca}{ii} \emph{h} and \emph{k} lines was first quantified by \citet{wilson_bappu_1957}. They showed how the logarithm of the \ion{Ca}{ii} line width increased linearly with decreasing absolute magnitude. This became known as the Wilson-Bappu Effect and in equation form is:
\begin{equation}
M = A + B \log W_{(k)}
\label{eqn:wilsonbappu}
\end{equation}
where $M$ is the absolute magnitude, $A$ and $B$ are coefficients of the fit and $W_{(k)}$ is the base width of the \emph{k} line. The relationship applies to late-type dwarf, giant, and supergiant stars \citep{stencel_2009}.

\citet{weiler_oegerle_1979} investigated the same effect using \ion{Mg}{ii} instead of \ion{Ca}{ii}. Their study used the {\sl Copernicus} satellite and the \emph{Catalogue of Bright Stars} \citep{hoffleit_1964} to come up with an analogous Wilson-Bappu type effect for \ion{Mg}{ii} \emph{k} emission lines.  \citet{scoville_menawerth_1998} recalibrated the Wilson-Bappu effect using {\sl IUE} archival data to measure the \ion{Mg}{ii} line widths.  It was expected that the linear regression fit would be improved upon with the addition of more accurate absolute magnitudes provided by {\sl Hipparcos}. This proved not to be the case and the scatter found by \citet{weiler_oegerle_1979}  appears to be intrinsic to the \ion{Mg}{ii} line. This empirical relationship can be used to estimate values of absolute magnitude from the observations. When applied to HD148349 and EG And using Equation \ref{eqn:wilsonbappu} with values of 34.93 and 15.15 for $A$ and $B$ from \citet{weiler_oegerle_1979}, absolute magnitudes of 0.07 and -1.40 were calculated for HD148349 and EG And respectively. Using the relationship between absolute magnitude $M$, apparent magnitude $m$ and distance $D$:
\begin{equation}
m = M - 5(1 - \log_{10}D)
\end{equation}
distances to both stars were estimated as $\sim$537pc for EG And and $\sim$110pc for HD148349. For EG And the distance is in good agreement with the value of $568\pm41$pc calculated in Section \ref{sec:extinction}. The different distances values are summarised in Table \ref{tab:dist_measures}.

\begin{table}
\caption[Distance Measurements for EG And and HD148349]{
Distance Measurements for EG And and HD148349. All distances are in parsecs. The second column lists the values of distance for EG And and HD148349 using {\sl Hipparcos} parallax measurements from \citet{vanleeuwen_2007}. The second column shows a more precise EG And distance calculated by comparing the G430L spectra to HD148349 and the last column shows the distance estimtaes using the Wilson-Bappu relationship.
\label{tab:dist_measures}}
\smallskip
\begin{center}
{\small
\begin{tabular}{cccc}
\hline
\noalign{\smallskip}
Targets & \citet{vanleeuwen_2007} & G430L Comparison & Wilson-Bappu  \\
\noalign{\smallskip}
\hline
\noalign{\smallskip}
EG And & $513\pm169$& $568\pm41$ & $\sim537$ \\
HD148349 & $177\pm12$ &- & $\sim110$ \\
\noalign{\smallskip}
\hline
\end{tabular}
}
\end{center}
\end{table}

Hipparcos distance to HD148349 and $B_{\mbox{\scriptsize HD/EG}}$ is the brightness ratio of HD148349 to EG And. The best fit values yielded E(B-V) of $0.05\pm0.02$ for EG And and $0.32\pm0.02$ for HD148349 and a new distance to EG And of $568\pm41$pc. EG And was found to be about $\sim$10\% as bright as HD148349 in the optical.

\section{Discussion and Conclusions}

A new method of obtaining low-resolution contemporaneous optical data from acquisition images taken during STIS observations was presented. This method allows coverage of the optical part of the spectrum while simultaneously observing in the ultraviolet. This is particularly useful in the case of EG And, as the effects of the dwarf (dominant in the UV) on the red giant (dominant in the optical) can be disentangled by having contemporaneous coverage in both spectral regions. Several separate approaches to determine the best extraction calibration values  converged on a  multiplicative correction factor of 1.35 to bring the pipeline produced spectrum to the absolute scale. This method has implications for all future small aperture observations with STIS, as a proven technique for obtaining useful contemporaneous optical data will benefit the study of symbiotic systems.

The presence of the white dwarf affects the photometry of the system, especially in the U band.  The variability can be accounted for using the period of orbit and a model of ellipsoidal distortion. When this has been accounted for, the remaining variability is on the scale of that expected for a red giant and so provides a good platform for extending this study to more isolated giants.

The variability of the spectral `standard' proved to be more of an issue than the symbiotic system. While HD148349 has been classed as a variable star in some catalogues, it remains a spectral standard in others. The range of magnitudes in the literature suggest that HD148349 does not vary by a large amount. The analysis shown here suggests that the observations of HD148349 took place when it was much fainter than preceding observations. Although there is no indication in the literature that it would be so faint, there are amateur ground-based AAVSO observations match values obtained here. Based on all of this evidence it would seem that HD148349 should be reclassified as a more variable star and should perhaps be demoted from the list of spectral standards.

The synthetic photometry values not only show the reliability of the red giant of EG And for further studies, it also provides a way of directly comparing UV to optical activity. In previous work on EG And \citep{cian_thesis}, the presence of ``higher-state'' UV observations has been an issue. If such phenomena (which are dealt with in Chapter 5) were tied to the red giant activity it would undermine its suitability for this work. However it can convincingly be shown that there is no definite correlation with brighter magnitudes in the optical regime.

Comparing the extracted G430L spectra of the two target stars shows how similar they are, at least in the optical. An extensive comparison to libraries of spectral types confirms both stars are M3 giants. This comparison also suggested that the M3/4 giant boundary is associated with the strong onset of TiO bands. The study also yielded E(B-V) values of $0.05\pm0.02$ for EG And and $0.32\pm0.02$ for HD148349 and a new distance to EG And of  $568\pm41$pc. EG And was found to be $\sim$10\% as bright as HD148349 in the optical.

The \ion{Mg}{ii} \emph{h} and \emph{k}  lines were used to show the behavior of the wind. Although mutilated by the ISM, terminal wind velocities of $\sim$70 km s$^{-1}$ and $\sim$15 km s$^{-1}$ were found for EG And and HD148349. The terminal wind velocity for HD148349 is similar to other M3 giants; $\alpha$ Aur, $\sigma$ Lib and $\mu$ Gem (18, 11 and 14 km s$^{-1}$ respectively). The higher velocity for EG And is indicative of binary stars (this will be discussed in Chapter 5).

To compare the chromospheres of the two stars the radiative losses can be investigated. Following the estimates of \citet{judge_stencel_1991_chromowinds} that \ion{Mg}{ii} \emph{h} and \emph{k}  losses account for $\sim$25\% of the total losses from cool giant chromospheres (other energy losses will occure from sources such as H Ly$\alpha$). The flux of the lines in Figure \ref{fig:mgii_asymm} can be used to calculate the total chromospheric radiative losses:
\begin{equation}
F_{\star}(\mbox{tot}) \simeq 4\times F_{\star}(\mbox{\ion{Mg}{ii}})
\label{eqn:f_chromo}
\end{equation}
where the surface flux of the \ion{Mg}{II} lines are related to the measured flux in Figure \ref{fig:mgii_asymm} by:
\begin{equation}
F_{\star}(\mbox{\ion{Mg}{ii}}) =  F_{\oplus}(\mbox{\ion{Mg}{ii}})\left(\frac{2}{\theta} \right)^2
\label{eqn:f_surface}
\end{equation}
where $\theta$ is the angular diameter of the star in radians. Angular diameters of $2.3\times10^{-8}$ and $6.2\times10^{-9}$ radians (4.85 and 1.27 milliarcseconds) for HD148349 and EG And were used (From Tables \ref{tab:target_params} and \ref{tab:target_params_hd}). Both targets were extinction-corrected and the measured flux of the \ion{Mg}{ii} \emph{h} and \emph{k} lines for HD148349 was $1.6\times10^{-11}$ erg cm$^{2}$ s$^{-1}$. The EG And \ion{Mg}{ii} lines were more difficult to measure. Using the same scaling and folding process that was used to determine the widths of the line, a flux of $1.9\times10^{-12}$ erg cm$^{-2}$ s$^{-1}$ was estimated for the fullest eclipse observation of EG And. When these angular diameters and fluxes were substituted into Equations \ref{eqn:f_chromo} and \ref{eqn:f_surface}, the total radiative losses of the chromosphere were estimated to be  $\sim 4.7\times10^{5}$ erg cm$^{-2}$ s$^{-1}$ for HD148349  and $\sim 7.9\times10^{5}$ erg cm$^{-2}$ s$^{-1}$ for Eg And. These values compare well to measurements by \citet{judge_stencel_1991_chromowinds} for similar M3III stars who show that $\pi$ Aur and $\mu$ Gem have total chromospheric radiative losses of  $2.0\times10^{5}$ and $1.3\times10^{5}$ erg cm$^{-2}$ s$^{-1}$, respectively. The mutilation of EG And's \ion{Mg}{ii} line profile and the subsequent correction value may have resulted in a slight overestimate of $F_{\star}(\mbox{tot})$ value.

New values of extinction, confirmation of spectral subclass, and refined astrometric values have provided a greater understanding of both EG And and HD148349 and their place among the general red giant population. Although this work has shown EG And's suitability for red giant study, the presence of the white dwarf cannot be ignored and will be considered in Chapter 5.

\chapter{EG Andromedae Wind Structure}
\label{chapter:results_egand_wind_structure}

In this chapter, a detailed analysis of the atmosphere of EG And, specifically the base conditions of the chromosphere is presented. This analysis is carried out using {\sl HST}/STIS observations of the system at various stages of eclipse spanning two {\sl HST} Observing Cycles. The importance of the \ion{C}{ii}] 2325\AA\ multiplet is discussed and the technique used to fit the lines is described in detail. Additionally, \ion{Al}{ii} features are also fitted to build up an ionisation picture of the symbiotic atmosphere. The results of those fits - emission line fluxes, line ratios, radial velocities, electron densities and opacity diagnostics are provided and discussed in the context of the physics of the emitting medium. Efforts to model the system are discussed and conclusions on the nature of the system are offered. The aim of this work is to try and identify whether the red giant chromosphere of EG And behaves similarly to an isolated giant star at any stage of its orbit, or whether the white dwarf distorts the giant's atmosphere to a significant extent. As described in Chapter 1, the advantages of using a symbiotic binary system holds potential for understanding giant star atmospheres and mass-loss. Extra significance is added to this work due to the finite lifetime of {\sl HST} and the lack of a natural replacement instrument meaning that this UV-based spectroscopic approach faces an uncertain future. 

\section{\ion{C}{ii]} 2325\AA\ Multiplet}

The boron-like \ion{C}{ii]} intercombination ($2s^{2}2p^{2}P^{0} -2s2p^{2\ 4}P$) UV 0.01 emission line multiplet at 2325\AA\  is the most important electron density ($n_e$) diagnostic in stellar chromospheres.  The emission lines of the multiplet are generated by electronic collisional excitation followed by radiative de-excitation. The multiplet is contained within a 5\AA\ range which lessens the effect of differential absorption due to dust. The narrow wavelength region also reduces the influence  of detector and spectrographic variations that become important over larger wavelengths. The line ratios are sensitive to changes in electron density ($n_e$) in the range $10^7  - 10^9$\ cm$^{-3}$ but are not sensitive to changing electron temperature because the collisional de-excitation rate is proportional to $T_{e}^{-\frac{1}{2}}$ \citep{harper_2006}. The diagnostics line ratios are shown in Table \ref{table:line_ratio_diagnostics}. Section \ref{sec:atomic_physics} describes in detail the atomic physics that gives rise to density-sensitive emission line diagnostics.

\citet{stencel_brown_carpenter_1980} and \citet{brown_ferraz_jordan_1981} offer some of the first attempts to diagnose the chromospheric electron density of a red giants. \citeauthor{stencel_brown_carpenter_1980} utilised {\sl IUE} observations of four cool giant and supergiant stars to determine chromospheric conditions of electron density and temperature. The importance of the \ion{C}{ii]} 2325\AA\ multiplet as a diagnostic in late-type stars was highlighted in \citet{stencel_linsky_brown_carpenter_1981} where a need for higher resolution observations and improved atomic data for the \ion{C}{ii} transitions was also stressed. Collision strengths were improved upon \citep{hayes_nussbaumer_1984_1, hayes_nussbaumer_1984_2} and a dichotomy emerged when comparing coronal stars to noncoronal stars \citep{brown_carpenter_1984}, as the emitting \ion{C}{ii} material appeared to come from a hotter region in coronal stars. \citet{carpenter_brown_1985} extended the method of using \ion{C}{ii} emitting lines to estimate the geometric extent of the chromospheric emission region, again making a distinction between coronal and noncoronal giants.  Further improvements to the atomic data were made by \citet{lennon_1985} and were implemented by \citet{byrne_1988} to show that previous values of electron density might have been too low on average by $0.75dex$ and the calculated geometric extent of the emitting region may have been too large. The new calibration showed better agreement with observed solar values. \citet{schroeder_reimers_carpenter_1988} discussed the \ion{C}{ii} 2325\AA\ diagnostic (incorrectly labelled ``2355\AA'' in that paper title) in terms of eclipsing binary $\zeta$ Aur systems, finding $n_e \sim 10^8$\ cm$^{-3}$. They found that while \ion{C}{ii} emission ratios are reliable for providing electron densities and geometrical extensions of chromospheres, they cannot be used to determine electron temperatures and ionization degrees. \citet{judge_1990} investigated the formation of emission lines in the UV region and urged caution when using the \ion{C}{ii} 2325/1335 ratio as a temperature diagnostic as the lines may arise in different parts of the chromosphere. Finally, with the launch of {\sl HST} the use of the \ion{C}{ii}] 2325\AA\ diagnostic was aided by higher resolution and high S/N data with good noise characteristics. GHRS provided high resolution observations of the \ion{C}{ii}] multiplet for  $\alpha$ Tau (K5III) \citep{carpenter_robinson_1991, judge_1994}. \citet{judge_carpenter_1998_basal_flux} used GHRS observations of the \ion{C}{ii}] multiplet in stars with outer sub-photospheric convection zones to show that the `basal' heating of  chromospheres could not be caused by acoustic wave propagation as had previously been claimed by \citet{ulm_1994}.  The full width at half maximum (FWHM) of the \ion{C}{ii}] lines in their study was  typically 20 - 25\ km\ s$^{-1}$ for cool giants with, for example, $n_e \sim 10^9$\ cm $^{-3}$ for $\gamma$ Cru (M3III). The succession of GHRS with STIS allowed even more detailed studies using the \ion{C}{ii}] multiplet, such as spatial scans of $\alpha$ Ori by analysed \citet{harper_2006} to  reveal the radial gradients of the electron density and turbulence.

\begin{figure}[ht!]
\centering
\includegraphics[width=\textwidth]{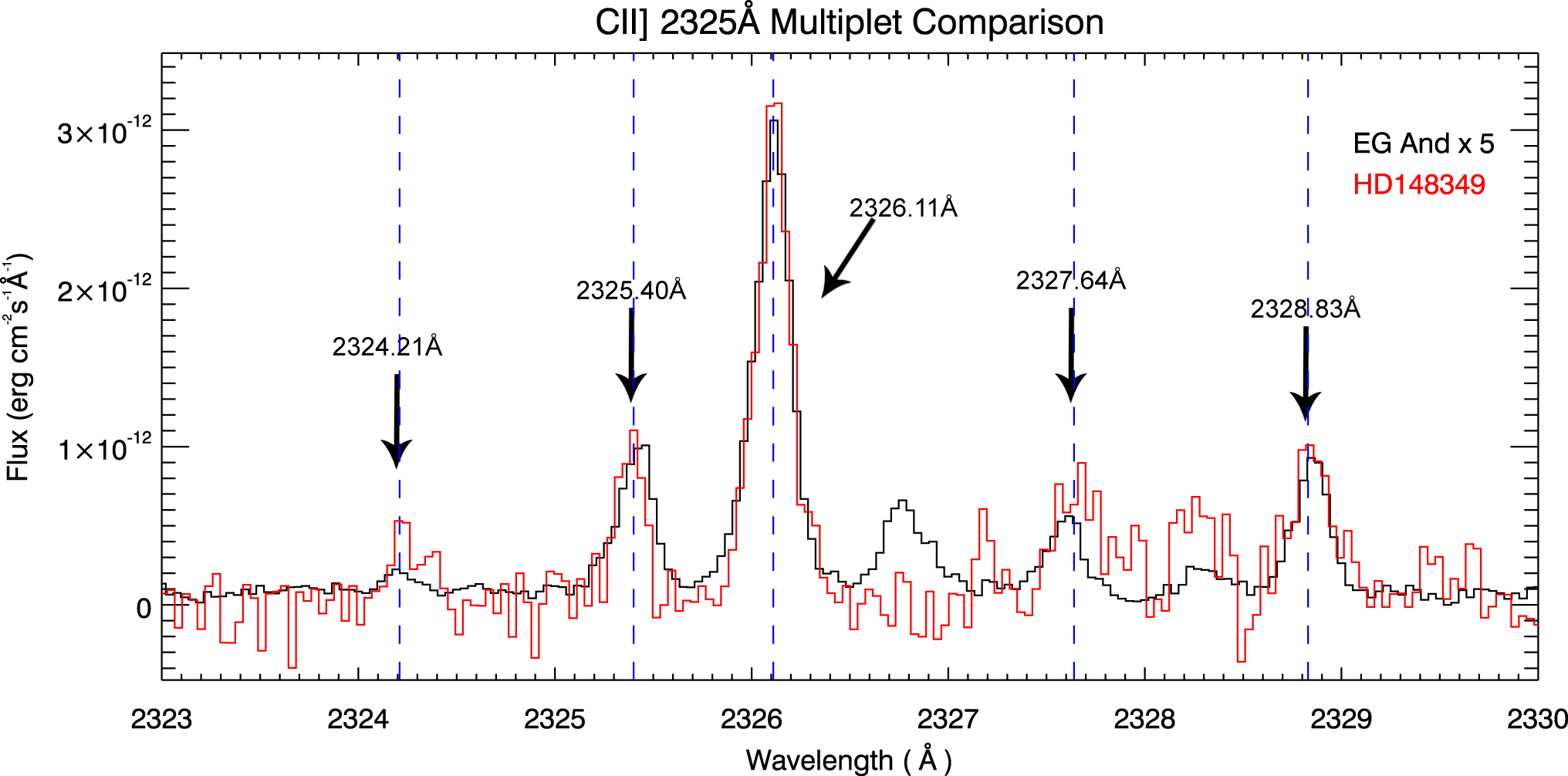}
\caption[\ion{C}{ii]} 2325\AA\ Multiplet Comparison]{{\sl HST}/STIS E230M observations of the \ion{C}{ii]} 2325\AA\ multiplet for EG And (black) and HD148349 (red). The HD148349 observation was taken on the 2nd September 2009. The locations of the five multiplet lines are marked by blue dashed lines. The EG And observation was taken during the eclipse of the white dwarf ($\phi=0.038$) on the 22nd December 2002, and should be dominated by the red giant's chromospheric emission. It has been scaled up by 5 for comparison with HD148349. The general similarity of the lines suggests that the emitting regions are broadly similar in nature, though there are differences in terms of the additional emission features between the multiplet components. Both stars have been corrected for interstellar extinction. HD148349 was corrected using E(B-V)=0.33, while EG And was corrected using E(B-V)=0.05.
\label{fig:cii_2325_egand_hd_comp}}
\end{figure}

Figure \ref{fig:cii_2325_egand_hd_comp} shows the multiplet in both EG And and HD148349. While the multiplet is less affected by blends than other diagnostics (i.e.\ \ion{Si}{ii]}) some of the emission lines are occasionally mutilated by \ion{Fe}{ii} and \ion{Co}{ii} line-blending. This will be taken into consideration in the analysis.

\begin{table}[ht!]
\caption[\ion{C}{ii]} Line Ratio Diagnostics]{\ion{C}{ii}] line ratios and diagnostics. The line vacuum wavelengths are from \citet{nist_book_values}. Unless otherwise stated, particle densities are always lowercase (e.g.\ $n_H$) and column densities are uppercase (e.g.\ $N_H$).}              
\label{table:line_ratio_diagnostics}      
\centering                                      
\begin{tabular}{c c c }          
\hline\hline                        
Ratio & Wavelength (\AA) & Diagnostic \\    
\hline                                   
R$_1$ & 2326.112 / 2328.837  & $n_e$ \\ 
R$_2$ & 2326.112 / 2327.644  & $n_e$ \\ 
R$_3$ & 2325.403 / 2327.644  & $n_e$ \\ 
R$_4$ & 2327.644 / 2324.214 & $N_{\mbox{\ion{C}{ii}}}$ \\ 
R$_5$ & 2328.837 / 2325.403  & $N_{\mbox{\ion{C}{ii}}}$ \\ 
\hline                                             
\end{tabular}
\end{table}

\section{MELF}\label{sec:melf}

A method of fitting the \ion{C}{ii]} 2325\AA\ multiplet in the E230M observations of HD148349 and EG And (Figure  \ref{fig:cii_2325_egand_hd_comp}) was developed. This method, referred to hereafter as MELF (\emph{\textbf{M}ultiplet \textbf{E}mission \textbf{L}ine \textbf{F}itting}), utilised MPCURVEFIT, a subroutine of the MPFIT IDL package \citep{markwardt_2009_mpfit}, to model and fit the emission line components. Unless otherwise noted, a gaussian profile was used to fit the emission lines. The modelled line fluxes were then used to estimate chromospheric characteristics. MELF required a purpose-built fitting procedure constructed to test the behaviour of the fitting routines and the whole fitting process in the presence of blending and different amounts of noise. The following items are supplied to MELF to fit a model to the STIS observations:

\begin{itemize}
\item An array of independent variable values (``X''), in this case the wavelength array associated with the emission feature.
\item An array of dependent variable  values (``Y''), the STIS measured flux values.
\item An array of weighting values (``WEIGHTS''), the STIS flux error values.
\item An IDL procedure which generates a function to calculate the Y values given X (``FUNC''), here the model can be defined as any function but usually based on some gaussian shape.
\item Initial guesses for all the parameters (``P'') called in the user-supplied FUNC.
\end{itemize}

Initial guesses of all the parameters were fed into MELF and for each set of parameters a model function was constructed using the following inputs:
\begin{enumerate}
\item X, the wavelength array.
\item P, the parameters of the model being adopted (See Equation \ref{eqn:melf_func} below).
\item FMOD, the variable array that will be the returned holding the computed model values at X.
\end{enumerate}

To fit all five emission lines of the multiplet simultaneously the function was given the form:
\begin{equation}
\label{eqn:melf_func}
\mbox{FMOD}(x) = \sum_{i=1}^{15}(P[i]e^{\frac{{(x - P[i+1])}^2}{2P[i+2]^2}}) + P[16]x  + P[17]
\end{equation}

In Equation \ref{eqn:melf_func} there are 17 parameters that can be varied in MELF. The first 15 of theses correspond to the values defining the gaussian shapes of the five emission lines (P[1]:P[15]). P[16] and P[17] define the continuum shape. If any of the parameters need to be constrained this can be done using the PARINFO keyword. At the end of the process the parameters that result in the best fit are returned, along with a measure of the best fit, $\chi ^2$.

\begin{figure}[ht!]
\centering
\includegraphics[width=\textwidth]{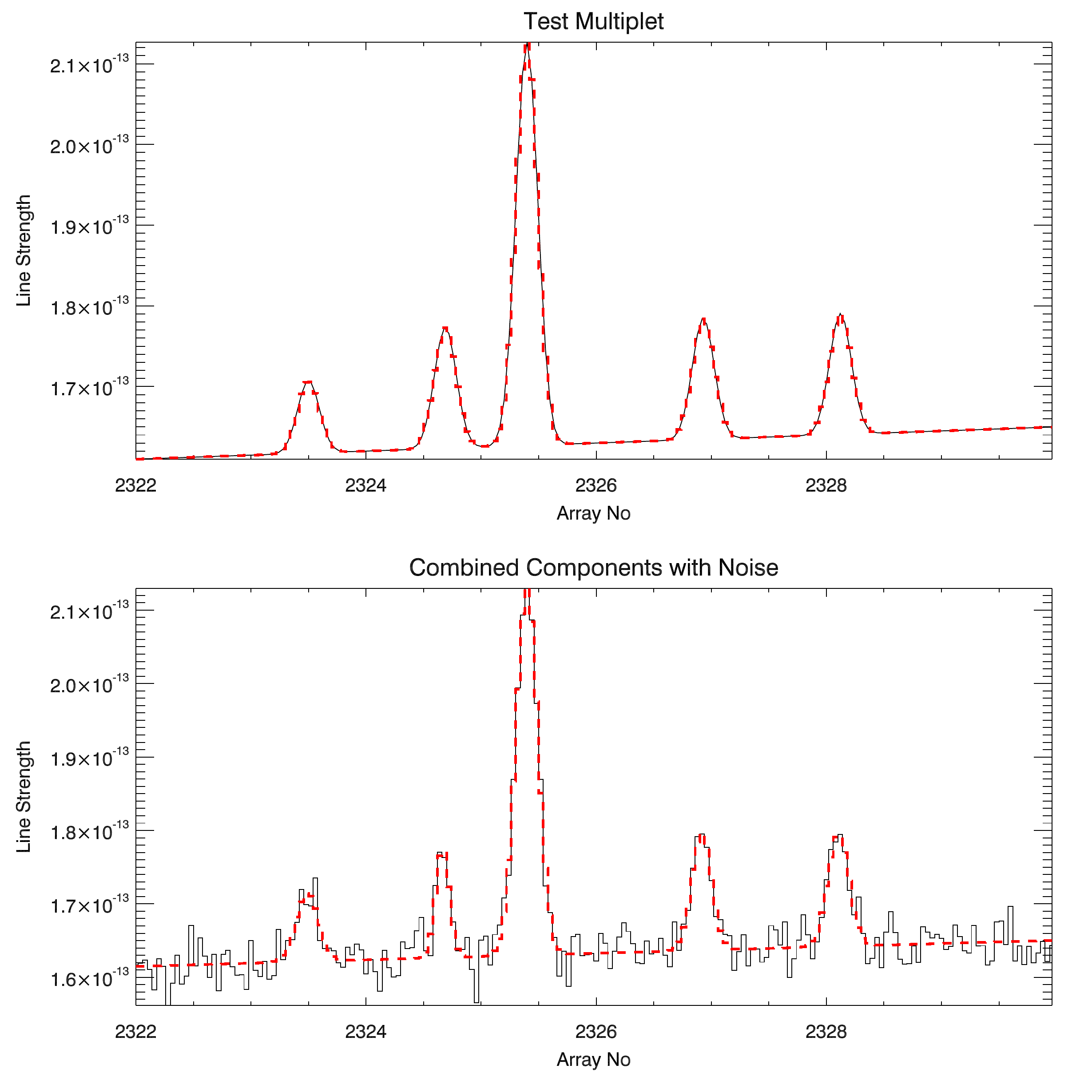}
\caption[MELF Test Fitting]{
In the top panel a simulation of an emission line multiplet is shown in solid black. Overplotted in dashed red is the MELF best fit. In the bottom panel, simulated data with a background signal-to-noise ratio of 4 is shown in black. The MELF best fit is again shown in dashed red. While the quality of the fit was reduced slightly by the addition of the noise to the model, it can be seen that MELF will still reliably produce a good quality fit.
\label{fig:mpfit_test_multiplet_data}}
\end{figure}

Several tests were carried out initially to measure the fitting strengths of the procedure. The idea was to generate a model of similar resolution and noise as would be expected in STIS data. Knowing the value of the parameters used to create the models it would be possible to measure the robustness of the MELF  fitting technique by comparing the best fit values returned. A multiplet with 5 emission lines was constructed with simulated wavelength and flux values close to those seen in the STIS data. The resolution was simulated to be around 0.04\AA\ per pixel. The top panel of Figure \ref{fig:mpfit_test_multiplet_data} shows the simulated data with the MELF best fit overplotted in red. Noise was then added to correspond to a signal-to-noise ratio of 4 (similar to the STIS SNR) and the procedure was again tested, as shown in the bottom panel of Figure \ref{fig:mpfit_test_multiplet_data}. 

Noise was added to the spectrum using:
\begin{equation}
I(\lambda)=I(\lambda)+\mbox{randomn}(\mbox{seed},\mbox{n\_elements}(\lambda))\times \frac{I(\lambda)}{S/N\times \sqrt{\frac{I(\lambda)}{\sigma}}}
\end{equation}
where $I(\lambda)$ was flux, $S/N$ was the desired signal-to-noise ratio of the noisy spectrum and $\sigma$ was the mean background continuum value. The IDL \textbf{randomn} function returns gaussian-distributed numbers with a mean of zero and a standard deviation of one, using the Box-Muller method \citep{numerical_recipes}. These values were then scaled by a value defined by the $S/N$ keyword to represent the signal-to-noise typically observed in the STIS E230M continuum. Where the signal increases to $N$ times the continuum level, the signal-to-noise ratio will improve by $\sqrt{N}$.  

MELF proved capable of obtaining fit parameters which matched closely the original parameters, even in the presence of a relatively large amount of noise. Reliable `by-eye' estimates were required for the procedure. If these initial guesses were replaced with random values, MELF struggled to find a good fit. In the above described test, MELF was given 17 input parameters which were allowed to vary to find the best fit. While some parameter constraints could be imposed on the fits to real data, for the purpose of testing no constraints were imposed. For each of the 5 emission lines there were parameters defining the location, width and strength of the line. This accounted for 15 of the 17 parameters. The remaining 2 parameters defined the slope and the offset of the continuum. In the test in the top panel of Figure \ref{fig:mpfit_test_multiplet_data} the percentage differences between the parameters of the simulated model and the MELF best fit parameters were less than 1\% for all 17 parameters. When noise was added in the bottom panel, the percentage difference between model and fit parameters increased but remained below 1\% apart from the 5 line strength parameters which differed by a maximum of 1.3\%.

\section{Error Masking and Fitting}\label{sec:error_mask_chi}

\begin{figure}[ht!]
\centering
\includegraphics[width=\textwidth]{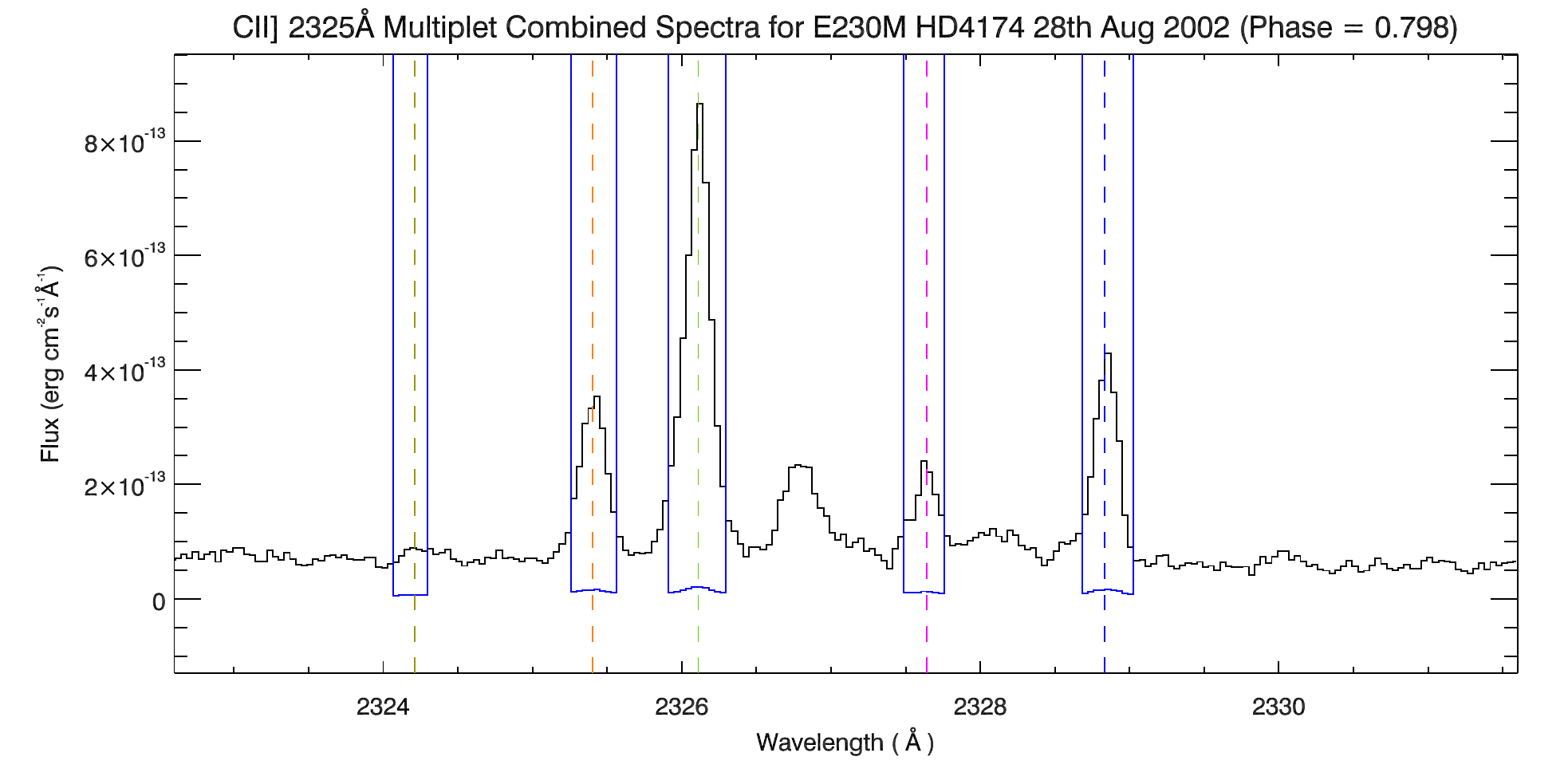}
\caption[MELF Masking Using Error Array]{An observation of EG And with the masked error array over-plotted in blue. All regions other than the narrow core of the emission lines have an error value sufficiently high to be given negligible weighting in the fit. The unmasked region is necessarily narrow as some of the broadened wings of the lines were asymmetric due to noisy data (and/or extra emission discussed in Section \ref{sec:rad_vel}) and could hamper the fitting. The point spread function of the telescope also contributed to a broadening at the base of the lines. The gaussian shape of the fitting function produced a good fit to most of the symmetric wings even though that part of the data was masked (see Figure \ref{fig:mpfit_multiplet}).
\label{fig:error_402010_402020}}
\end{figure}

The best-fit model of the multiplet emission was generated by simultaneously fitting all five of the \ion{C}{ii}] lines. To ensure that the absorption features and noise between the lines did not reduce the quality of the fit, these regions were masked from the fitting process. This was done by setting the entire error array to a sufficiently high value except for the regions encompassing the narrow emission lines. These narrow regions kept their original error values. By setting the error array to a high value elsewhere these regions are effectively given no weighting in the fit. Figure \ref{fig:error_402010_402020} shows  the error array overplotted on a masked region. It can be seen that outside the emission lines the error values rise drastically. To ensure the error value was set to a high enough value, the $\chi^2$ statistic was calculated for the whole region and compared to the value calculated from solely looking at the narrow emission line regions. Both calculations produced the same result proving that manually setting the error of the surrounding regions to a high value renders them negligible in the fitting process.

MELF calculates the goodness-of-fit from the $\chi^2$ statistic - the summed total of the square of the weighted residuals. In the line-fitting code the value was calculated from:
\begin{equation}\label{eqn:chisquared}
\chi^2 = \sum_{\lambda} \left( \frac{I(\lambda) - I_{\mbox{\tiny{FIT}}}(\lambda)}{\sigma(\lambda)}\right)^2
\end{equation}
In this case $I(\lambda)$ was the flux array for the region of the spectrum that is being fitted, $I_{\mbox{\tiny{FIT}}}(\lambda)$ was the fitted flux array and $\sigma(\lambda)$ was the flux uncertainty for the observation. The best fit was then taken as the set of parameters which produced the lowest value using Equation \ref{eqn:chisquared}. To see the quality of the fit the ``reduced'' $\chi^2$ statistic can be calculated by dividing the best-fit $\chi^2$ statistic by the number of degrees of freedom. The degrees of freedom was given by subtracting the number of fitting parameters from the number of data points in the fit. The error-masked data points were not counted as they are ignored in the fitting process, leaving $\sim30$ data points. These calculations follow the guidelines suggested in \citet{numerical_recipes} and \citet{bevington_data_reduction}. 

\begin{figure}[ht!]
\centering
\includegraphics[width=\textwidth]{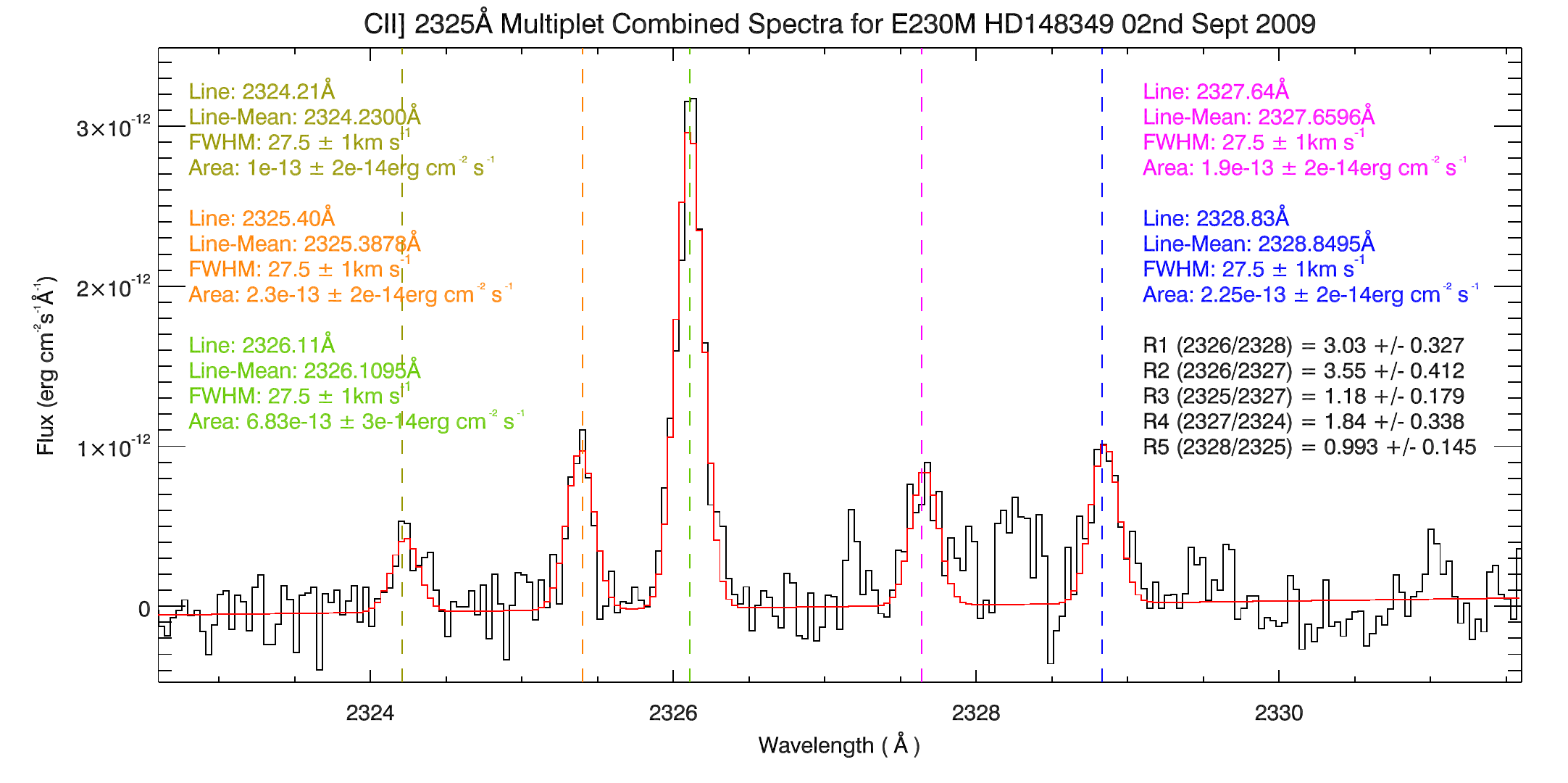}
\caption[MELF Fit to HD148349]{E230M {\sl HST}/STIS observation of HD148349 on the 2nd of September 2009. The \ion{C}{ii]} 2325\AA\ multiplet is shown by the solid black line, with the best fit overplotted in red. The data has been dereddened using E(B-V) = 0.33. \label{fig:mpfit_multiplet}}
\end{figure}

After testing the fitting procedure for cases similar to the data, the procedure was extended to the E230M {\sl HST}/STIS observation of HD148349 on the 2nd of September 2009. The model defined as the user-supplied input function for MELF was built around Equation \ref{eqn:melf_func} with 5 emission lines.  The slope and offset values of the continuum (parameters P[3] and P[4] in Equation \ref{eqn:melf_func}) were chosen by fitting a straight line between featureless regions of the spectrum either side of the \ion{C}{ii]} 2325\AA\ multiplet. This constraint was added as it speeded up the computational times and eliminated the codes tendency to occasionally choose wildly inaccurate continuum values. One of the advantages of MELF calling the MPCURVEFIT routine is the selection of fitting options available to the user through parameter constraints. The fitting parameters can be fixed to a certain value, bound between certain limits, or parameters can be tied to one another. The desired constraints are fed into a structure called PARFINO which is passed as a keyword into the code. Extensive testing with the available constraints led to the best fit shown in Figure \ref{fig:mpfit_multiplet}. The wavelengths of the multiplet emission lines are reasonably well known, however it was found that fixing these values over-constrained the fit. Instead, the location of the line centres was determined by permitting narrow limits on the parameters to take account of the radial velocity variations with orbit, and also to take account of any small offsets due to additional motion of the emitting material. The wavelength limits imposed gave a range of $\pm$0.02\AA. In the case of the line widths, the parameters were tied such that the same line width was used for all lines of the multiplet. This resulted in effectively one line width value that was allowed to vary to find the best-fitting width for all five emission lines, which is a fair assumption for optically thing lines like  \ion{C}{ii]}. The fluxes of the individual lines were unconstrained. The continuum slope and offset were fixed to the values from the straight-line fit mentioned above.

\section{Issues with the 2324 and 2327\AA\ Lines}\label{sec:prob_lines_3sig}

These two lines proved the most problematic throughout the MELF process. The 2327.64\AA\ line is particularly difficult to fit. Due to the presence of underlying absorption (\ion{Fe}{ii}) 2328.11\AA\ (see Section \ref{sec:absorption_component}) the \ion{C}{ii]} line suffers mutilation by absorption, especially close to eclipse. There is also a marked increase in the amount of absorption in the recent observations with the 2009 $\phi=0.070$ and $\phi=0.078$ observations demonstrating particularly high absorption (See plots in Appendix \ref{app:cii_2325_multiplet_fits}). These epochs also show higher-than-expected continuum levels, so the observed changes may be related, and these data need to be considered separately. One possibility is that the higher continuum is a reflection of the increased luminosity of the white dwarf which, consequently, has affected the physical conditions in the outer atmosphere and wind. Another possibility is distortion of the red giant wind, and support for this comes from the slightly different radial velocities found from the more recent data (See Section \ref{sec:rad_vel}). 

To get an estimate of the flux in these mutilated lines, the range of wavelengths that the MELF process fits is manually restricted by use of the error-masking technique described in Section \ref{sec:error_mask_chi}. Although this helps to estimate the line flux (sometimes using only a portion of the blueward, unabsorbed wing of the line and the line-width parameter constrained by the other lines) the errors are necessarily high after this process. As a result it is difficult to place faith in diagnostic ratios that are dependent on this value i.e.\ $R_2$ and $R_3$ (See Table \ref{table:line_ratio_diagnostics}). The $R_4$ opacity diagnostic is also ignored as it relies on the 2324.21\AA\ \ion{C}{ii]} line. While this line does not suffer from obvious absorption mutilation, the weakness of the line makes it less reliable. Typically this line comprises only 2-5\% of the flux that is seen in the strongest 2326.11\AA\ line. In the $\phi=0.497$ observation the high level of continuum flux swamps the 2324.21\AA\ line, so the MELF process struggles to obtain a fit. Although a minimum in $\chi^2$-space was obtained, this value is not trustworthy as there seems to be extra emission during the  observation \citep{cian_thesis}. An upper-limit was also calculated to see how much flux from the 2324.21\AA\ line could be realistically hidden in the spectrum. To calculate this amount of flux, a series of model lines were produced and the $\chi^2$ statistic was calculated for each one.  

\begin{figure}[ht!]
\centering
\includegraphics[width=\textwidth]{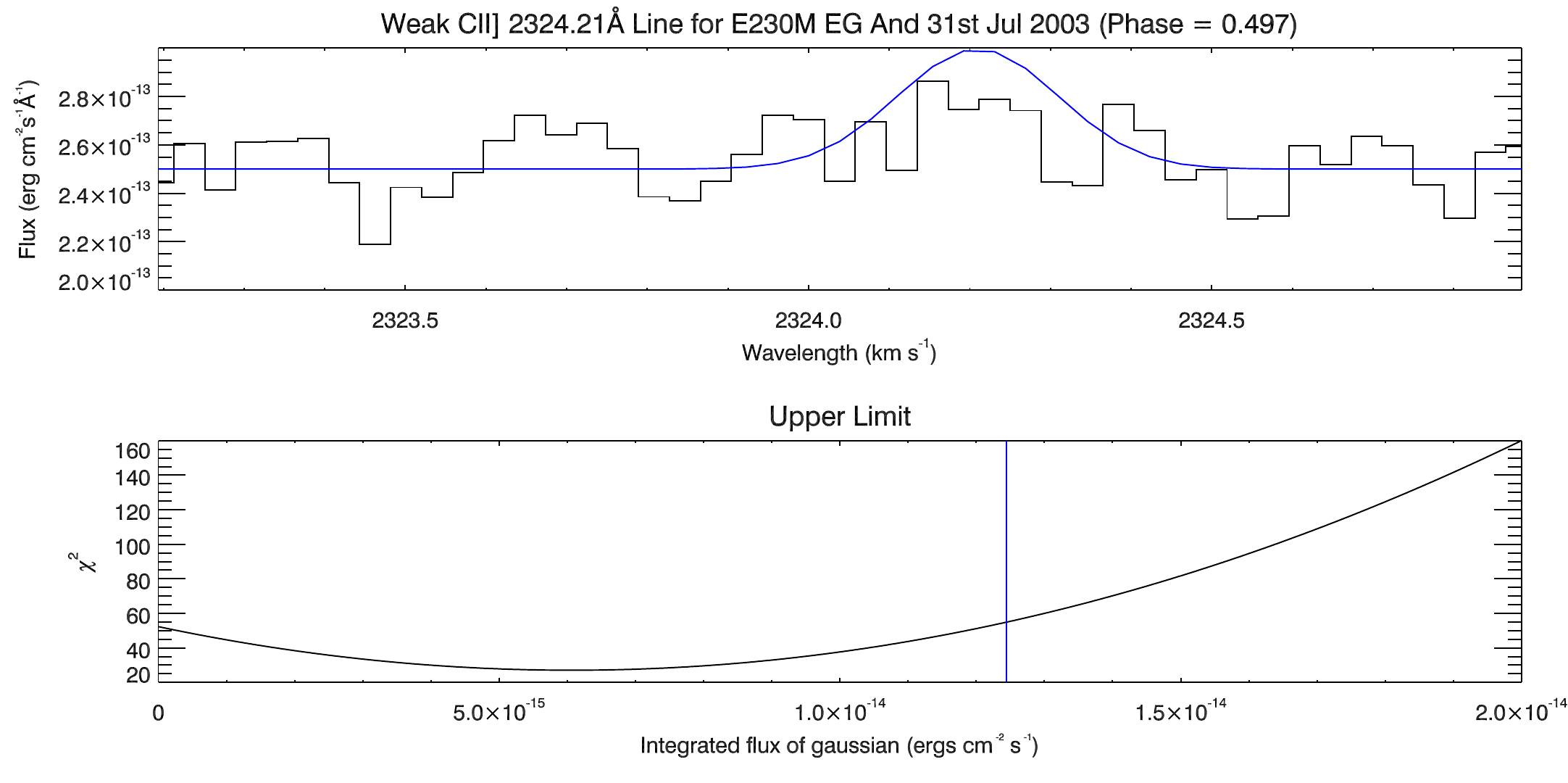}
\caption[Upper-Limit on Weak \ion{C}{ii]} 2324.21\AA\ line]{Upper-limit for the weak \ion{C}{ii]} 2324.21\AA\ line. The top panel shows the weak 2324.41\AA\ line in the  observation of EG And on the 31st July 2003. As the location and width of the gaussian are already known from the other \ion{C}{ii]} lines, all that is left to find is the area. For each successive gaussian the $\chi^{2}$ statistic is calculated to find the best fit to the data. Overplotted in blue is the gaussian which produces a $\chi^2$ value of +2.71 greater than the zero flux model (the case of no line existing). This value is shown as the solid vertical line on the bottom plot which corresponds to the upper limit on the line flux ($1.2\times10^{-14}$erg\ cm$^{-2}$\ s$^{-1}$). \label{fig:sig3_401010_401020}}
\end{figure}

Figure \ref{fig:sig3_401010_401020} shows how gaussians were produced from the known mean (line location) and sigma with only the area of the line varying. The line location is known in wavelength space and the width of the line is known from the other lines in the spectrum. The models were calculated from a zero-flux reference point up to a gaussian of $2\times10^{-14}$\ erg\ cm$^{-2}$\ s$^{-1}$. To select the location of the zero-flux reference, the median of the two regions of $\sim1$\AA\ either side of the proposed feature location was calculated and this constant was used as the zero-flux level. The median was chosen instead of the mean to negate the effect of single bad pixels,  as outlined in \citet{der_snr}. An upper-limit was found by selecting the flux that was closest to $\chi^2_{ref}+ 2.71$ value, where $\chi^2_{ref}$ was the value for the zero flux model (the case where no lines exists).  2.71 is the value in a $\chi^2$ distribution that gives 90\% confidence for one degree of freedom \citep{chi_table}. In this case the upper-limit was calculated to be $1.2\times10^{-14}$erg\ cm$^{-2}$\ s$^{-1}$.

\section{Emission Line Fluxes}

The fits for all 11 {\sl HST}/STIS E230M observations of EG And from Cycle 11 and Cycle 17 can be seen in Appendix \ref{app:cii_2325_multiplet_fits}.  For each observation the data have been fit with the MELF technique described in Section \ref{sec:melf}. Each plot provides the best-fit value, together with the relevant error for the main line parameters, as well as the values of the five diagnostic line ratios, together with their errors. Table \ref{tab:cii_2325_lines} shows the integrated line fluxes for all of the observations.

\begin{table*}
\caption[Integrated Fluxes for \ion{C}{ii]} 2325\AA\ Multiplet Lines]{Integrated line fluxes and errors for \ion{C}{ii]} 2325\AA\ multiplet emission lines. These fluxes were measured using the MELF technique described in Section \ref{sec:melf}. All fluxes are quoted in units of $10^{-15}$erg\ cm$^{-2}$\ s$^{-1}$. The first seven rows of the table correspond to Cycle 11 observations while the remaining four are Cycle 17 observations. The $\phi$ column indicates the orbital phase of the observation. The EG And observations have been de-reddened using an E(B-V) value of 0.05, while the HD148349 observation has been de-reddened using a value of 0.33. The asterisk marks an upper-limit where a flux could not be accurately measured. In this case the upper-limit was found by selecting the flux equal to $\chi^2_{ref}+ 2.71$, where $\chi^2_{ref}$ was the value for the case where no lines exists. The date of observation (HJD) is given by JD - 2,450,000. Target names are shortened to E (EG And) and H (HD148349).
\label{tab:cii_2325_lines}}
\centering
\begin{tabular}{cccccccc}
\hline\hline
HJD & Target  & $\phi$ & 2324.21 & 2325.40 & 2326.11 & 2327.64 & 2328.83\\
\hline
     2515 & E  & 0.798 &            5 $\pm$            2 &           94 $\pm$            5 &          249 $\pm$            6 &           51 $\pm$            4 &          108 $\pm$            5 \\
     2564 & E  & 0.899 &            8 $\pm$            2 &           67 $\pm$            4 &          174 $\pm$            4 &           35 $\pm$            3 &           62 $\pm$            3 \\
     2631 & E  & 0.038 &            6 $\pm$            1 &           44 $\pm$            2 &          137 $\pm$            3 &           24 $\pm$            2 &           39 $\pm$            2 \\
     2658 & E  & 0.095 &            4 $\pm$            1 &           45 $\pm$            2 &          133 $\pm$            2 &           28 $\pm$            2 &           43 $\pm$            2 \\
     2677 & E  & 0.133 &            9 $\pm$            1 &           75 $\pm$            3 &          196 $\pm$            3 &           39 $\pm$            3 &           65 $\pm$            2 \\
     2687 & E  & 0.153 &            5 $\pm$            1 &           69 $\pm$            2 &          175 $\pm$            2 &           35 $\pm$            2 &           56 $\pm$            2 \\
     2852 & E  & 0.497 &            12* &          138 $\pm$            6 &          427 $\pm$            8 &           72 $\pm$            5 &          173 $\pm$            7 \\
     5059 & E  & 0.070 &            8 $\pm$            3 &           79 $\pm$            6 &          190 $\pm$            6 &           40 $\pm$           12 &           82 $\pm$            5 \\
     5063 & E  & 0.078 &            9 $\pm$            5 &           65 $\pm$            8 &          180 $\pm$            9 &           41 $\pm$           20 &           78 $\pm$            8 \\
     5109 & E  & 0.173 &            5 $\pm$            4 &          112 $\pm$            7 &          272 $\pm$            9 &           64 $\pm$            6 &          134 $\pm$            7 \\
\hline
     5077 & H  & - &          105 $\pm$           15 &          227 $\pm$           24 &          683 $\pm$           26 &          192 $\pm$           21 &          225 $\pm$           23 \\
\hline     
\end{tabular}
\end{table*}

\begin{figure}[ht!]
\centering
\includegraphics[width=\textwidth]{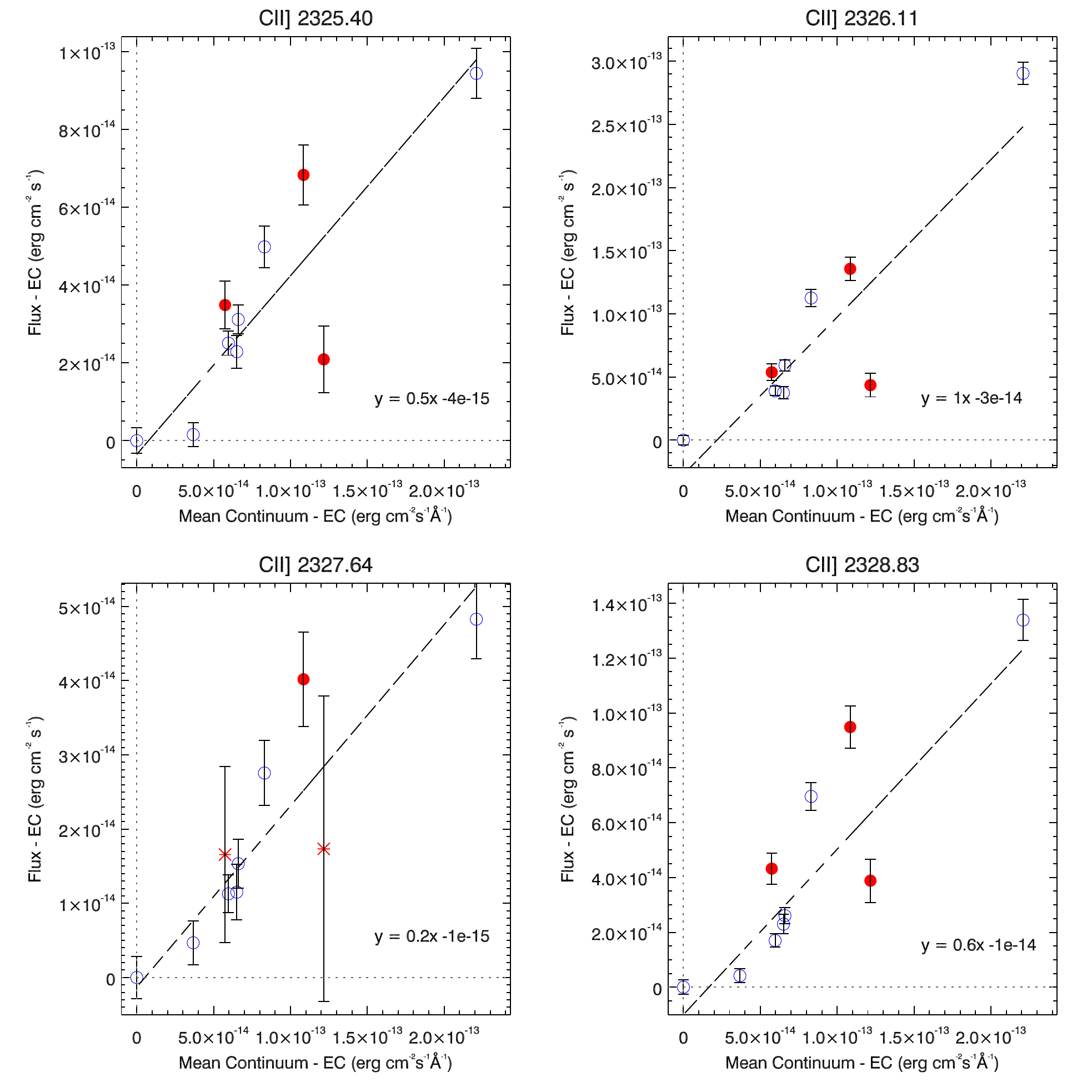}
\caption[Fluxes of \ion{C}{ii} 2325\AA\ Multiplet vs Continuum]{Fluxes of each emission line of the \ion{C}{ii]} 2325\AA\ multiplet against the mean continuum at that region. Each line and continuum value have had the eclipsed component ($\phi=0.038$ phase) values subtracted. In the plot ``EC'' stands for ``Eclipsed Component''. The filled red circles correspond to the Cycle 17 observations while the empty blue circles correspond to the Cycle 11 observations. The red asterisks indicate fluxes that were gathered from poor fits. A fit to the points has been included showing the linear relationship of the two parameters. The slope of the lines can be seen to be different for each fit as would be expected if the electron density is changing. \label{fig:cii2325_combo_lineflux_vs_meancontin_minphase}}
\end{figure}

\begin{figure}[ht!]
\centering
\includegraphics[width=\textwidth]{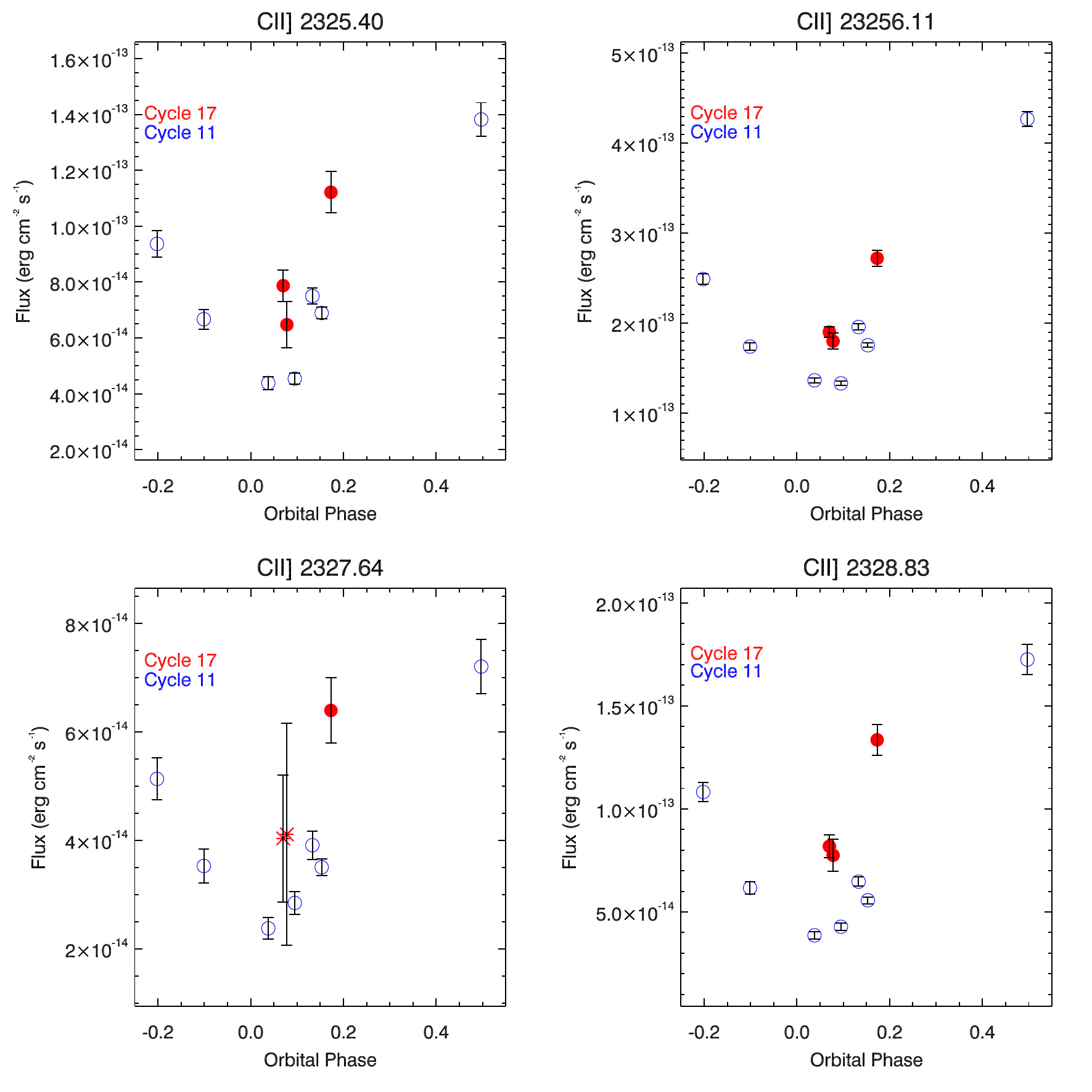}
\caption[Fluxes of \ion{C}{ii} 2325\AA\ Multiplet vs Phase]{Fluxes of each emission line of the \ion{C}{ii]} 2325\AA\ multiplet against orbital phase. The filled red circles correspond to the Cycle 17 observations while the empty blue circles correspond to the Cycle 11 observations. The red asterisks indicate fluxes that were gathered from poor fits.  \label{fig:cii2325_combo_line_fluxes_vs_phase}}
\end{figure}

Figure \ref{fig:cii2325_combo_lineflux_vs_meancontin_minphase} shows how the flux from each of the four strongest lines from the \ion{C}{ii]} 2325\AA\ multiplet varies with mean continuum. The mean continuum in this case has been calculated from emission-free regions either side of the multiplet. In each case, the flux and mean continuum from the most-eclipsed observation ($\phi = 0.038$) have been subtracted. This effectively removes the levels of chromospheric flux that would be expected from the red giant. By analysing the excess flux in the other observations (outside eclipse), the effect of the white dwarf on the system can be assessed. Straight-line fits, which take the error values into account, illustrate how the emission for each line varies with continuum level. While in general the fluxes appear to increase linearly with mean continuum, the higher-state observations (Cycle 17 and $\phi = 0.497$) do not appear to obey the trend as much as the other observations. Figure \ref{fig:cii2325_combo_line_fluxes_vs_phase} shows how the flux from the four strongest lines of the multiplet varies with orbital phase. It can be seen again that both the Cycle 17 observations display higher flux levels than their corresponding orbital phase observation from the earlier cycle. This proves that not only is the white dwarf responsible for excess emission outside of eclipse, but that amount of emission it contributes varies for different observing cycles.

\section{Line Ratio Diagnostics}

\begin{figure}[ht!]
\centering
\includegraphics[width=\textwidth]{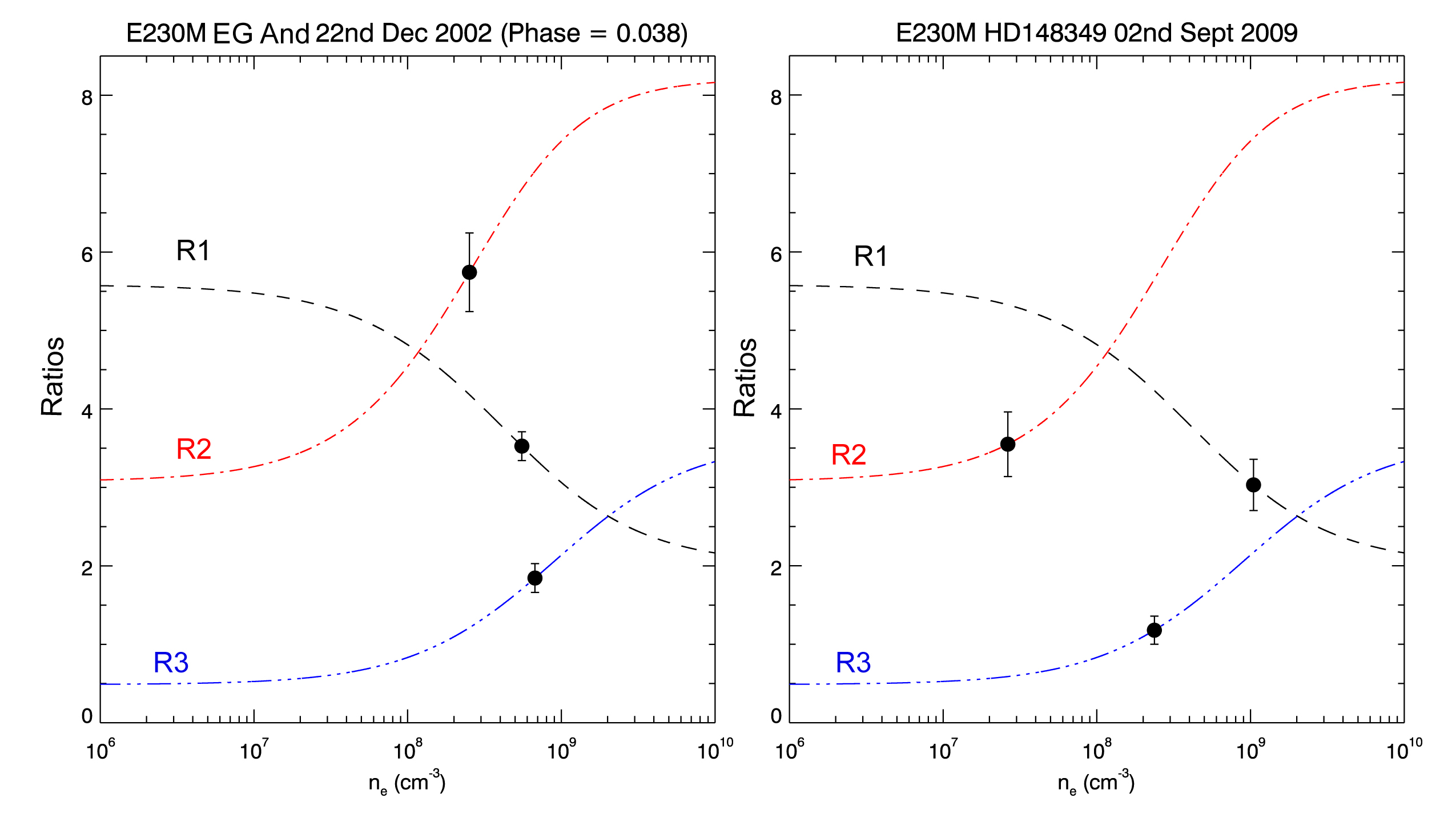}
\caption[R$_1$, R$_2$ and R$_3$ \ion{C}{ii]} Electron Density Diagnostics]{The line ratio (R$_1$, R$_2$ and R$_3$) \ion{C}{ii]} electron density diagnostics for EG And and HD148349. The plot on the left shows the three line ratios for EG And, while HD148349 is shown on the right. In both plots, the line ratios R$_1$, R$_2$ and R$_3$ are shown in black, red and blue respectively.
\label{fig:elec_dens_406010_406020_406030}}
\end{figure}

As noted by \citet{harper_2006} the \ion{C}{ii]}  lines have an advantage over the more commonly used \ion{Si}{ii]} lines because they have lower overall opacity, and hence are useful electron density probes for cool star atmospheres. Table \ref{table:line_ratio_diagnostics} shows that the five line ratios used are R$_{1}$, R$_{2}$ and R$_{3}$ (Electron Density Diagnostics) and R$_{4}$ and R$_{5}$ (Opacity Diagnostics). The three electron density diagnostics are shown in Figure \ref{fig:elec_dens_406010_406020_406030} for EG And and HD148349. The EG And observations give an average electron density of $\sim7\times10^8$\ cm$^{-3}$. As there is only a single HD148349 observation the average value from the three ratios of  $\sim4\times10^8$\ cm$^{-3}$ is less dependable. While the electron density values seem to converge for the three EG And line ratios, there is an obvious spread in values for HD148349. This discrepancy is discussed in \citet{stencel_linsky_brown_carpenter_1981}. More accurate collision strength values and improved $A$-values are suggested by Stencel as ways to help constrain the calculations. In this case the uncertainty in the flux values contributing to the line ratios is the  most likely factor in the disparity. Multiple observations of HD148349 would improve the accuracy. The R$_{4}$ diagnostic ratios the 2327.64\AA\ line (the line of the multiplet that is most affected by absorption) to the weakest 2324.21\AA\ line, the ratio itself is the least reliable of all five of the line ratios, so the R$_5$ ratio is chosen instead, which ratios the less absorbed 2328.83\AA\ and 2325.40\AA\ features.

\begin{figure}[ht!]
\centering
\includegraphics[width=\textwidth]{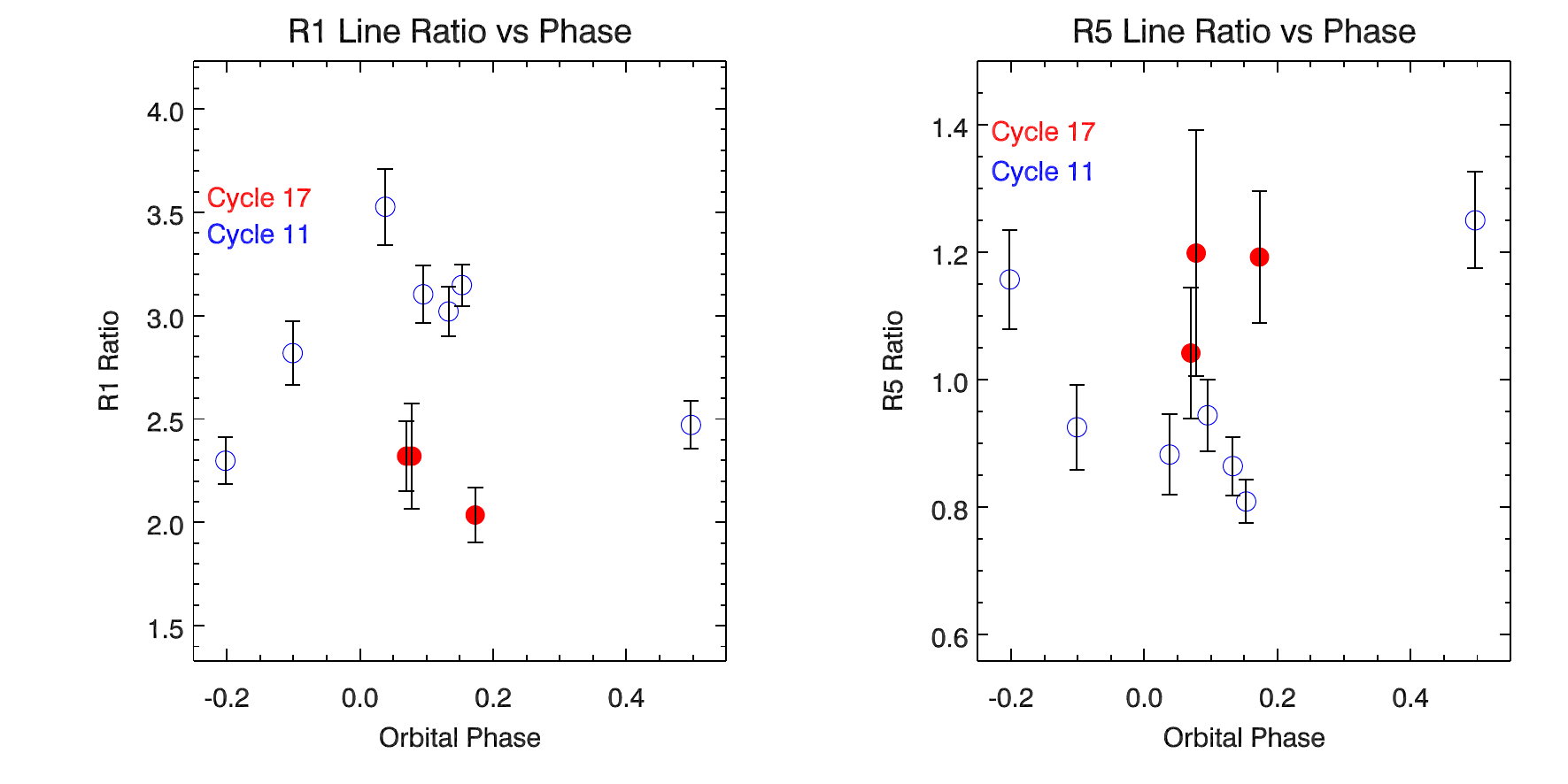}
\caption[R$_1$ and R$_5$ \ion{C}{ii]} Ratios against Phase ]{R$_1$ and R$_5$ line ratios plotted against phase. The R$_1$ ratio on the left shows how the electron density decreases (corresponding to an increasing R$_1$ value) outside of eclipse. The electron density also appears to have increased between observing cycles 11 and 17. The right plot shows how, within error, there is little change in the R$_5$ ratio.  
\label{fig:cii2325_combo_line_ratios_1_5_vs_phase}}
\end{figure}

Figure \ref{fig:cii2325_combo_line_ratios_1_5_vs_phase} shows the line ratios that have the least uncertainty associated with them, R$_1$ and R$_5$. These ratios do not make use of a common transition. Excluding the high-state observations in the right panel of Figure \ref{fig:cii2325_combo_line_ratios_1_5_vs_phase}, there is evidence for a decrease in R$_1$ which implies an overall increase in density outside eclipse. The R$_5$\ plot can be interpreted in different ways. Again, ignoring the high-state observations (the $\phi=0.497$\ data from Cycle 11, together with the Cycle 17 observations), the measurements for phases around mid-eclipse $\pm 0.2$\ do not differ within errors, indicating a uniform medium is observed when the white dwarf is eclipsed.  Compared with the out-of-eclipse (high-state) observations, there is evidence for a decrease in opacity. This indicates  a change in the emitting gas, or a dilution with lower-density material.

\section{Radial Velocities and Asymmetry}\label{sec:rad_vel}

\begin{figure}[ht!]
\centering
\includegraphics[width=\textwidth]{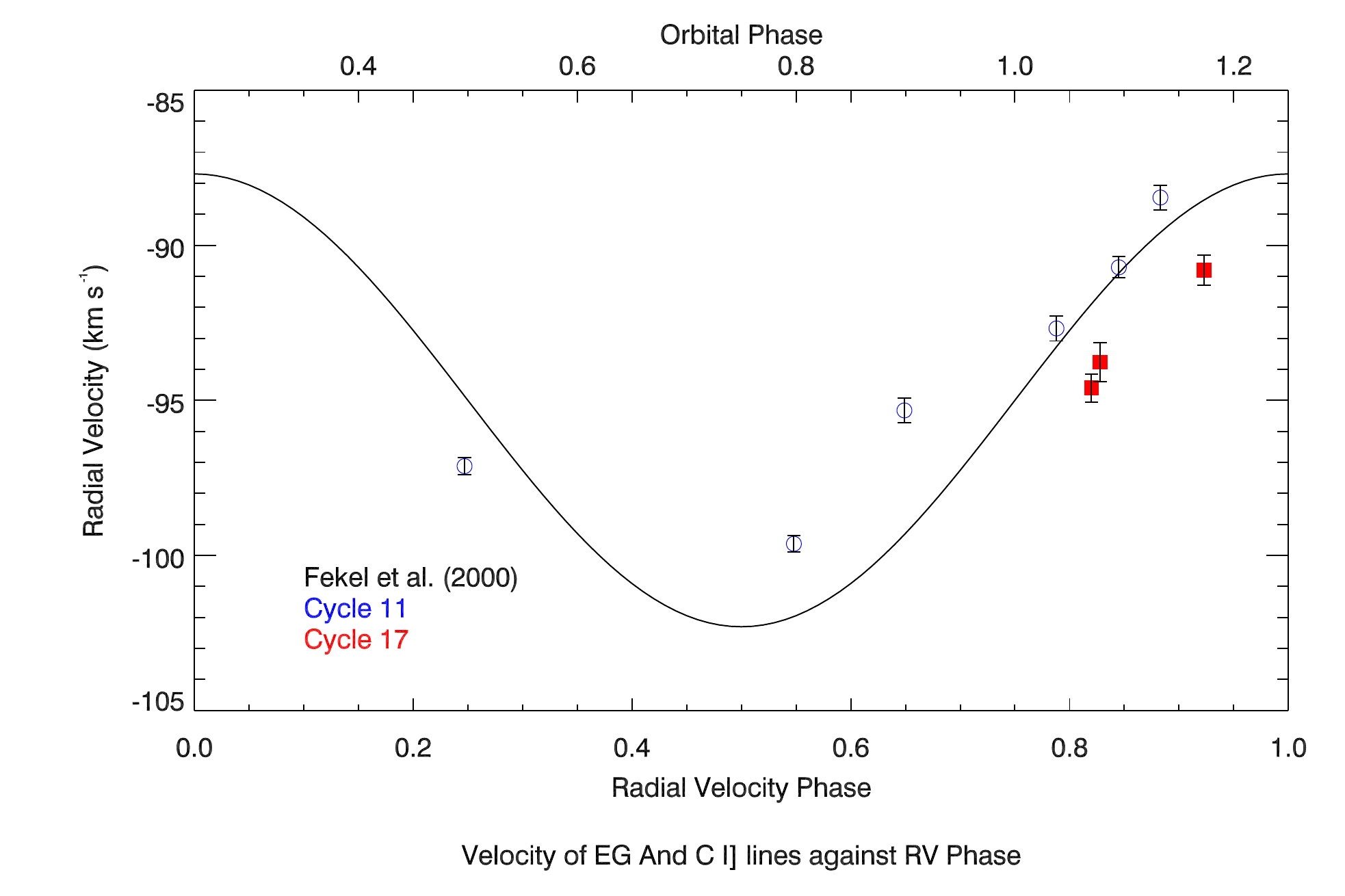}
\caption[EG And \ion{C}{ii]} Radial Velocities]{Radial velocities of EG And \ion{C}{ii]} lines. The black solid-line curve is based on the best-fit red giant velocity curve calculated by \citet{fekel_part1} based on infrared photospheric radial velocity measurements of EG And. Note that the best-fit curve is circular. The filled blue circles represent Cycle 11 observations, while the filled red squares represent Cycle 17 observations. 
\label{fig:velocity_vs_rv_phase}}
\end{figure}

To measure the radial velocity of EG And (and HD148349) at each observational phase, the \ion{C}{ii}] 2325\AA\ multiplet in each {\sl HST}/STIS E230M observation was fit using MELF to obtain the best-fit radial velocity shift for each observation. Comparing the resulting shifts between the \ion{C}{ii}] peak locations of the best fit and the wavelengths from \citet{nist_book_values} resulted in a velocity measurement at each phase. The velocities are shown in Table \ref{table:rvs}. These values were compared to those compiled from several sources in \citet{fekel_part1} in Figure \ref{fig:velocity_vs_rv_phase}.

\begin{table}
\caption[EG And Radial Velocities]{Radial Velocities for EG And and HD148349 based on \ion{C}{ii]} 2325\AA\ line locations. The date of observation is given by JD - 2,450,000. Target names are shortened to E (EG And) and H (HD148349).}              
\label{table:rvs}      
\centering                                      
\begin{tabular}{c c c c c c}          
\hline\hline                        
HJD &Target & Cycle & Orbital Phase& Radial Velocity Phase & Velocity\\    
& &  & ($\phi$) & ($\phi - 0.25$) & ($km\ s^{-1}$) \\ 
\hline                                   
2515 &E & 11 & 0.798 & 0.548  & -99.6 $\pm$ 0.3 \\ 
2564 &E & 11 & 0.899 & 0.649  & -95.3 $\pm$ 0.4 \\  
2631 &E & 11& 0.038 & 0.788  & -92.7 $\pm$ 0.4 \\
2658 &E & 11& 0.095 & 0.845  & -90.1$\pm$ 0.3 \\  
2677 &E & 11& 0.133 & 0.883  & -88.4 $\pm$ 0.4 \\  
2687 &E & 11& 0.153 & 0.903  & -84.3 $\pm$ 0.3 \\ 
2852 &E & 11 & 0.497 & 0.247  & -97.1 $\pm$ 0.3 \\ 
5059 &E & 17 & 0.070 & 0.820  & -94.6 $\pm$ 0.5 \\ 
5063 &E & 17 & 0.078 & 0.828  & -93.8 $\pm$ 0.6 \\
5109 &E & 17 &  0.173 & 0.923  & -90.8 $\pm$ 0.5 \\ 
\hline                                             
5077 &H & 17 & - & -  & +102.3 $\pm$ 0.4 \\ 
\hline
\end{tabular}
\end{table}

\begin{figure}[ht!]
\includegraphics[width=\textwidth]{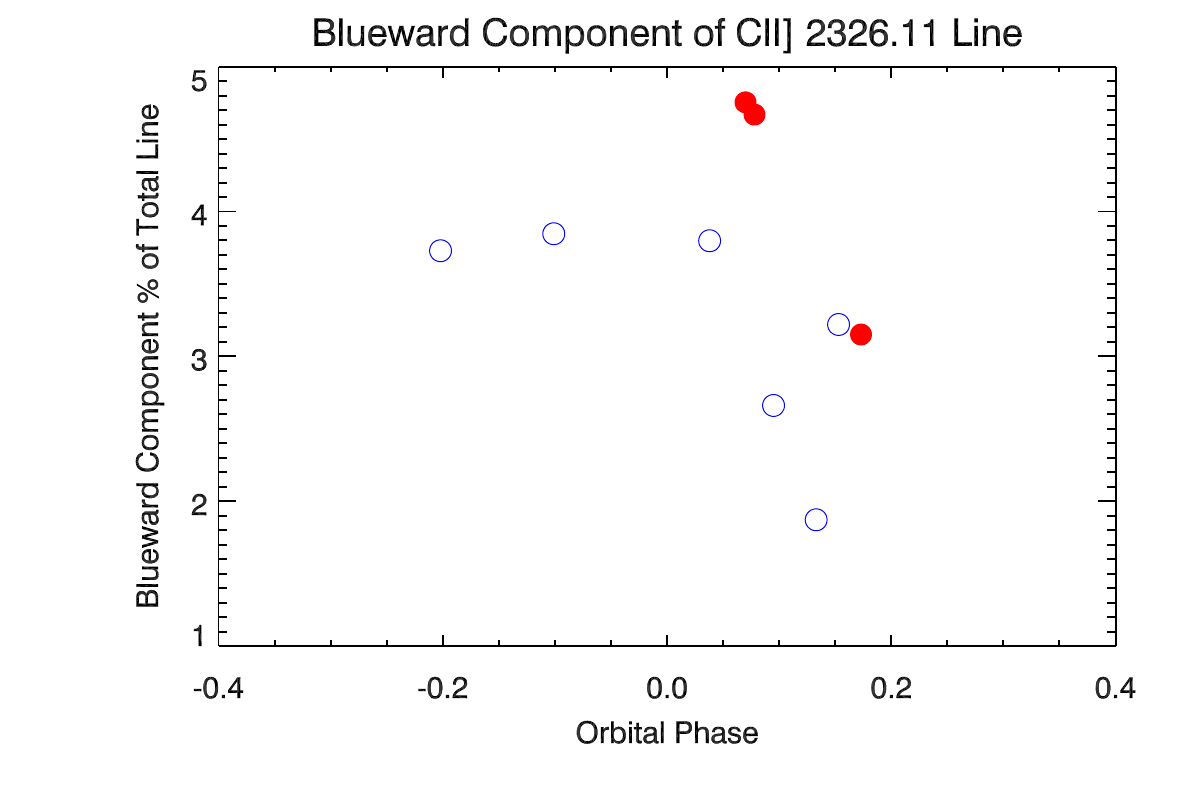}
\caption[Blueward Asymmetry in \ion{C}{ii]} Profiles]{Asymmetrical scattering flux as a percentage of the total line profile. The filled red circles represent the Cycle 17 observations, while the empty blue circles show Cycle 11. Just as the Cycle 17 appears to have stronger flux levels than corresponding phases from Cycle 11, so too the asymmetrical flux in the \ion{C}{ii]} line profiles is stronger in Cycle 17.
\label{fig:cii_blue_asym_c2326}}
\end{figure}

While the shape of the \ion{C}{ii]} lines can be described generally by a gaussian function (see fits in Appendix \ref{app:cii_2325_multiplet_fits}), it can be seen that there is a blue asymmetry evident in the base of the strongest lines, which could be either due to electron scattering (if present in all lines), or a wind component (if strongest in only some). For details of electron scattering by a hot gas see \citet{shields_mckee_1981}. In this case it can be seen that the scattering is asymmetrical in nature. If the model fit is compared to the data, it can be seen that on the red-wing the line profile is fit quite well by the model gaussian. The blue-wing, however, has additional flux when compared to the model. The additional flux reaches out to -50\ km\ s$^{-1}$ from the line centre around eclipse, while the red-wing of the line profile is consistently 30\ km\ s$^{-1}$ from the centre. By subtracting the model from the data, the percentage of the additional flux relative to the flux of the entire line can be measured. Figure \ref{fig:cii_blue_asym_c2326} shows how the asymmetry is strongest around eclipse and not apparent at all in the uneclipsed observations or in the comparison star, HD148349. 

\section{Fitting \ion{Al}{ii} Features}

Using the fitting technique applied to the \ion{C}{ii]} 2325\AA\ multiplet, some \ion{Al}{ii} features were identified in the data and fit in a similar manner. As outlined in \citet{johnson_smith_1986} and \citet{keenan_espey_1999} several \ion{Al}{ii} lines can be utilised as diagnostics of electron density and temperature. In particular, [\ion{Al}{ii]} 2661\AA\ and \ion{Al}{ii]} 2669\AA\ are especially useful. Reference wavelengths for the \ion{Al}{ii} line locations in vacuum were taken from \citet{al_lines_ref_1991} as 2669.951\AA\ and 2661.147\AA.

\begin{figure}[ht!]
\centering
\includegraphics[width=\textwidth]{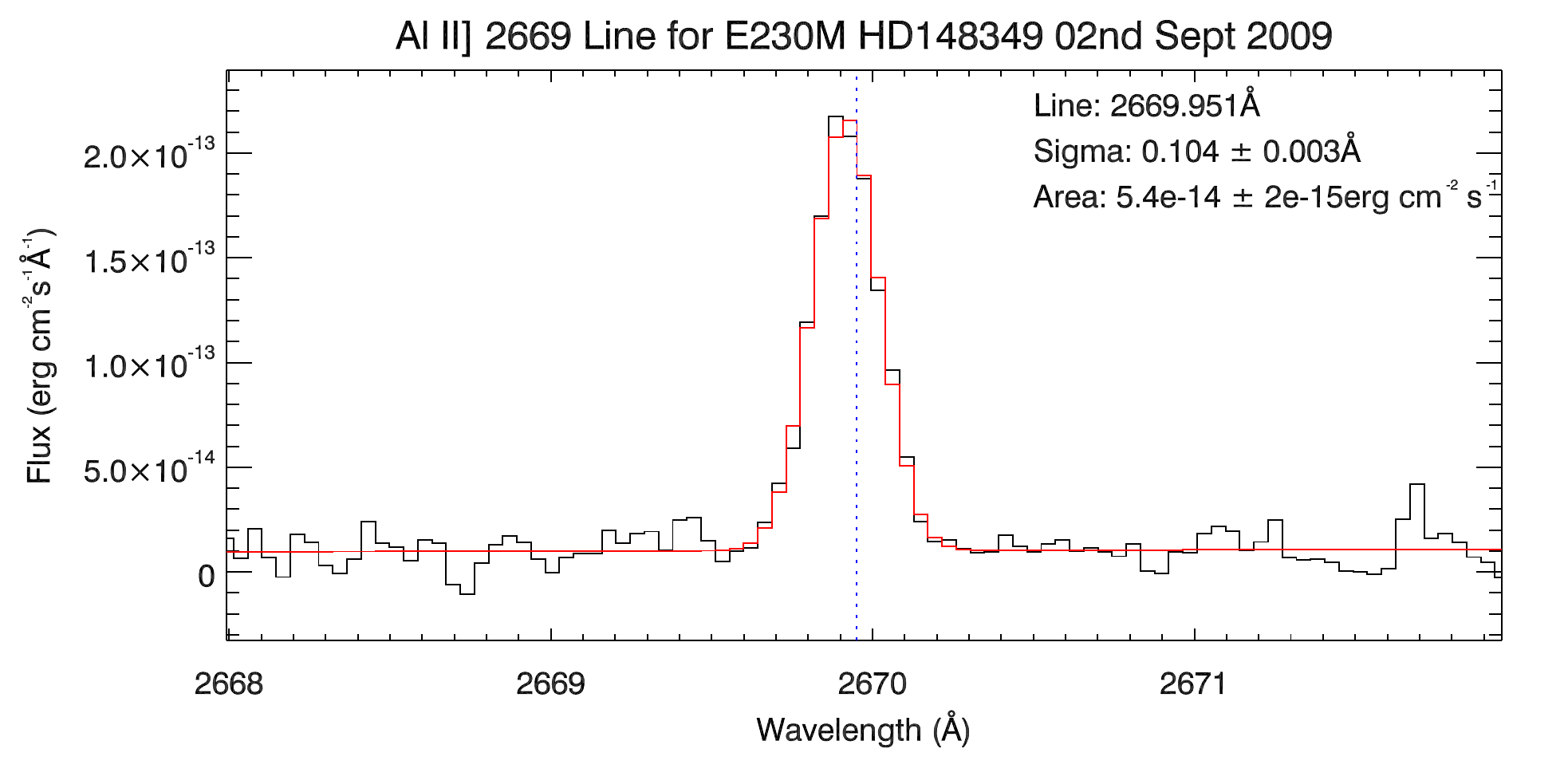}
\includegraphics[width=\textwidth]{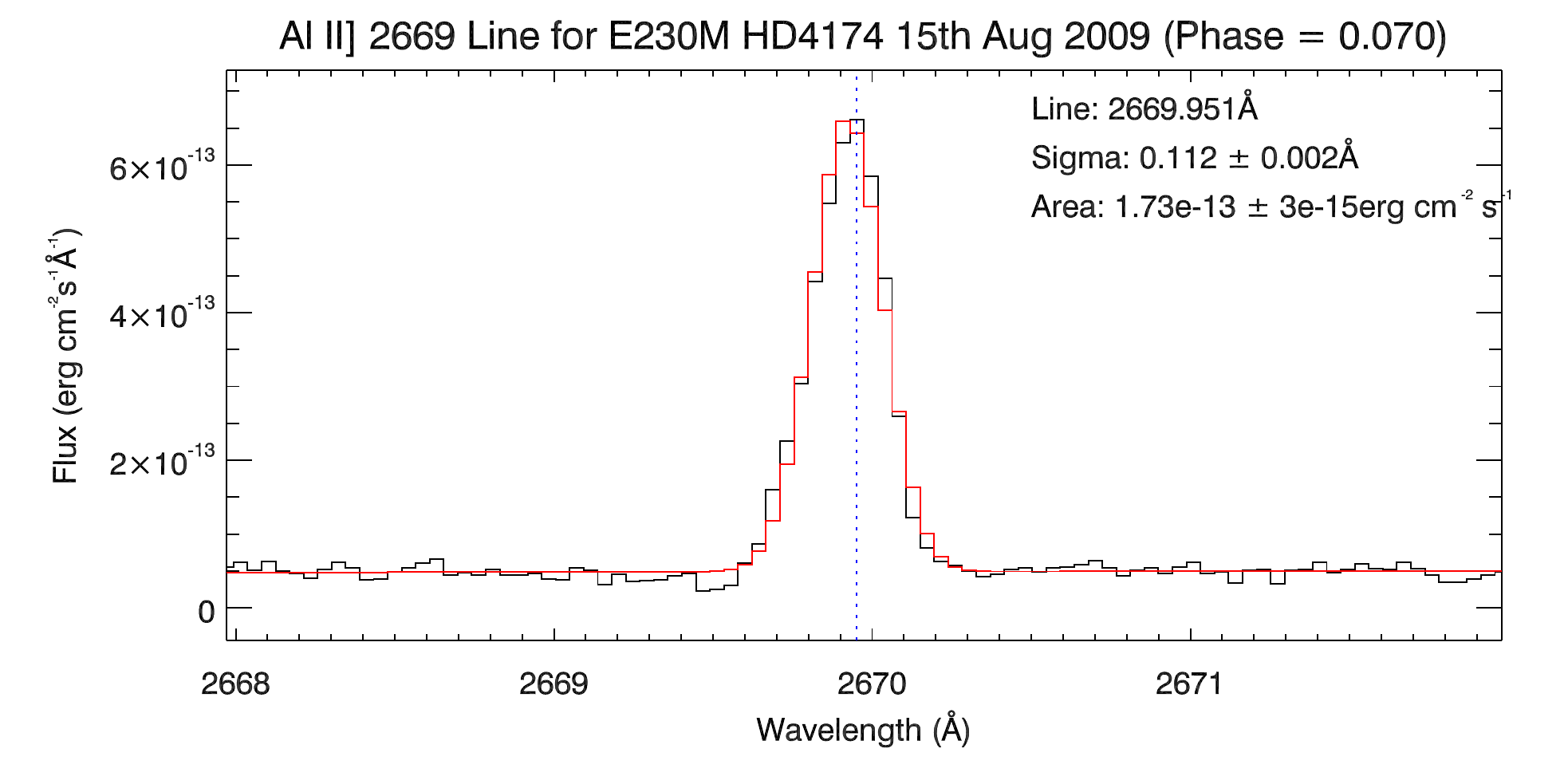}
\caption[\ion{Al}{ii]} 2669\AA\ Line Profiles]{\ion{Al}{ii]} 2669\AA\ lines in both HD148349 (top panel) and EG And (bottom panel). In both cases the data is in black with the model fit in red. The integrated line fluxes are annotated on both plots. 
\label{fig:al_ii_2669_line_fits}}
\end{figure}

\begin{figure}[ht!]
\centering
\includegraphics[width=\textwidth]{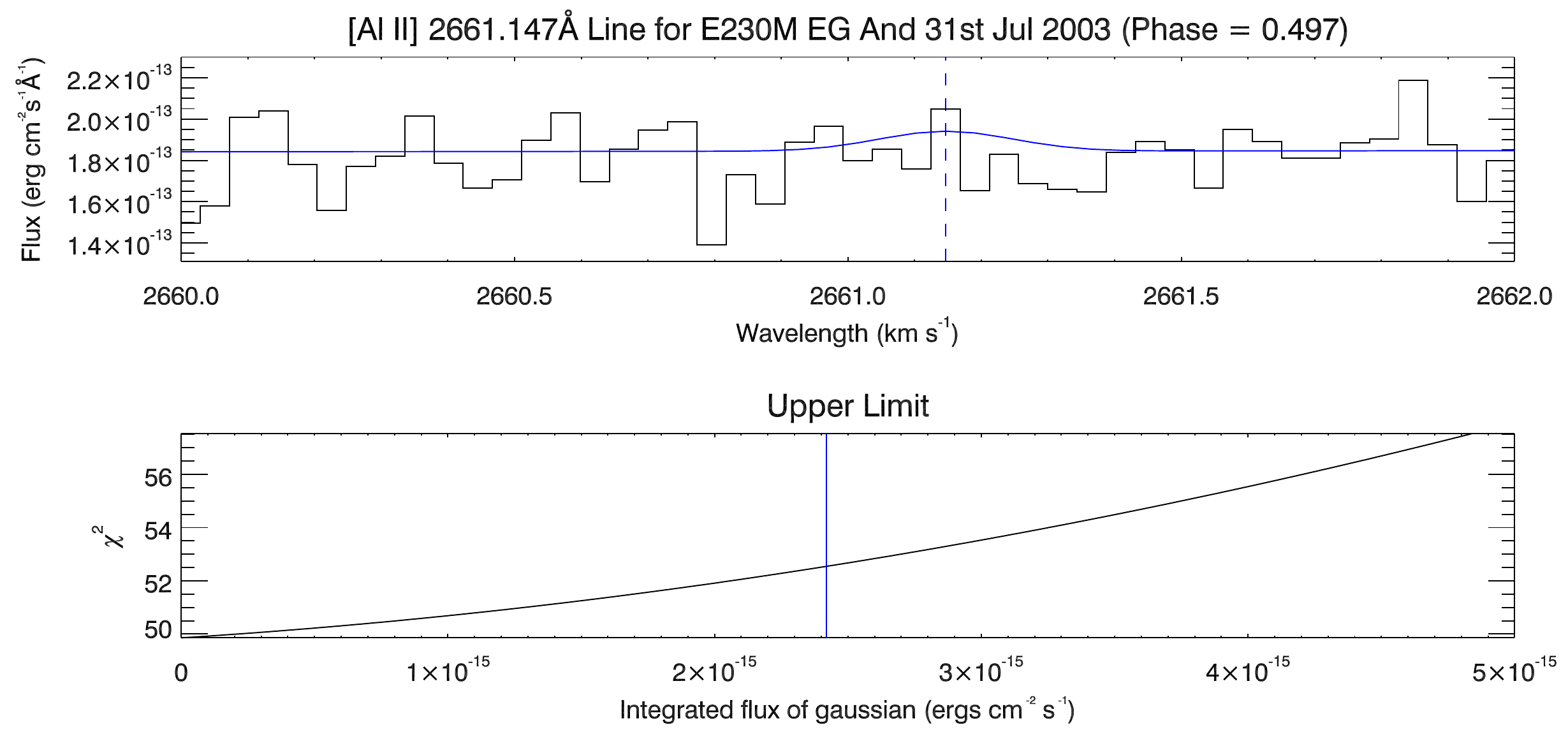}
\caption[[\ion{Al}{ii]} 2661\AA\ Upper-Limit]{Upper-limit for the [\ion{Al}{ii]} 2661\AA\ line. In the top panel, successive models were compared to the data by calculating the $\chi^{2}$ statistic. The bottom panel shows the $\chi^2$ values for the models. Overplotted in blue (in the top panel) is the gaussian which produces a $\chi^2$ value of +2.71 greater than the zero flux model (the case of no line existing). In this case the value was found to be $2.42\times10^{-15}$erg\ cm\ $^{-2}$\ s$^{-1}$.  \label{fig:sig3_alii_2661_401010_401020}}
\end{figure}

The 3s$^2$ $^1$S - 3s3p $^3$P$_1$ \ion{Al}{ii]} 2669\AA\ intercombination line is a strong emission line  of the \ion{Al}{ii} features to fit. It can be seen in Figure \ref{fig:al_ii_2669_line_fits} that the line is in a relatively featureless part of the spectrum and can be fit with a gaussian. While the forbidden 3s$^2$ $^1$S - 3s3p $^3$P$_2$  [\ion{Al}{ii]}  2661.147\AA\ line has been identified in some symbiotic nebulae, such as RR Tel \citep{keenan_espey_1999}, little evidence was found for its presence in either EG And or HD148349. Instead, upper-limits were obtained to see how much flux could be hidden at this wavelength. A similar method of calculation as that shown in Section \ref{sec:prob_lines_3sig} was utilised. Figure \ref{fig:sig3_alii_2661_401010_401020} shows a sample upper-limit for the [\ion{Al}{ii]} line. 

\begin{figure}[ht!]
\centering
\includegraphics[width=\textwidth]{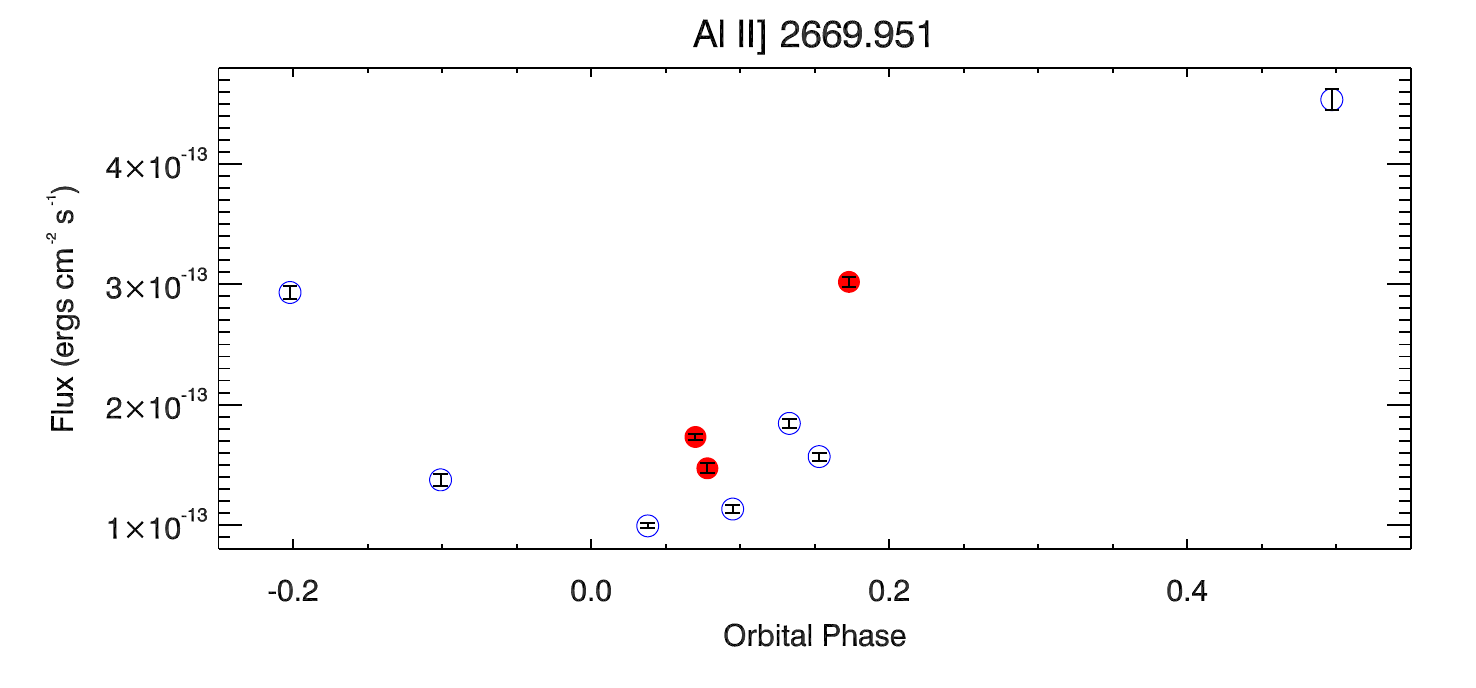}
\caption[\ion{Al}{ii]} 2669\AA\ vs Phase]{Integrated line fluxes for \ion{Al}{ii]} 2669\AA\ lines against orbital phase. The filled red circles represent the Cycle 17 observations while the empty blue circles show the Cycle 11 observations. The error in the line fluxes is of the order of the symbol size.
\label{fig:al_ii_2669_phase}}
\end{figure}

\begin{figure}[ht!]
\centering
\includegraphics[width=\textwidth]{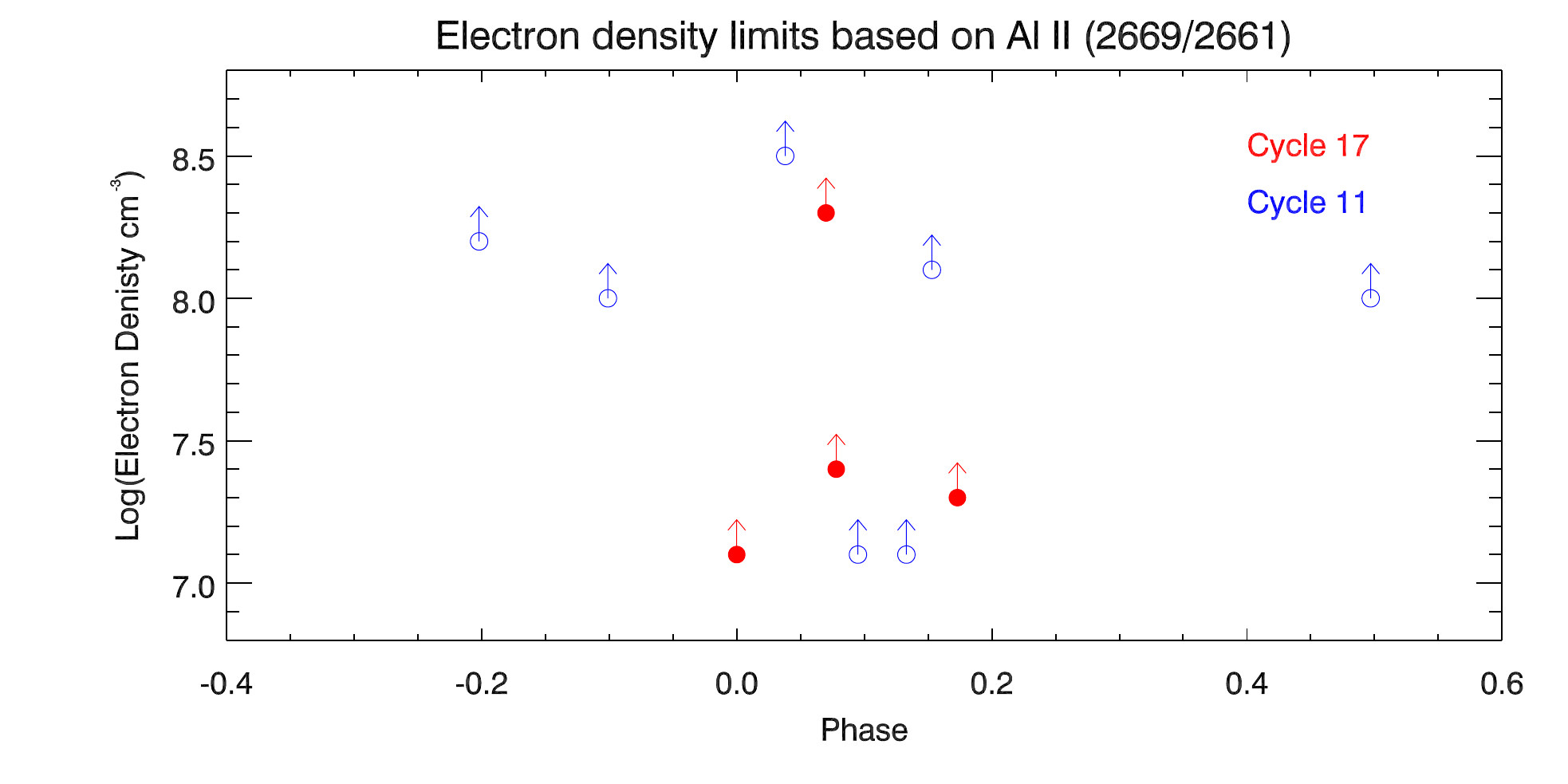}
\caption[\ion{Al}{ii} 2669/2661Electron Density Lower-Limits]{Electron density limits, based on the ratio of the \ion{Al}{ii} lines 2669/2661. As the integrated fluxes for the 2661\AA\ line are all upper-limits, the resulting electron densities are all lower-limits. The electron density values are measured using the values in \citet{keenan_espey_1999}. \label{fig:al_ii_2661_2669}}
\end{figure}

Figure \ref{fig:al_ii_2669_phase} shows that the variation of integrated flux from the \ion{Al}{ii]} 2669\AA\ line varies with orbital phase in a similar manner to the \ion{C}{ii]} lines in Figure \ref{fig:cii2325_combo_line_fluxes_vs_phase}. Once again it appears that the higher-state Cycle 17 and $\phi=0.497$ observations do not follow the same trend as the other observations. The [\ion{Al}{ii}] 2661\AA\ upper-limits can be used to obtain lower-limits of the electron density by calculating the ratio to the \ion{Al}{ii]} 2669\AA\ line. The resulting lower-limits of electron density are show in Figure \ref{fig:al_ii_2661_2669}. These limits reinforce the electron density values calculated from the \ion{C}{ii]} 2325\AA\ density diagnostic as they provide lower-limits to the values obtained from that analysis.

\section{Constructing \ion{C}{ii]} Models}\label{cii_3comp_models}

By combining the emission line fits from MELF (section \ref{sec:melf}) with a model of the absorption due to the giants chromosphere and lower wind using Atomicspec, a software package developed by \citet{cian_thesis}, a model of the \ion{C}{ii]} 2325\AA\ multiplet was developed to show how the dual effects of emission and absorption shape the spectrum at this region. For all modelling, vacuum wavelengths were adopted. The models were constructed by combining 3 separate components corresponding to different regions of the system: a continuum component, emission component and an absorption component.

\subsubsection*{Continuum Component}

The continuum due to the white dwarf dominates the UV. At redder wavelengths, nebular recombination continuum and continuum from the giant are seen. As a comparison the isolated red giant (Figure \ref{fig:mpfit_multiplet}) shows very little UV continuum. To estimate the continuum component of the 3-component model, the mean continuum for the 2320 - 2330\AA\ region of the spectrum is used. This was calculated as part of the MELF process by finding a mean continuum value in a featureless region either side of the \ion{C}{ii} 2325\AA\ multiplet and fitting a straight-line between them. For each observation that was modelled a corresponding constant mean continuum value was adopted (the mid-point of the continuum line described above) instead of a line-fit that varied with wavelength. This meant that any slope in the data continuum will not be fit accurately but this was considered an appropriate simplification, at least initially.

\subsubsection*{Emission Component}

The emission component of the model is taken from the profile fits using the described MELF technique. As the base model, the emission line model that was the best fit to the most-eclipsed observation of EG And ($\phi=0.038$) was chosen. This model should correspond to the emission line flux caused by collisional excitation of the \ion{C}{ii} in the chromosphere with no additional influence from the white dwarf. As it happens, some continuum in the $\phi=0.038$ observation can be seen and this is taken into account by the MELF fitting - resulting in a model with a small amount of continuum. This low-level continuum could be caused by scattered light from the dwarf through the red giant atmosphere, due to partial covering of the nebular continuum region around the white dwarf, or an additional, more extended continuum component. This additional continuum in the emission component is taken into account in constructing model below.

\subsubsection*{Absorption Component}\label{sec:absorption_component}

\begin{figure}[ht!]
\centering
\includegraphics[width=\textwidth]{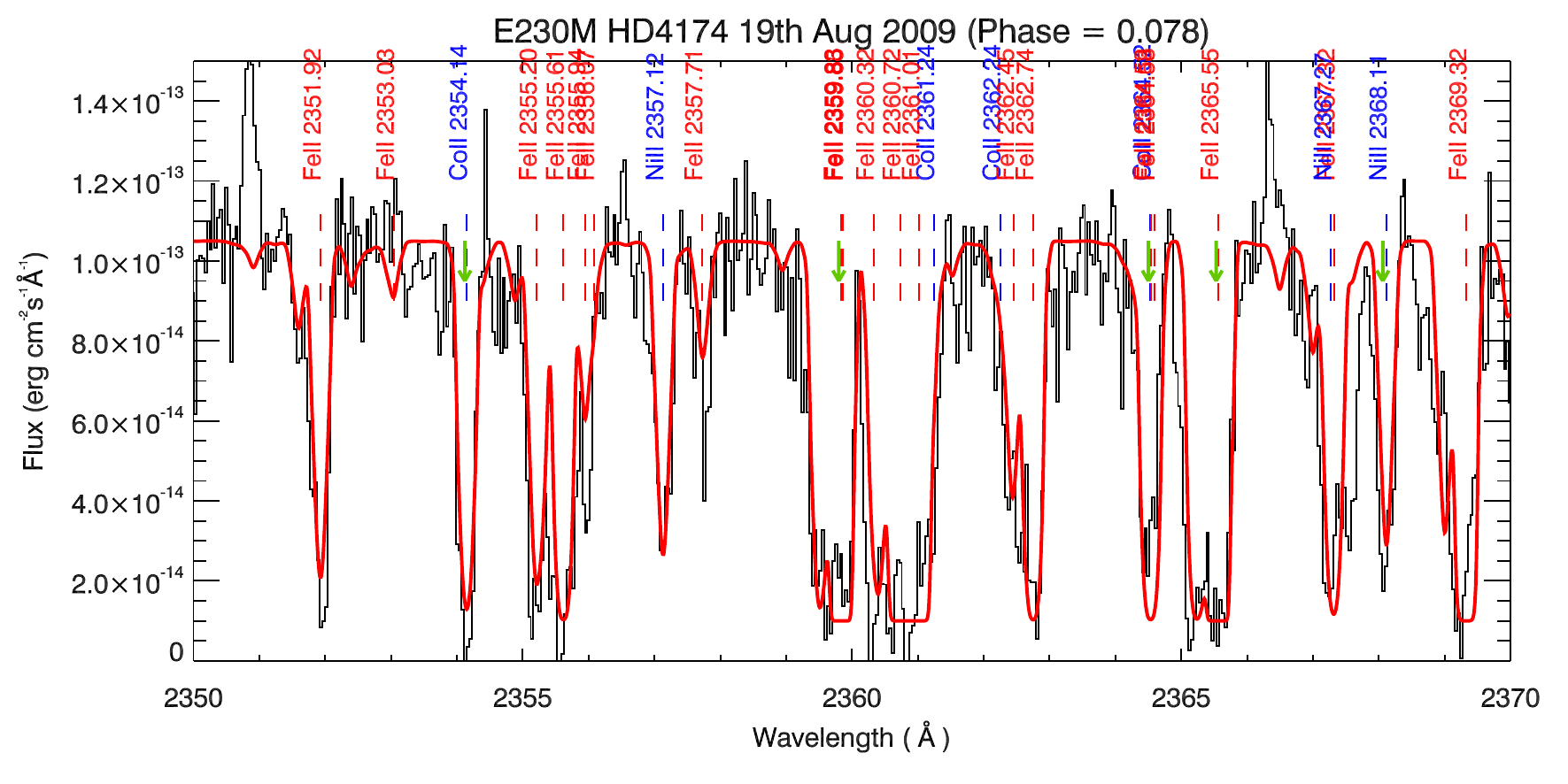}
\includegraphics[width=\textwidth]{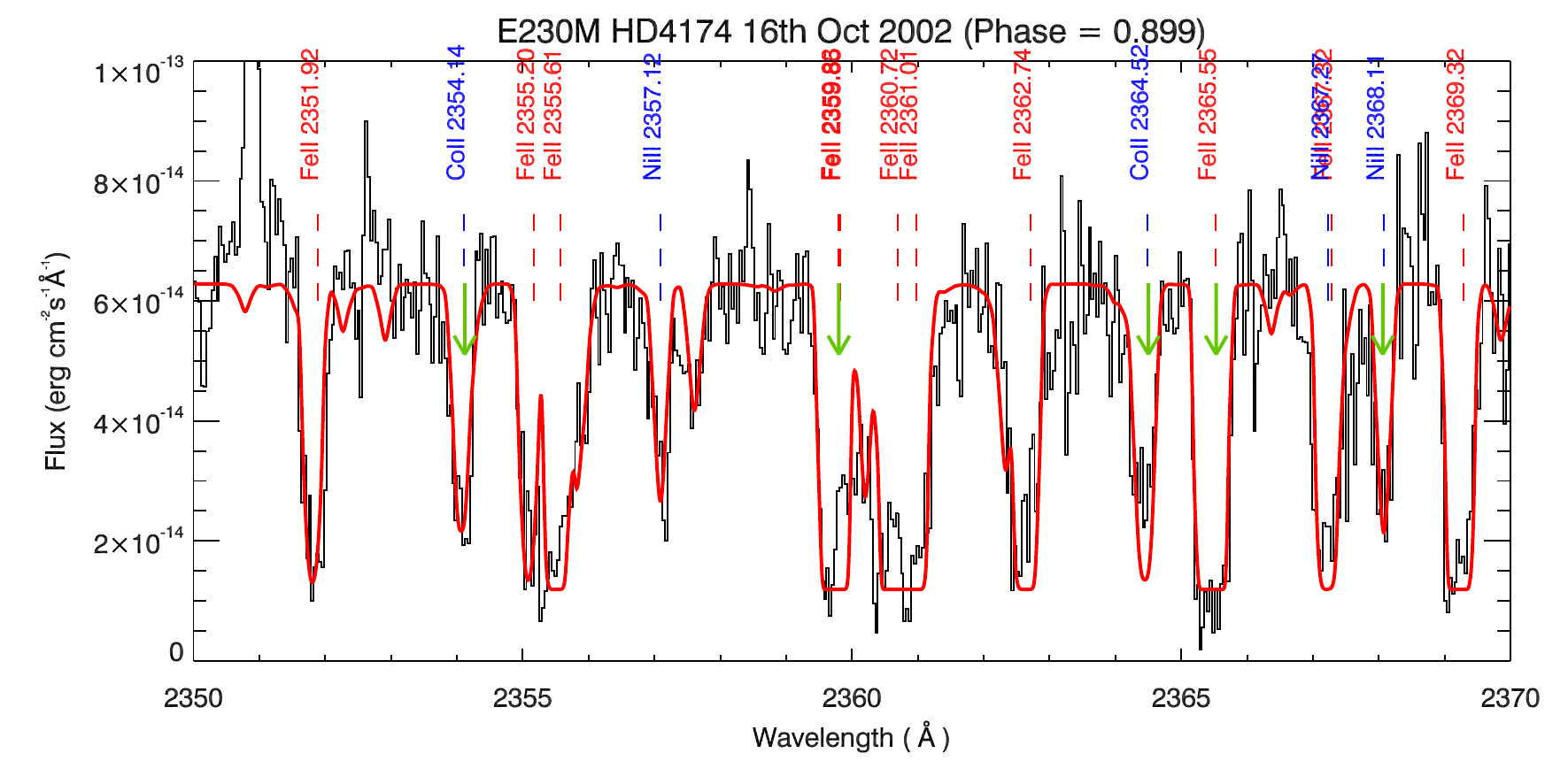}
\caption[Absorption Model Fits]{
The region of the spectrum that was used to help construct the absorption models by fitting \ion{Fe}{ii}, \ion{Co}{ii} and \ion{Ni}{ii} in a largely emission free region. The top panel is an observation from Cycle 17, while the bottom panel is an observation from Cycle 11. The data is in black with the best Atomicspec generated absorption model overplotted in red. The absorption species are labelled where known. The green arrows indicate the lines used to determine the column densities of the absorption species in the \ion{C}{ii]} 2325\AA\ multiplet region.
\label{fig:abs_models}}
\end{figure}

The white dwarf continuum undergoes absorption as its light passes through the red giant upper atmosphere and wind. This absorption component was modelled using Atomicspec, a spectral modelling code developed by \citet{cian_thesis}. This code populates the atoms thermally using Boltzmann's distributions and and input values of temperature and column density. Although this model assumes local-thermodynamic equilibrium, in the absence of detailed observations to form a more complicated model the is a good approximation to the observations. When generating absorption models it is necessary to consider the \ion{Co}{ii}, \ion{Ni}{ii} and \ion{Fe}{ii} species, which have lines around the location of the \ion{C}{ii}] multiplet. To disentangle this absorption from the \ion{C}{ii}] multiplet components in the 2320 - 2330\AA\ region, the column densities of the individual species are estimated by using lines from similar multiplet terms, but in the 2350 - 2370\AA\ region, which is emission-line free (see Figure \ref{fig:abs_models}). In the case of \ion{Co}{ii}, the lines at 2354.14\AA\ and 2364.55\AA\ correspond to multiplet(s) in the \ion{C}{ii}] region, while for \ion{Fe}{ii}, lines at 2359.83\AA\ and 2365.55\AA\ match. Finally, for \ion{Ni}{ii}, the line at 2368.107\AA\ can be used. Temperature was varied to see the effect on the absorption component of the models. Having discerned no appreciable sensitivity to temperature within  $\pm$5000\ K, a temperature of 8000\ K was used, which was the approximate value for the wind component at all phases, as discussed in \citet{cian_2007}.

\section{Underlying Absorption}\label{sec:test_under_abs}

\begin{figure}[ht!]
\centering
\includegraphics[width=\textwidth]{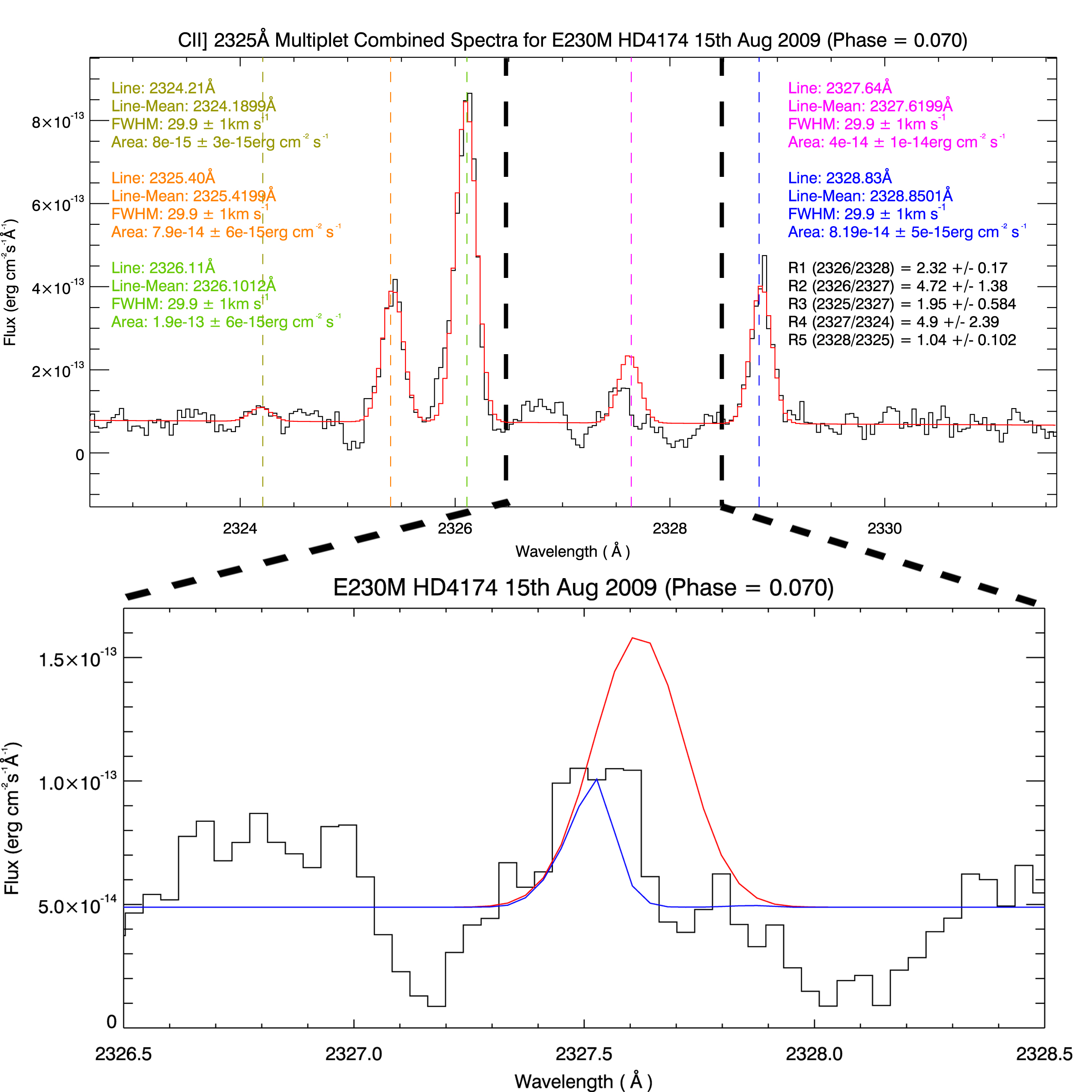}
\caption[Absorption Affecting \ion{C}{ii]} Lines]{Flux estimates of the mutilated 2327.64\AA\ line. The top panel shows the entire multiplet for EG And on the 15th August 2009, while the bottom panel zooms in on the 2327.64\AA\ line. While absorption either side of the line is self-evident, it is difficult to gauge how much absorption underlies the \ion{C}{ii]} emission feature. In the bottom panel, the red line shows the estimate of the flux that could well be present if there was no absorption component. This estimate is based on fitting the unaffected blue-wing of the line and constraining the width of the line to match that of the other \ion{C}{ii]} lines. The blue shows how that estimate is reduced when subject to the absorption model. Up to 60\% of the emission flux appears hidden due to underlying absorption. 
\label{fig:cii_model_3010_abs}}
\end{figure}

To test the effect of underlying absorption on the \ion{C}{ii}] region, the MELF emission line models can be combined with the constructed absorption models and the impact of  absorption on the integrated line fluxes can be calculated. For four of the five lines the effect of the underlying absorption on the integrated flux is found to be negligible. In the case of the 2327.64\AA\ line the integrated flux is significantly changed. Figure \ref{fig:cii_model_3010_abs} shows a close-up of the \ion{C}{ii]} 2327.64\AA\ line.

In the bottom panel, overplotted in red is the estimated best fit. While the other lines in the multiplet are fitted by finding the best fit to the core of the full line (excluding broad base features), in this case only the blue wing of the line profile was used for the fitting and hence is considered a best guess. The rest of the line is clearly mutilated by absorption. When the emission line model is subjected to the suspected absorption  taking place, it results in the blue gaussian line on the plot. This shows that the absorption is significantly reducing the flux that is seen from the emission line. Assuming that the original estimated flux (in red) is a good approximation of the actual emission line flux, and comparing that amount of flux to what is actually observed in the data, up to 60\% of the line flux has been lost due to underlying absorption. While this observation is an extreme example, it can be seen  from the line profiles that all of the observations close to minimum phase suffer absorption that affects the 2327.64\AA\ line. For this reason, the line ratio diagnostics that are dependent on the 2327.64\AA\ line are not used.

\section{Testing \ion{C}{ii]} Models}

Putting together the components outlined in Section \ref{cii_3comp_models} in different ways allows the structure of the system to be tested. Intuitively, the white dwarf continuum might be expected to be absorbed by the cool giant gas along the line of sight. The \ion{C}{ii} emission that is observed could be from different parts of the system. It is known that such emission is visible in isolated giants, and also in the phases when the white dwarf is eclipsed, so there is an underlying chromospheric component. An increase in the flux of the \ion{C}{ii}] lines outside of eclipse is also observed, together with changes in the diagnostic parameters derived, which indicates that there is additional material outside the normal chromospheric regions.

\begin{figure}[ht!]
\centering
\includegraphics[width=\textwidth]{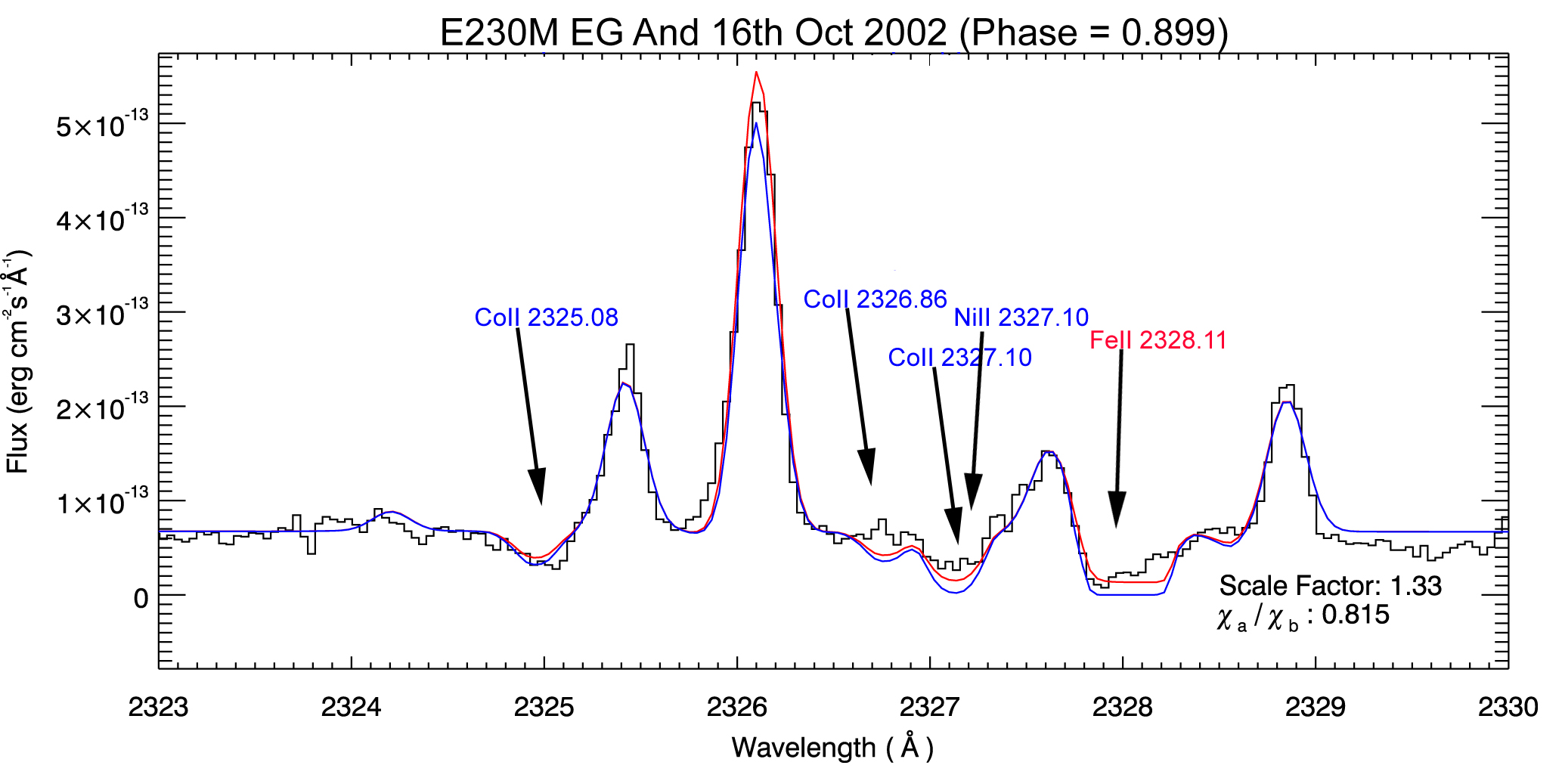}
\caption[\ion{C}{ii]} 2325\AA\ 3-Component Model Fits]{
\ion{C}{ii]} 2325\AA\ 3-Component Model Fits. The components of continuum, emission and absorption were combined in different orders to try and match the observations. At 2327.1\AA\ there are overlapping \ion{Ni}{ii} and \ion{Co}{ii} absorption features.
\label{fig:cii_model_4010}}
\end{figure}

Based on this, the fitting of the \ion{C}{ii}] region needs to be approached in two complementary ways: one in which only the white dwarf continuum is absorbed, and the other in which some (or all) of the \ion{C}{ii}] emission is absorbed too. By multiplying the white dwarf continuum component by the absorption component it is possible to simulate the white dwarf light passing through the red giant wind. A scaled version of the low-level eclipsed component continuum can be added on top of the absorption lines to represent the (unabsorbed) chromospheric emission, perhaps enhanced through the effects of the illuminating white dwarf. The other basic set of models assumes that any additional emission is absorbed by the same absorbing material as affects the white dwarf continuum. Figure \ref{fig:cii_model_4010} shows the models generated from the three components of continuum, emission and absorption. The blue line corresponds to adding the emission component to the continuum component and then multiplying by the absorption component. The red is the continuum multiplied by the absorption component with the emission component added in at the end.

The two versions of the 3-component models are:

\begin{itemize}
\item Red: (continuum$\times$absorption)$+$emission
\item Blue: (continuum$+$emission)$\times$absorption
\end{itemize}

It has been shown (in Section \ref{sec:test_under_abs}) that the multiplet emission lines of interest as diagnostics are largely unaffected by the presence of the underlying absorption lines. 

\begin{figure}[ht!]
\centering
\includegraphics[width=\textwidth]{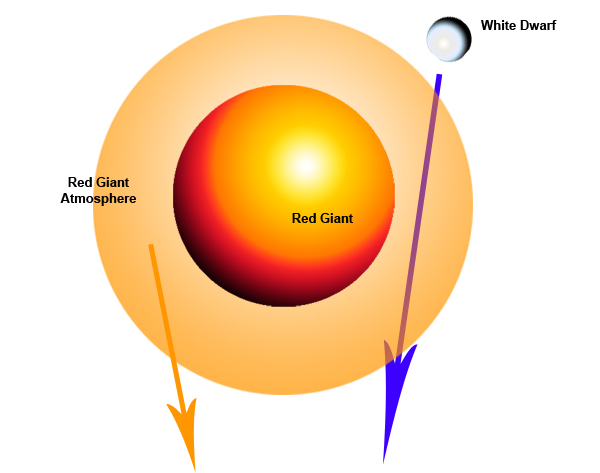}
\caption[EG And Schematic Diagram]{
EG And Schematic Diagram. The UV continuum observed in EG And comes from the white dwarf (blue arrow). Depending on the orbital phase this continuum might then be attenuated as it passes through the giant atmosphere. The chromospheric emission lines could arise from a separate part of the system and be unaffected by the dwarf (orange arrow). Alternatively, the dwarf could cause excess chromospheric emission which might then be subject to atmospheric absorption too.
\label{fig:egand_schematic}}
\end{figure}

\section{Modelling the White Dwarf Effect}

\begin{figure}[ht!]
\centering
\includegraphics[width=\textwidth]{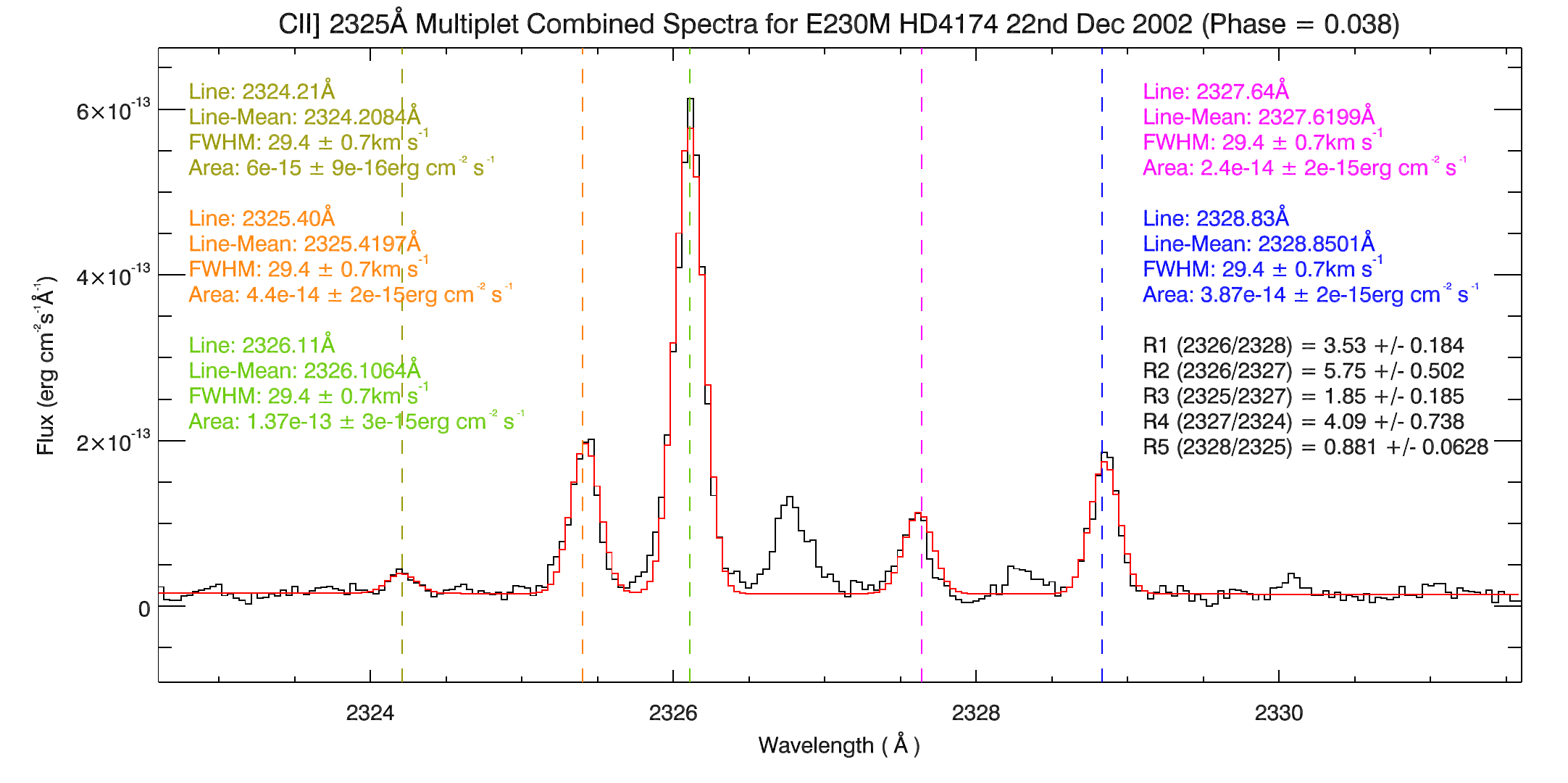}
\caption[Most-eclipsed \ion{C}{ii]} 2325\AA\ Observation]{\ion{C}{ii]} 2325\AA\ multiplet for an orbital phase of $\phi = 0.038$. This spectrum was produced by combining three separate observations of EG And on the 22nd December 2003. 
\label{fig:mpfit_cycle11_combo_406010_406020_406030}}
\end{figure}

The \ion{C}{ii]} 2325\AA\ multiplet comprises chromospheric emission lines  produced by collisional excitation followed by radiative de-excitation. At eclipse, the white dwarf's effect on the amount of \ion{C}{ii]} emission is minimal and this flux component is referred to as the `eclipsed component' (see Figure \ref{fig:mpfit_cycle11_combo_406010_406020_406030}).

As the white dwarf travels out of eclipse it causes more emission in these lines due to photoexcitation. To characterise this additional flux component with phase, a model of the white dwarf irradiating a growing area of the giant might be assumed. A geometrical Sobolev law \citep{sobolev_1975} with Lambertian scattering could thus be employed to model how the additional flux due to the white dwarf varies with orbital phase using an expression of the form: 

\begin{equation}\label{eqn:power_law}
f(\theta) = a[(Sin(\theta) + (\pi - \theta)Cos(\theta))/\pi]
\end{equation}
where $\theta$ is the phase angle and $a$ defines the strength of the curve. 
  
 \begin{figure}[ht!]
\includegraphics[width=\textwidth]{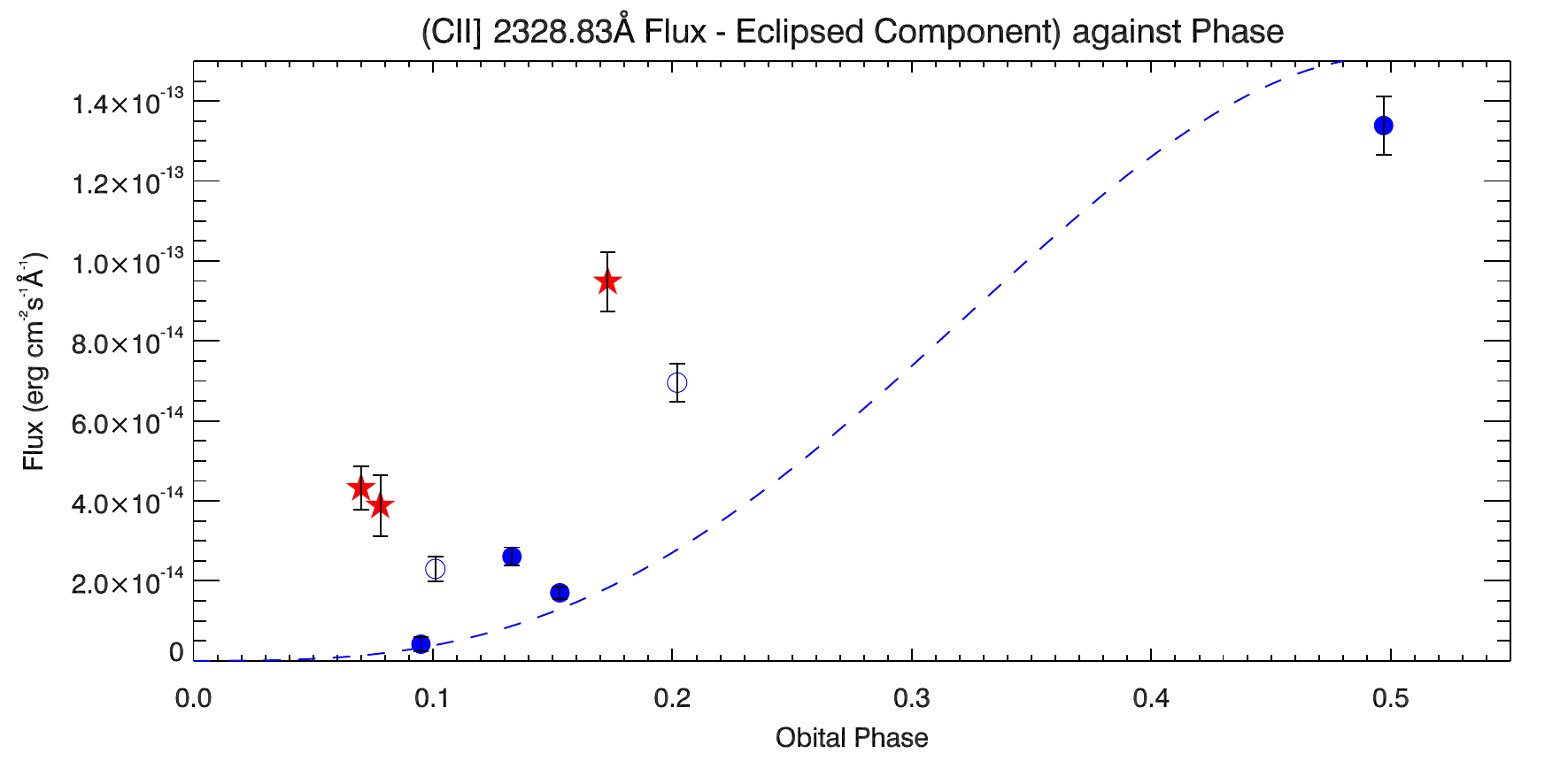}
\caption[EG And Simple Geometric Flux Model]{Additional flux (i.e. flux after subtraction of eclipsed component $\sim3.9\times10^{-14}$erg\ cm$^{-2}$\ s${-1}$). The filled blue circles show egress Cycle 11 observations, while the empty blue circles are the same observing cycle but at ingress. The red stars are Cycle 17 observations. The blue dashed-line shows the best fit form of Equation \ref{eqn:power_law} fit to the Cycle 11 egress values.
\label{fig:cii_model_power}}
\end{figure}

Figure \ref{fig:cii_model_power} shows the additional flux due to the white dwarf as a function of orbital phase. It is apparent that the effect of the white dwarf must be treated separately for different observing cycles (comparing the star symbols to the circles) and possibly even between ingress and egress (comparing the empty to the filled circles). When a curve of the form described in Equation \ref{eqn:power_law} was fit to the egress observations from Cycle 11, the best fit occurred when $a=1.5\times10^{-13}$. While the Cycle 17 data appears to have high additional flux components, this could be explained by the fact that the eclipsed component that was subtracted corresponded to the Cycle 11 eclipse. If the system changed with time (as appears to be the case) then the eclipsed component could possibly be higher during Cycle 17.

\section{Discussion and Conclusions}

The \ion{C}{ii}] 2325\AA\ multiplet was discussed in terms of its significance in characterizing giant chromospheres. MELF, a method of fitting all five lines of the multiplet was described and values of integrated flux, line width and line ratios were listed. This analysis represents the most in-depth  exploration of the \ion{C}{ii}] 2325\AA\ multiplet in any symbiotic system to date. The analysis spans two observing cycles and portrays the changing parameters of the chromosphere and wind. The average best-fit wavelengths of the five lines were 2324.213, 2325.413, 2326.106, 2327.634 and 2328.856\AA. While these vary by a maximum of 0.02\AA\ (2.6\ km\ s$^{-1}$) compared to the accepted NIST values from \citet{nist_book_values}, they are closer to \citet{young_2011_cii_wavelengths} values which were derived using {\sl HST}/STIS observations of  the symbiotic nova RR Telescopii. This shows that the \ion{C}{ii} plasma was at rest with respect to the M giant.

Estimates of the flux of each line of the \ion{C}{ii}] 2325\AA\ multiplet were used to calculate diagnostic line ratios and also to chart the changing flux levels for each line as the system goes through eclipse. Values of $n_e \sim7\times10^{8}$\ cm$^{-3}$ and  $n_e \sim4\times10^{8}$\ cm$^{-3}$ were measured for EG And and HD148349 respectively. These are similar to the expected values  of $n_e \sim 10^{8} - 10^{9}$\ cm$^{-3}$ found for $\gamma$ Dra (K5III), $\gamma$ Cru (M3III), $\alpha$ Ori (M2Iab) and RR Tel by \citet{judge_carpenter_1998_basal_flux}. 

The flux from each line of the multiplet is seen to increase linearly with the mean continuum of that observation. However the slope of each linear fit is different, suggesting that the line fluxes do not increase at the same rate, indicative of a change in electron density with phase. The change in the system between the two observing cycles is investigated. When characterising the EG And wind structure, the Cycle 17 and $\phi=0.497$ higher-state fluxes need to be handled separately. The additional \ion{C}{ii}] flux due to the white dwarf was modelled as a geometrical Sobolev law. The photometric variations shown in Chapter 4 indicate that EG And undergoes semi-regular pulsations. This is supported by the fact that the different observing cycle observations have to be handled separately as the system changes considerably over timescales of the order of observing cycles. Higher density and different radial velocities suggest slightly different emitting regions between observing cycles. 

The \ion{C}{ii]} models show that the underlying absorption lines do not affect the important flux lines significantly and the density diagnostics can be used. They also illustrate where the hot and cool components of gas arise geometrically in the system. Absorption line profiles were modelled by varying the column density, width and number of line components. Initial, single-component fits, did not fully describe the line profiles at all epochs, and it was found that that a second, weaker component sometimes had to be added in at an offset between -10 and -50\ km\ s$^{-1}$. This is evidence for flocculent wind structures similar to those found previously by \citet{cian_2007}. Since it was not necessary to include this additional component for all of the observations, this hints at the dynamic nature of the wind structure.

Following the discussion of radiative heat loss in the chromosphere (see Chapter 4), the energy budget for the outer atmosphere can be revisited.  The flux energy of the wind itself can now be included in the discussion. In the higher gravity cool stars, like the Sun, comparable parts of the wind energy go into expanding the atmosphere out of the stellar gravitational field and accelerating it to the  terminal wind velocity ($v_\infty \sim v_{esc}$). For  lower gravity cool stars almost all of the driving energy of the wind goes into lifting the expanding atmosphere out of the gravitational field and $v_\infty << v_{esc}$ \citep{holzer_macgregor_1988_massloss}:
\begin{equation}
v_{esc} = \sqrt{\frac{2GM}{R}}
\label{eqn:escape_vel}
\end{equation}
Using the parameters outlined in Chapter 2 for EG And ($R = 75R_{\odot}$, $M = 1.5M_{\odot}$) and HD148359  ($R = 83R_{\odot}$, $M = 2M_{\odot}$), escape velocities of 87 and 96\ km\ s$^{-1}$ respectively, are estimated. The total energy flux through the outer atmosphere must equal the wind flux $F_W$ plus the total energy flux radiated by the outer atmosphere above the photosphere, $F_{\star}(tot)$ \citep{judge_stencel_1991_chromowinds}. The latter term was calculated in Chapter 4.  $F_W$ is the energy flux at the atmospheric base required to drive the wind alone. Assuming a gravitationally bound expanding stellar atmosphere, this term is given by \citet{holzer_macgregor_1988_massloss} as:
\begin{equation}
F_W = \frac{1}{2} \dot{M} \left(v_{\infty} ^2 +  v_{esc}^2 \right) \frac{1}{4\pi R_{\star}^2}
\label{eqn:holzer_massloss1}
\end{equation}
Assuming that there are initially no radiative losses and that the flow is subsonic at $r=R_{\star}$ and supersonic at $r=\infty$, \citeauthor{holzer_macgregor_1988_massloss} show:
\begin{equation}
F_W \simeq 3.3\times10^3 \left(  \frac{\dot{M}_{\star}}{10^{-7}} \right)  \left(  \frac{M_{\star}}{M_{\odot}}  \right)  \left(  \frac{400R_{\odot}}{R_{\star}}  \right)^3   \left( 1 +  \frac{v_{\infty} ^2}{v_{esc}^2}  \right)
\label{eqn:holzer_massloss}
\end{equation}
Using an upper-limit of $\dot{M}_{\star} = 7.5\times10^{-8}M_{\odot}$ yr$^{-1}$ \citep{cian_thesis} in Equation \ref{eqn:holzer_massloss} gives a value of $F_W = 4.7\times10^{5}$\ erg\ cm$^{-2}$\ s$^{-1}$ for EG And which is less than the $ F_{\star}(tot)$ value of $7.9\times10^{5}$\ erg\ cm$^{-2}$\ s$^{-1}$ from Chapter 4. Compared to other stars in \citet{judge_stencel_1991_chromowinds}, the trend of slightly larger $F_{\star}(tot)$ compared to $F_W$ is maintained. $F_W$ is most-likely larger compared to other stars as an upper-limit was used for $\dot{M}_{\star}$.

By calculating the response of the outer atmosphere to the passage of mechanical energy fluxes, \citet{hartmann_mcgregor_1980} suggest a way of explaining the different atmospheric phenomena seen across the cool star population. A mechanical flux density, $F_m$, can be represented as the product of the characteristic energy density $\epsilon_m$ and a characteristic speed of propagation $v_{prop}$:
\begin{equation}
F_m \simeq \epsilon_m v_{prop}
\label{eqn:flux_dens}
\end{equation}
The energy density can be expressed as:
\begin{equation}
\epsilon_m = \frac{1}{2}C \rho v_{rms}^2
\label{eqn:energy_dens}
\end{equation}
where $\rho$ is the density of the outer atmosphere, $v_{rms}$ is the characteristic wave amplitude velocity, and $C$ is a constant of order unity that depends on the wave geometry. The uncertainty in this value is negligible compared to the size of the uncertainty in the density. For the purpose of these calculations $\frac{1}{2}Cv_{rms}^2 \sim v_{mp}^2$, where $v_{mp}$ is the most probable velocity. If the density is taken to be:
\begin{equation}
\rho = n_H m_p
\label{eqn:dens_h}
\end{equation}
where $n_H$ is the hydrogen particle density and $m_p$ is the proton mass, Equation \ref{eqn:flux_dens} becomes:
\begin{equation}
F_m \simeq  v_{prop}  n_H m_p v_{mp}^2
\label{eqn:flux_dens_mech}
\end{equation}
There are two alternatives for $v_{prop}$. The speed of propagation may be described by the sound wave speed $v_S$ or the Alfv\'{e}n wave speed $v_A$:
\begin{equation}
v_{S} = 12.85\sqrt{\frac{T}{10^4}}
\label{eqn:sound_vel}
\end{equation}
\begin{equation}
v_{A} = \frac{B}{\sqrt{4\pi \rho}}
\label{eqn:alf_vel}
\end{equation}
The characteristic temperature of the lower chromosphere is taken to be 8000\ K \citep{cian_thesis}, yielding $v_S = 1.15\times10^6$\ cm\ s$^{-1}$. Subbing this into Equation \ref{eqn:flux_dens_mech} as the value of $v_{prop}$, the mechanical flux energy provided by sound waves, $F_{M_{S}}$, can be calculated.  $v_{mp}$ can be found from the FWHM value of 29\ km\ s$^{-1}$ found from the \ion{C}{ii}] 2325\AA\ multiplet lines. The lines have a gaussian profile\footnote{As shown in Equation \ref{eqn:gauss_profile}.} and because of this, $v_{mp}$ can be found from the line widths:
\begin{equation}
\mbox{FWHM }= 2\sqrt{ln2}v_{mp} =1.66v_{mp}
\label{eqn:line_widths}
\end{equation} 
as shown by \citet{judge_carpenter_1998_basal_flux}.  $n_H$ can be estimated using the $n_e$ values calculated from the density diagnostic study.  To do this, a measure of the ionisation balance ${n_e}/{n_H}$ is needed. This can be taken to be 0.01\footnote{This is based on ionisation balances from studies of $\alpha$ Tau (K4III), $\alpha$ Ori (M2Iab) and G Her (M6III) \citet{mcmurry_1999, hartmann_avrett_1984, luttermoser_1994}.} and leads to $n_H \sim 10^{11}$\ cm$^{-3}$, similar to values shown by \citet{cian_thesis}. This results in $F_{M_{S}} = 7.5\times10^{5}$\ erg\ cm$^{-2}$\ s$^{-1}$. 

To estimate the mechanical flux energy if the speed of propagation is the Alfv\'{e}n wave speed, it is necessary to estimate the strength of the magnetic field. For Alfv\'{e}nic waves $v_{mp} = \delta v_A$, where $\delta v_A$ is the tangential Alfv\'{e}n wave speed. To avoid non-linear damping, $\delta v_A \leq \frac{1}{2} v_A$. Using $v_{mp} =1.75\times10^{6}$\ cm\ s$^{-1}$, $v_A = 3.5\times10^{6}$\ cm\ s$^{-1}$. Substituting this value into Equation \ref{eqn:alf_vel} gives $B\sim6$ Gauss, similar to values estimated by \citet{hartmann_mcgregor_1980}. Solving Equation \ref{eqn:flux_dens_mech} using the Alfv\'{e}n speed gives a  mechanical flux energy of $F_{M_{A}} = 2.3\times10^{6}$\ erg\ cm$^{-2}$\ s$^{-1}$. 

The energy budget for the outer atmosphere of EG And can be found by adding the calculated values of the total chromospheric radiative losses and the energy required to drive a wind:
\begin{equation}
F_{\star}(tot) + F_W \simeq 1.3\times10^{6}\mbox{ erg\ cm$^{-2}$\ s$^{-1}$}
\label{eqn:total_energy_flux_combined}
\end{equation}
This value can be directly compared to the two values of mechanical flux energy produced by sound and Alfv\'{e}n waves, $F_{M_{S}} = 7.5\times10^{5}$\ erg\ cm$^{-2}$\ s$^{-1}$ and $F_{M_{A}} = 2.3\times10^{6}$\ erg\ cm$^{-2}$\ s$^{-1}$. While sound waves cannot be completely ruled out as an energy mechanism, they do not seem to produce enough energy to heat the chromosphere and drive the wind. Conversely, Alfv\'{e}n waves appear to provide more than the required energy. This fits in with the theory of  \citet{hartmann_mcgregor_1980} that while acoustic waves dissipating through shock formation in the low chromosphere can account for the temperature and density structure, their damping lengths are far too short to effectively drive mass-loss. On the other hand, damped Alfv\'{e}n waves are efficient drivers of mass-loss and can account for the observed terminal wind velocities. Frictional dissipation of Alfv\'{e}n waves in higher gravity stars could lead to the coronal heating and fast winds while in lower gravity stars it could result in slower, more massive mass-loss. The role of photospheric convection in producing these energetic waves is discussed in the next chapter.

\newpage
\thispagestyle{plain}
\mbox{}	
\chapter{Photospheric Convection in Red Giants}
\label{chapter:results_granulation}

Along with the complexity of EG And's atmosphere outlined in the previous chapter, the mechanism that initiates and drives its wind is not yet clear. This ambiguity is endemic of red giant stars. In a bid to comprehend what the underlying mass-loss driving mechanism might be, a theory tentatively mentioned in \citet{schwarzschild_1975} was investigated. Schwarzschild hypothesized that \emph{``it appears plausible - though far from certain - that this mass ejection is triggered by photospheric convection''}. This idea was also suggested by Sun Kwok \citep{kwok_press_relase} and \citet{schroder_2005}. In this chapter the photospheric motion of giants is discussed along with previous efforts to understand the asymmetry of optical spectral lines. LAST, a tomographic technique for cross-correlating low-resolution spectra to measure the asymmetry of optical features, was developed and applied to archival {\sl ELODIE} data. Initial testing of these techniques are discussed in \citet{roche_2009_cs15poster}. While the results do not prove that photospheric convection could provide the initial impetus to instigate mass-loss in red giant stars, they do impart several consequences on the nature of line asymmetries in photospheric spectral features. The convective-granulation motions of the deep photosphere  will be transmitted to any magnetic field lines extending outwards into the chromosphere and wind. This leads to magnetic Alfv\'{e}n waves being induced and could possibly be the mechanism behind the heating of the outer atmosphere of cool stars \citep{gray_photosphere_book}.

\section{Photospheric Motions and Line Asymmetry}\label{sec:photo_motion}

\begin{figure} 
\centering
\includegraphics[width=\textwidth]{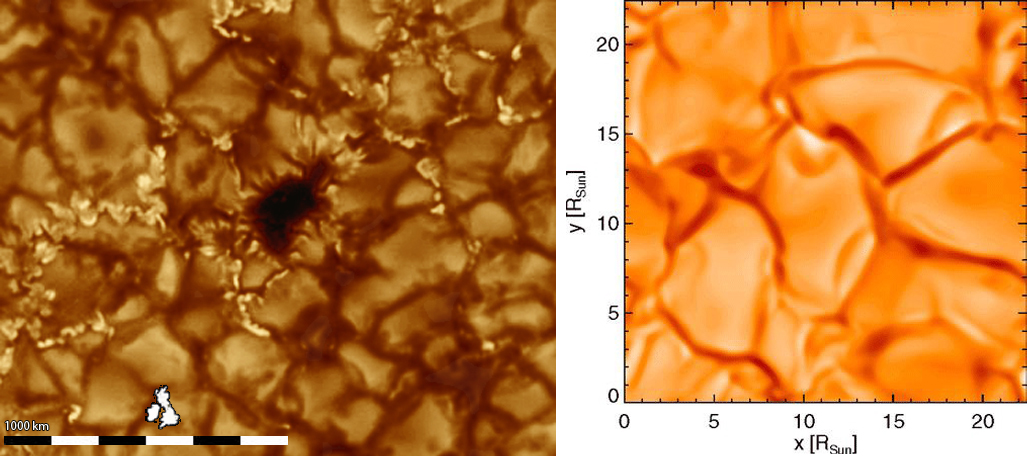}
\caption[Solar Granulation and Giant Granules]{
The image on the left shows the scale of solar granulation. The image on the right shows a 3D hydrodynamical model of convection in an M3 giant from \citet{phoenix_gran} using the stellar atmosphere modelling code, PHOENIX \citep{phoenix_1997}. While the granular surface cannot yet be directly imaged on giants, the models now being produced show granular patterns to those observed on the Sun. 
\label{fig:solar_gran}}
\end{figure}

The photosphere, the visible surface of a star, is the region at which the plasma of the star becomes transparent to photons of light. At this point, any energy generated in the stellar interior becomes free to propagate out into space. Most stars having $T_e$ less than about 10,000\ K undergo convection in their visible photospheric layers \citep{landstreet_2007}. The convective bubbling on the surface of the Sun provided the first case of convection being observable in stellar atmospheres. These solar granules are well described by B\'enard Cells -  convection cells that appear spontaneously in a liquid/gas layer when heat is applied from below. Visually they have the appearance of hot rising bubbles, as shown in the left panel of Figure \ref{fig:solar_gran}. This figure shows a comparison of granule size for the Sun and that expected for an M3 giant. The solar disk is covered by approximately a million cells. They have a lifetime of about 10 minutes and an average size of about $1.3\times10^{8}$\ cm (larger cells can stretch to 2$\times10^8$\ cm). It is estimated that the temperature difference between the hotter rising material in the middle of the cell and cooler/darker infalling material in the surrounding lanes is about 100\ K, corresponding to a 25\% brightness difference \citep{gray_photosphere_book}.

\begin{figure}[ht!]
\centering
\includegraphics[width=\textwidth]{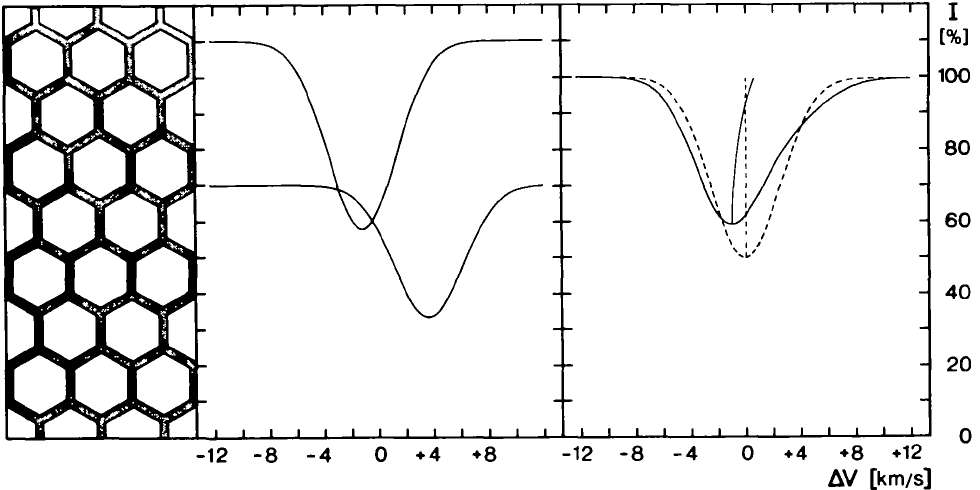}
\caption[Origin of Spectral Line Asymmetry]{
The origin of spectral line asymmetry from  \citet{dravins_1981_solar_gran_asym}. The left panel shows an idealised depiction of a stellar surface where $\sim75$\% is covered with bright granules while the rest is  darker intergranular lanes. The center panel shows a high-resolution spectral line profile for both regions of the surface. Hot material rising from the granule will give a blueward Doppler shift to its spectrum while the material falling away in the cool lane will give a redward Doppler shift. The right panel displays the resulting profile with low spatial resolution to average over many granules. The line bisector is show as a slight ``c-shape'' and is blue-shifted. The dashed overplotted profile shows what a symmetric profile would look like if there were no velocity fields on the surface.
\label{fig:dravins_solar_asym}}
\end{figure}

\begin{figure}[ht!]
\centering
\includegraphics[width=\textwidth]{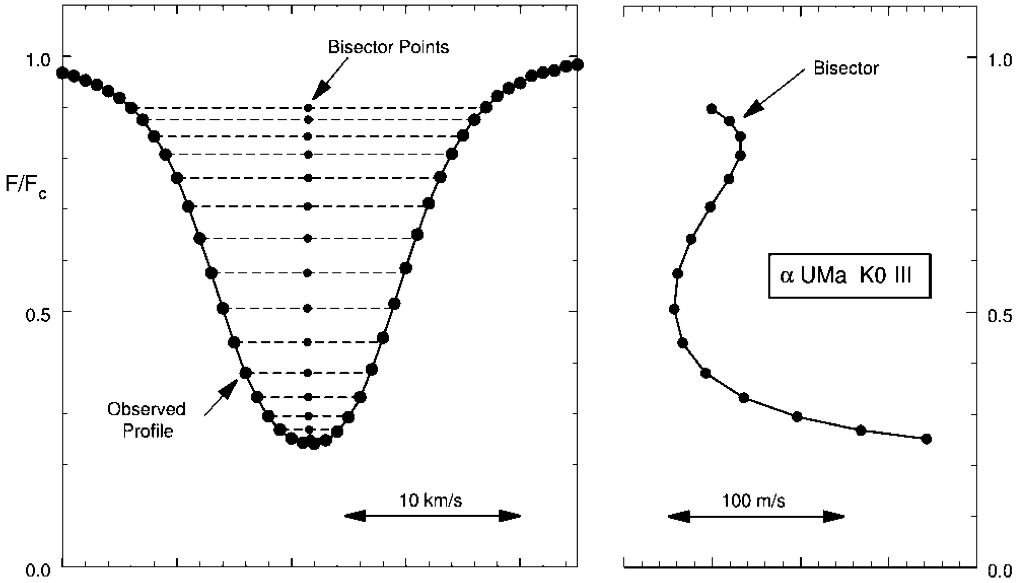}
\caption[Line Bisector Construction]{
Line bisector construction from \citet{gray_2005}. 
The left panel shows how the midpoints of the widths of several horizontal line segments (dashed lines) are used to make up a bisector. The right panel is an expansion of the bisector which is necessary to appreciate its shape and the asymmetry of the line profile. 
\label{fig:bisector}}
\end{figure}

Stellar convection dominates the shaping of spectral lines in cool stars \citep{gray_review_1992} as a convecting atmosphere will not have equal regions of upward, horizontal and downward flows. Figure \ref{fig:dravins_solar_asym} shows that the upward flow has a higher temperature and will be blue-shifted, while the sinking region will be red-shifted, as seen in the solar case. The weaker contribution will tend to form a depressed wing on the stronger contributor \citep{gray_2005}. A way of differentiating between the different velocity fields and the asymmetry present became known as the line-bisector method. It involves joining the midpoints of several horizontal lines across a spectral line profile. The construction of such a bisector is shown in Figure \ref{fig:bisector}. This line bisector method was first applied to solar spectral lines by \cite{voigt_1956_solarasym} before being applied to other stars by \cite{gray_1980_sunarcasym, gray_1981_procyonasym, gray_1982_bisect}. Unlike the case of the Sun, convective structure in the atmospheres of other stars cannot be directly imaged. \citet{landstreet_2007} provides a review of some of the observational clues concerning photospheric velocity fields and, along with \citet{gray_2009_third_sig, gray_2010_line_bisectors}, cites the main sources of velocity fields in the photosphere as macroturbulence (broadening due to the velocity of the distribution of granulation) and asymmetry arising from the rising and falling motion of the granules and their subsequent Doppler shifts. \citet{schwarzschild_1975} suggests that by comparing the physical parameters of granules on the Sun to observations of giant and supergiant stars that the dominant convective elements might  be so large as to only a number a few.

\begin{figure} 
\centering
\includegraphics[width=\textwidth]{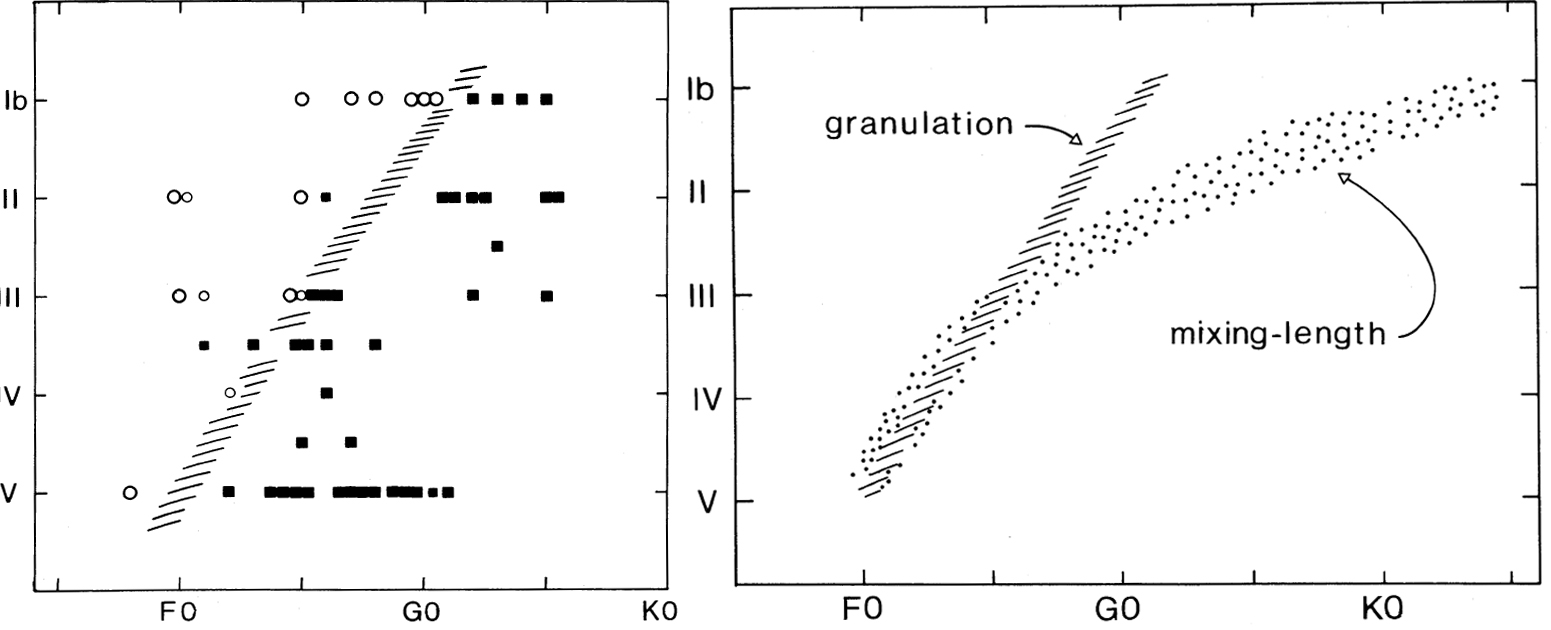}
\caption[Granulation and Convection Boundaries]{
The left panel shows luminosity class plotted against temperature to show the position of the granulation boundary. The filled squares denote normal granulation while the inverse bisectors are indicated by open circles \citep{gray_1989_granubound}. This boundary is compared to the convection boundary in the right panel, which \citeauthor{gray_1989_granubound} label ``mixing-length''.
\label{fig:gran_bound}}
\end{figure}

As well as the coronal/non-coronal divide between stars, there are other empirical boundaries across the HR Diagram.  The granulation boundary runs from F0V to G1Ib stars \citep{1990_granb}. This boundary is defined by bisector shapes such as those in Figure \ref{fig:bisector}. The bisector inclination changes direction from one side of the boundary to the other. The right (cool) side of the boundary shows bisectors typically expected from granulation while the left (hot) side display a bisector indicative of some other type of velocity field. The granulation boundary was discovered  by \citet{gray_1989_granubound} and is shown in Figure \ref{fig:gran_bound}. Granulation is considered to be the top of the convective envelope in stars and the region defining the onset of convection on the HR Digram will be associated with the boundary. The right panel of Figure  \ref{fig:gran_bound} shows how they are close enough to be deemed connected, except at higher luminosities where the mixing length is not as clearly defined. Another empirical boundary is the rotation boundary (not shown). It parallels the granulation boundary and separates the slow rotating cool stars on the right of the HR Diagram with fast rotating hot stars on the left of the HR Diagram. Doppler shifts caused by  stellar rotation can affect spectral line profiles, especially for fast rotating hot stars. Since this thesis is mainly concerned with cool slow rotating giants, the minor effects of rotational broadening will be ignored for the rest of the analysis.

\section{Tomographic Technique and Data Masks}\label{sec:tomog_tech}

\citet{dravins_1987_stellargran} points out that when observing solar granulation in optical line profiles, the resulting asymmetries are on the order of 1\% of the line width. To search for line asymmetries in stellar photospheres, very high spectral resolution is required. Even in cases where such high resolution is attainable, analysing individual line profiles may not be an efficient method of identifying velocity fields in the photosphere. If many spectral lines could be inspected simultaneously, efficiency would be improved and the need for high resolution data could be lessened.   \citet{alvarez_tomog_1} describes the need for such a method and this leads to the tomographic\footnote{The technique is described as \emph{``Tomographic''}  in the classical sense - due to the process of taking cuts or slices of a spectrum. It should not to be confused with the broader astronomical use of the word meaning a reconstruction of a configuration using projections taken under different angles.} technique employed in \citet{alvarez_tomog_2}. Although originally implemented to follow the propagation of shock waves across the photosphere of long-period variable (LPV) stars, there is scope for applying this method to non-LPV giants.

While the bisector technique (Figure \ref{fig:bisector}) is sufficient for observing asymmetry in a single feature, more persuasive methods are needed for detecting asymmetry across a series of lines. Red giant spectra are extremely crowded in the optical domain. The tomographic technique relies on the correlation of the observed spectrum with numerical masks probing layers of different optical depths. The masks are numerical arrays of one's and zero's that allow spectral values through the ``holes'' (the one's) while blocking out the rest of the spectrum (the zero's). The mask is stepped across the spectrum and at each step an average value of the unmasked parts of the spectrum is calculated. It is therefore possible to extract relevant information from crowded optical spectra by forming a composite profile of the lines formed at the same depth. The value of the composite at each wavelength step is given in \citet{jorissen_2003_tomog} as a convolution:
\begin{equation}
\mbox{Comp}(\Delta\lambda) = \int_{\lambda_1}^{\lambda2} s(\lambda - \Delta\lambda)m(\lambda)d\lambda 
\end{equation}
where $s(\lambda)$ is the observed spectrum, $m(\lambda)$ is the mask template and $\lambda_1$ and $\lambda_2$ are the boundaries of the spectral range covered by the observed spectrum.

The masks themselves are constructed using reliable synthetic spectra of late-type giant stars to identify the depth of formation of any given spectral line. The computed synthetic spectra originate from static models of red giant stars in spherical symmetry  by \citet{plez_1992_2} and \citet{plez_1992_1}. These synthetic spectra were computed for the spectral range 3850 - 6900\AA\ with a resolution of $\Delta\lambda=0.03$\AA. The spectral synthesis used linelists from \citet{plez_1998}, \citet{bessell_plez_1998} and \citet{alvalrez_plez_1998} which address the TiO and VO bands that can be prevalent in the optical spectra of giants (as seen in Chapter 4). 

\begin{figure}[ht!]
\includegraphics[width=\textwidth]{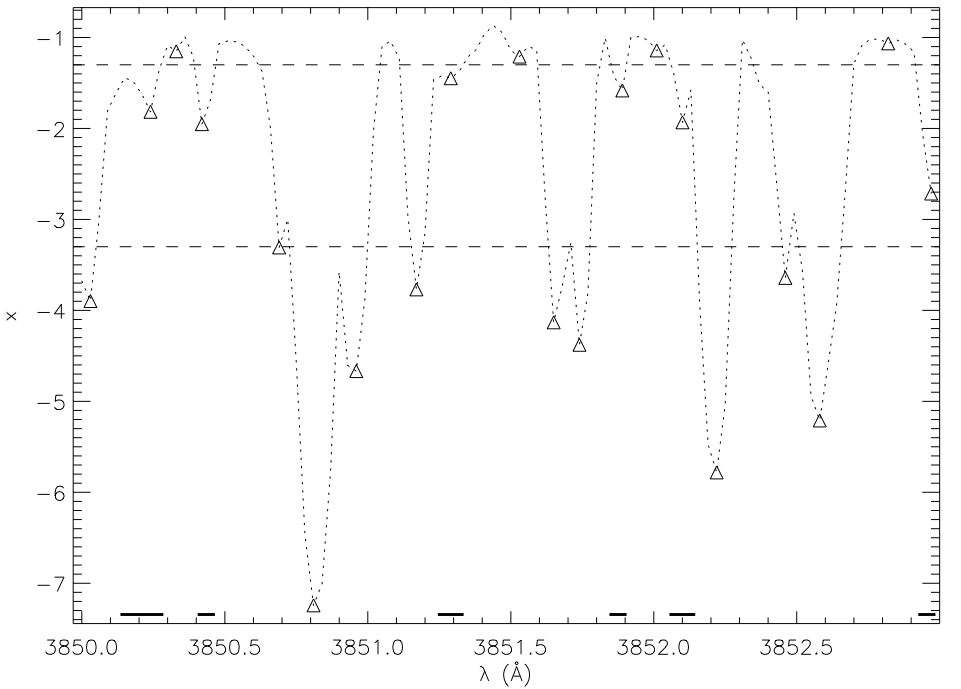}
\caption[Tomographic Mask Construction]{
Tomographic mask construction showing how the ``holes'' (thick dashed lines at the bottom of the blot)  correspond to lines formed in the range selected (triangles falling within the thin dashed lines). The range chosen was for  lines forming at depths $x$ such that $-1.3 \leq x < -3.3$ and a large $\Delta x$ value was used for clarity.
\label{fig:alvarex_masks_selection}}
\end{figure}

\begin{figure}[ht!]
\includegraphics[width=\textwidth]{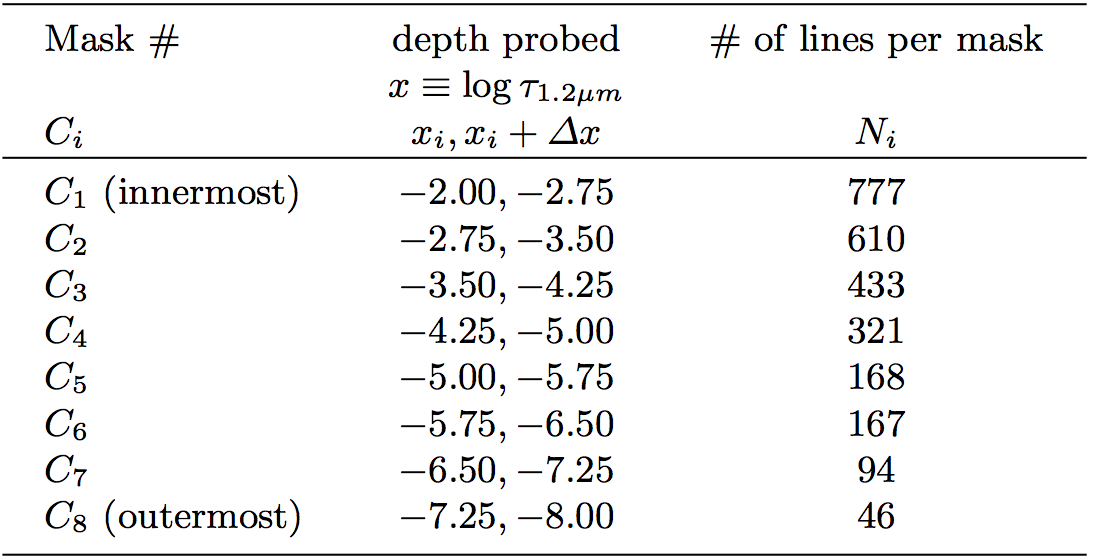}
\caption[Numerical Masks]{
Numerical masks from \citet{alvarez_tomog_2} and \citet{jorissen_2003_tomog}. The masks were constructed from synthetic templates with log(g) of 0.9 and a temperature of 3500\ K. 
\label{fig:c_masks}}
\end{figure}

For simplicity, the geometrical depth corresponding to monochromatic optical depth $\tau_{\lambda} = 2/3$ was found at each considered wavelength, $\lambda$. This function should not differ much from the average depth of formation for sufficiently strong lines \citep{magain_1986_linedepths}.  For the masks used in this study the depth function is:
\begin{equation}
x =log\tau_0
\end{equation}
where $\tau_0$ is the optical depth at the reference wavelength of $1.2\mu m$.  Different masks are then constructed from the collection of  $N$ wavelengths  $\lambda _{i,j}(1 \leq j \leq N)$ such that $x_{i} \leq x(\tau _{\lambda _{i,j}}=2/3) < x_{i+1}  = x_{i} + \Delta x$ where $\Delta x$ is some constant optimized to keep enough lines in any given mask without losing too much resolution in terms of geometrical depth. Figure \ref{fig:alvarex_masks_selection} shows how the masks are formed. The triangles mark the minima of the depth function. The range, or depth, that is being considered is marked by dashed lines. When a minima falls in this range, a corresponding ``hole'' is created in the mask as shown by the thick dashes at the bottom of the plot. The details of the masks used in this study are shown in Figure \ref{fig:c_masks}. The masks were constructed from synthetic templates with log(g) of 0.9 and a temperature of 3500\ K. 

\section{Line Asymmetries with Spectral Tomography}

While there have been several attempts to apply the tomographic technique to LPV and supergiant stars \citep{alvarez_tomog_1, alvarez_tomog_2, alvarez_tomog_3, jorissen_2003_tomog, josselin_2007} it has not been rigorously tested on a sample of low and intermediate mass non-LPV giants. To investigate the photospheric motions of the cool giants like EG And and HD148349, a method was developed to analyze existing optical spectra of giant stars. This method will hereafter be referred to as LAST (\emph{\textbf{L}ine \textbf{A}symmetries with \textbf{S}pectral \textbf{T}omography}). LAST is an original IDL procedure written for this study in order to incorporate several processes. These processes included reading both archival spectra and numerical masks and formatting them in a manner that allowed them to be compared. LAST then applies the masks to the data and forms composite spectral features. Finally, a measurement of the skewness of the resultant feature is carried out for each mask to determine how the asymmetry changes throughout the photosphere.

\begin{figure}[ht!]
\centering
\includegraphics[width=\textwidth]{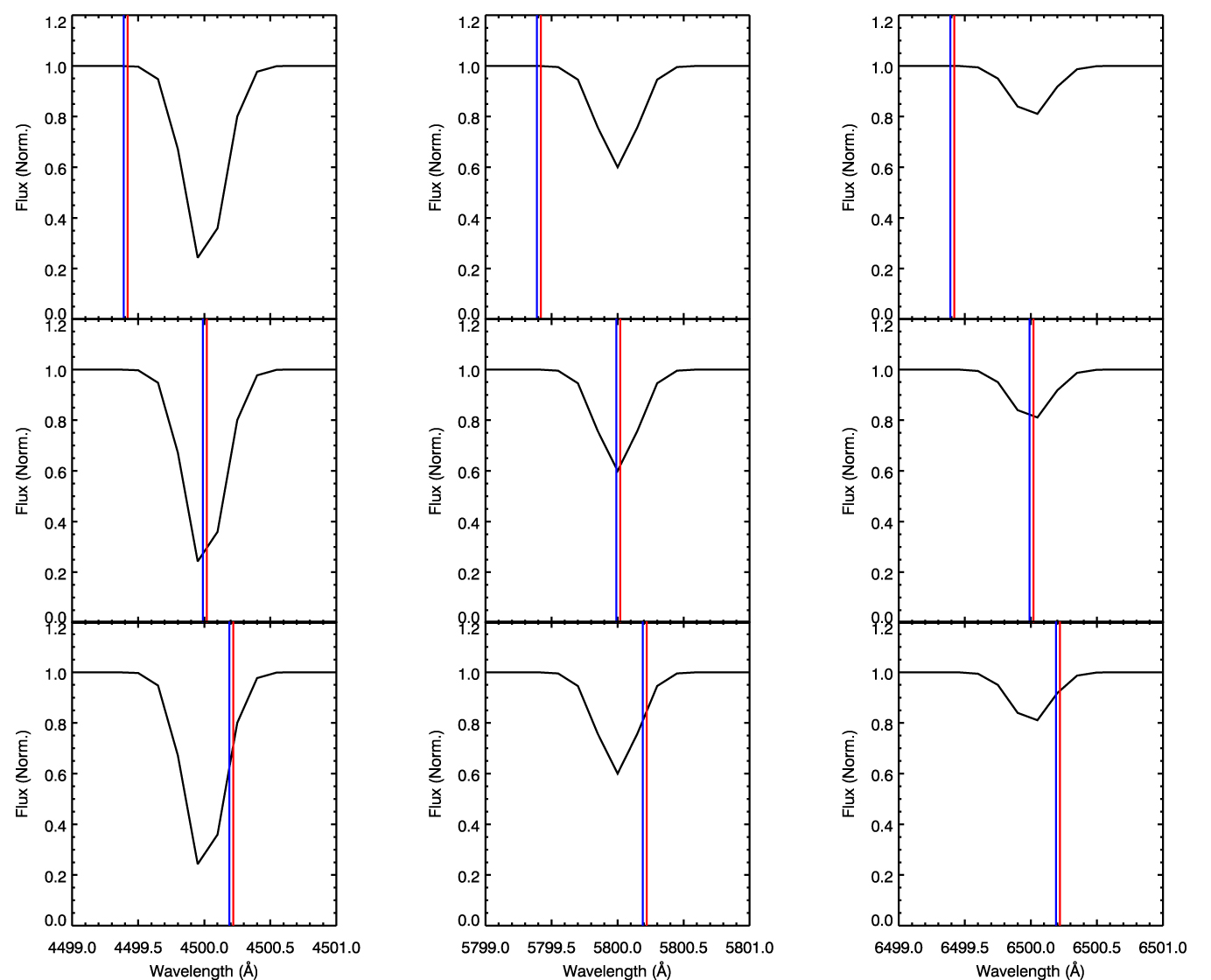}
\caption[LAST Simulated Data and Masks]{
Simulated data to test LAST. The resolution is comparable to that of {\sl ELODIE} with three absorption features of similar dimensions to those appearing in {\sl ELODIE} spectra. A numerical mask has also been simulated with holes that correspond to the size of those from \citet{alvarez_tomog_2}. The blue and red vertical lines show the start and end points respectively of the mask hole. The columns display the three simulated absorption lines, while the rows show the a mask hole as it is stepped along the spectrum. The resulting composite spectral feature is show in Figure \ref{fig:blast_test_ccf}. 
\label{fig:blast_test_data}}
\end{figure}

\begin{figure}[ht!]
\centering
\includegraphics[width=\textwidth]{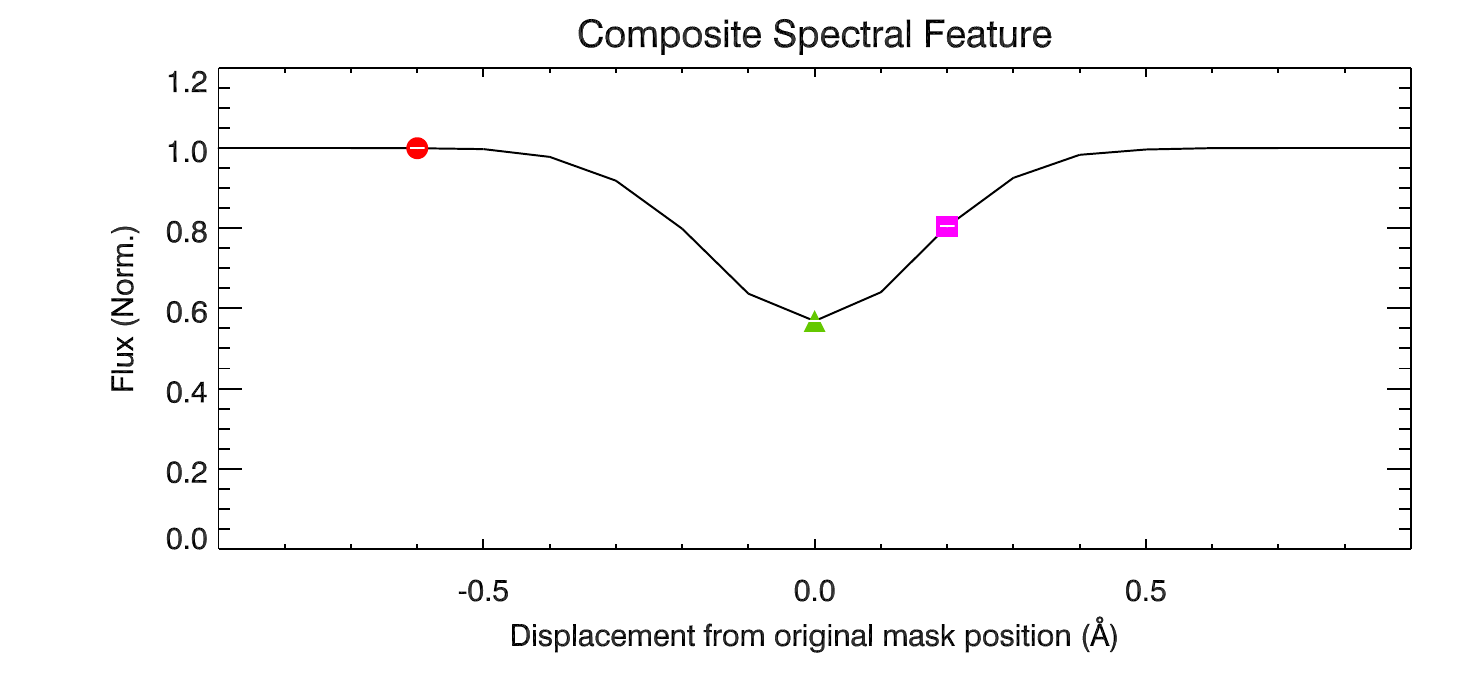}
\caption[LAST Test Composite Feature]{
As the numerical mask is stepped across a spectrum it will form a composite spectral feature. Using the simulated mask and data shown in Figure \ref{fig:blast_test_data} the above composite spectral feature was formed. The red circle, green triangle and purple square correspond to the three steps shown in  Figure \ref{fig:blast_test_data}.
\label{fig:blast_test_ccf}}
\end{figure}

To test LAST, a simplified version of the data and masks were modelled. The data was simulated as three gaussian lines and a continuum at a resolution similar to that of the {\sl ELODIE} data. A numerical mask was also simulated with holes that correspond to the size of those from \citet{alvarez_tomog_2}. The holes lined up with the three simulated absorption lines. The top row of Figure \ref{fig:blast_test_data} shows close-ups of the three simulated absorption lines. The mask holes are shown as the narrow gaps between the blue and red vertical lines. The two subsequent rows of  Figure \ref{fig:blast_test_data} correspond to the mask being stepped along the spectrum. The composite spectral feature formed from this process is shown in Figure \ref{fig:blast_test_ccf}.  It can be seen that lowest point of the feature is at the zero displacement point as expected due to to the mask holes lining up with the deepest, central region of the spectral lines. 

\begin{figure}[ht!]
\centering
\includegraphics[width=\textwidth]{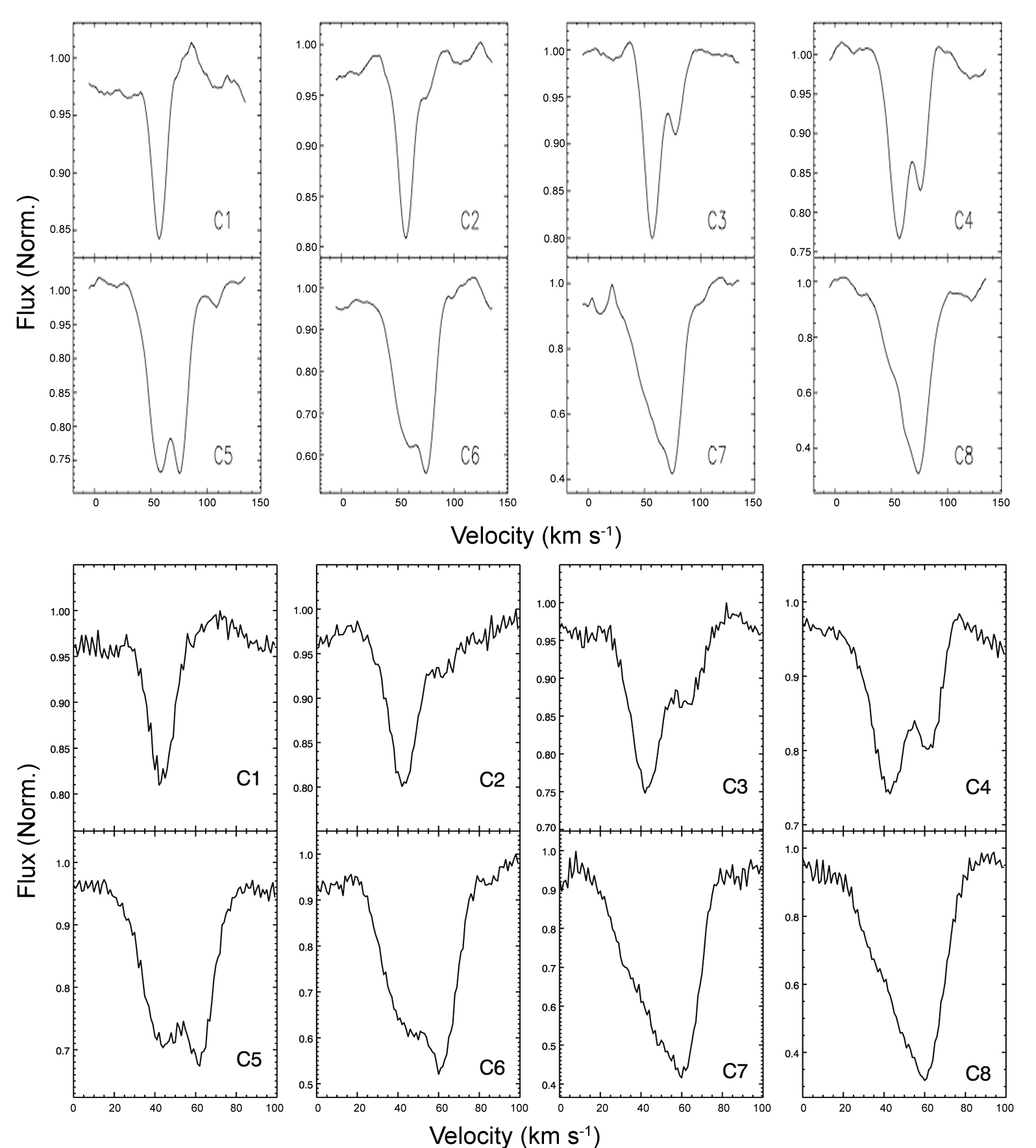}
\caption[LAST Composite for V Tau]{
LAST composite comparison for V Tau. The top panel shows  the results obtained in \citet{alvarez_tomog_2}, while the bottom panel shows the LAST output for the same target. The numerical masks from Figure \ref{fig:c_masks} are labelled C1 to C8. The features have been flux normalised and the the x-axis is -20 to 150\ km\ s$^{-1}$ for the \citeauthor{alvarez_tomog_2} plots, and 0 to 100\ km\ s$^{-1}$ for each LAST plot.
\label{fig:ccf_vel_v_tau}}
\end{figure}

After confirming the composite formation was successful for simulated data, LAST was applied to an {\it ELODIE} observation of  the mira V Tau and compared to the results obtained in \citet{alvarez_tomog_2} for the same target. The composite spectral features matched up almost perfectly for all 8 numerical masks as shown in Figure \ref{fig:ccf_vel_v_tau}.

\begin{figure}[ht!]
\centering
\includegraphics[width=\textwidth]{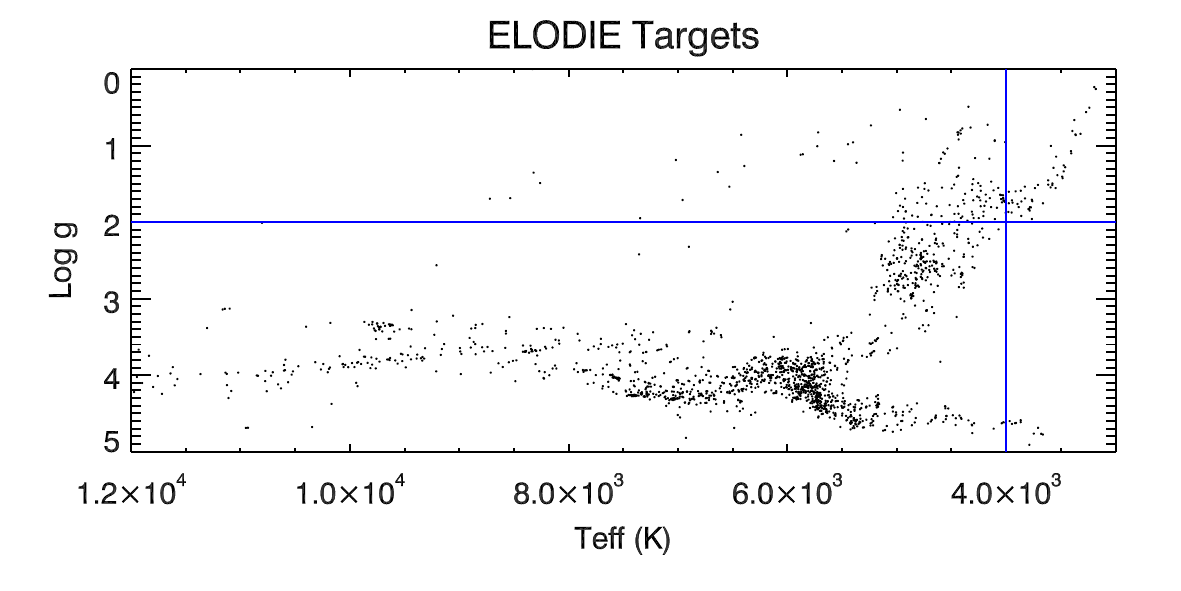}
\caption[{\sl ELODIE} Targets]{
All the archival {\sl ELODIE} targets. The cool giants analysed for this study have a Log(g) less than 2 and $T_{e}$ less than 4000\ K. These limits are shown as blue lines and confine the study to the targets in the top right-hand corner of the plot. 
\label{fig:logg_teff_all_lined}}
\end{figure}

Next it was necessary to identify which targets could be examined with LAST.  The tomographic technique was originally developed for stars that were observed with {\sl ELODIE}. {\sl ELODIE}, as discussed in Chapter 2, covered a spectral range 3850 - 6800\AA\ with a spectral resolution of 42,000 \citep{elodie_baranne}. Archival data were obtained in FITS format through the CDS database. All the spectra were reduced to the rest-frame and normalized to their pseudo-continuum \citep{elodie_archive_2001}.  This made the library ideal for applying a code to analyse all its contents in a consistent manner. There are 709 unique targets in the {\sl ELODIE} archive that have high enough spectral resolution for the masks to be applied. These targets are shown in Figure \ref{fig:logg_teff_all_lined}. As the masks were constructed with models using a Log(g) of 0.9 and a temperature of 3500\ K, LAST was applied to the 51 targets that have Log(g) less than 2 and a temperature of less than 4000\ K. These targets are highlighted in the top right-hand corner of Figure \ref{fig:logg_teff_all_lined}.

\begin{figure}[ht!] 
\centering
\includegraphics[width=\textwidth]{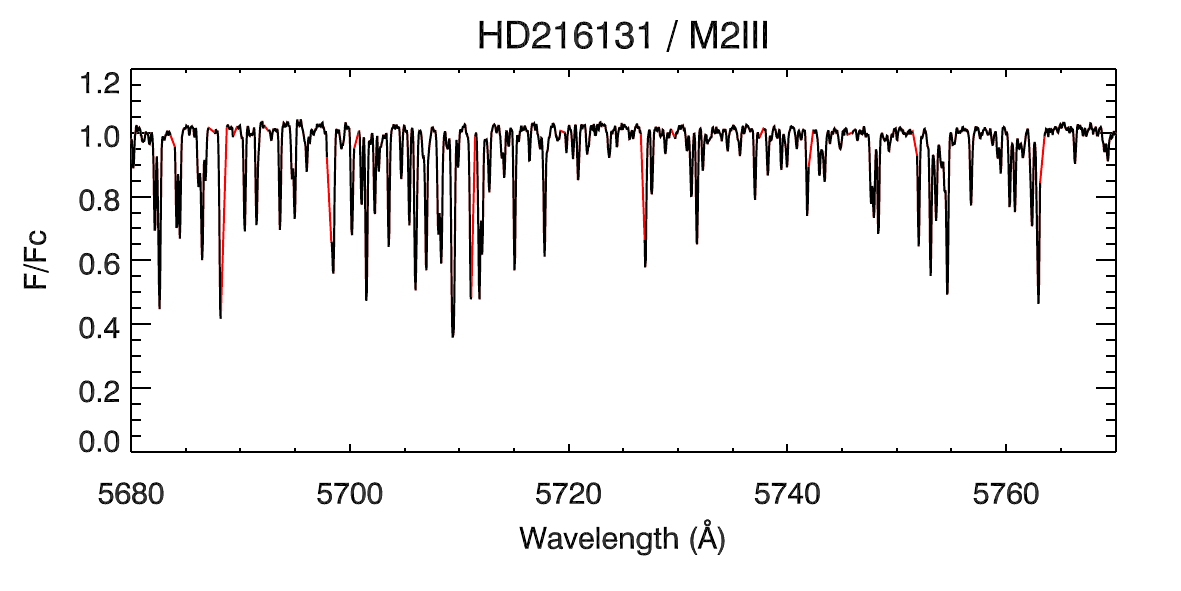}
\caption[Typical ELODIE Spectrum]{
A typical {\sl ELODIE} Spectrum for an M2 giant. As with all {\sl ELODIE} library stars, the spectrum has been corrected to the rest frame and normalised to the pseudo-continuum.  The black line shows the original data with the gaps indicating \emph{NaN}s are still present. The underplotted red spectrum is the \emph{NaN}-free spectrum.
\label{fig:typical_elodie_spectrum}}
\end{figure}

When LAST  is applied to a spectrum, it initially reads in a flux array directly from the {\sl ELODIE} FITS file for that star. The corresponding wavelength array is then  generated from parameters contained in the header file. Before applying the numerical masks to the spectrum, bad pixel points must be removed. These points are labelled as \emph{NaN} values (i.e.\ ``\emph{Not a Number}'') in IDL and are the cause of the gaps in the spectrum shown in Figure \ref{fig:typical_elodie_spectrum}. While the removal of these points may result in some of the spectral lines having an unrealistic shape, the number of lines affected are very small and can be considered negligible. The region shown in Figure \ref{fig:typical_elodie_spectrum} has a higher than normal number of \emph{NaN}'s and was chosen to highlight their presence,  but overall  \emph{NaN}'s make up less than 3\% of the flux points in each spectrum.

\begin{figure}[ht!]
\centering
\includegraphics[width=\textwidth]{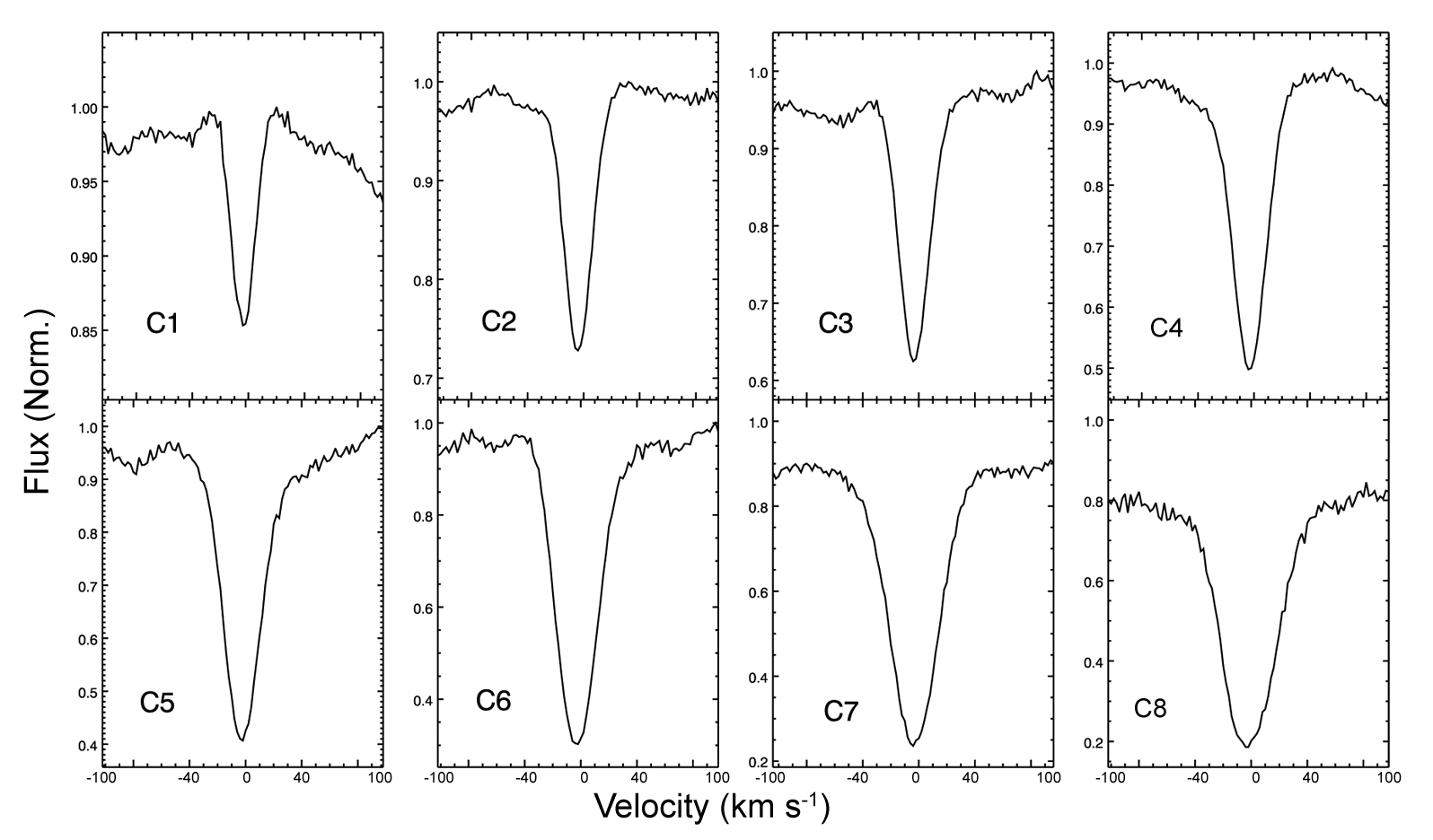}
\caption[LAST Composite Feature for an {\sl ELODIE} M2 Giant]{
LAST Composite Feature for an {\sl ELODIE} M2 Giant. All 8 numerical masks are shown. In each case the x-axis show velocity from $-100$\ km\ s$^{-1}$ to $+100$\ km\ s$^{-1}$, while the y-axis shows the continuum-normalised flux.
\label{fig:ccf_HD219734}}
\end{figure}

\begin{figure}[ht!]
\centering
\includegraphics[width=\textwidth]{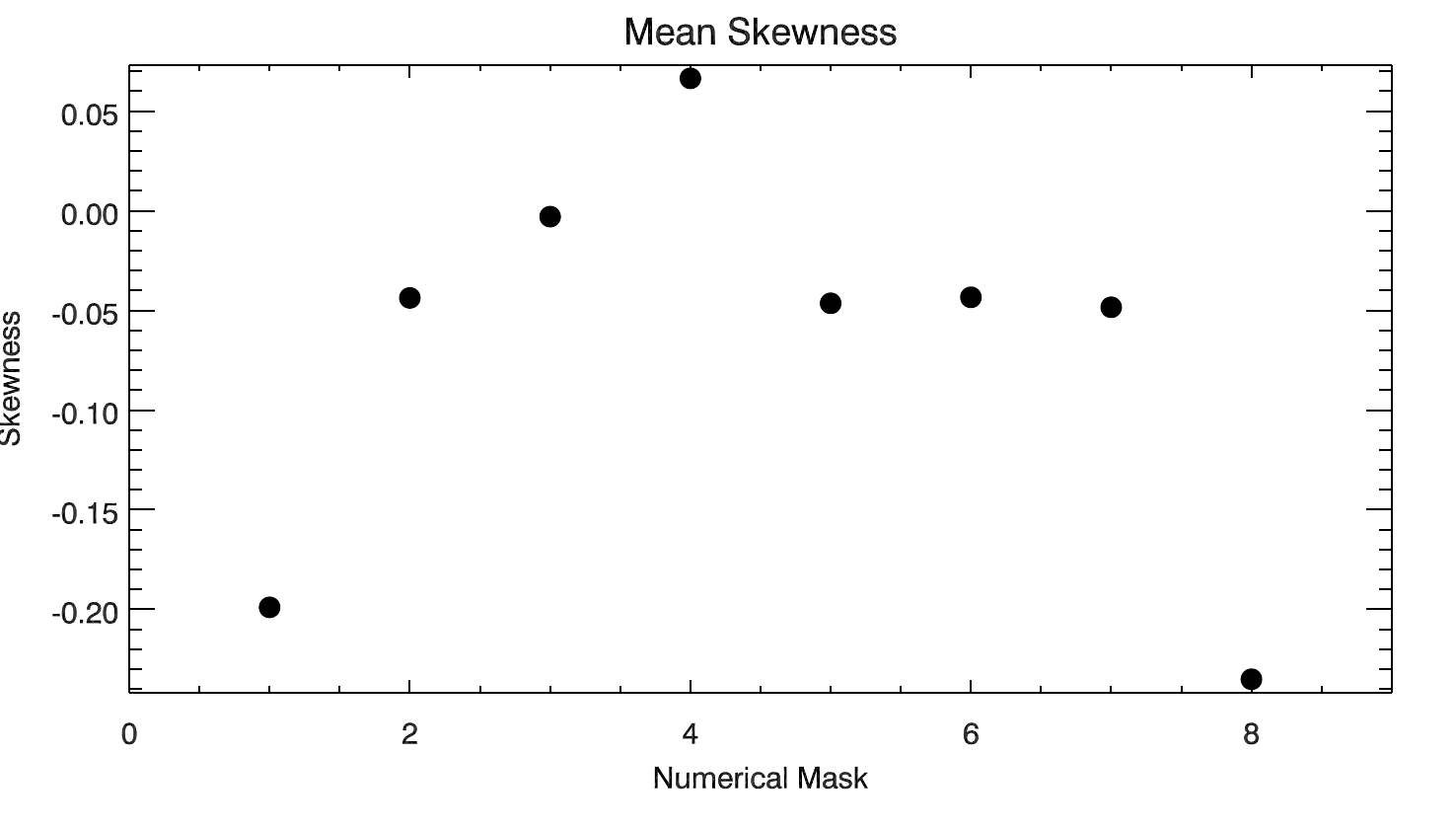}
\caption[LAST Mean Skewness Across Masks]{
The mean skewness for all of the ELODIE targets at each numerical mask. While the absolute values will remain difficult to tie down (both for the layers of the photosphere and the asymmetry values themselves) the values of skewness through the photospheric layers relative to each other implies that the peak occurs when the most material can be considered uprising.
\label{figccf_skew_mean}}
\end{figure}

Once the \emph{NaN}'s have been removed,  the numerical masks are applied to the spectrum and the mean of the unmasked fluxes is calculated. The mask is then stepped along the spectrum and at each step a new mean is calculated. As the mask was generated by finding the optical lines formed at certain depths in a synthetic spectrum, a composite spectral line should emerge from the steps, with the lowest point coinciding with the point at which the mask holes line up with the middle of the spectral lines. The code repeats this process for all eight of the numerical masks shown in in Figure \ref{fig:c_masks}. The result is a series of composite spectral features formed from lines at different depths in the photosphere. A sample plot is shown in Figure \ref{fig:ccf_HD219734}. 
The skewness of each composite spectral feature is measured. Finally, the above steps are carried out for all 51 cool giants in the {\sl ELODIE} archive. When the skewness for at each layer (i.e.\ for each mask) was averaged over all 51 targets a trend in the asymmetry of the optical spectral lines was uncovered. This trend is shown in Figure \ref{figccf_skew_mean}. A positive skewness suggests a blueshifted velocity field (upward motion) while a negative skewness would imply a redshifted velocity field. While it is not possible to determine absolute geometrical depths from the mask layers, the value of skewness of each layer relative to each other indicates an upwelling motion similar to what we see in solar granulation. When these composite features were folded back across their centre points and subtracted from each other, the peaks in their difference gave an indication of the velocity fields involved. In the central layers (C3 - C5) velocity fields of around $\sim 5$\ km\ s$^{-1}$ were measured, while the innermost and outermost layers showed little evidence of velocity differences.

\section{Discussion and Conclusions}

A method of measuring the asymmetry of spectral lines formed in the photosphere was developed using the numerical masks of \citet{alvarez_tomog_2, jorissen_2003_tomog} and a tomographic technique to obtain composite spectral features from the optical lines forming at different layers in the photosphere. An original IDL code, LAST,  was written to process more than 50 {\sl ELODIE} giants that had log(g) less than 2 and $T_{e}$ less than 4000 K. 

This analysis represented the first attempt to determine the variation of spectral line asymmetry across the photosphere of red giant stars. While it is difficult to determine quantitatively the actual value of the asymmetry, just as it is difficult to assign the masks with real geometrical depths in the photosphere \citep{jorissen_2003_tomog}, the change in the asymmetry of the layers relative to each other would suggest an upward moving velocity field that increases in strength and then subsides towards the outer layers of the photosphere. This image seems to fit in with the expected picture of a stellar  convecting atmosphere (Section \ref{sec:photo_motion}) not having equal regions of upward, horizontal and downward photospheric motions.

The lack of progress in this field is highlighted by the fact that \citet{schwarzschild_1975} is the most frequently cited work for analysis in this field despite the ``tentative'' nature of the  conclusions in that study. To estimate an energy flux from the photospheric motions requires the use of an even older study.  \citet{schwarzschild_1948} argued that  the energy responsible for heating the outer atmosphere is a stream of mechanical noise produced by granulation and transported into the chromosphere. This convective energy flux is given by:
\begin{equation}
F_{con} = \rho_{\star} \omega^2 v_S
\label{eqn:schwarz_gran}
\end{equation}
where $\rho_{\star}$ is the photospheric density, $\omega$ is the material velocity and $v_s$ is the isothermal sound speed. To calculate $\rho_{\star}$, the ideal gas law was used:
\begin{equation}
\rho_{\star} = \frac{P_{\star} \mu m_H}{kT_{\star}}
\label{eqn:ideal_gas}
\end{equation}
where $P_{\star}$ is the photospheric pressure, $\mu$ is the average particle mass, $m_H$ is the mass of hydrogen, $k$ is the Boltzmann constant and $T_{\star}$ is the effective temperature (given a value of 3500\ K for the cool giants in this study). To estimate $P_{\star}$, a pressure-gravity relationship from \citet{gray_photosphere_book} was used:
\begin{equation}
P_{\star} = P_{\odot}\left(\frac{g_{\star}}{g_{\odot}} \right)^{0.6}
\label{eqn:pressure_grav}
\end{equation}
Adopting a $g_{\star}$ value of 10\ cm\ s$^{-2}$ (appropriate for the evolved cool giants in this study), and solar values of $P_{\odot} = 3.8\times10^{4}$\ g\ cm$^{-1}$\ s$^{-2}$ and $g_{\odot} = 2.74\times10^{4}$\ cm\ s$^{-1}$ results in $P_{\star} = 3.3\times10^{2}$\ g\ cm$^{-1}$\ s$^{-2}$.  $v_S$ is given by:
\begin{equation}
v_S = \sqrt{\frac{k T}{\mu m_H}}
\label{eqn:sound_vel}
\end{equation}
where $k$ is the Boltzmann constant, $\mu$ is the mean particle mass and $m_H$ is the mass of hydrogen. This works out at $v_S\sim 5.5$\ km\ s$^{-1}$. $\omega$ can be estimated as the velocity of the rising granules taken from the LAST code $\sim 5$\ km\ s$^{-1}$, which is similar to the radialtangential macrotrubulence velocities found by \citet{1985gray}. Substituting all three values into Equation \ref{eqn:schwarz_gran} gives $F_{con} = 1.7\times10^{8}$\ erg\ cm$^{-2}$\ s$^{-1}$. Following \cite{schwarzschild_1948}, this value is divided by 10 (as at any moment only one tenth of the surface will be occupied by rising granules) to give an upper-limit for the energy transport of the steam of noise, $L_{noise} <  1.7\times10^{7}$\ erg\ cm$^{-2}$\ s$^{-1}$. When compared with the energy budget discussions from the previous chapters, this value appears high enough to satisfy the energy requirements of heating the chromosphere and driving the wind.  
	
\newpage
\thispagestyle{plain}
\mbox{}	

\chapter{Conclusions and Future Work} 
\label{chapter:conclusions_and_future_work}

In this thesis, the advantages of using EG Andromedae as a resource for studying red giant chromospheres has been demonstrated. Unlike other binary systems (i.e.\ $\zeta$ Aur and VV Cep), the relative size of the white dwarf compared to the red giant means it can act as a pencil-beam UV backlight to spatially probe the giant atmosphere. Recent  {\sl HST}/STIS ultraviolet observations of both EG And and the M3III spectral standard HD148349 from Observing Cycle 17 are presented here  for the first time. This dataset is from the {\sl HST} proposal No.11690, PI: Crowley \citep{hst_prop}. These observations are part of an ongoing program that also encompasses EG And observations from Observing Cycle 13, {\sl HST} proposal No.9487, PI: Espey \citep{espey_2002_prop}.

As well as the UV data, new optical spectra of both EG And and HD148349 were recovered from acquisition images taken during both observing cycles. These spectra were uncovered by developing a new technique for extracting contemporaneous low-resolution optical spectra from previously unused acquisition images. These spectra were then incorporated into a synthetic photometry study to measure the variability of both targets. While EG And was found to be quite well-behaved when the binarity of the system was accounted for, HD148349 was found to be less bright than previous observations. Although there had been no published values as low as the magnitudes found in this thesis, there is a wealth of amateur observations that cluster around the values presented here. It seems necessary that HD148349 should be further studied and perhaps demoted from the catalogues of spectral standards. Through the photometric measurements and comparisons to all relevant archival small aperture STIS observations, a calibration correction factor was derived that makes the optical extraction technique applicable to all past and future small aperture observations with {\sl HST}/STIS.

The optical spectra from both targets were also compared to  libraries of M giant spectra. This comparison confirmed that both are currently M3 giants, but also unearthed an onset point for the strong TiO  bands in M class giants between the M3 and M4 subclasses. New values of extinction were calculated and the flux ratio of the two stars was used to calculate an improved distance to EG And.

This distance was supported by using the Wilson-Bappu relationship between luminosity and the width of the \ion{Mg}{ii} \emph{h} and \emph{k} lines. This relationship was examined, despite the line profiles being mutilated by the interstellar medium, by exploiting the symmetry of the emission lines. Both EG And HD148349 showed terminal wind velocities expected of intermediate size cool giants with EG And's higher speed possibly a symptom of the binarity the system. A similar effect is seen in $\zeta$ Aurigae binaries.

Another aspect of the analysis was the \ion{C}{ii}] 2325\AA\ multiplet. This thesis represents the most detailed study of these important electron density diagnostic lines in any symbiotic system.  MELF, a fitting procedure to simultaneously fit all the lines of the multiplet and return values of integrated line flux, line width and radial velocity was developed. The fitting routine also led to new values of the wavelength of the emission lines that differ slightly from the accepted wavelengths from \citet{nist_book_values} and are closer to the new values presented by \citet{young_2011_cii_wavelengths}. Underlying absorption from line blends was tested to determine which line ratio diagnostics were most trust-worthy.  These ratios were used as electron density diagnostics and both targets were found to show similar values to well-studied isolated cool giants. The effect of the white dwarf on the emission lines in the chromosphere was modelled using a geometric Sobolev light-scattering planetary atmosphere law. The additional flux seen at uneclipsed phases suggests that additional material is being irradiated by the dwarf. This could be due to ancillary structure in the atmosphere caused by pulsations lifting material out of the Roche potential.

The idea of the giant losing mass through energetic mechanisms in the photosphere was investigated. LAST, a tomographic technique to form composite spectral profiles from low-resolution optical lines using numerical masks developed by \citet{alvarez_tomog_2} and \citet{jorissen_2003_tomog} was created and used to form composites and measure the asymmetry across the photospheric layers of archival cool giant stars. The resulting trend in asymmetry confirmed the granular motion expected of red giant stars. The energy mechanism imparting the heat in the outer atmosphere could well originate from the convective envelope and a combination of acoustic and Alfv\'{e}n waves.

\begin{figure}[ht]
\centering
\includegraphics[width=\textwidth]{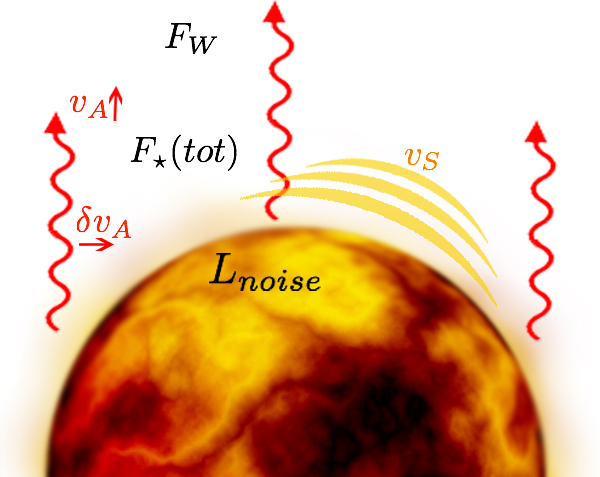}
\caption[Energy Schematic of Red Giant]{
Author's energy schematic of a red giant showing the different parts of the atmosphere where the energy mechanisms can be found. $L{noise}$ originates in the photosphere while $F_{\star}(tot)$ and $F_W$ arise in the outer atmosphere.  Alfv\'{e}n waves are shown in red while acoustic waves are yellow.
\label{fig:rg_energy.jpg}}
\end{figure}

The energy budget further into the atmosphere was easier to determine. Using the \ion{Mg}{ii} lines as a proxy for the radiative losses in the chromosphere, along with the stellar parameters determined for the system,  the energy flux densities for the atmosphere and wind were calculated.  Evidence was found to support the theory of  Alfv\'{e}n waves driving the wind and perhaps heating the chromosphere. Table \ref{tab:energy_budget} summarises the energy fluxes calculated in this study and Figure \ref{fig:rg_energy.jpg} shows where they arise in the red giant atmosphere. It has been shown that EG And is a well-behaved system and once the effects of the white dwarf are taken into  account it can be used as an insight into red giant chromospheres. 

\begin{table}
\begin{center}
\begin{tabular}{ll}
\hline
\textbf{Energy Flux}  & \textbf{Calculated Value}   \\
& (erg cm$^{-2}$ s$^{-1}$) \\
\hline
\hline 
$L_{noise}$ &  $1.7\times10^{7}$ \\   
$F_{\star}(tot)$ & $7.9\times10^{5}$ \\
$F_W$ & $4.7\times10^{5}$ \\
$F_{M_{S}}$ &  $7.5\times10^{5}$ \\
$F_{M_{A}}$ & $2.3\times10^{6}$ \\
\hline
\hline
\end{tabular}
\caption[Red Giant Atmospheric Energy Budget]{
Red giant atmospheric energy budget values calculated in this study.}
\label{tab:energy_budget}
\end{center}
\end{table}

\section{Future Work}\label{sec:future_work}

In this thesis, the importance of UV observations of symbiotic stars for addressing the unknown structure of red giant chromospheres has been demonstrated. An obvious concern for this type of work is the lack of a natural successor to {\sl HST}. Having outlived its expected lifetime and no longer being considered for any future servicing missions, the future of ultraviolet spectroscopy looks bleak. Hypothetically a new ultraviolet instrument could conduct long-scale observing programs of symbiotic stars to accurately determine their stellar parameters. Their use as probes of red giant chromospheres would then become even more wide-ranging as the effects of different types of binary companions could be explored. To continue this work without UV capabilities, the resolution and sensitivity offered by the radio observatories EVLA and ALMA could be used to probe even more distant systems. Previous radio observations by \cite{taylor_seaqusit_1984, seaquist_taylor_1990} of the emission from photoionised regions of symbiotic systems helped to estimate binary separation, wind density, and ionising photon luminosity. Exploring the radio thermal free-free continuum would help probe the same chromospheric plasma as the \ion{C}{ii]} 2325\AA\ and \ion{Mg}{ii} \emph{h} and \emph{k} lines.

The new method of extracting optical data from STIS G430L acquisition images is based on IRAF scripts. This procedure could be automated so that optical spectra could be easily gleaned from all similar STIS observations in the future. A potential stumbling block  in this area is the necessary calibration scale factor. In this study the value of this correction factor was estimated by several methods. A way of estimating the value in an automated manner could prove to be non-trivial. Automating software is a logical and necessary next step with several aspects of this work. Along with the optical spectra extraction technique from Chapter 4, the splicing and weighting of the data outlined in Chapter 3 has several manual steps. MELF, the multiplet fitting tool described in Chapter 5 needs sensible initial estimates of all the fitting parameters. Replacing these manual steps with automated ones would make the reduction and analysis tools developed here more applicable to future (and archival) datasets. At that point the tools could easily be made available to the community online.

The tomographic technique could be extended and applied to other targets. The numerical masks themselves could be extended to make use of  2D and 3D convective models of the photosphere, such as those from \citet{asplund_2000}, to help interpret the asymmetries caused by convective velocity fields. The wealth of archival data could be processed with these improved numerical masks and a broad view of photospheric motions across the HR Diagram could be established.

\newpage
\thispagestyle{plain}
\mbox{}	

\appendix






\chapter{\ion{C}{ii]} 2325\AA\ Multiplet Fits}\label{app:cii_2325_multiplet_fits}

This appendix shows the \ion{C}{ii]} 2325\AA\ linefits for all 11 {\sl HST}/STIS E230M Cycle 11 and Cycle 17 observations. In each case the data is in black with the fit overplotted in red. Annotated on each plot are the line centre (mean) the FWHM in terms of km\ s$^{-1}$ and the area of each gaussian fit. Also annotated are the 5 line ratios.

\begin{figure}[ht]
\centering
\includegraphics[width=\textwidth]{writeup_images/mpfit_cycle17_1010.pdf}
\includegraphics[width=\textwidth]{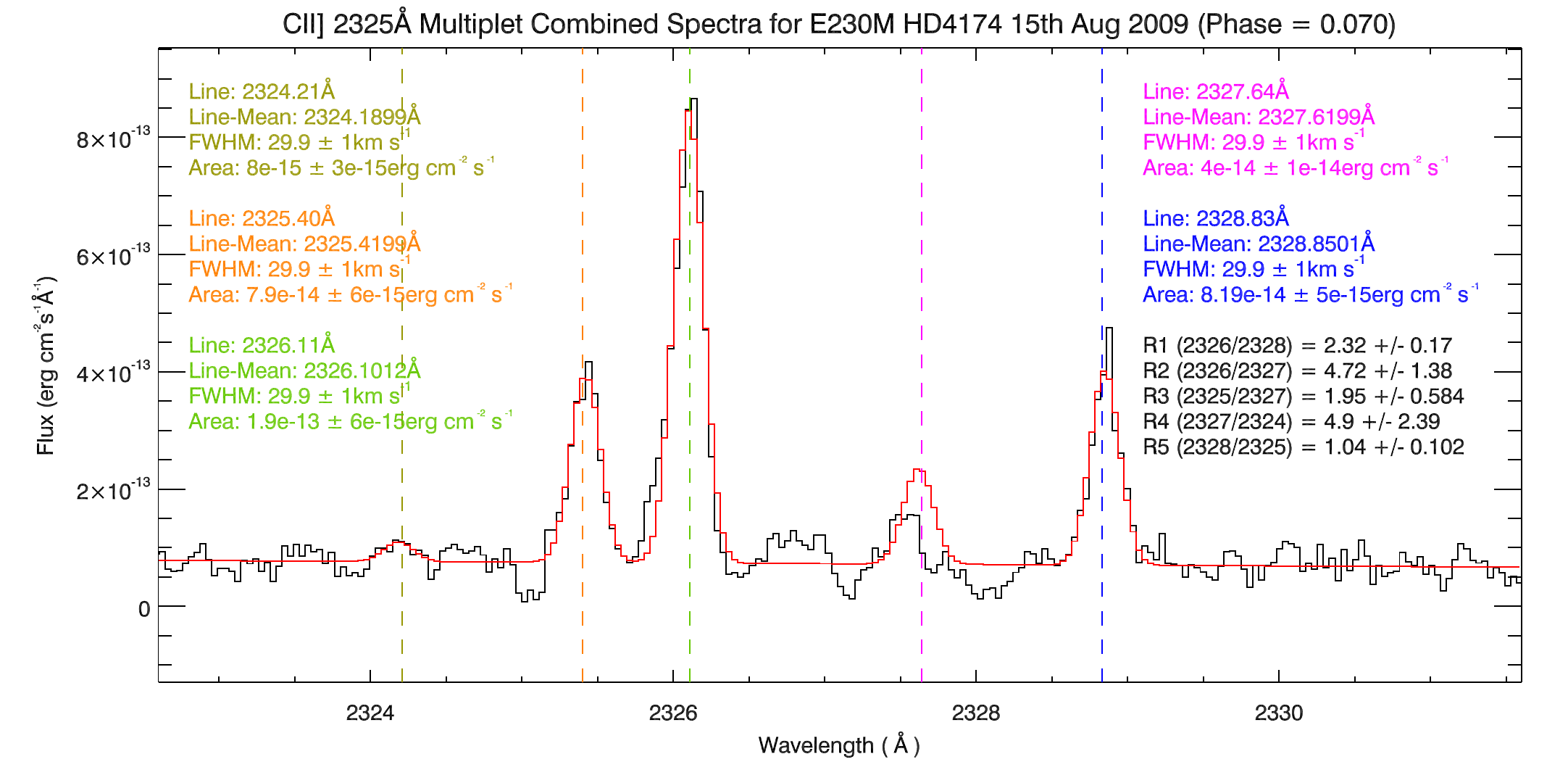}
\end{figure}
\begin{figure}[ht]
\centering
\includegraphics[width=\textwidth]{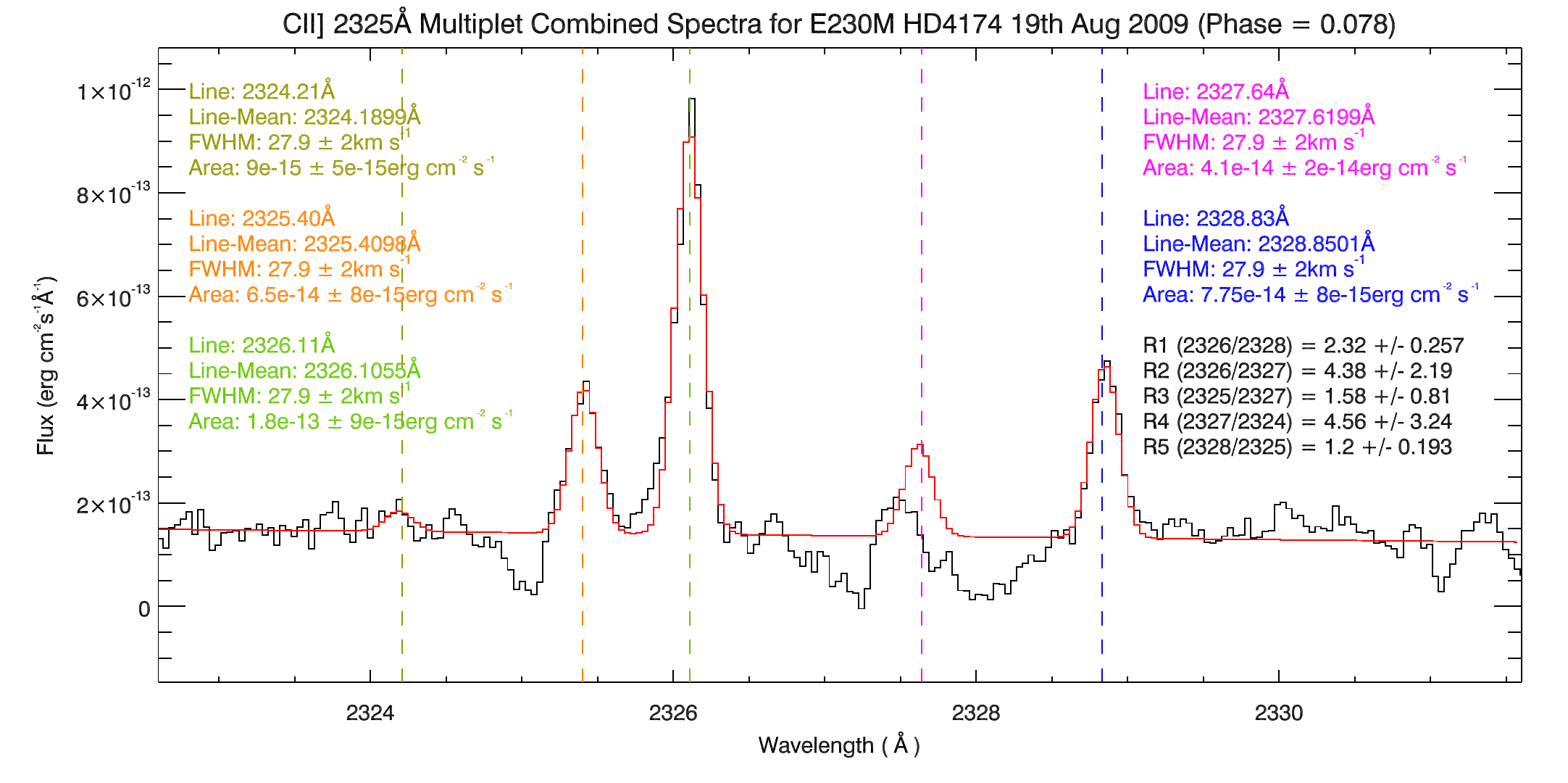}
\includegraphics[width=\textwidth]{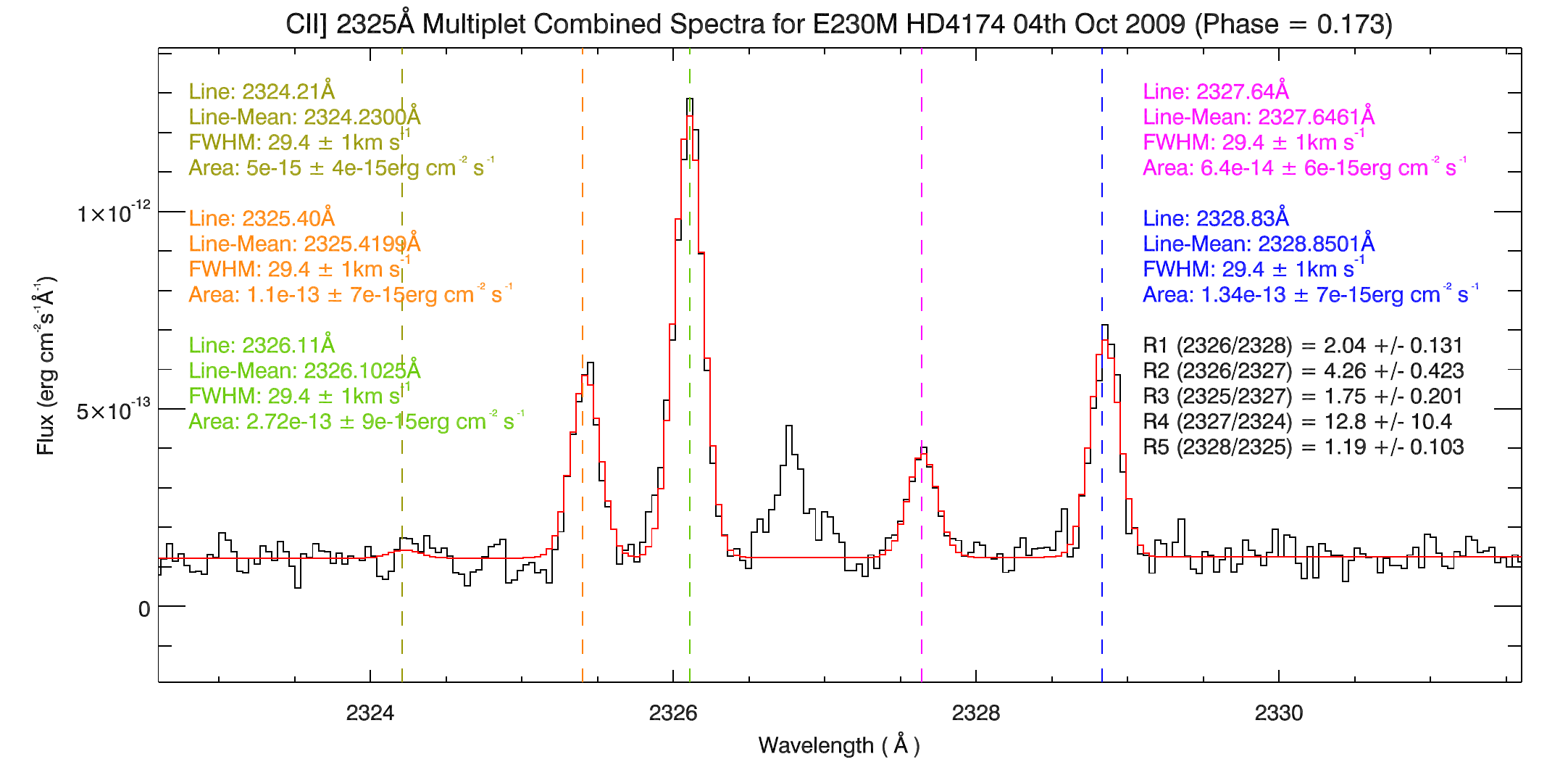}
\end{figure}
\begin{figure}[ht]
\centering
\includegraphics[width=\textwidth]{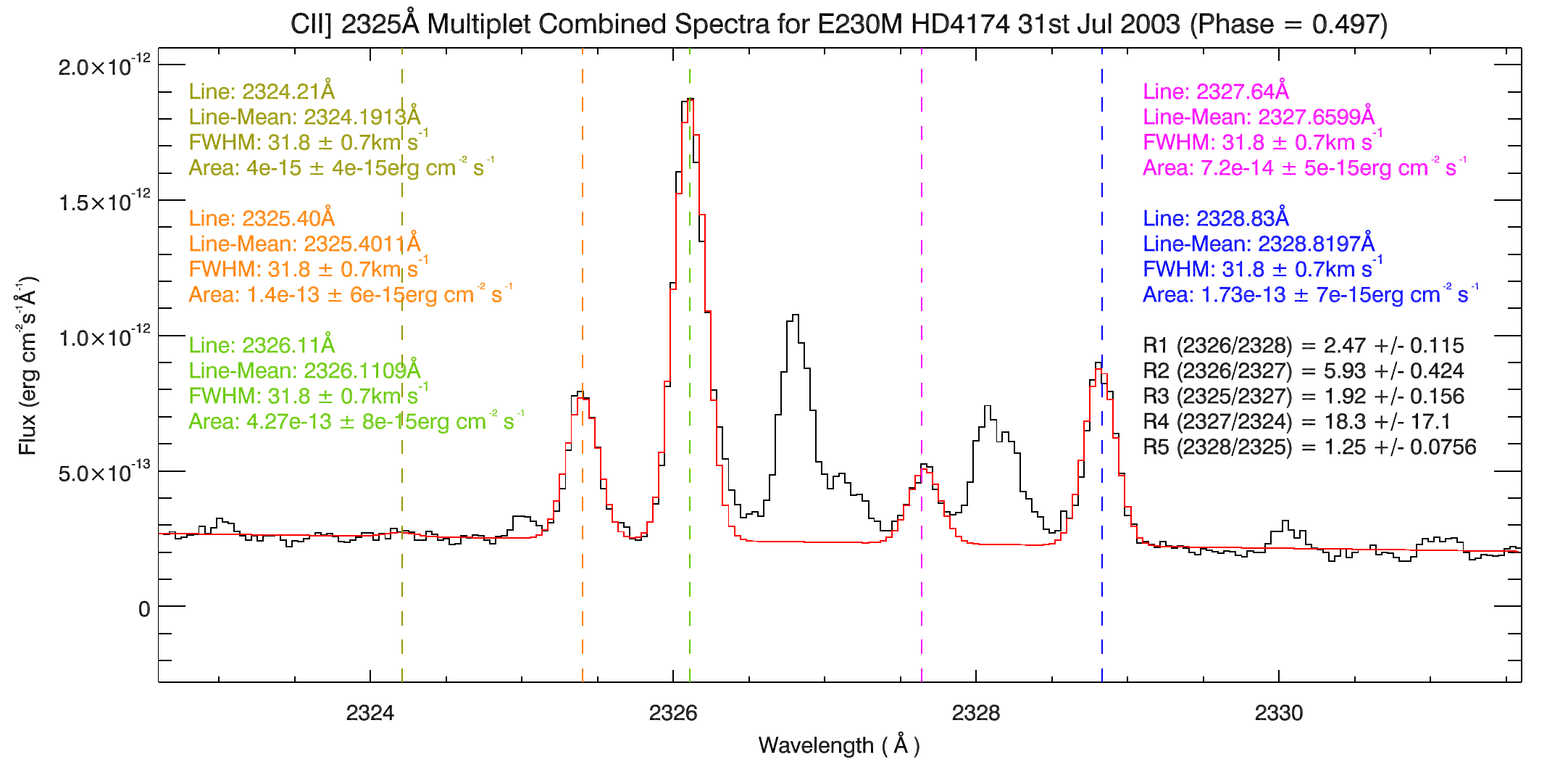}
\includegraphics[width=\textwidth]{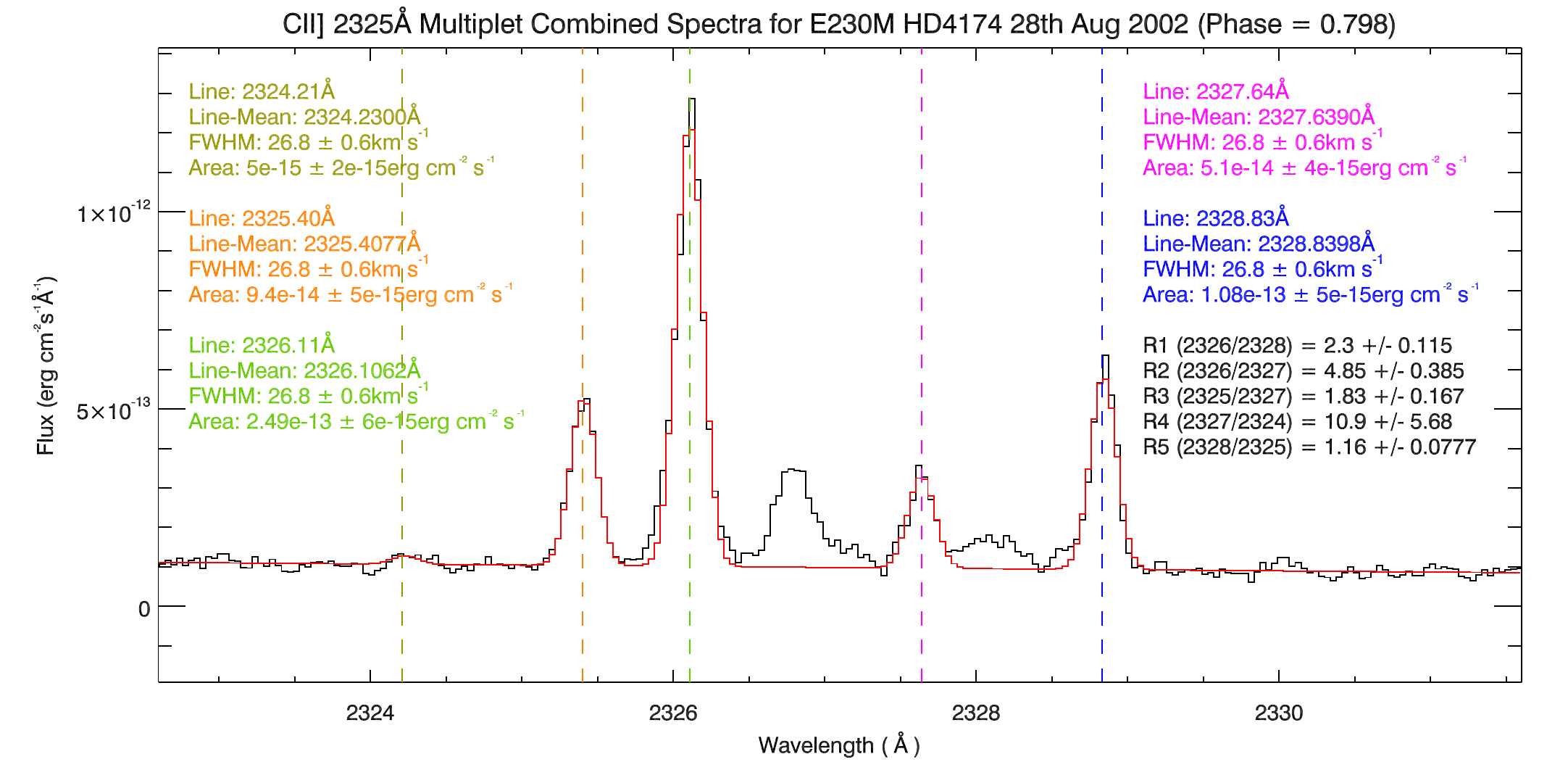}
\end{figure}
\begin{figure}[ht]
\centering
\includegraphics[width=\textwidth]{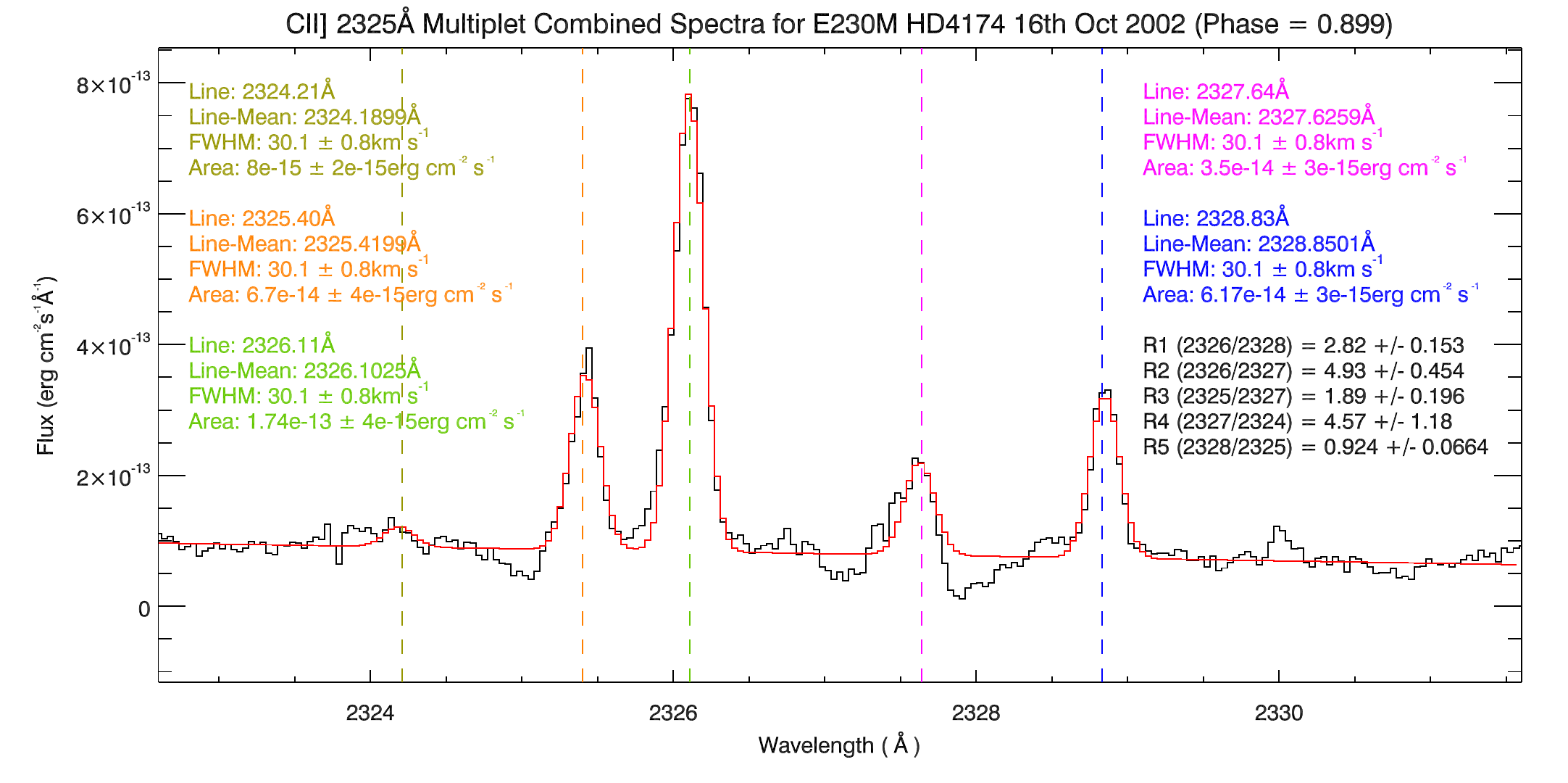}
\includegraphics[width=\textwidth]{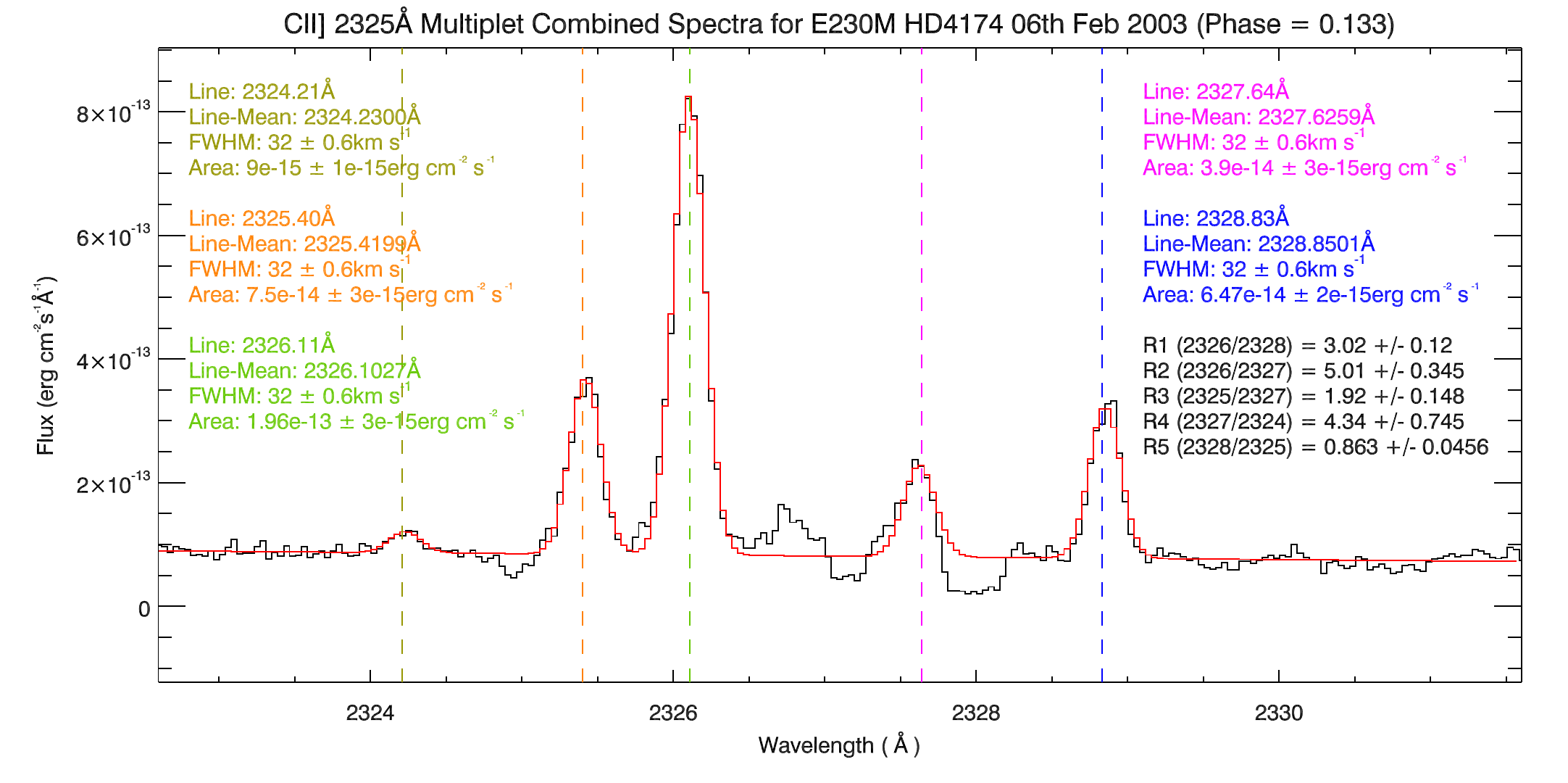}
\end{figure}
\begin{figure}[ht]
\centering
\includegraphics[width=\textwidth]{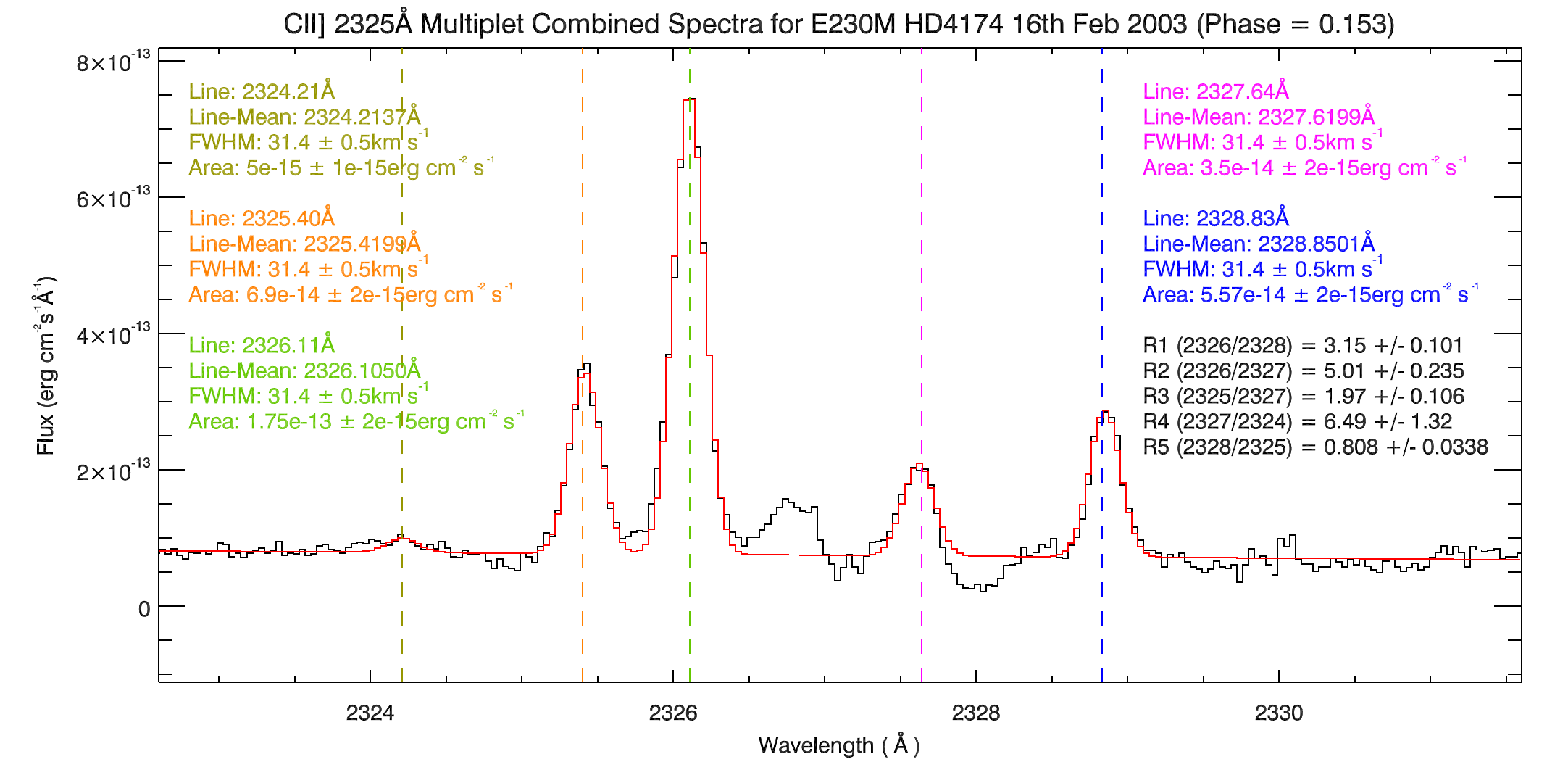}
\includegraphics[width=\textwidth]{writeup_images/mpfit_cycle11_combo_406010_406020_406030.pdf}
\end{figure}


\begin{footnotesize} 
\bibliographystyle{Latex/Classes/jmb}
\renewcommand{\bibname}{References} 
\bibliography{mybib} 

\begin{thebibliography}{216}
\expandafter\ifx\csname natexlab\endcsname\relax\def\natexlab#1{#1}\fi

\bibitem[{{Aller}(1991)}]{aller_atoms_stars_nebulae}
{\sc {Aller}, L.H.} (1991). {\em {Atoms, Stars, and Nebulae}\/}. Cambridge
  University Press.

\bibitem[{{Alvarez} \& {Plez}(1998)}]{alvalrez_plez_1998}
{\sc {Alvarez}, R. \& {Plez}, B.} (1998). {Near-infrared narrow-band photometry
  of M-giant and Mira stars: models meet observations}. {\em \aap\/}, {\bf
  330}, 1109--1119.

\bibitem[{{Alvarez} {\em et~al.\/}(2000){Alvarez}, {Jorissen}, {Plez}, {Gillet}
  \& {Fokin}}]{alvarez_tomog_1}
{\sc {Alvarez}, R., {Jorissen}, A., {Plez}, B., {Gillet}, D. \& {Fokin}, A.}
  (2000). {Envelope tomography of long-period variable stars. I. The
  Schwarzschild mechanism and the Balmer emission lines}. {\em \aap\/}, {\bf
  362}, 655--665.

\bibitem[{{Alvarez} {\em et~al.\/}(2001{\natexlab{a}}){Alvarez}, {Jorissen},
  {Plez}, {Gillet}, {Fokin} \& {Dedecker}}]{alvarez_tomog_2}
{\sc {Alvarez}, R., {Jorissen}, A., {Plez}, B., {Gillet}, D., {Fokin}, A. \&
  {Dedecker}, M.} (2001{\natexlab{a}}). {Envelope tomography of long-period
  variable stars II. Method}. {\em \aap\/}, {\bf 379}, 288--304.

\bibitem[{{Alvarez} {\em et~al.\/}(2001{\natexlab{b}}){Alvarez}, {Jorissen},
  {Plez}, {Gillet}, {Fokin} \& {Dedecker}}]{alvarez_tomog_3}
{\sc {Alvarez}, R., {Jorissen}, A., {Plez}, B., {Gillet}, D., {Fokin}, A. \&
  {Dedecker}, M.} (2001{\natexlab{b}}). {Envelope tomography of long-period
  variable stars III. Line-doubling frequency among Mira stars}. {\em \aap\/},
  {\bf 379}, 305--322.

\bibitem[{{Asplund} {\em et~al.\/}(2000){Asplund}, {Ludwig}, {Nordlund} \&
  {Stein}}]{asplund_2000}
{\sc {Asplund}, M., {Ludwig}, H.G., {Nordlund}, {\AA}. \& {Stein}, R.F.}
  (2000). {The effects of numerical resolution on hydrodynamical surface
  convection simulations and spectral line formation}. {\em \aap\/}, {\bf 359},
  669--681.

\bibitem[{{Ayres}(2010)}]{ayres_2010_stis_wrinkles}
{\sc {Ayres}, T.R.} (2010). {Ironing Out the Wrinkles in STIS}. In {\em 2010
  Space Telescope Science Institute Calibration Workshop - Hubble after SM4\/}.

\bibitem[{{Babcock}(1950)}]{babcock_1950}
{\sc {Babcock}, H.W.} (1950). {Magnetic Field of a Late-Type Star, HD 4174}.
  {\em \pasp\/}, {\bf 62}, 277.

\bibitem[{Bally \& Reipurth(2006)}]{bally2006birth}
{\sc Bally, J. \& Reipurth, B.} (2006). {\em The birth of stars and planets\/}.
  Cambridge University Press.

\bibitem[{{Baranne} {\em et~al.\/}(1996){Baranne}, {Queloz}, {Mayor},
  {Adrianzyk}, {Knispel}, {Kohler}, {Lacroix}, {Meunier}, {Rimbaud} \&
  {Vin}}]{elodie_baranne}
{\sc {Baranne}, A., {Queloz}, D., {Mayor}, M., {Adrianzyk}, G., {Knispel}, G.,
  {Kohler}, D., {Lacroix}, D., {Meunier}, J.P., {Rimbaud}, G. \& {Vin}, A.}
  (1996). {ELODIE: A spectrograph for accurate radial velocity measurements.}
  {\em \aaps\/}, {\bf 119}, 373--390.

\bibitem[{{Belczy{\'n}ski} {\em et~al.\/}(2000){Belczy{\'n}ski},
  {Miko{\l}ajewska}, {Munari}, {Ivison} \& {Friedjung}}]{belczy_2000}
{\sc {Belczy{\'n}ski}, K., {Miko{\l}ajewska}, J., {Munari}, U., {Ivison}, R.J.
  \& {Friedjung}, M.} (2000). {A catalogue of symbiotic stars}. {\em \aaps\/},
  {\bf 146}, 407--435.

\bibitem[{{Berman}(1932)}]{berman_1932}
{\sc {Berman}, L.} (1932). {The Spectrum and Temperature of T Coronae (Nova
  1866)}. {\em \pasp\/}, {\bf 44}, 318.

\bibitem[{{Bessell}(1990)}]{bessell_1990_ubvri_passbands}
{\sc {Bessell}, M.S.} (1990). {UBVRI passbands}. {\em \pasp\/}, {\bf 102},
  1181--1199.

\bibitem[{{Bessell}(2000)}]{bessell_2000_hipparcos_passbands}
{\sc {Bessell}, M.S.} (2000). {The Hipparcos and Tycho Photometric System
  Passbands}. {\em \pasp\/}, {\bf 112}, 961--965.

\bibitem[{{Bessell} {\em et~al.\/}(1998){Bessell}, {Castelli} \&
  {Plez}}]{bessell_plez_1998}
{\sc {Bessell}, M.S., {Castelli}, F. \& {Plez}, B.} (1998). {Erratum: Model
  atmospheres broad-band colors, bolometric corrections and temperature
  calibrations for O-M stars}. {\em \aap\/}, {\bf 337}, 321.

\bibitem[{{Bevington} \& {Robinson}(2003)}]{bevington_data_reduction}
{\sc {Bevington}, P.R. \& {Robinson}, D.K.} (2003). {\em {Data reduction and
  error analysis for the physical sciences}\/}. {McGraw-Hill.}

\bibitem[{{Boggess} {\em et~al.\/}(1978){Boggess}, {Carr}, {Evans}, {Fischel},
  {Freeman}, {Fuechsel}, {Klinglesmith}, {Krueger}, {Longanecker} \&
  {Moore}}]{boggess_1978_iue}
{\sc {Boggess}, A., {Carr}, F.A., {Evans}, D.C., {Fischel}, D., {Freeman},
  H.R., {Fuechsel}, C.F., {Klinglesmith}, D.A., {Krueger}, V.L., {Longanecker},
  G.W. \& {Moore}, J.V.} (1978). {The IUE spacecraft and instrumentation}. {\em
  \nat\/}, {\bf 275}, 372--377.

\bibitem[{{Bohlin}(1998)}]{bohin_1998_isr}
{\sc {Bohlin}, R.} (1998). {Absolute Flux Calibration For Prime STIS Echelle
  Modes With The 0.2x0.2'' Slit}. Tech. rep., {Space Telescope Science
  Institute}.

\bibitem[{{Bohlin}(1996)}]{bohlin_1996_stis_calib}
{\sc {Bohlin}, R.C.} (1996). {Spectrophotometric Standards From the Far-UV to
  the Near-IR on the White Dwarf Flux Scale}. {\em \aj\/}, {\bf 111}, 1743.

\bibitem[{{Brown} \& {Carpenter}(1984)}]{brown_carpenter_1984}
{\sc {Brown}, A. \& {Carpenter}, K.G.} (1984). {The temperature of C II
  emission-line formation regions in cool stars}. {\em \apjl\/}, {\bf 287},
  L43--L46.

\bibitem[{{Brown} {\em et~al.\/}(1981){Brown}, {Ferraz} \&
  {Jordan}}]{brown_ferraz_jordan_1981}
{\sc {Brown}, A., {Ferraz}, M. \& {Jordan}, C.} (1981). {The structure of
  chromospheres around late-type giants and supergiants}. In {R.~D.~Chapman},
  ed., {\em NASA Conference Publication\/}, vol. 2171 of {\em NASA Conference
  Publication\/}, 297--302.

\bibitem[{{Buchholz} \& {Ulmschneider}(1994)}]{ulm_1994}
{\sc {Buchholz}, B. \& {Ulmschneider}, P.} (1994). {Theoretical Basal Flux
  Limits from Acoustic Wave Calculations}. In {J.-P.~Caillault}, ed., {\em Cool
  Stars, Stellar Systems, and the Sun\/}, vol.~64 of {\em Astronomical Society
  of the Pacific Conference Series\/}, 363.

\bibitem[{{Byrne} {\em et~al.\/}(1988){Byrne}, {Murphy}, {Dufton}, {Kingston}
  \& {Lennon}}]{byrne_1988}
{\sc {Byrne}, P.B., {Murphy}, H.M., {Dufton}, P.L., {Kingston}, A.E. \&
  {Lennon}, D.J.} (1988). {Electron densities in late-type stars}. {\em
  \apj\/}, {\bf 197}, 205--208.

\bibitem[{{Campbell}(1892)}]{campbell_1892}
{\sc {Campbell}, W.W.} (1892). {The spectrum of Nova Aurigae in February and
  March, 1892.} {\em Astronomy and Astro-Physics\/}, {\bf 11}, 799--811.

\bibitem[{{Cardelli} {\em et~al.\/}(1989){Cardelli}, {Clayton} \&
  {Mathis}}]{cardelli_1989}
{\sc {Cardelli}, J.A., {Clayton}, G.C. \& {Mathis}, J.S.} (1989). {The
  relationship between infrared, optical, and ultraviolet extinction}. {\em
  \apj\/}, {\bf 345}, 245--256.

\bibitem[{{Carpenter}(1991)}]{carpenter_1991_prop}
{\sc {Carpenter}, K.} (1991). {Physical Conditions and Velocity Structures in
  the Red Giant Winds in the Binaries CI CYG and EG and}. In {\em HST
  Proposal\/}, 4251.

\bibitem[{{Carpenter}(1998)}]{carpenter_1998_coolstars}
{\sc {Carpenter}, K.G.} (1998). {Cool Stars}. In {J.~C.~Brandt, T.~B.~Ake, \&
  C.~C.~Petersen}, ed., {\em The Scientific Impact of the Goddard High
  Resolution Spectrograph\/}, vol. 143 of {\em Astronomical Society of the
  Pacific Conference Series\/}, 67.

\bibitem[{{Carpenter} {\em et~al.\/}(1985){Carpenter}, {Brown} \&
  {Stencel}}]{carpenter_brown_1985}
{\sc {Carpenter}, K.G., {Brown}, A. \& {Stencel}, R.E.} (1985). {The geometric
  extent of C II (UV 0.01) emitting regions around luminous, late-type stars}.
  {\em \apj\/}, {\bf 289}, 676--680.

\bibitem[{{Carpenter} {\em et~al.\/}(1991){Carpenter}, {Robinson}, {Wahlgren},
  {Ake}, {Ebbets}, {Linsky}, {Brown} \& {Walter}}]{carpenter_robinson_1991}
{\sc {Carpenter}, K.G., {Robinson}, R.D., {Wahlgren}, G.M., {Ake}, T.B.,
  {Ebbets}, D.C., {Linsky}, J.L., {Brown}, A. \& {Walter}, F.M.} (1991). {First
  results from the Goddard High-Resolution Spectrograph - The chromosphere of
  Alpha Tauri}. {\em \apjl\/}, {\bf 377}, L45--L48.

\bibitem[{{Clampin} {\em et~al.\/}(1996){Clampin}, {Hartig}, {Baum}, {Kraemer},
  {Kinney}, {Kutina}, {Pitts} \& {Balzano}}]{stis_acq_isr_1996}
{\sc {Clampin}, M., {Hartig}, G., {Baum}, S., {Kraemer}, S., {Kinney}, E.,
  {Kutina}, R., {Pitts}, R. \& {Balzano}, V.} (1996). {STIS Target Acquisitions
  I: CCD Point Source Acquisitions}. Tech. rep., STScI.

\bibitem[{{Cousins}(1964)}]{cousins_1964_hd_photom}
{\sc {Cousins}, A.W.J.} (1964). {Photometric Data for Stars in the Equatorial
  Zone (Seventh List)}. {\em Monthly Notes of the Astronomical Society of South
  Africa\/}, {\bf 23}, 175.

\bibitem[{{Cousins}(1971)}]{cousins_1971_photom}
{\sc {Cousins}, A.W.J.} (1971). {Photometric standard stars}. {\em Royal
  Observatory Annals\/}, {\bf 7}.

\bibitem[{{Cranmer} \& {van Ballegooijen}(2005)}]{cranmer_2005_alfenwaves}
{\sc {Cranmer}, S.R. \& {van Ballegooijen}, A.A.} (2005). {On the Generation,
  Propagation, and Reflection of Alfv{\'e}n Waves from the Solar Photosphere to
  the Distant Heliosphere}. {\em \apjs\/}, {\bf 156}, 265--293.

\bibitem[{{Crowley}(2006)}]{cian_thesis}
{\sc {Crowley}, C.} (2006). {\em {Red Giant Mass-Loss: Studying Evolved Stellar
  Winds with FUSE and HST/STIS}\/}. Ph.D. thesis, School of Physics, Trinity
  College Dublin, Dublin 2, Ireland.

\bibitem[{{Crowley} \& {Espey}(2010)}]{cian_2010_passadena}
{\sc {Crowley}, C. \& {Espey}, B.R.} (2010). {Cool Evolved Winds: Use of the
  Eclipsing Technique}. In {C.~Leitherer, P.~D.~Bennett, P.~W.~Morris, \&
  J.~T.~Van Loon}, ed., {\em Hot and Cool: Bridging Gaps in Massive Star
  Evolution\/}, vol. 425 of {\em Astronomical Society of the Pacific Conference
  Series\/}, 191.

\bibitem[{{Crowley} {\em et~al.\/}(2007){Crowley}, {Espey} \&
  {McCandliss}}]{cian_2007}
{\sc {Crowley}, C., {Espey}, B.R. \& {McCandliss}, S.R.} (2007). {EG And: FUSE
  and HST/STIS Monitoring of an Eclipsing Symbiotic Binary}. {\em ArXiv
  e-prints\/}.

\bibitem[{{Crowley} {\em et~al.\/}(2008{\natexlab{a}}){Crowley}, {Espey},
  {Roche}, {McCandliss} \& {Suzuki}}]{hst_prop}
{\sc {Crowley}, C., {Espey}, B., {Roche}, J., {McCandliss}, S. \& {Suzuki}, S.}
  (2008{\natexlab{a}}). {EG And: Providing the Missing Link Required for
  Modelling Red Giant Mass-loss}. In {\em HST Proposal\/}, 11690.

\bibitem[{{Crowley} {\em et~al.\/}(2008{\natexlab{b}}){Crowley}, {Espey} \&
  {McCandliss}}]{cian_2008}
{\sc {Crowley}, C., {Espey}, B.R. \& {McCandliss}, S.R.} (2008{\natexlab{b}}).
  {EG And: Far Ultraviolet Spectroscopic Explorer and Hubble Space Telescope
  STIS Monitoring of an Eclipsing Symbiotic Binary}. {\em \apj\/}, {\bf 675},
  711--722.

\bibitem[{{De Jager}(1990)}]{1990_granb}
{\sc {De Jager}, C.} (1990). {An explanation of the 'granulation boundary' in
  the HR diagram}. {\em \solphys\/}, {\bf 126}, 201--205.

\bibitem[{{Deutsch}(1956)}]{deutsch_1956}
{\sc {Deutsch}, A.J.} (1956). {The Circumstellar Envelope of Alpha Herculis.}
  {\em \apj\/}, {\bf 123}, 210.

\bibitem[{{Dravins}(1987)}]{dravins_1987_stellargran}
{\sc {Dravins}, D.} (1987). {Stellar granulation. I - The observability of
  stellar photospheric convection}. {\em \aap\/}, {\bf 172}, 200--224.

\bibitem[{{Dravins} {\em et~al.\/}(1981){Dravins}, {Lindegren} \&
  {Nordlund}}]{dravins_1981_solar_gran_asym}
{\sc {Dravins}, D., {Lindegren}, L. \& {Nordlund}, A.} (1981). {Solar
  granulation - Influence of convection on spectral line asymmetries and
  wavelength shifts}. {\em \aap\/}, {\bf 96}, 345--364.

\bibitem[{{Dreiling} \& {Bell}(1980)}]{dreiling_bell_1980_vega}
{\sc {Dreiling}, L.A. \& {Bell}, R.A.} (1980). {The chemical composition,
  gravity, and temperature of VEGA}. {\em \apj\/}, {\bf 241}, 736--758.

\bibitem[{{Dressel}(2007)}]{dhb}
{\sc {Dressel}, L.}, ed. (2007). {\em {STIS Data Handbook, Version 5.0}\/}.

\bibitem[{{Dumm} \& {Schild}(1998)}]{dumm_schild_1998}
{\sc {Dumm}, T. \& {Schild}, H.} (1998). {Stellar radii of M giants}. {\em New
  Astronomy\/}, {\bf 3}, 137--156.

\bibitem[{{Dupree} \& {Reimers}(1987)}]{dupree_reimers_1987}
{\sc {Dupree}, A.K. \& {Reimers}, D.} (1987). {Mass loss from cool stars}. In
  {Y.~Kondo}, ed., {\em Exploring the Universe with the IUE Satellite\/}, vol.
  129 of {\em Astrophysics and Space Science Library\/}, 321--353.

\bibitem[{{Eggleton}(1983)}]{eggleton_1983_rochelobe}
{\sc {Eggleton}, P.P.} (1983). {Approximations to the radii of Roche lobes}.
  {\em \apj\/}, {\bf 268}, 368.

\bibitem[{{Espey}(2002)}]{espey_2002_prop}
{\sc {Espey}, B.} (2002). {UV Sounding of the M-Giant Atmosphere in the
  Symbiotic Binary EG-AND}. In {\em HST Proposal\/}, 9487.

\bibitem[{{Evans}(1967)}]{evans_1967_hdradvel}
{\sc {Evans}, D.S.} (1967). {The Revision of the General Catalogue of Radial
  Velocities}. In {A.~H.~Batten \& J.~F.~Heard}, ed., {\em Determination of
  Radial Velocities and their Applications\/}, vol.~30 of {\em IAU
  Symposium\/}, 57.

\bibitem[{{Fekel} {\em et~al.\/}(2000){Fekel}, {Joyce}, {Hinkle} \&
  {Skrutskie}}]{fekel_part1}
{\sc {Fekel}, F.C., {Joyce}, R.R., {Hinkle}, K.H. \& {Skrutskie}, M.F.} (2000).
  {Infrared Spectroscopy of Symbiotic Stars. I. Orbits for Well-Known S-Type
  Systems}. {\em \aj\/}, {\bf 119}, 1375--1388.

\bibitem[{{Fitzpatrick}(1999)}]{fitzpatrick_1999_extinct}
{\sc {Fitzpatrick}, E.L.} (1999). {Correcting for the Effects of Interstellar
  Extinction}. {\em \pasp\/}, {\bf 111}, 63--75.

\bibitem[{{Fluks} {\em et~al.\/}(1994){Fluks}, {Plez}, {The}, {de Winter},
  {Westerlund} \& {Steenman}}]{fluks_1994}
{\sc {Fluks}, M.A., {Plez}, B., {The}, P.S., {de Winter}, D., {Westerlund},
  B.E. \& {Steenman}, H.C.} (1994). {M giants spectra and photometry (Fluks,
  1994)}. {\em VizieR Online Data Catalog\/}, {\bf 4105}, 50311.

\bibitem[{{Garcia}(1989)}]{1989_hdstand}
{\sc {Garcia}, B.} (1989). {A list of MK standard stars}. {\em Bulletin
  d'Information du Centre de Donnees Stellaires\/}, {\bf 36}, 27.

\bibitem[{{Gillett} {\em et~al.\/}(1971){Gillett}, {Merrill} \&
  {Stein}}]{gillett_1971_vega}
{\sc {Gillett}, F.C., {Merrill}, K.M. \& {Stein}, W.A.} (1971). {Observations
  of Infrared Radiation from Cool Stars}. {\em \apj\/}, {\bf 164}, 83.

\bibitem[{{Gray}(1976)}]{gray_photosphere_book}
{\sc {Gray}, D.F.} (1976). {\em {The observation and analysis of stellar
  photospheres}\/}. {Wiley.}

\bibitem[{{Gray}(1980)}]{gray_1980_sunarcasym}
{\sc {Gray}, D.F.} (1980). {Measurements of spectral line asymmetries for
  Arcturus and the sun}. {\em \apj\/}, {\bf 235}, 508--514.

\bibitem[{{Gray}(1981)}]{gray_1981_procyonasym}
{\sc {Gray}, D.F.} (1981). {Asymmetries in the spectral lines of Procyon}. {\em
  \apj\/}, {\bf 251}, 583.

\bibitem[{{Gray}(1982)}]{gray_1982_bisect}
{\sc {Gray}, D.F.} (1982). {Observations of spectral line asymmetries and
  convective velocities in F, G, and K stars}. {\em \apj\/}, {\bf 255},
  200--209.

\bibitem[{{Gray}(1992)}]{gray_review_1992}
{\sc {Gray}, D.F.} (1992). {Stellar Convection: The Observations (Invited
  Review)}. In M.S. {Giampapa} \& J.A. {Bookbinder}, eds., {\em Cool Stars,
  Stellar Systems, and the Sun\/}, vol.~26 of {\em Astronomical Society of the
  Pacific Conference Series\/}, 127.

\bibitem[{{Gray}(2005)}]{gray_2005}
{\sc {Gray}, D.F.} (2005). {Shapes of Spectral Line Bisectors for Cool Stars}.
  {\em PASP\/}, {\bf 117}, 711--720.

\bibitem[{{Gray}(2009)}]{gray_2009_third_sig}
{\sc {Gray}, D.F.} (2009). {The Third Signature of Stellar Granulation}. {\em
  \apj\/}, {\bf 697}, 1032--1043.

\bibitem[{{Gray}(2010)}]{gray_2010_line_bisectors}
{\sc {Gray}, D.F.} (2010). {Empirical Decoding of the Shapes of Spectral-Line
  Bisectors}. {\em \apj\/}, {\bf 710}, 1003--1008.

\bibitem[{{Gray} \& {Nagar}(1985)}]{1985gray}
{\sc {Gray}, D.F. \& {Nagar}, P.} (1985). {The rotational discontinuity shown
  by luminosity class IV stars}. {\em \apj\/}, {\bf 298}, 756--760.

\bibitem[{{Gray} \& {Nagel}(1989)}]{gray_1989_granubound}
{\sc {Gray}, D.F. \& {Nagel}, T.} (1989). {The granulation boundary in the H-R
  diagram}. {\em \apj\/}, {\bf 341}, 421--426.

\bibitem[{{Harper}(1996)}]{harper_1996_massloss}
{\sc {Harper}, G.} (1996). {Mass loss and winds from cool giants}. In
  {R.~Pallavicini \& A.~K.~Dupree}, ed., {\em Cool Stars, Stellar Systems, and
  the Sun\/}, vol. 109 of {\em Astronomical Society of the Pacific Conference
  Series\/}, 481.

\bibitem[{{Harper} \& {Brown}(2006)}]{harper_2006}
{\sc {Harper}, G.M. \& {Brown}, A.} (2006). {Electron Density and Turbulence
  Gradients within the Extended Atmosphere of the M Supergiant Betelgeuse
  ({$\alpha$} Orionis)}. {\em \apj\/}, {\bf 646}, 1179--1202.

\bibitem[{{Harrison}(1949)}]{harrison_1949_echelle}
{\sc {Harrison}, G.R.} (1949). The production of diffraction gratings: The
  design of echelle gratings and spectrographs. {\em J. Opt. Soc. Am.\/}, {\bf
  39}, 522--527.

\bibitem[{{Hartmann} \& {Avrett}(1984)}]{hartmann_avrett_1984}
{\sc {Hartmann}, L. \& {Avrett}, E.H.} (1984). {On the extended chromosphere of
  Alpha Orionis}. {\em \apj\/}, {\bf 284}, 238--249.

\bibitem[{{Hartmann} \& {MacGregor}(1980)}]{hartmann_mcgregor_1980}
{\sc {Hartmann}, L. \& {MacGregor}, K.B.} (1980). {Momentum and energy
  deposition in late-type stellar atmospheres and winds}. {\em \apj\/}, {\bf
  242}, 260--282.

\bibitem[{{Hartmann} {\em et~al.\/}(1980){Hartmann}, {Dupree} \&
  {Raymond}}]{hartmann_1980}
{\sc {Hartmann}, L., {Dupree}, A.K. \& {Raymond}, J.C.} (1980). {Hybrid
  atmospheres and winds in supergiant stars}. {\em \apjl\/}, {\bf 236},
  L143--L147.

\bibitem[{{Hauschildt} {\em et~al.\/}(1997){Hauschildt}, {Baron} \&
  {Allard}}]{phoenix_1997}
{\sc {Hauschildt}, P.H., {Baron}, E. \& {Allard}, F.} (1997). {Parallel
  Implementation of the PHOENIX Generalized Stellar Atmosphere Program}. {\em
  \apj\/}, {\bf 483}, 390.

\bibitem[{{Hayashi}(1961)}]{hayashi_1961}
{\sc {Hayashi}, C.} (1961). {Stellar evolution in early phases of gravitational
  contraction.} {\em \pasj\/}, {\bf 13}, 450--452.

\bibitem[{{Hayes} \&
  {Nussbaumer}(1984{\natexlab{a}})}]{hayes_nussbaumer_1984_1}
{\sc {Hayes}, M.A. \& {Nussbaumer}, H.} (1984{\natexlab{a}}). {The C II 2325 A
  multiplet and the 158 micron transition}. {\em \aap\/}, {\bf 139}, 233--236.

\bibitem[{{Hayes} \&
  {Nussbaumer}(1984{\natexlab{b}})}]{hayes_nussbaumer_1984_2}
{\sc {Hayes}, M.A. \& {Nussbaumer}, H.} (1984{\natexlab{b}}). {The C II
  infrared and ultraviolet lines}. {\em \aap\/}, {\bf 134}, 193--197.

\bibitem[{{Henyey} {\em et~al.\/}(1955){Henyey}, {Lelevier} \&
  {Lev{\'e}e}}]{henyey_1955}
{\sc {Henyey}, L.G., {Lelevier}, R. \& {Lev{\'e}e}, R.D.} (1955). {The Early
  Phases of Stellar Evolution}. {\em \pasp\/}, {\bf 67}, 154.

\bibitem[{{Hodge} {\em et~al.\/}(1998{\natexlab{a}}){Hodge}, {Baum}, {McGrath},
  {Hulbert} \& {Christensen}}]{wavecal_1998_isr}
{\sc {Hodge}, P., {Baum}, S., {McGrath}, M., {Hulbert}, S. \& {Christensen},
  J.} (1998{\natexlab{a}}). {Calstis4, calstis11, calstis12: Wavecal Processing
  in the STIS Calibration Pipeline}. Tech. rep., {Space Telescope Science
  Institute}.

\bibitem[{{Hodge} {\em et~al.\/}(1998{\natexlab{b}}){Hodge}, {Baum}, {McGrath},
  {Shaw}, {Busko}, {Christensen}, {Goudfrooij}, {Hsu}, {Hulbert} \&
  {Katsanis}}]{hodge_1998_calstis0}
{\sc {Hodge}, P., {Baum}, S., {McGrath}, M., {Shaw}, D., {Busko}, I.,
  {Christensen}, J., {Goudfrooij}, P., {Hsu}, J.C., {Hulbert}, S. \&
  {Katsanis}, R.} (1998{\natexlab{b}}). {Calstis0: Pipeline Calibration of STIS
  Data-A Detailed View}. Tech. rep., {Space Telescope Science Institute}.

\bibitem[{{Hodge} {\em et~al.\/}(1998{\natexlab{c}}){Hodge}, {Hulbert},
  {Lindler}, {Busko}, {Hsu}, {Baum}, {McGrath}, {Goudfrooij}, {Shaw},
  {Katsanis}, {Keener} \& {Bohlin}}]{hodge_stis_1998}
{\sc {Hodge}, P.E., {Hulbert}, S.J., {Lindler}, D., {Busko}, I., {Hsu}, J.,
  {Baum}, S., {McGrath}, M., {Goudfrooij}, P., {Shaw}, R., {Katsanis}, R.,
  {Keener}, S. \& {Bohlin}, R.} (1998{\natexlab{c}}). {Pipeline Calibration for
  STIS}. In {R.~Albrecht, R.~N.~Hook, \& H.~A.~Bushouse}, ed., {\em
  Astronomical Data Analysis Software and Systems VII\/}, vol. 145 of {\em
  Astronomical Society of the Pacific Conference Series\/}, 316.

\bibitem[{{Hoffleit}(1964)}]{hoffleit_1964}
{\sc {Hoffleit}, D.} (1964). {\em {Catalogue of bright stars}\/}. {New Haven}.

\bibitem[{{Hogg}(1936)}]{hogg_1936}
{\sc {Hogg}, F.S.} (1936). {Recent variations in spectrum of Z Andromedae}. In
  {\em Publications of the American Astronomical Society\/}, vol.~8 of {\em
  Publications of the American Astronomical Society\/}, 14.

\bibitem[{{Holzer} \& {MacGregor}(1985)}]{holzer_macgregor_1988_massloss}
{\sc {Holzer}, T.E. \& {MacGregor}, K.B.} (1985). {Mass loss mechanisms for
  cool, low-gravity stars}. In {M.~Morris \& B.~Zuckerman}, ed., {\em Mass Loss
  from Red Giants\/}, vol. 117 of {\em Astrophysics and Space Science
  Library\/}, 229--255.

\bibitem[{{Holzer} {\em et~al.\/}(1983){Holzer}, {Fla} \& {Leer}}]{alfven_1983}
{\sc {Holzer}, T.E., {Fla}, T. \& {Leer}, E.} (1983). {Alfven waves in stellar
  winds}. {\em \apj\/}, {\bf 275}, 808--835.

\bibitem[{{Jeans}(1902)}]{jeans_1902_instability}
{\sc {Jeans}, J.H.} (1902). {The Stability of a Spherical Nebula}. {\em Royal
  Society of London Philosophical Transactions Series A\/}, {\bf 199}, 1--53.

\bibitem[{{Johnson} {\em et~al.\/}(1986){Johnson}, {Smith} \&
  {Parkinson}}]{johnson_smith_1986}
{\sc {Johnson}, B.C., {Smith}, P.L. \& {Parkinson}, W.H.} (1986). {Transition
  probability of the AL II 2669 intersystem line}. {\em \apj\/}, {\bf 308},
  1013--1017.

\bibitem[{{Johnson} \& {Harris}(1954)}]{johnson_harris_1954_photom}
{\sc {Johnson}, H.L. \& {Harris}, D.L.} (1954). {Three-Color Observations of
  108 Stars Intended for Use as Photometric Standards.} {\em \apj\/}, {\bf
  120}, 196.

\bibitem[{{Johnson} \& {Morgan}(1953)}]{johnson_morgan_1953_photom}
{\sc {Johnson}, H.L. \& {Morgan}, W.W.} (1953). {Fundamental stellar photometry
  for standards of spectral type on the revised system of the Yerkes spectral
  atlas}. {\em \apj\/}, {\bf 117}, 313.

\bibitem[{{Jorissen} {\em et~al.\/}(2003){Jorissen}, {Dedecker}, {Plez},
  {Gillet} \& {Fokin}}]{jorissen_2003_tomog}
{\sc {Jorissen}, A., {Dedecker}, M., {Plez}, B., {Gillet}, D. \& {Fokin}, A.}
  (2003). {Tomography of the atmosphere of long-period variable stars}. {\em
  ArXiv Astrophysics e-prints\/}.

\bibitem[{{Josselin} \& {Plez}(2007)}]{josselin_2007}
{\sc {Josselin}, E. \& {Plez}, B.} (2007). {Atmospheric dynamics and the mass
  loss process in red supergiant stars}. {\em \aap\/}, {\bf 469}, 671--680.

\bibitem[{{Joye} \& {Mandel}(2003)}]{joye_2003_ds9}
{\sc {Joye}, W.A. \& {Mandel}, E.} (2003). {New Features of SAOImage DS9}. In
  {H.~E.~Payne, R.~I.~Jedrzejewski, \& R.~N.~Hook}, ed., {\em Astronomical Data
  Analysis Software and Systems XII\/}, vol. 295 of {\em Astronomical Society
  of the Pacific Conference Series\/}, 489.

\bibitem[{{Judge}(1990)}]{judge_1990}
{\sc {Judge}, P.G.} (1990). {On the interpretation of chromospheric emission
  lines}. {\em \apj\/}, {\bf 348}, 279--296.

\bibitem[{{Judge}(1994)}]{judge_1994}
{\sc {Judge}, P.G.} (1994). {The 'monochromatic density diagnostic' technique:
  First detection of multiple density components in the chromosphere of
  $\alpha$ Tauri}. {\em \apj\/}, {\bf 430}, 351--359.

\bibitem[{{Judge} \& {Carpenter}(1998)}]{judge_carpenter_1998_basal_flux}
{\sc {Judge}, P.G. \& {Carpenter}, K.G.} (1998). {On Chromospheric Heating
  Mechanisms of ``Basal Flux'' Stars}. {\em \apj\/}, {\bf 494}, 828.

\bibitem[{{Judge} \& {Stencel}(1991)}]{judge_stencel_1991_chromowinds}
{\sc {Judge}, P.G. \& {Stencel}, R.E.} (1991). {Evolution of the chromospheres
  and winds of low- and intermediate-mass giant stars}. {\em \apj\/}, {\bf
  371}, 357--379.

\bibitem[{{Karakla}(2007)}]{hst_primer}
{\sc {Karakla}, D.}, ed. (2007). {\em {Hubble Space Telescope Primer for Cycle
  17}\/}.

\bibitem[{Kaufman \& Martin(1991{\natexlab{a}})}]{kaufman1991wavelengths}
{\sc Kaufman, V. \& Martin, W.} (1991{\natexlab{a}}). {\em Wavelengths and
  energy level classifications of magnesium spectra for all stages of
  ionization (Mg I through Mg XII)\/}. Journal of physical and chemical
  reference data. Reprint, American Chemical Society.

\bibitem[{Kaufman \& Martin(1991{\natexlab{b}})}]{al_lines_ref_1991}
{\sc Kaufman, V. \& Martin, W.C.} (1991{\natexlab{b}}). Wavelength and energy
  level classifications for the spectra of aluminum (al~i through al~xiii).
  {\em J. Phys. Chem. Ref. Data\/}, {\bf 20}, 775--858.

\bibitem[{{Keenan} {\em et~al.\/}(1999){Keenan}, {Espey}, {Mathioudakis},
  {Aggarwal}, {Crawford}, {Feibelman} \& {McKenna}}]{keenan_espey_1999}
{\sc {Keenan}, F.P., {Espey}, B.R., {Mathioudakis}, M., {Aggarwal}, K.M.,
  {Crawford}, F.L., {Feibelman}, W.A. \& {McKenna}, F.C.} (1999). {[Alii] in
  the ultraviolet spectrum of the symbiotic star RR Telescopii}. {\em
  \mnras\/}, {\bf 309}, 195--198.

\bibitem[{{Keenan} \& {McNeil}(1989)}]{keenan_1989}
{\sc {Keenan}, P.C. \& {McNeil}, R.C.} (1989). {The Perkins catalog of revised
  MK types for the cooler stars}. {\em \apjs\/}, {\bf 71}, 245--266.

\bibitem[{{Kenyon}(1994)}]{kenyon_1994_symb_evo}
{\sc {Kenyon}, S.J.} (1994). {Formation and evolution of symbiotic stars}. {\em
  Memorie della Societa Astronomia Italiana\/}, {\bf 65}, 135.

\bibitem[{{Kenyon} \& {Fernandez-Castro}(1987)}]{kenyon_fernandez_castro_1987}
{\sc {Kenyon}, S.J. \& {Fernandez-Castro}, T.} (1987). {The cool components of
  symbiotic stars. I - Optical spectral types}. {\em \aj\/}, {\bf 93},
  938--949.

\bibitem[{{Kenyon} \& {Webbink}(1984)}]{1984_kenyon_webbink}
{\sc {Kenyon}, S.J. \& {Webbink}, R.F.} (1984). {The nature of symbiotic
  stars}. {\em \apj\/}, {\bf 279}, 252--283.

\bibitem[{{Kenyon} {\em et~al.\/}(1993){Kenyon}, {Mikolajewska},
  {Mikolajewski}, {Polidan} \& {Slovak}}]{kenyon_1993_ag_peg}
{\sc {Kenyon}, S.J., {Mikolajewska}, J., {Mikolajewski}, M., {Polidan}, R.S. \&
  {Slovak}, M.H.} (1993). {Evolution of the symbiotic binary system AG Pegasi -
  The slowest classical nova eruption ever recorded}. {\em \aj\/}, {\bf 106},
  1573--1598.

\bibitem[{{Keyes} \& {Preblich}(2004)}]{kreyes_preblich_2004}
{\sc {Keyes}, C.D. \& {Preblich}, B.} (2004). {Spectral and Luminosity
  Classification of Symbiotic Star Cool Components with Near-Infrared
  Photometry}. {\em \aj\/}, {\bf 128}, 2981--2987.

\bibitem[{{Kimble} {\em et~al.\/}(1998){Kimble}, {Woodgate}, {Bowers},
  {Kraemer}, {Kaiser}, {Gull}, {Heap}, {Danks}, {Boggess}, {Green},
  {Hutchings}, {Jenkins}, {Joseph}, {Linsky}, {Maran}, {Moos}, {Roesler},
  {Timothy}, {Weistrop}, {Grady}, {Loiacono}, {Brown}, {Brumfield}, {Content},
  {Feinberg}, {Isaacs}, {Krebs}, {Krueger}, {Melcher}, {Rebar}, {Vitagliano},
  {Yagelowich}, {Meyer}, {Hood}, {Argabright}, {Becker}, {Bottema}, {Breyer},
  {Bybee}, {Christon}, {Delamere}, {Dorn}, {Downey}, {Driggers}, {Ebbets},
  {Gallegos}, {Garner}, {Hetlinger}, {Lettieri}, {Ludtke}, {Michika},
  {Nyquist}, {Rose}, {Stocker}, {Sullivan}, {van Houten}, {Woodruff}, {Baum},
  {Hartig}, {Balzano}, {Biagetti}, {Blades}, {Bohlin}, {Clampin}, {Doxsey},
  {Ferguson}, {Goudfrooij}, {Hulbert}, {Kutina}, {McGrath}, {Lindler}, {Beck},
  {Feggans}, {Plait}, {Sandoval}, {Hill}, {Collins}, {Cornett}, {Fowler},
  {Hill}, {Landsman}, {Malumuth}, {Standley}, {Blouke}, {Grusczak}, {Reed},
  {Robinson}, {Valenti} \& {Wolfe}}]{kimble_1998_stis}
{\sc {Kimble}, R.A., {Woodgate}, B.E., {Bowers}, C.W., {Kraemer}, S.B.,
  {Kaiser}, M.E., {Gull}, T.R., {Heap}, S.R., {Danks}, A.C., {Boggess}, A.,
  {Green}, R.F., {Hutchings}, J.B., {Jenkins}, E.B., {Joseph}, C.L., {Linsky},
  J.L., {Maran}, S.P., {Moos}, H.W., {Roesler}, F., {Timothy}, J.G.,
  {Weistrop}, D.E., {Grady}, J.F., {Loiacono}, J.J., {Brown}, L.W.,
  {Brumfield}, M.D., {Content}, D.A., {Feinberg}, L.D., {Isaacs}, M.N.,
  {Krebs}, C.A., {Krueger}, V.L., {Melcher}, R.W., {Rebar}, F.J., {Vitagliano},
  H.D., {Yagelowich}, J.J., {Meyer}, W.W., {Hood}, D.F., {Argabright}, V.S.,
  {Becker}, S.I., {Bottema}, M., {Breyer}, R.R., {Bybee}, R.L., {Christon},
  P.R., {Delamere}, A.W., {Dorn}, D.A., {Downey}, S., {Driggers}, P.A.,
  {Ebbets}, D.C., {Gallegos}, J.S., {Garner}, H., {Hetlinger}, J.C.,
  {Lettieri}, R.L., {Ludtke}, C.W., {Michika}, D., {Nyquist}, R., {Rose}, D.M.,
  {Stocker}, R.B., {Sullivan}, J.F., {van Houten}, C.N., {Woodruff}, R.A.,
  {Baum}, S.A., {Hartig}, G.F., {Balzano}, V., {Biagetti}, C., {Blades}, J.C.,
  {Bohlin}, R.C., {Clampin}, M., {Doxsey}, R., {Ferguson}, H.C., {Goudfrooij},
  P., {Hulbert}, S.J., {Kutina}, R., {McGrath}, M., {Lindler}, D.J., {Beck},
  T.L., {Feggans}, J.K., {Plait}, P.C., {Sandoval}, J.L., {Hill}, R.S.,
  {Collins}, N.R., {Cornett}, R.H., {Fowler}, W.B., {Hill}, R.J., {Landsman},
  W.B., {Malumuth}, E.M., {Standley}, C., {Blouke}, M., {Grusczak}, A., {Reed},
  R., {Robinson}, R.D., {Valenti}, J.A. \& {Wolfe}, T.} (1998). {The On-Orbit
  Performance of the Space Telescope Imaging Spectrograph}. {\em \apjl\/}, {\bf
  492}, L83.

\bibitem[{{Knill} {\em et~al.\/}(1993){Knill}, {Dgani} \& {Vogel}}]{knill_1993}
{\sc {Knill}, O., {Dgani}, R. \& {Vogel}, M.} (1993). {A new approach to Abel's
  integral operator and its application to stellar winds}. {\em \aap\/}, {\bf
  274}, 1002.

\bibitem[{{Koninx} \& {Pijpers}(1992)}]{koninx_1992_soundwinds}
{\sc {Koninx}, J.P.M. \& {Pijpers}, F.P.} (1992). {The applicability of the
  linearized theory of sound-wave driven winds}. {\em \aap\/}, {\bf 265},
  183--194.

\bibitem[{{Kraemer} {\em et~al.\/}(1997){Kraemer}, {Downes}, {Katsanis},
  {Crenshaw}, {McGrath} \& {Robinson}}]{kraemer_1997_stic_acq}
{\sc {Kraemer}, S., {Downes}, R., {Katsanis}, R., {Crenshaw}, M., {McGrath}, M.
  \& {Robinson}, R.} (1997). {STIS target acquisition}. In {S.~Casertano,
  R.~Jedrzejewski, T.~Keyes, \& M.~Stevens}, ed., {\em The 1997 HST Calibration
  Workshop with a New Generation of Instruments\/}, 39.

\bibitem[{{Kucinskas} {\em et~al.\/}(2006){Kucinskas}, {Ludwig} \&
  {Hauschildt}}]{phoenix_gran}
{\sc {Kucinskas}, A., {Ludwig}, H.G. \& {Hauschildt}, P.H.} (2006). {Convection
  and observable properties of late-type giants}. In P.~{Whitelock},
  M.~{Dennefeld} \& B.~{Leibundgut}, eds., {\em The Scientific Requirements for
  Extremely Large Telescopes\/}, vol. 232 of {\em IAU Symposium\/}, 498--501.

\bibitem[{{Kudritzki}(2002)}]{kudritzki_2002_linedrivenwinds}
{\sc {Kudritzki}, R.P.} (2002). {Line-driven Winds, Ionizing Fluxes, and
  Ultraviolet Spectra of Hot Stars at Extremely Low Metallicity. I. Very
  Massive O Stars}. {\em \apj\/}, {\bf 577}, 389--408.

\bibitem[{{Lamers} \& {Cassinelli}(1999)}]{lamers_book}
{\sc {Lamers}, H.J.G.L.M. \& {Cassinelli}, J.P.} (1999). {\em {Introduction to
  Stellar Winds}\/}. Cambridge University Press.

\bibitem[{{Landstreet}(2007)}]{landstreet_2007}
{\sc {Landstreet}, J.D.} (2007). {Observing convection in stellar atmospheres}.
  In F.~{Kupka}, I.~{Roxburgh} \& K.~{Chan}, eds., {\em IAU Symposium\/}, vol.
  239 of {\em IAU Symposium\/}, 103--112.

\bibitem[{{Leedj{\"a}rv}(2006)}]{leed_2006}
{\sc {Leedj{\"a}rv}, L.} (2006). {Symbiotic Stars in the Context of Binary
  Evolution and Metallicity}. {\em \aaps\/}, {\bf 304}, 105--107.

\bibitem[{{Lennon} {\em et~al.\/}(1985){Lennon}, {Dufton}, {Hibbert} \&
  {Kingston}}]{lennon_1985}
{\sc {Lennon}, D.J., {Dufton}, P.L., {Hibbert}, A. \& {Kingston}, A.E.} (1985).
  {C II emission lines formed in optically thin plasmas}. {\em \apj\/}, {\bf
  294}, 200--206.

\bibitem[{{Lim}(1998)}]{kwok_press_relase}
{\sc {Lim}, J.} (1998). {VLA Shows ``Boiling'' in Atmosphere of Betelgeuse}. In
  {\em NRAO Press Release, 04/1998\/}, 1.

\bibitem[{{Lindgren} \& {McElrath}(1959)}]{chi_table}
{\sc {Lindgren}, B.W. \& {McElrath}, G.W.} (1959). {\em {Introduction to
  Probability and Statistics}\/}. {Macmillan New York.}

\bibitem[{{Linsky} \& {Haisch}(1979)}]{linsky_haisch_1979}
{\sc {Linsky}, J.L. \& {Haisch}, B.M.} (1979). {Outer atmospheres of cool
  stars. I - The sharp division into solar-type and non-solar-type stars}. {\em
  \apjl\/}, {\bf 229}, L27--L32.

\bibitem[{{Lunt}(1918)}]{lunt_1918}
{\sc {Lunt}, J.} (1918). {The radial velocities of 60 southern stars.} {\em
  \apj\/}, {\bf 47}, 201--205.

\bibitem[{{Luttermoser} {\em et~al.\/}(1994){Luttermoser}, {Johnson} \&
  {Eaton}}]{luttermoser_1994}
{\sc {Luttermoser}, D.G., {Johnson}, H.R. \& {Eaton}, J.} (1994). {The
  chromospheric structure of the cool giant star G Herculis}. {\em \apj\/},
  {\bf 422}, 351--365.

\bibitem[{{Magain}(1986)}]{magain_1986_linedepths}
{\sc {Magain}, P.} (1986). {Contribution functions and the depths of formation
  of spectral lines}. {\em \aap\/}, {\bf 163}, 135--139.

\bibitem[{Maoz(2007)}]{maoz_2007_astrophysics}
{\sc Maoz, D.} (2007). {\em Astrophysics in a Nutshell\/}. In a nutshell,
  Princeton University Press.

\bibitem[{Mariska(1992)}]{mariska1992solar}
{\sc Mariska, J.} (1992). {\em The Solar Transition Region\/}. Cambridge
  Astrophysics Series, Cambridge University Press.

\bibitem[{{Markwardt}(2009)}]{markwardt_2009_mpfit}
{\sc {Markwardt}, C.B.} (2009). {Non-linear Least-squares Fitting in IDL with
  MPFIT}. In {D.~A.~Bohlender, D.~Durand, \& P.~Dowler}, ed., {\em Astronomical
  Data Analysis Software and Systems XVIII\/}, vol. 411 of {\em Astronomical
  Society of the Pacific Conference Series\/}, 251.

\bibitem[{{McGrath} {\em et~al.\/}(1999){McGrath}, {Busko} \&
  {Hodge}}]{stis_isr_calstis_1999}
{\sc {McGrath}, M.A., {Busko}, I. \& {Hodge}, P.} (1999). {Calstis6: Extraction
  of 1-D Spectra in the STIS Calibration Pipeline}. Tech. rep., STScI.

\bibitem[{{McMurry}(1999)}]{mcmurry_1999}
{\sc {McMurry}, A.D.} (1999). {The outer atmosphere of Tau - I. A new
  chromospheric model}. {\em \mnras\/}, {\bf 302}, 37--47.

\bibitem[{{Mermilliod}(1986)}]{mermilliod_1986_hd_photom}
{\sc {Mermilliod}, J.C.} (1986). {Compilation of Eggen's UBV data, transformed
  to UBV (unpublished)}. {\em Catalogue of Eggen's UBV data\/}, 10.

\bibitem[{{Merrill}(1944)}]{merrill_1944_symb}
{\sc {Merrill}, P.W.} (1944). {Spectroscopic Observations of AX Persei, RW
  Hydrae, CI Cygni, and Z Andromedae.} {\em \apj\/}, {\bf 99}, 15.

\bibitem[{{Merrill}(1950)}]{merrill_1950_symb}
{\sc {Merrill}, P.W.} (1950). {Measurements in the Combination Spectra of RW
  Hydrae, BF Cygni, and CI Cygni.} {\em \apj\/}, {\bf 111}, 484.

\bibitem[{{Merrill}(1958)}]{merrill_1958}
{\sc {Merrill}, P.W.} (1958). {Symbiosis in Astronomy : introductory report.}
  {\em 8eme Colloque Intern.~d'Astrophys.~a Liege, 20, 436 (1958)\/}, {\bf 20},
  436.

\bibitem[{{Mihalas}(1978)}]{mihalis_book}
{\sc {Mihalas}, D.} (1978). {\em {Stellar Atmospheres (2nd edition)}\/}. W. H.
  Freeman.

\bibitem[{{Mikolajewska}(2002)}]{mikolajewska_2002}
{\sc {Mikolajewska}, J.} (2002). {Orbital and stellar parameters of symbiotic
  stars}. {\em ArXiv Astrophysics e-prints\/}.

\bibitem[{{Miko{\l}ajewska}(2007)}]{mikolajewska_2007}
{\sc {Miko{\l}ajewska}, J.} (2007). {Symbiotic Stars: Continually Embarrassing
  Binaries}. {\em Baltic Astronomy\/}, {\bf 16}, 1--9.

\bibitem[{{Monet} {\em et~al.\/}(2003){Monet}, {Levine}, {Canzian}, {Ables},
  {Bird}, {Dahn}, {Guetter}, {Harris}, {Henden}, {Leggett}, {Levison},
  {Luginbuhl}, {Martini}, {Monet}, {Munn}, {Pier}, {Rhodes}, {Riepe}, {Sell},
  {Stone}, {Vrba}, {Walker}, {Westerhout}, {Brucato}, {Reid}, {Schoening},
  {Hartley}, {Read} \& {Tritton}}]{monet_2003_vega_mags}
{\sc {Monet}, D.G., {Levine}, S.E., {Canzian}, B., {Ables}, H.D., {Bird}, A.R.,
  {Dahn}, C.C., {Guetter}, H.H., {Harris}, H.C., {Henden}, A.A., {Leggett},
  S.K., {Levison}, H.F., {Luginbuhl}, C.B., {Martini}, J., {Monet}, A.K.B.,
  {Munn}, J.A., {Pier}, J.R., {Rhodes}, A.R., {Riepe}, B., {Sell}, S., {Stone},
  R.C., {Vrba}, F.J., {Walker}, R.L., {Westerhout}, G., {Brucato}, R.J.,
  {Reid}, I.N., {Schoening}, W., {Hartley}, M., {Read}, M.A. \& {Tritton},
  S.B.} (2003). {The USNO-B Catalog}. {\em \aj\/}, {\bf 125}, 984--993.

\bibitem[{{Moos} {\em et~al.\/}(2000){Moos}, {Cash}, {Cowie}, {Davidsen},
  {Dupree}, {Feldman}, {Friedman}, {Green}, {Green}, {Gry}, {Hutchings},
  {Jenkins}, {Linsky}, {Malina}, {Michalitsianos}, {Savage}, {Shull},
  {Siegmund}, {Snow}, {Sonneborn}, {Vidal-Madjar}, {Willis}, {Woodgate},
  {York}, {Ake}, {Andersson}, {Andrews}, {Barkhouser}, {Bianchi}, {Blair},
  {Brownsberger}, {Cha}, {Chayer}, {Conard}, {Fullerton}, {Gaines}, {Grange},
  {Gummin}, {Hebrard}, {Kriss}, {Kruk}, {Mark}, {McCarthy}, {Morbey},
  {Murowinski}, {Murphy}, {Oegerle}, {Ohl}, {Oliveira}, {Osterman}, {Sahnow},
  {Saisse}, {Sembach}, {Weaver}, {Welsh}, {Wilkinson} \&
  {Zheng}}]{moos_2000_fuse}
{\sc {Moos}, H.W., {Cash}, W.C., {Cowie}, L.L., {Davidsen}, A.F., {Dupree},
  A.K., {Feldman}, P.D., {Friedman}, S.D., {Green}, J.C., {Green}, R.F., {Gry},
  C., {Hutchings}, J.B., {Jenkins}, E.B., {Linsky}, J.L., {Malina}, R.F.,
  {Michalitsianos}, A.G., {Savage}, B.D., {Shull}, J.M., {Siegmund}, O.H.W.,
  {Snow}, T.P., {Sonneborn}, G., {Vidal-Madjar}, A., {Willis}, A.J.,
  {Woodgate}, B.E., {York}, D.G., {Ake}, T.B., {Andersson}, B.G., {Andrews},
  J.P., {Barkhouser}, R.H., {Bianchi}, L., {Blair}, W.P., {Brownsberger}, K.R.,
  {Cha}, A.N., {Chayer}, P., {Conard}, S.J., {Fullerton}, A.W., {Gaines}, G.A.,
  {Grange}, R., {Gummin}, M.A., {Hebrard}, G., {Kriss}, G.A., {Kruk}, J.W.,
  {Mark}, D., {McCarthy}, D.K., {Morbey}, C.L., {Murowinski}, R., {Murphy},
  E.M., {Oegerle}, W.R., {Ohl}, R.G., {Oliveira}, C., {Osterman}, S.N.,
  {Sahnow}, D.J., {Saisse}, M., {Sembach}, K.R., {Weaver}, H.A., {Welsh}, B.Y.,
  {Wilkinson}, E. \& {Zheng}, W.} (2000). {Overview of the Far Ultraviolet
  Spectroscopic Explorer Mission}. {\em \apjl\/}, {\bf 538}, L1--L6.

\bibitem[{{Muerset} {\em et~al.\/}(1991){Muerset}, {Nussbaumer}, {Schmid} \&
  {Vogel}}]{muerset_1991}
{\sc {Muerset}, U., {Nussbaumer}, H., {Schmid}, H.M. \& {Vogel}, M.} (1991).
  {Temperature and luminosity of hot components in symbiotic stars}. {\em
  \aap\/}, {\bf 248}, 458--474.

\bibitem[{{Munari}(1989)}]{munari_1989}
{\sc {Munari}, U.} (1989). {Studies of symbiotic stars. I - Location of the UV
  emitting regions in 6 S-type systems monitored by the IUE satellite}. {\em
  \aap\/}, {\bf 208}, 63--68.

\bibitem[{{Munari} \& {Zwitter}(2002)}]{munari_zwitter_atlas_2002}
{\sc {Munari}, U. \& {Zwitter}, T.} (2002). {Spectrophotometric atlas of
  symbiotic stars (Munari+, 2002)}. {\em VizieR Online Data Catalog\/}, {\bf
  3383}, 30188.

\bibitem[{{Munari} {\em et~al.\/}(1988){Munari}, {Margoni}, {Iijima} \&
  {Mammano}}]{munari_1988}
{\sc {Munari}, U., {Margoni}, R., {Iijima}, T. \& {Mammano}, A.} (1988). {The
  spectroscopic orbit of the symbiotic star EG Andromedae}. {\em \aap\/}, {\bf
  198}, 173--178.

\bibitem[{{Oliversen} {\em et~al.\/}(1985){Oliversen}, {Anderson}, {Slovak} \&
  {Stencel}}]{oliverson_1985}
{\sc {Oliversen}, N.A., {Anderson}, C.M., {Slovak}, M.H. \& {Stencel}, R.E.}
  (1985). {Observational studies of the symbiotic stars. III High-dispersion
  IUE and H-alpha observations of EG Andromedae}. {\em \apj\/}, {\bf 295},
  620--627.

\bibitem[{Palmer \& Loewen(2000)}]{palmer2000diffraction}
{\sc Palmer, C. \& Loewen, E.} (2000). {\em Diffraction Grating Handbook\/}.
  Richardson Grating Laboratory.

\bibitem[{{Parker}(1958)}]{parker_1958}
{\sc {Parker}, E.N.} (1958). {Suprathermal Particle Generation in the Solar
  Corona}. {\em \apj\/}, {\bf 128}, 677.

\bibitem[{{Pasinetti Fracassini} {\em et~al.\/}(2001){Pasinetti Fracassini},
  {Pastori}, {Covino} \& {Pozzi}}]{angdiam_2001}
{\sc {Pasinetti Fracassini}, L.E., {Pastori}, L., {Covino}, S. \& {Pozzi}, A.}
  (2001). {Catalogue of Apparent Diameters and Absolute Radii of Stars (CADARS)
  - Third edition - Comments and statistics}. {\em \aap\/}, {\bf 367},
  521--524.

\bibitem[{Percy(2007)}]{percy2007understanding}
{\sc Percy, J.} (2007). {\em Understanding variable stars\/}. Cambridge
  University Press.

\bibitem[{{Percy} \& {Harrett}(2004)}]{percy_harrett_2004}
{\sc {Percy}, J.R. \& {Harrett}, A.} (2004). {Self-Correlation Analysis of the
  Brightness Variability of Symbiotic Stars: A Pilot Project}. {\em Journal of
  the American Association of Variable Star Observers (JAAVSO)\/}, {\bf 33},
  34--41.

\bibitem[{{Perryman} {\em et~al.\/}(1997){Perryman}, {Lindegren}, {Kovalevsky},
  {Hoeg}, {Bastian}, {Bernacca}, {Cr{\'e}z{\'e}}, {Donati}, {Grenon}, {van
  Leeuwen}, {van der Marel}, {Mignard}, {Murray}, {Le Poole}, {Schrijver},
  {Turon}, {Arenou}, {Froeschl{\'e}} \& {Petersen}}]{hipparcos_catalogue_1997}
{\sc {Perryman}, M.A.C., {Lindegren}, L., {Kovalevsky}, J., {Hoeg}, E.,
  {Bastian}, U., {Bernacca}, P.L., {Cr{\'e}z{\'e}}, M., {Donati}, F., {Grenon},
  M., {van Leeuwen}, F., {van der Marel}, H., {Mignard}, F., {Murray}, C.A.,
  {Le Poole}, R.S., {Schrijver}, H., {Turon}, C., {Arenou}, F.,
  {Froeschl{\'e}}, M. \& {Petersen}, C.S.} (1997). {The HIPPARCOS Catalogue}.
  {\em \aap\/}, {\bf 323}, L49--L52.

\bibitem[{Phillips(1999)}]{phillips1999physics}
{\sc Phillips, A.} (1999). {\em The physics of stars\/}. Manchester physics
  series, John Wiley.

\bibitem[{{Pickles}(1998)}]{pickles_1998_stellar_library}
{\sc {Pickles}, A.J.} (1998). {A Stellar Spectral Flux Library: 1150-25000
  {\AA}}. {\em \pasp\/}, {\bf 110}, 863--878.

\bibitem[{{Plez}(1992)}]{plez_1992_2}
{\sc {Plez}, B.} (1992). {Spherical opacity sampling model atmospheres for
  M-giants and supergiants. II - A grid}. {\em \aap\/}, {\bf 94}, 527--552.

\bibitem[{{Plez}(1998)}]{plez_1998}
{\sc {Plez}, B.} (1998). {A new TiO line list}. {\em \aap\/}, {\bf 337},
  495--500.

\bibitem[{{Plez} {\em et~al.\/}(1992){Plez}, {Brett} \&
  {Nordlund}}]{plez_1992_1}
{\sc {Plez}, B., {Brett}, J.M. \& {Nordlund}, A.} (1992). {Spherical opacity
  sampling model atmospheres for M-giants. I - Techniques, data and
  discussion}. {\em \aap\/}, {\bf 256}, 551--571.

\bibitem[{{Press} {\em et~al.\/}(1992){Press}, {Teukolsky}, {Vetterling} \&
  {Flannery}}]{numerical_recipes}
{\sc {Press}, W.H., {Teukolsky}, S.A., {Vetterling}, W.T. \& {Flannery}, B.P.}
  (1992). {\em {Numerical recipes in C. The art of scientific computing}\/}.
  {Cambridge}.

\bibitem[{Prialnik(2000)}]{prialnik2000introduction}
{\sc Prialnik, D.} (2000). {\em An introduction to the theory of stellar
  structure and evolution\/}. Cambridge University Press.

\bibitem[{{Proffitt}(2010)}]{ihb}
{\sc {Proffitt}, C.}, ed. (2010). {\em {STIS Instrument Handbook, Version
  9.0}\/}.

\bibitem[{Prothero \& Buell(2007)}]{prothero2007evolution}
{\sc Prothero, D. \& Buell, C.} (2007). {\em Evolution: what the fossils say
  and why it matters\/}. Columbia University Press.

\bibitem[{{Prugniel} \& {Soubiran}(2001)}]{elodie_archive_2001}
{\sc {Prugniel}, P. \& {Soubiran}, C.} (2001). {A database of high and
  medium-resolution stellar spectra}. {\em \aap\/}, {\bf 369}, 1048--1057.

\bibitem[{{Przybylski} \& {Kennedy}(1965)}]{przybylski_1965_hd_photom}
{\sc {Przybylski}, A. \& {Kennedy}, P.M.} (1965). {Radial velocities and
  three-colour photometry of 166 southern stars}. {\em \mnras\/}, {\bf 131},
  95.

\bibitem[{Reader {\em et~al.\/}(1980)Reader, Corliss, Wiese \&
  Martin}]{nist_book_values}
{\sc Reader, J., Corliss, C.H., Wiese, W.L. \& Martin, G.A.} (1980). {\em
  Wavelengths and Transition Probabilities for Atoms and Atomic Ions,
  Part.~I.~Wavelengths, Part~II.~Transition Probabilities, Nat. Stand. Ref.
  Data Ser., NSRDS-NBS~68\/}. U.S. Government Printing Office, Washington,
  D.C., nIST compilation.

\bibitem[{{Reimers}(1975)}]{reimers_1975}
{\sc {Reimers}, D.} (1975). {\em {Circumstellar envelopes and mass loss of red
  giant stars}\/}, 229--256. Problems in stellar atmospheres and envelopes.

\bibitem[{{Reimers}(1987)}]{reimers_1987_binary}
{\sc {Reimers}, D.} (1987). {What do binaries teach us about mass-loss from
  late-type stars?} In {I.~Appenzeller \& C.~Jordan}, ed., {\em Circumstellar
  Matter\/}, vol. 122 of {\em IAU Symposium\/}, 307--318.

\bibitem[{{Robinson} \& {Carpenter}(1995)}]{robinson_carpenter_1995}
{\sc {Robinson}, R.D. \& {Carpenter}, K.G.} (1995). {MG II H and K profiles in
  high-liminosity, late-type stars}. {\em \apj\/}, {\bf 442}, 328--336.

\bibitem[{{Roche} {\em et~al.\/}(2009){Roche}, {Espey} \&
  {Crowley}}]{roche_2009_cs15poster}
{\sc {Roche}, J., {Espey}, B.R. \& {Crowley}, C.} (2009). {Exploring the Origin
  of Red Giant Winds}. In {E.~Stempels}, ed., {\em American Institute of
  Physics Conference Series\/}, vol. 1094 of {\em American Institute of Physics
  Conference Series\/}, 888--891.

\bibitem[{{Roche} {\em et~al.\/}(2011){Roche}, {Espey} \&
  {Crowley}}]{roche_2011_cs16poster}
{\sc {Roche}, J., {Espey}, B.R. \& {Crowley}, C.} (2011). {Symbiotic Stars and
  the Origin of Red Giant Winds}. In C.~{Johns-Krull}, M.K. {Browning} \& A.A.
  {West}, eds., {\em Astronomical Society of the Pacific Conference Series\/},
  vol. 448 of {\em Astronomical Society of the Pacific Conference Series\/},
  713.

\bibitem[{{Ruban} {\em et~al.\/}(2006){Ruban}, {Alekseeva}, {Arkharov},
  {Hagen-Thorn}, {Galkin}, {Nikanorova}, {Novikov}, {Pakhomov} \&
  {Puzakova}}]{ruban_2006}
{\sc {Ruban}, E.V., {Alekseeva}, G.A., {Arkharov}, A.A., {Hagen-Thorn}, E.I.,
  {Galkin}, V.D., {Nikanorova}, I.N., {Novikov}, V.V., {Pakhomov}, V.P. \&
  {Puzakova}, T.Y.} (2006). {Spectrophotometric observations of variable
  stars}. {\em Astronomy Letters\/}, {\bf 32}, 604--607.

\bibitem[{Sagan(1997)}]{sagan1997demon}
{\sc Sagan, C.} (1997). {\em The demon-haunted world: science as a candle in
  the dark\/}. Ballantine Books.

\bibitem[{{Salaris} \& {Cassisi}(2005)}]{salaris_2005_stellar_evo_book}
{\sc {Salaris}, M. \& {Cassisi}, S.} (2005). {\em {Evolution of Stars and
  Stellar Populations}\/}. {John Wiley \& Sons Ltd}.

\bibitem[{{Salpeter}(1955)}]{salpeter_1955_inital_mass_function}
{\sc {Salpeter}, E.E.} (1955). {The Luminosity Function and Stellar Evolution.}
  {\em \apj\/}, {\bf 121}, 161.

\bibitem[{Schreier(2009)}]{schreier_2009}
{\sc Schreier, F.} (2009). Comments on Òa common misunderstanding about the
  voigt line profile. {\em Journal of the Atmospheric Sciences\/}, {\bf 66},
  1860--1864.

\bibitem[{Schrijver \& Zwaan(2008)}]{schrijver2008solar}
{\sc Schrijver, C. \& Zwaan, C.} (2008). {\em Solar and Stellar Magnetic
  Activity\/}. Cambridge Astrophysics, Cambridge University Press.

\bibitem[{{Schr{\"o}der} \& {Connon Smith}(2008)}]{schroder_smith_2008}
{\sc {Schr{\"o}der}, K. \& {Connon Smith}, R.} (2008). {Distant future of the
  Sun and Earth revisited}. {\em \mnras\/}, {\bf 386}, 155--163.

\bibitem[{{Schr{\"o}der} \& {Cuntz}(2005)}]{schroder_2005}
{\sc {Schr{\"o}der}, K.P. \& {Cuntz}, M.} (2005). {A New Version of Reimers'
  Law of Mass Loss Based on a Physical Approach}. {\em \apj\/}, {\bf 630},
  L73--L76.

\bibitem[{{Schr{\"o}der} \& {Sedlmayr}(2001)}]{schroder_2001_gal}
{\sc {Schr{\"o}der}, K.P. \& {Sedlmayr}, E.} (2001). {The galactic mass
  injection from cool stellar winds of the 1 to 2.5 M$_{sun}$ stars in the
  solar neighbourhood}. {\em \aap\/}, {\bf 366}, 913--922.

\bibitem[{{Schr{\"o}der} {\em et~al.\/}(1988){Schr{\"o}der}, {Reimers},
  {Carpenter} \& {Brown}}]{schroeder_reimers_carpenter_1988}
{\sc {Schr{\"o}der}, K.P., {Reimers}, D., {Carpenter}, K.G. \& {Brown}, A.}
  (1988). {What does C II lambda 2325 A emission tell us about chromospheres of
  red supergiants? - A critical test using Zeta Aurigae-type K supergiants}.
  {\em \aap\/}, {\bf 202}, 136--142.

\bibitem[{{Schwarzschild}(1948)}]{schwarzschild_1948}
{\sc {Schwarzschild}, M.} (1948). {On Noise Arising from the Solar
  Granulation.} {\em \apj\/}, {\bf 107}, 1.

\bibitem[{{Schwarzschild}(1975)}]{schwarzschild_1975}
{\sc {Schwarzschild}, M.} (1975). {On the scale of photospheric convection in
  red giants and supergiants}. {\em \apj\/}, {\bf 195}, 137--144.

\bibitem[{{Scoville} \& {Mena-Werth}(1998)}]{scoville_menawerth_1998}
{\sc {Scoville}, F. \& {Mena-Werth}, J.} (1998). {Recalibration of the
  Wilson-Bappu Effect Using the Singly Ionized Magnesium K Line}. {\em
  \pasp\/}, {\bf 110}, 794--803.

\bibitem[{{Seaquist} \& {Taylor}(1990)}]{seaquist_taylor_1990}
{\sc {Seaquist}, E.R. \& {Taylor}, A.R.} (1990). {The collective radio
  properties of symbiotic stars}. {\em \apj\/}, {\bf 349}, 313--327.

\bibitem[{{Sedlmayr} \& {Dominik}(1995)}]{sedlmayr_1995_dustdriven}
{\sc {Sedlmayr}, E. \& {Dominik}, C.} (1995). {Dust Driven Winds}. {\em
  \ssr\/}, {\bf 73}, 211--272.

\bibitem[{{Shields} \& {McKee}(1981)}]{shields_mckee_1981}
{\sc {Shields}, G.A. \& {McKee}, C.F.} (1981). {Electron scattering by hot gas
  in QSOs}. {\em \apjl\/}, {\bf 246}, L57--L60.

\bibitem[{{Skopal} {\em et~al.\/}(1988){Skopal}, {Chochol}, {Vittone} \&
  {Mammano}}]{skopal_1988}
{\sc {Skopal}, A., {Chochol}, D., {Vittone}, A. \& {Mammano}, A.} (1988). {\em
  {Photometric and Spectroscopic Variations of the Symbiotic Star EG
  Andromedae}\/}, 289. IAU.

\bibitem[{{Skopal} {\em et~al.\/}(2002){Skopal}, {Vanko}, {Pribulla}, {Wolf},
  {Semkov} \& {Jones}}]{skopal_2002_egand_photom}
{\sc {Skopal}, A., {Vanko}, M., {Pribulla}, T., {Wolf}, M., {Semkov}, E. \&
  {Jones}, A.} (2002). {Photometry of symbiotic stars. X. EG And, Z And, BF
  Cyg, CH Cyg, V1329 Cyg, AG Dra, RW Hya, AX Per and IV Vir}. {\em
  Contributions of the Astronomical Observatory Skalnate Pleso\/}, {\bf 32},
  62--78.

\bibitem[{{Smak}(1964)}]{smak_1964_hd_photom}
{\sc {Smak}, J.} (1964). {Photometry and Spectrophotometry of Long-Period
  Variables.} {\em \apjs\/}, {\bf 9}, 141.

\bibitem[{{Smith}(1980)}]{smith_1980}
{\sc {Smith}, S.E.} (1980). {The symbiotic star HD 4174}. {\em \apj\/}, {\bf
  237}, 831--839.

\bibitem[{{Sobolev}(1975)}]{sobolev_1975}
{\sc {Sobolev}, V.V.} (1975). {\em {Light scattering in planetary
  atmospheres}\/}. {Pergamon Press.}

\bibitem[{Spitzer(1974)}]{lst_history_1974}
{\sc Spitzer, L.} (1974). {\em {Large Space Telescopes - A New Tool for
  Science}\/}. University of Michigan.

\bibitem[{Spitzer \& Ostriker(1997)}]{spitzer1997dreams}
{\sc Spitzer, L. \& Ostriker, J.} (1997). {\em Dreams, stars, and electrons:
  selected writings of Lyman Spitzer, Jr\/}. Princeton University Press.

\bibitem[{{Spitzer}(1990)}]{spitzer_1990_hst_idea}
{\sc {Spitzer}, L., Jr.} (1990). {REPORT TO PROJECT RAND: The Astronomical
  Advantages of an Extra-terrestrial Observatory (1 September 1946)}. {\em
  Astronomy Quarterly\/}, {\bf 7}, 131.

\bibitem[{{Stencel}(1984)}]{stencel_1984}
{\sc {Stencel}, R.E.} (1984). {Changes in the ultraviolet spectrum of EG
  Andromedae}. {\em \apjl\/}, {\bf 281}, L75--L77.

\bibitem[{{Stencel}(2009)}]{stencel_2009}
{\sc {Stencel}, R.E.} (2009). {The Wilson-Bappu Effect - 50 Years Later}. In
  {D.~G.~Luttermoser, B.~J.~Smith, \& R.~E.~Stencel}, ed., {\em Astronomical
  Society of the Pacific Conference Series\/}, vol. 412 of {\em Astronomical
  Society of the Pacific Conference Series\/}, 251.

\bibitem[{{Stencel} \& {Mullan}(1980)}]{stencel_mullan_1980apj}
{\sc {Stencel}, R.E. \& {Mullan}, D.J.} (1980). {Detection of mass loss in
  stellar chromospheres}. {\em \apj\/}, {\bf 238}, 221--228.

\bibitem[{{Stencel} \& {Sahade}(1980)}]{stencel_sahade_1980}
{\sc {Stencel}, R.E. \& {Sahade}, J.} (1980). {IUE observations of the peculiar
  M giant HD 4174}. {\em \apj\/}, {\bf 238}, 929--934.

\bibitem[{{Stencel} {\em et~al.\/}(1980){Stencel}, {Jordan}, {Wing}, {Linsky},
  {Carpenter}, {Brown} \& {Czyzak}}]{stencel_brown_carpenter_1980}
{\sc {Stencel}, R.E., {Jordan}, C., {Wing}, R.F., {Linsky}, J.L., {Carpenter},
  K.G., {Brown}, A. \& {Czyzak}, S.J.} (1980). {Chromospheric Densities and
  Geometrical Extensions of Red Giants and Supergiants using C II Lines as
  Diagnostics}. In {\em Bulletin of the American Astronomical Society\/},
  vol.~12 of {\em Bulletin of the American Astronomical Society\/}, 806.

\bibitem[{{Stencel} {\em et~al.\/}(1981){Stencel}, {Linsky}, {Brown}, {Jordan},
  {Carpenter}, {Wing} \& {Czyzak}}]{stencel_linsky_brown_carpenter_1981}
{\sc {Stencel}, R.E., {Linsky}, J.L., {Brown}, A., {Jordan}, C., {Carpenter},
  K.G., {Wing}, R.F. \& {Czyzak}, S.} (1981). {Density sensitive C II lines in
  cool stars of low gravity}. {\em \mnras\/}, {\bf 196}, 47P--53P.

\bibitem[{{Stoehr} {\em et~al.\/}(2008){Stoehr}, {White}, {Smith}, {Kamp},
  {Thompson}, {Durand}, {Freudling}, {Fraquelli}, {Haase}, {Hook}, {Kimball},
  {K{\"u}mmel}, {Levay}, {Lombardi}, {Micol} \& {Rogers}}]{der_snr}
{\sc {Stoehr}, F., {White}, R., {Smith}, M., {Kamp}, I., {Thompson}, R.,
  {Durand}, D., {Freudling}, W., {Fraquelli}, D., {Haase}, J., {Hook}, R.,
  {Kimball}, T., {K{\"u}mmel}, M., {Levay}, K., {Lombardi}, M., {Micol}, A. \&
  {Rogers}, T.} (2008). {DER\_SNR: A Simple \& General Spectroscopic
  Signal-to-Noise Measurement Algorithm}. In {R.~W.~Argyle, P.~S.~Bunclark, \&
  J.~R.~Lewis}, ed., {\em Astronomical Data Analysis Software and Systems
  XVII\/}, vol. 394 of {\em Astronomical Society of the Pacific Conference
  Series\/}, 505.

\bibitem[{{Swade} {\em et~al.\/}(2001){Swade}, {Hopkins} \&
  {Swam}}]{swade_2001_hst_archive}
{\sc {Swade}, D.A., {Hopkins}, E. \& {Swam}, M.S.} (2001). {HST Data Flow with
  On-The-Fly Reprocessing}. In {F.~R.~Harnden Jr., F.~A.~Primini, \&
  H.~E.~Payne}, ed., {\em Astronomical Data Analysis Software and Systems X\/},
  vol. 238 of {\em Astronomical Society of the Pacific Conference Series\/},
  295.

\bibitem[{{Taylor} \& {Seaquist}(1984)}]{taylor_seaqusit_1984}
{\sc {Taylor}, A.R. \& {Seaquist}, E.R.} (1984). {Radio emission from symbiotic
  stars - A binary model}. {\em \apj\/}, {\bf 286}, 263--268.

\bibitem[{{Thiering} {\em et~al.\/}(1990){Thiering}, {Baade}, {Schr{\"o}der},
  {Erhorn} \& {Huensch}}]{baade_1990_binary}
{\sc {Thiering}, I., {Baade}, R., {Schr{\"o}der}, K.P., {Erhorn}, G. \&
  {Huensch}, M.} (1990). {A binary star technique for investigating the
  extended atmospheres of cool stars}. In {G.~Wallerstein}, ed., {\em Cool
  Stars, Stellar Systems, and the Sun\/}, vol.~9 of {\em Astronomical Society
  of the Pacific Conference Series\/}, 243--245.

\bibitem[{{Treffers} {\em et~al.\/}(1995){Treffers}, {Filippenko}, {van Dyk},
  {Paik} \& {Richmond}}]{treffers_1995_bait}
{\sc {Treffers}, R.R., {Filippenko}, A.V., {van Dyk}, S.D., {Paik}, Y. \&
  {Richmond}, M.W.} (1995). {The Berkeley Automatic Imaging Telescope: an
  Update}. In {G.~W.~Henry \& J.~A.~Eaton}, ed., {\em Robotic Telescopes.
  Current Capabilities, Present Developments, and Future Prospects for
  Automated Astronomy\/}, vol.~79 of {\em Astronomical Society of the Pacific
  Conference Series\/}, 86.

\bibitem[{{Turon} \& {Arenou}(2008)}]{turon_hipparcos_2008}
{\sc {Turon}, C. \& {Arenou}, F.} (2008). {The Hipparcos Catalogue: 10th
  anniversary and its legacy}. In {W.~J.~Jin, I.~Platais, \&
  M.~A.~C.~Perryman}, ed., {\em IAU Symposium\/}, vol. 248 of {\em IAU
  Symposium\/}, 1--7.

\bibitem[{{Ulmschneider}(1996)}]{ulms_1996}
{\sc {Ulmschneider}, P.} (1996). {Chromosphere and coronal heating mechanisms}.
  In {R.~Pallavicini \& A.~K.~Dupree}, ed., {\em Cool Stars, Stellar Systems,
  and the Sun\/}, vol. 109 of {\em Astronomical Society of the Pacific
  Conference Series\/}, 71.

\bibitem[{{Van Buren} {\em et~al.\/}(1994){Van Buren}, {Dgani} \&
  {Noriega-Crespo}}]{van_buren_1994}
{\sc {Van Buren}, D., {Dgani}, R. \& {Noriega-Crespo}, A.} (1994). {The wind of
  EG Andromedae is not dust driven}. {\em \aj\/}, {\bf 108}, 1112--1114.

\bibitem[{{Van Leeuwen}(2007)}]{vanleeuwen_2007}
{\sc {Van Leeuwen}, F.} (2007). {Validation of the new Hipparcos reduction}.
  {\em \aap\/}, {\bf 474}, 653--664.

\bibitem[{{Vernazza} {\em et~al.\/}(1973){Vernazza}, {Avrett} \&
  {Loeser}}]{vernazza_1973_val}
{\sc {Vernazza}, J.E., {Avrett}, E.H. \& {Loeser}, R.} (1973). {Structure of
  the Solar Chromosphere. Basic Computations and Summary of the Results}. {\em
  \apj\/}, {\bf 184}, 605--632.

\bibitem[{{Vogel}(1991)}]{vogel_1991}
{\sc {Vogel}, M.} (1991). {Empirical velocity laws for cool giants. I - The
  symbiotic binary EG Andromedae}. {\em \aap\/}, {\bf 249}, 173--180.

\bibitem[{{Vogel} {\em et~al.\/}(1992){Vogel}, {Nussbaumer} \&
  {Monier}}]{vogel_nussbaumer_1992}
{\sc {Vogel}, M., {Nussbaumer}, H. \& {Monier}, R.} (1992). {Absolute radius of
  an M giant}. {\em \aap\/}, {\bf 260}, 156--160.

\bibitem[{{Voigt}(1956)}]{voigt_1956_solarasym}
{\sc {Voigt}, H.H.} (1956). {``Drei-Strom-Modell'' der Sonnenphotosph{\"a}re
  und Asymmetrie der Linien des infraroten Sauerstoff-Tripletts. Mit 12
  Textabbildungen}. {\em {\it Zeitschrift fŸr Astrophysik}\/}, {\bf 40}, 157.

\bibitem[{{Wallerstein}(1981)}]{wallerstein_1981_olddisk}
{\sc {Wallerstein}, G.} (1981). {Symbiotic stars as an old disk population}.
  {\em The Observatory\/}, {\bf 101}, 172--174.

\bibitem[{{Wedemeyer-B{\"o}hm} {\em et~al.\/}(2009){Wedemeyer-B{\"o}hm}, {Lagg}
  \& {Nordlund}}]{wedemeyer_2009}
{\sc {Wedemeyer-B{\"o}hm}, S., {Lagg}, A. \& {Nordlund}, {\AA}.} (2009).
  {Coupling from the Photosphere to the Chromosphere and the Corona}. {\em
  \ssr\/}, {\bf 144}, 317--350.

\bibitem[{{Weiler} \& {Oegerle}(1979)}]{weiler_oegerle_1979}
{\sc {Weiler}, E.J. \& {Oegerle}, W.R.} (1979). {A Copernicus survey of MG II
  emission in late-type stars}. {\em \apjs\/}, {\bf 39}, 537--547.

\bibitem[{{Wells} {\em et~al.\/}(1981){Wells}, {Greisen} \&
  {Harten}}]{wells_1981_fits}
{\sc {Wells}, D.C., {Greisen}, E.W. \& {Harten}, R.H.} (1981). {FITS - a
  Flexible Image Transport System}. {\em \aaps\/}, {\bf 44}, 363.

\bibitem[{{Whitelock}(1987)}]{whitelock_1987_symb_mira}
{\sc {Whitelock}, P.A.} (1987). {Symbiotic Miras}. {\em \pasp\/}, {\bf 99},
  573--591.

\bibitem[{{Wilson} \& {Vainu Bappu}(1957)}]{wilson_bappu_1957}
{\sc {Wilson}, O.C. \& {Vainu Bappu}, M.K.} (1957). {H and K Emission in
  Late-Type Stars: Dependence of Line Width on Luminosity and Related Topics.}
  {\em \apj\/}, {\bf 125}, 661.

\bibitem[{{Wilson}(1950)}]{wilson_1950}
{\sc {Wilson}, R.E.} (1950). {The Spectrum of HD 4174}. {\em \pasp\/}, {\bf
  62}, 14.

\bibitem[{{Wilson}(1953)}]{wilson_1953_egandradvel}
{\sc {Wilson}, R.E.} (1953). {General catalogue of stellar radial velocities.}
  {\em Carnegie Institute Washington D.C.~Publication\/}, 0.

\bibitem[{{Wilson} \& {Vaccaro}(1997)}]{wilson_vaccaro_1997_egand_photom}
{\sc {Wilson}, R.E. \& {Vaccaro}, T.R.} (1997). {Ellipsoidal variation in EG
  Andromedae}. {\em \mnras\/}, {\bf 291}, 54--58.

\bibitem[{{Woodgate} {\em et~al.\/}(1998){Woodgate}, {Kimble}, {Bowers},
  {Kraemer}, {Kaiser}, {Danks}, {Grady}, {Loiacono}, {Brumfield}, {Feinberg},
  {Gull}, {Heap}, {Maran}, {Lindler}, {Hood}, {Meyer}, {Vanhouten},
  {Argabright}, {Franka}, {Bybee}, {Dorn}, {Bottema}, {Woodruff}, {Michika},
  {Sullivan}, {Hetlinger}, {Ludtke}, {Stocker}, {Delamere}, {Rose}, {Becker},
  {Garner}, {Timothy}, {Blouke}, {Joseph}, {Hartig}, {Green}, {Jenkins},
  {Linsky}, {Hutchings}, {Moos}, {Boggess}, {Roesler} \&
  {Weistrop}}]{woogdate_stis_1998}
{\sc {Woodgate}, B.E., {Kimble}, R.A., {Bowers}, C.W., {Kraemer}, S., {Kaiser},
  M.E., {Danks}, A.C., {Grady}, J.F., {Loiacono}, J.J., {Brumfield}, M.,
  {Feinberg}, L., {Gull}, T.R., {Heap}, S.R., {Maran}, S.P., {Lindler}, D.,
  {Hood}, D., {Meyer}, W., {Vanhouten}, C., {Argabright}, V., {Franka}, S.,
  {Bybee}, R., {Dorn}, D., {Bottema}, M., {Woodruff}, R., {Michika}, D.,
  {Sullivan}, J., {Hetlinger}, J., {Ludtke}, C., {Stocker}, R., {Delamere}, A.,
  {Rose}, D., {Becker}, I., {Garner}, H., {Timothy}, J.G., {Blouke}, M.,
  {Joseph}, C.L., {Hartig}, G., {Green}, R.F., {Jenkins}, E.B., {Linsky}, J.L.,
  {Hutchings}, J.B., {Moos}, H.W., {Boggess}, A., {Roesler}, F. \& {Weistrop},
  D.} (1998). {The Space Telescope Imaging Spectrograph Design}. {\em \pasp\/},
  {\bf 110}, 1183--1204.

\bibitem[{{Young} {\em et~al.\/}(2011){Young}, {Feldman} \&
  {Lobel}}]{young_2011_cii_wavelengths}
{\sc {Young}, P.R., {Feldman}, U. \& {Lobel}, A.} (2011). {Forbidden and
  Intercombination Lines of RR Telescopii: Wavelength Measurements and Energy
  Levels}. {\em \apjs\/}, {\bf 196}, 23.

\bibitem[{Zeilik {\em et~al.\/}(1992)Zeilik, Gregory \&
  Smith}]{zeilik_1992_introductory}
{\sc Zeilik, M., Gregory, S. \& Smith, E.} (1992). {\em Introductory astronomy
  and astrophysics\/}. Saunders golden sunburst series, Saunders College Pub.

\end{thebibliography}
\end{footnotesize}

\end{document}